\documentclass[%
 reprint,
 showpacs,
 amsmath,amssymb,
 aps,
prstab,
floatfix,
]{revtex4-1}

\usepackage{graphicx}
\usepackage{dcolumn}
\usepackage{bm}
\usepackage{textcomp}
\usepackage{txfonts}  
\usepackage{booktabs}

\newcommand{\pubtitle}[1]{#1}

\begin{document}


\title{Commissioning experience and beam physics measurements \\ 
at the SwissFEL Injector Test Facility}

 
\author{T.~Schietinger}   
\email{thomas.schietinger@psi.ch} 
\author{M.~Pedrozzi}      
 
\author{M.~Aiba}          
\author{V.~Arsov}         
\author{S.~Bettoni}       
\author{B.~Beutner}       
\author{M.~Calvi}         
\author{P.~Craievich}     
\author{M.~Dehler}        
\author{F.~Frei}          
\author{R.~Ganter}        
\author{C.~P.~Hauri}      
\author{R.~Ischebeck}     
\author{Y.~Ivanisenko}    
\author{M.~Janousch}      
\author{M.~Kaiser}        
\author{B.~Keil}          
\author{F.~L\"ohl}        
\author{G.~L.~Orlandi}    
\author{C.~Ozkan~Loch}    
\author{P.~Peier}         
\author{E.~Prat}          
\author{J.-Y.~Raguin}     
\author{S.~Reiche}        
\author{T.~Schilcher}     
\author{P.~Wiegand}       
\author{E.~Zimoch}        
 
\author{D.~Anicic}        
\author{D.~Armstrong}     

\author{M.~Baldinger}     
\author{R.~Baldinger}     
\author{A.~Bertrand}      
\author{K.~Bitterli}      
\author{M.~Bopp}          
\author{H.~Brands}        
\author{H.~H.~Braun}      
\author{M.~Br\"onnimann}  
\author{I.~Brunnenkant}   
 
\author{P.~Chevtsov}      
\author{J.~Chrin}         
\author{A.~Citterio}      
\author{M.~Csatari Divall}
 
\author{M.~Dach}          
\author{A.~Dax}           
\author{R.~Ditter}        
\author{E.~Divall}        
\author{A.~Falone}        
\author{H.~Fitze}         

\author{C.~Geiselhart}    
\author{M.~W.~Guetg}      
 
\author{F.~H\"ammerli}    
\author{A.~Hauff}         
\author{M.~Heiniger}      
\author{C.~Higgs}         
\author{W.~Hugentobler}   
\author{S.~Hunziker}      
 
\author{G.~Janser}        
 
\author{B.~Kalantari}     
\author{R.~Kalt}          
\author{Y.~Kim}           
\author{W.~Koprek}        
\author{T.~Korhonen}      
\author{R.~Krempaska}     
 
\author{M.~Laznovsky}     
\author{S.~Lehner}        
\author{F.~Le Pimpec}     
\author{T.~Lippuner}      
\author{H.~Lutz}          
 
\author{S.~Mair}          
\author{F.~Marcellini}    
\author{G.~Marinkovic}    
\author{R.~Menzel}        
\author{N.~Milas}         
 
\author{T.~Pal}           
\author{P.~Pollet}        
\author{W.~Portmann}      
 
\author{A.~Rezaeizadeh}   
\author{S.~Ritt}          
\author{M.~Rohrer}        
 
\author{M.~Sch\"ar}       
\author{L.~Schebacher}    
\author{St.~Scherrer}     
\author{V.~Schlott}       
\author{T.~Schmidt}       
\author{L.~Schulz}        
\author{B.~Smit}          
\author{M.~Stadler}       
\author{B.~Steffen}       
\author{L.~Stingelin}     
\author{W.~Sturzenegger}  
 
\author{D.~M.~Treyer}     
\author{A.~Trisorio}      
\author{W.~Tron}          
 
\author{C.~Vicario}       
 
\author{R.~Zennaro}       
\author{D.~Zimoch}        
 
\affiliation{Paul Scherrer Institut, CH-5232 Villigen PSI, Switzerland}

\date{\today}

\begin{abstract}
The SwissFEL Injector Test Facility operated at the Paul Scherrer Institute
between 2010 and 2014, serving as a pilot plant and testbed for the development
and realization of SwissFEL, the X-ray Free-Electron Laser facility under 
construction at the same institute.
The test facility consisted of a laser-driven rf electron gun followed by an 
S-band booster linac, a magnetic bunch compression chicane and a diagnostic
section including a transverse deflecting rf cavity.
It delivered electron bunches of up to 200~pC charge and up to 250~MeV beam
energy at a repetition rate of 10~Hz.
The measurements performed at the test facility not only demonstrated the beam 
parameters required to drive the first stage of an FEL facility, but also 
led to significant advances in instrumentation technologies, beam 
characterization methods and the generation, transport and compression of 
ultra-low-emittance beams.
We give a comprehensive overview of the commissioning experience of the 
principal subsystems and the beam physics measurements performed during the
operation of the test facility, including the results of the test of an 
in-vacuum undulator prototype generating radiation in the vacuum ultraviolet
and optical range.
\end{abstract}


\maketitle

\newpage \mbox{} \newpage \mbox{} \newpage                              
\tableofcontents
\newpage 

\section{\label{sec:intro}Introduction}

The Paul Scherrer Institute (PSI) is currently constructing SwissFEL, an X-ray 
Free-Electron Laser (FEL) designed to produce coherent, ultra-bright and
ultra-short photon pulses in the wavelength range between 0.1 and 7~nm, with
first pilot experiments scheduled for late 2017 \cite{SwissFEL,Pat10}.
During the preparation phase of SwissFEL, PSI designed, built and operated
a 250-MeV photoinjector, the SwissFEL Injector Test Facility (SITF)~\cite{SITF}.
Its main purposes consisted in the detailed study of the generation, 
acceleration and time compression of high-brightness electron beams needed to 
drive a compact SASE (self-amplified spontaneous emission) FEL and in providing
a platform for the development and test of accelerator and undulator components 
or systems planned for SwissFEL.

Several FEL test facilities with varying objectives have operated at other
labs, going back to the TESLA Test Facility (TTF) at DESY~\cite{Ros96,Bre97}, 
the photoinjector test facility at DESY, Zeuthen (PITZ)~\cite{Ste10},
and the SLAC Gun Test Facility (GTF)~\cite{Sch02}, 
followed by 
the SPARC photoinjector~\cite{Ale03,Ale04} at INFN Frascati, 
the Shanghai Deep Ultraviolet Free-Electron Laser source 
(SDUV-FEL)~\cite{Zha04}, 
the SPring–8 Compact SASE Source (SCSS) test accelerator at 
RIKEN~\cite{Tan06,Shi08}, or, more recently, the
PAL-XFEL injector test facility at the Pohang Accelerator Laboratory
(PAL)~\cite{Par13,Han14}, 
and further facilities are being constructed in this context, e.g., 
the CLARA test facility at Daresbury~\cite{Cla14}.
Moreover, SLAC's Next Linear Collider Test Accelerator (NLCTA)~\cite{Rut93}
has been used extensively for FEL studies (see, e.g., Ref.~\cite{Xia10}), and
reports on commissioning experience have been given for the injector of the 
Linac Coherent Light Source (LCLS) at SLAC~\cite{Akr08} and the 
FERMI project at Elettra Sincrotrone Trieste~\cite{Pen13}. 
The primary focus of the SITF lied on the systematic characterization and 
optimization of the electron source up to and including the first bunch 
compression stage. 
The beam optimization was accomplished through the tuning of all relevant 
parameters including the gun laser, cathode properties, radiofrequency (rf) 
acceleration and beam optics setup.
The aim of these efforts was to generate, and preserve after magnetic bunch 
compression, a high-brightness beam fulfilling the stringent requirements for 
driving an FEL of compact and thus cost-effective design like SwissFEL.
As the verification of such a high-brightness beam requires advanced beam 
instrumentation as well as excellent beam characterization methods, much
emphasis was put on both of these areas at the SITF.

The SITF operated between March 2010 and October 2014, with installation
and repair breaks in between, in various configurations allowing for a wide 
range of beam measurements and component tests.
In the present report we give a review of the commissioning experience of the
main subsystems and summarize the beam physics measurements performed at the 
test facility.
We describe the design and general layout of the facility in 
Sec.~\ref{sec:layout}, followed by a chronological account 
of the various operating phases in Sec.~\ref{sec:operation}.
Section~\ref{sec:subsystems} contains commissioning and performance reviews
of the main subsystems, with particular emphasis on the gun laser, the 
rf systems and the electron beam diagnostics.
In Sec.~\ref{sec:diag-dev} we present some diagnostics developments performed
at the SITF in view of the realization of SwissFEL.
In the following two sections we report on our beam development work, 
describing our methods and procedures (Sec.~\ref{sec:bd-procs}) as well as 
presenting and discussing the results of our measurements 
(Sec.~\ref{sec:bd-results}).
Section~\ref{sec:undulator} is dedicated to the undulator prototype test 
performed at the SITF. 
In Sec.~\ref{sec:summary} we summarize our work and give a brief outlook on 
SwissFEL.

\begin{figure*}[t]  
  \includegraphics*[width=1\linewidth]{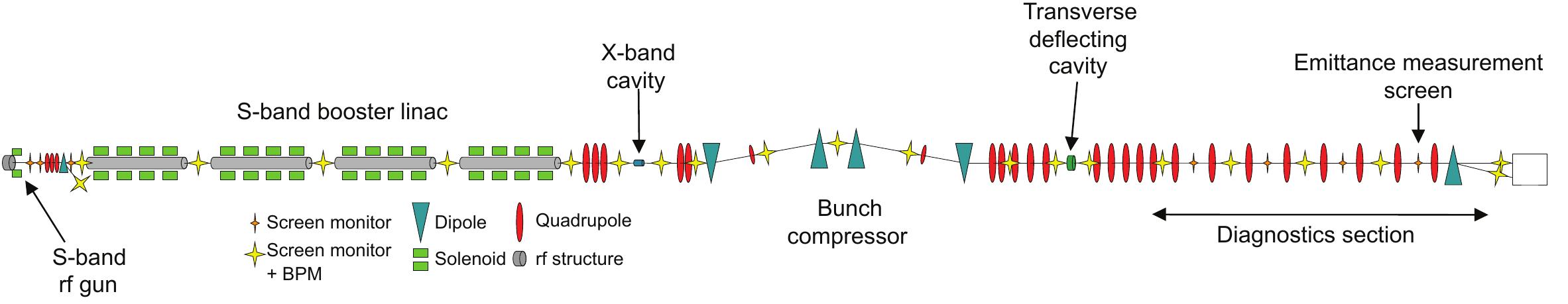}
  \caption{Schematic layout of the SwissFEL Injector Test Facility 
  (original configuration, not to scale). 
  The total length of the beam line is approximately 60~m.}
  \label{fig:layout}
\end{figure*}

\section{\label{sec:layout}Machine layout and design considerations}

The SITF was conceived as a split rf photoinjector followed by a booster linac,
a magnetic bunch compression chicane and a diagnostic section. 
The facility was housed in a dedicated building at PSI, the laboratory hall 
WLHA, which was constructed in the years 2008 and 2009 and includes a 
shielding tunnel for the beam line.

The design of the facility~\cite{Kim08} benefited from several conceptual and 
experimental insights in injector design gained in the last three decades. 
In particular the design is based on the concepts of emittance 
compensation~\cite{Car89} and invariant envelope matching~\cite{Ser97}, as well
as the effective working point found by Ferrario et al.~\cite{Fer00} during the
redesign of the Linac Coherent Light Source (LCLS) and later confirmed by 
measurements performed at the SPARC photoinjector~\cite{Fer07} and, eventually,
by the successful operation of the LCLS facility~\cite{Akr08,Emm10}.
Further guiding principles were given by the need for an azimuthally symmetric
beam at low energy to provide an equal space-charge density in the vertical
and horizontal plane, the preference for a relatively low bunch charge
(compared to earlier X-ray FEL designs), allowing for a smaller laser spot size 
on the cathode and thus resulting in a smaller thermal emittance contribution, 
the minimization of nonlinearities and nonuniformities in the bunch slice 
parameters (good slice matching) and the aim for a small $\beta$-function to 
minimize chromatic effects in the focusing quadrupole magnets.

A schematic representation of the layout of the facility in its initial 
configuration is given in Fig.~\ref{fig:layout}.
The electron source is a laser-driven S-band (3~GHz) rf gun.
During the first four years of operation, until April 2014, electrons were
produced with an rf gun originally built for the CLIC test facility (CTF) at 
CERN, the CTF gun No.~V~\cite{Bos95,Bos96}, kindly lent by CERN to PSI.
Later, the PSI-built SwissFEL rf gun~\cite{Rag12} was installed and used. 
In both cases, the electrons were extracted by a laser pulse from an
exchangeable copper cathode plug (at later stages in some cases coated with 
cesium telluride) and accelerated to an initial momentum of 7.1~MeV/$c$.
From 2013 onwards, the exchange of cathodes was facilitated by a load-lock 
chamber mounted onto the rf gun.
Immediately after the gun, a movable solenoid magnet provides initial focusing
and is used to optimize the emittance implementing the emittance
compensation technique along the following 3~m long drift.
The gun solenoid includes two weak, individually tunable quadrupole magnets
(normal and skew) to correct potential quadrupole terms in the solenoid 
field~\cite{Sch05} (see also Ref.~\cite{Dow13} for a theoretical description
of the correction scheme).
The ensuing drift accommodates various low-energy beam diagnostics including a 
spectrometer arm at 30$^\circ$ deflection angle for the measurement of beam 
energy and energy spread.
The main accelerating section consists of four S-band traveling-wave
structures boosting the beam energy to a nominal value of 270~MeV.
Each of the structures is surrounded by four solenoid magnets for additional 
symmetric transverse focusing.
The solenoids are each 75~cm long and provide magnetic fields up to 100~mT.
The two downstream S-band structures are simultaneously used to generate, by 
off-crest acceleration, the energy chirp necessary for the longitudinal 
compression of the electron bunches in the magnetic four-dipole chicane after 
the booster.
Since the energy chirp generated in this way results in a curved longitudinal
phase-space distribution, a fourth-harmonic cavity (X-band, 12~GHz) is operated
upstream of the bunch compressor for the linearization of the longitudinal 
phase space, thereby suppressing the development of peaks in the head and tail 
of the current profile of the compressed bunch.
This linearization includes the compensation of second-order terms arising from
the bunch compression process. 
Under normal conditions, the harmonic cavity has a net decelerating effect on
the bunch centroid, reducing the beam energy to the nominal final energy of
250~MeV.
The matching of the beam optics to the design optics is accomplished with five
quadrupole magnets upstream of the chicane.
After the bunch compressor the beam enters an extensive diagnostic section
optimized for bunch length and emittance measurements.
An array of quadrupole magnets allows the implementation of numerous optics 
schemes, in particular to perform various kinds of optics-based measurements of
emittance and Twiss parameters.
The initial configuration of the diagnostics section, as depicted in 
Fig.~\ref{fig:layout}, included a Focus-Drift-Defocus-Drift (FODO) channel for the 
measurement of emittance using multiple screens inside the FODO cells, as an 
alternative to single or multiple quadrupole scan techniques.
The FODO option was lost in the course of the modification of the diagnostics 
beam line to accommodate an undulator prototype in early 2014. 
At this point, however, the quadrupole scan methods had already been firmly 
established as being both more robust and more precise.
A vertically deflecting S-band rf cavity located after a few more matching 
quadrupoles beyond the bunch compressor can be used to streak the bunch, thus 
enabling the measurement of bunch length and slice parameters with a screen at 
a suitable distance. 
Immediately before the beam dump, a high-energy spectrometer, consisting of a 
deflecting dipole magnet (6$^\circ$ bending angle), a straight and a dispersive 
section allows momentum and momentum spread of the final electron beam to be 
measured. 
It can also be used in combination with the transverse deflecting cavity 
for the detailed imaging of the longitudinal phase space.
At various positions along the entire beam line, transverse beam profiles can 
be obtained by imaging the beam with insertable screens, consisting of 
crystals (YAG:Ce or LuAG:Ce) emitting scintillation light or thin metal foils 
emitting optical transition radiation (OTR). 
The beam orbit and charge transmission is monitored by a series of resonant 
stripline beam position monitors (BPMs).
Additional cavity-based BPMs were installed for test purposes along the 
straight section of the high-energy spectrometer.
A wall current monitor in the gun section provides a measurement of the absolute
bunch charge with an uncertainty of 10--15\% and serves as a calibration for
the BPM charge measurements.
Additional information on the bunch charge can be obtained from a few prototype
integrating current transformers installed along the beam line.

\begin{figure*}[t]  
  \includegraphics*[width=0.7\linewidth]{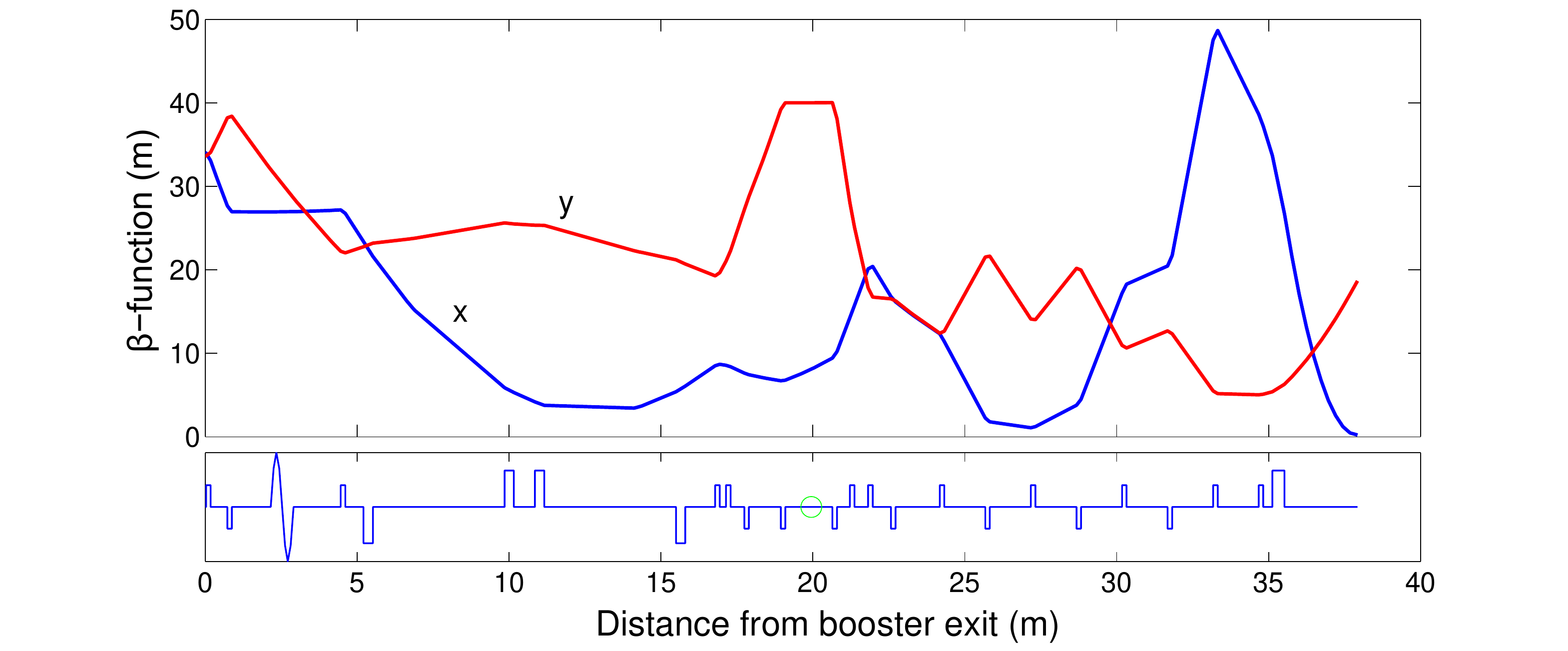}
  \caption{Horizontal (blue) and vertical (red) $\beta$-functions for a 
  typical beam optics setup, covering the beam line from the S-band booster 
  exit down to the profile monitor at the high-energy spectrometer (upper 
  panel). 
  The corresponding magnetic lattice is shown in the lower panel, where the 
  larger rectangular
  shapes indicate the locations of the main dipole (bending) magnets and the 
  smaller those of quadrupole (focusing) magnets.
  The wave-like curve at 2.5~m represents the position of the X-band cavity.
  The green circle at 20~m marks the location of the rf deflector.} 
  \label{fig:optics}
\end{figure*}

Figure~\ref{fig:optics} shows a typical beam optics setting, as it was applied 
for the main measurements related to beam characterization and optimization 
described in Sec.~\ref{sec:bd-results}.
The main design aims leading to this choice of optics are:
\begin{itemize}
\item
small $\beta$-functions ($\beta_{x,y} < 50$~m) throughout the lattice;
\item
a small horizontal $\beta$-function between the third and fourth dipole magnet 
of the bunch compression chicane to limit potential detrimental effects from 
coherent synchrotron radiation;
\item
adequate time resolution for longitudinal bunch measurements with the rf 
deflector at a profile monitor in the high-energy spectrometer;
\item
appropriate energy resolution at the high-energy spectrometer;
\item
proper optics for the emittance measurement (in this case based on a 
single-quadrupole scan, see Sec.~\ref{sec:bd-trans} for details).
\end{itemize}

The SITF was designed for two standard bunch charge modes, a high-charge mode 
with 200~pC bunch charge, and a low-charge mode with 10~pC, in accordance
with the planned SwissFEL bunch charge range of 10--200~pC.
Table~\ref{tab:pars-high} summarizes the main design beam parameters for the 
high-charge mode and compares them to typically achieved values.
The corresponding numbers for the low-charge mode, where different from the 
high-charge mode, are listed in Table~\ref{tab:pars-low}. 
Note that systematic measurements with compressed beams have only been done
for the high-charge mode.

\begin{table}[tb]
  \caption{Main design parameters of the SwissFEL Injector Test Facility for 
    the high-charge mode (200~pC bunch charge), with comparison to typically 
    realized values.}
  \begin{ruledtabular}
    \begin{tabular}{lcc}
      Parameter & Design  & Realized  \\
      \colrule
      Final beam energy (MeV)                    &    250 &    260\footnotemark[1] \\
      Repetition rate (Hz)                       &     10 &     10 \\
      Cathode quantum efficiency                 &    5$\times$10$^{-5}$ (Cu) &  3$\times$10$^{-4}$ (Cu) \\  
                                                 &                           &  6$\times$10$^{-2}$ (Cs$_2$Te) \\  
      Peak rf gun field at cathode (MV/m)        &    100 &     100\footnotemark[2] \\
      Intrinsic emittance contribution (\textmu m) &    0.195 &    $\approx$0.1 \\   
      \textit{Gun laser parameters}              &        &                                       \\
      Spot diameter on cathode (mm)              &    \multicolumn{2}{c}{1}                    \\
      Pulse length (FWHM) (ps)                   &    \multicolumn{2}{c}{9.9}       \\
      Wavelength (nm)                            &    \multicolumn{2}{c}{257--280}            \\
      \colrule
      \multicolumn{3}{l}{\textit{Before compression:}}                     \\
      Bunch length (rms) (ps)                    &    2.8 &    3.8 \\      
      Peak current (A)                           &     22 &    20  \\
      Normalized projected emittance (\textmu m)     &    0.35 & $\approx$0.30 \\
      Normalized core slice emittance (\textmu m)    &    0.32 & $\approx$0.20 \\
      \colrule
      \multicolumn{3}{l}{\textit{After compression:}}                      \\
      Bunch length (rms) (ps)                    &    0.193 &   0.3--0.6   \\
      Peak current (A)                           &    352 &   120--240     \\
      Normalized projected emittance (\textmu m)     &   0.379 &   $\approx$0.40\footnotemark[3] \\
      Normalized core slice emittance (\textmu m)    &    0.33 &   $\approx$0.20\footnotemark[3] 
    \end{tabular}
    \footnotetext[1]{Uncompressed beam only.}
    \footnotetext[2]{85~MV/m for the CTF photocathode gun.}
    \footnotetext[3]{Measured for a peak current of 150~A.}
  \end{ruledtabular}
  \label{tab:pars-high}
\end{table}

\begin{table}[bt]
  \caption{Design parameters of the SwissFEL Injector Test Facility for 
    the low-charge mode (10~pC bunch charge), as far as different from the 
    high-bunch-charge mode, with comparison to typically realized values.
    (Systematic compression studies have only been carried out for the 
    high-charge case.)}
  \begin{ruledtabular}
    \begin{tabular}{lcc}
      Parameter                                      & Design   & Realized          \\
      \colrule
      Intrinsic emittance contribution (\textmu m)   &   0.072  & $\approx$0.04     \\   
      Bunch length (rms) (ps)                        &    1.05  & 0.7               \\     
      Peak current (A)                               &     3.0  & 4.0                 \\
      Normalized projected emittance (\textmu m)         &   0.095  & $\approx$0.15     \\
      Normalized core slice emittance (\textmu m)        &   0.078  & $\approx$0.10     \\ 
      \textit{Gun laser parameters}                  &          &                   \\
      Spot diameter on cathode (\textmu m)           &    \multicolumn{2}{c}{400}   \\
      Pulse length (FWHM) (ps)                       &     3.7  &  2.5              \\
    \end{tabular}
  \end{ruledtabular}
  \label{tab:pars-low}
\end{table}

The SITF typically operated at a single-bunch repetition rate of 10~Hz.
In contrast SwissFEL will operate at a repetition rate of 100~Hz.
The SITF was deliberately designed to run at 10~Hz, which allowed the 
demonstration of the required beam parameters at somewhat relaxed requirements 
on hardware and shielding components.
Moreover, the overall cooling layout of the CTF rf gun did not allow for 
repetition rates higher than 10~Hz.
The SwissFEL gun by contrast is designed to run at up to 400~Hz and was 
tested at a rate of 100~Hz at the SITF, albeit as an isolated system only.
Several of the other rf stations were also operated individually at 100~Hz 
during dedicated test runs without beam.

\section{\label{sec:operation}Operation overview}

In this section we give a brief overview of the four main operation phases of 
the SITF, dating from March 2010 until October 2014.
Driven by the progressive evolution of the machine, each operating phase had
its own primary focus.

\subsection{Phase I: gun commissioning}

The first phase of commissioning, Phase I, was dedicated to the commissioning
of the rf gun and the low-energy beam diagnostics~\cite{Sch10}.
It started immediately after the completion of the component installation in 
the gun section, with first photo-electrons generated on 12 March 2010.
The booster linac was still undergoing installation during this phase and was
separated from the gun area by a concrete shielding wall.
For this initial measuring period, the first girder of the S-band booster was 
replaced by a temporary diagnostics girder with additional profile monitors for
direct measurements at the future booster entrance.
After the removal of the shielding wall at the end of May 2010, gun 
commissioning continued for another month, during which beam operation was
limited to evening and night hours due to the ongoing installation of the
booster hardware during working days.

\subsection{Phase II: basic booster commissioning}

For Phase II the entire beam line down to the beam dump was available.
In place of the bunch compressor, which was still under construction at the
time, a simple tube was installed to connect the booster to the diagnostics 
line downstream.
On the booster side, only two out of the four structures were available for
acceleration, thus limiting the final beam energy to about 165 MeV in this 
commissioning phase.
For most measurements the first structure was operated at a reduced gradient
to fulfill the optimal conditions for emittance compensation, giving a final
beam energy of about 130~MeV.
Phase II started in early August 2010, focusing first on preparing the machine
for an official inauguration ceremony held on the 24th of the month.
At this occasion the beam was successfully accelerated to 165 MeV 
for demonstration purposes, but many systems were still incomplete.
Therefore a further installation period of several weeks followed immediately 
after the inauguration.
Beam operation under limited conditions recommenced in mid-October, alternating
with more installation work on a weekly basis.
The initial set of installation tasks was completed by the end of the year, 
such that starting from early 2011, the injector was fully available for beam 
development work.
The first weeks of 2011 were dedicated to setting up (matching) and verifying
the beam optics for various configurations and to first optimizations of the
transverse beam emittance.
In mid-March, the transverse deflecting cavity was put into operation for the
first time, allowing the first bunch length measurements. 
Phase II commissioning lasted until the end of May 2011. 
For most of the beam development work in Phases I and II, the robust and
reliable Nd:YLF laser system was used for the generation of the electrons. 
The considerably more complex Ti:sapphire laser system was only used towards the
end of Phase II (see Sec.~\ref{sec:laser}).

\subsection{Phase III: bunch compression and phase-space linearization}

During the summer of 2011, the movable magnetic chicane intended for bunch
compression was installed between the booster and the diagnostics sections of 
the injector beam line.
The following beam commissioning period, Phase III, in principle started in 
September of the same year, but suffered considerably from several problems with
the rf systems, which severely hampered further beam development.
As a consequence, the following months were mainly dedicated to the 
consolidation of the rf systems, with limited beam operation making use of the 
gun and the third booster structure only.
During another shutdown in February 2012, the repaired fourth booster structure
was reintegrated. 
The same shutdown was used to add the requisite infrastructure for the THz
diagnostics system at the bunch compressor exit.

By April 2012, the rf repair work had been completed and the injector could
be run with all accelerating structures for the first time.
With the refurbished rf system in operation, it did not take long to achieve 
several important milestones within the same month: the design energy of 
250 MeV was attained, the first slice emittance measurements were performed
and bunch compression was demonstrated for the first time.
Since the X-band cavity had not yet been installed, only nonlinear compression
could be realized at this time, with a bunch length reduction from 3.6 ps 
to 0.2 ps (rms). 
The further study of the compressed beam was postponed to the time when the
linearizing cavity would be available.

In the period from May to July 2012 the focus was on the systematic 
optimization of the uncompressed beam, which resulted in emittance values
well below the SwissFEL design figures (see Sec.~\ref{sec:bd-uncomp}).
In this context, a number of beam characterization methods were developed or
refined, as will be described in detail in Sec.~\ref{sec:bd-procs}.
The X-band cavity was installed for a first time in another shutdown in 
August, but had to be removed again due to a design flaw of the mover system.
In the time after the summer shutdown the injector test facility was used for
systematic measurements of the intrinsic emittance of copper cathodes and for
the testing and development of various diagnostics systems, in particular 
charge monitors and bunch arrival-time monitors. 
These and similar activities based on uncompressed beam were resumed after the 
successful re-integration of the X-band cavity in January 2013 (with a 
re-dimensioned mover system), since the cavity could not yet be powered due to 
delays with the preamplifier system.
During the same shutdown in January, the rf gun was equipped with a load-lock
chamber to facilitate the exchange of cathodes. 

In June 2013 the X-band cavity was finally ready for beam operation and the 
experimental investigation of linearly compressed bunches could commence.
The initial measurements revealed a few surprises, such as a growth in
slice emittance and a fragmentation of the longitudinal phase space with 
progressive compression, and triggered a series of systematic investigations
of these phenomena (see Sec.~\ref{sec:bd-comp}).
The compressed bunches were also used to test concepts of bunching monitors
based on THz radiation emitted by the shortened bunches in the last dipole of
the magnetic chicane.

Towards the end of Phase III, a modification of the Nd:YLF laser opened up the 
possibility to accelerate two electron bunches in the same rf pulse, separated 
by 28~ns.
These two-bunch tests were carried out in the context of diagnostics 
developments in view of the final configuration of SwissFEL, where two undulator
lines will be fed simultaneously at 100 Hz.
In the last week of Phase III operations, at the end of October 2013, a 
copper cathode coated with cesium telluride (Cs$_2$Te) was successfully tested.
The significantly higher quantum efficiency allowed checks of diagnostics 
components at bunch charges exceeding 1~nC, but at a reduced beam energy to
keep the beam power within limits.

In the course of Phase III operations, several campaigns to measure ambient 
radiation levels were carried out to evaluate shielding and instrumentation
options for SwissFEL~\cite{Hoh14,Hoh14a}.

\subsection{\label{sec:phaseIV}Phase IV: undulator experiment and new SwissFEL 
gun}

The fourth and last commissioning phase (Phase IV) was mainly dedicated to the 
real-conditions test of two crucial hardware components for SwissFEL: 
an undulator prototype and the new SwissFEL electron gun.

The installation of the undulator prototype in the diagnostics section of the 
injector (at the location of the former FODO channel) was carried out in the 
last weeks of 2013.
Recommissioning of the machine started on January 14, 2014, still using the
CTF rf gun for the generation of electrons.
After a swift commissioning sequence, ultraviolet FEL radiation from the 
undulator exhibiting SASE characteristics was observed within two days
(see Sec.~\ref{sec:undulator} for details).
The distorted photon beam profile measured at the undulator exit, however, 
hinted at a photon beam alignment problem in this vicinity, which prevented a 
detailed characterization of the radiation.
In addition, a failure of the klystron driving the last two S-band structures
soon compromised booster performance.
A brief shutdown was consequently scheduled to rearrange the beam line after 
the undulator and to repair the defective klystron.
While the first effort was successful the repair of the klystron took longer
than initially expected.
For the remainder of the undulator experiment, the beam energy was therefore
limited to a maximum of about 130~MeV, thereby shifting the FEL radiation into 
the optical range.
After some basic FEL characterization studies, the undulator setup was mainly
used for testing alignment schemes and to measure the beam kicks at the 
undulator entrance and exit.
These undulator measurements were alternating with further diagnostics 
developments and more intrinsic emittance measurements, in particular a 
systematic comparison between pure copper and cesium-telluride coated 
cathodes.

To spare the undulator prototype unnecessary radiation exposure in the test
facility, it was again removed during the following shutdown in April-May
2014.
At the same occasion, the CTF gun No.~V was replaced by the newly manufactured 
SwissFEL rf gun, capable of running at 100~Hz repetition rate.
After a brief conditioning period it reached its nominal power without
difficulty.

The first characterization of the new gun under operating conditions, at
repetition rates of both 10 and 100~Hz, yielded very satisfying results, among 
them a significant reduction in dark current compared to the previously 
installed gun, as described in Sec.~\ref{sec:rf-gun}.
The tests at 100~Hz were performed in stand-alone mode, since the injector as
a whole was not designed to run at that rate, as stated earlier.

The remaining beam dynamics and diagnostics measurements could once again 
profit from the full beam energy, as the klystron driving the two last booster 
structures had been replaced by a new one.
Measurements were, however, impeded by the limited bunch charge provided by the 
installed copper cathode, which exhibited a low quantum efficiency.
The problem was exacerbated by the fact that the cathode could no longer be
easily exchanged owing to a design flaw in the load-lock chamber, which only 
became evident with the new rf gun.
Another brief shutdown in mid-July 2014 was used to apply a temporary fix to the
load-lock chamber in addition to the exchange of the bunch compressor dipole
magnets (see Sec.~\ref{sec:bc}) and the installation or replacement of various
diagnostics components.
Unfortunately, the regained ability to exchange cathodes only revealed that all
remaining copper cathodes exhibited quickly deteriorating quantum efficiencies,
leveling at rather low values.
To have access to the full range of bunch charges, the remaining program was
therefore carried out with cesium-telluride-coated cathodes.
The primary goal of the last weeks of operation was the investigation and 
possible mitigation of the emittance growth in conjunction with compression
(Sec.~\ref{sec:bd-comp}).
These beam dynamics measurements were again performed in alternation with 
diagnostics developments and some final intrinsic emittance studies.

Injector operations came to an end in the early hours of October 13, 2014.

\section{\label{sec:subsystems}Subsystems and operational experience}

We describe the main technical subsystems constituting the SITF and summarize 
the experience gained during the time of operation of the test facility.
Where applicable, lessons learnt for the design and construction of SwissFEL
are presented and elaborated on.

Apart from the bunch compressor dipole magnets, which are discussed in 
Sec.~\ref{sec:bc}, we will not further describe the various magnets used for 
beam manipulation at the SITF, since in most cases standard electromagnet 
solutions have been chosen~\cite{San10,Neg12}.

\subsection{\label{sec:laser}Laser system}

Two different laser systems based on Ti:sapphire and Nd:YLF laser technology
were employed for operating the photo-electron guns at the SITF.
Both the Ti:sapphire and the Nd:YLF systems start with oscillators running at
83.275~MHz and actively synchronized within less than 50~fs~\cite{Div14} to an 
external radio-frequency master clock running at 1\,498.956~MHz (the 7th 
harmonic of the master clock, see Sec.~\ref{sec:timsynch}).
They act as seed for the corresponding amplifier stages delivering mJ pulses
at a repetition rate of up to 100~Hz (10~Hz for normal injector operation).
The requirements and performance for the SwissFEL gun laser are listed in 
Table~\ref{tab:laser}.

\begin{table*}[hbt]
  \centering
  \caption{SwissFEL gun laser requirements and achieved performance of the
  Ti:sapphire and Nd:YLF amplifier systems used at the SITF.} 
  \begin{ruledtabular}
    \begin{tabular}{lllll}
       Parameter & Unit & Requirement & Ti:sapphire & Nd:YLF \\ 
                 &      & (SwissFEL)  &             &        \\ 
       \colrule
           Pulse duration & ps & 3--10 & 3--10 & 8 \\
           Pulse rise-time & ps & $<$0.7 & 0.5--1 & 5 \\
           Temporal pulse shape &  & flat-top & flat-top-like, stacked, & Gaussian \\
           & & & \quad $\approx$10\% modulation & \\
           Pulse energy at cathode & \textmu J & up to 30 & $<$40 & $<$20 \\
           Central wavelength & nm & 250--300 & 257--280 & 262 \\
           UV energy stability at cathode & \% (rms) & $<$0.5 & 1--2 & 0.9--1.2 \\
           Transverse profile & & uniform & truncated Gaussian, & truncated Gaussian, \\
           & & & \quad $\approx$10\% modulation & \quad $\approx$10\% modulation \\
           Laser-to-rf jitter & fs & $<$100 & $<$30 at oscillator, & $<$40 at oscillator, \\
           & & & \quad $\approx$160 at cathode & \quad $\approx$130 at cathode \\
           Beam pointing stability at & \% (peak- & $<$1 & 0.15 & 0.15 \\
           \quad cathode (relative to laser beam size) & \quad to-peak) & & & \\
    \end{tabular}
  \end{ruledtabular}
  \label{tab:laser}
\end{table*}

\begin{figure}[hbt]  
  \includegraphics*[width=1\linewidth]{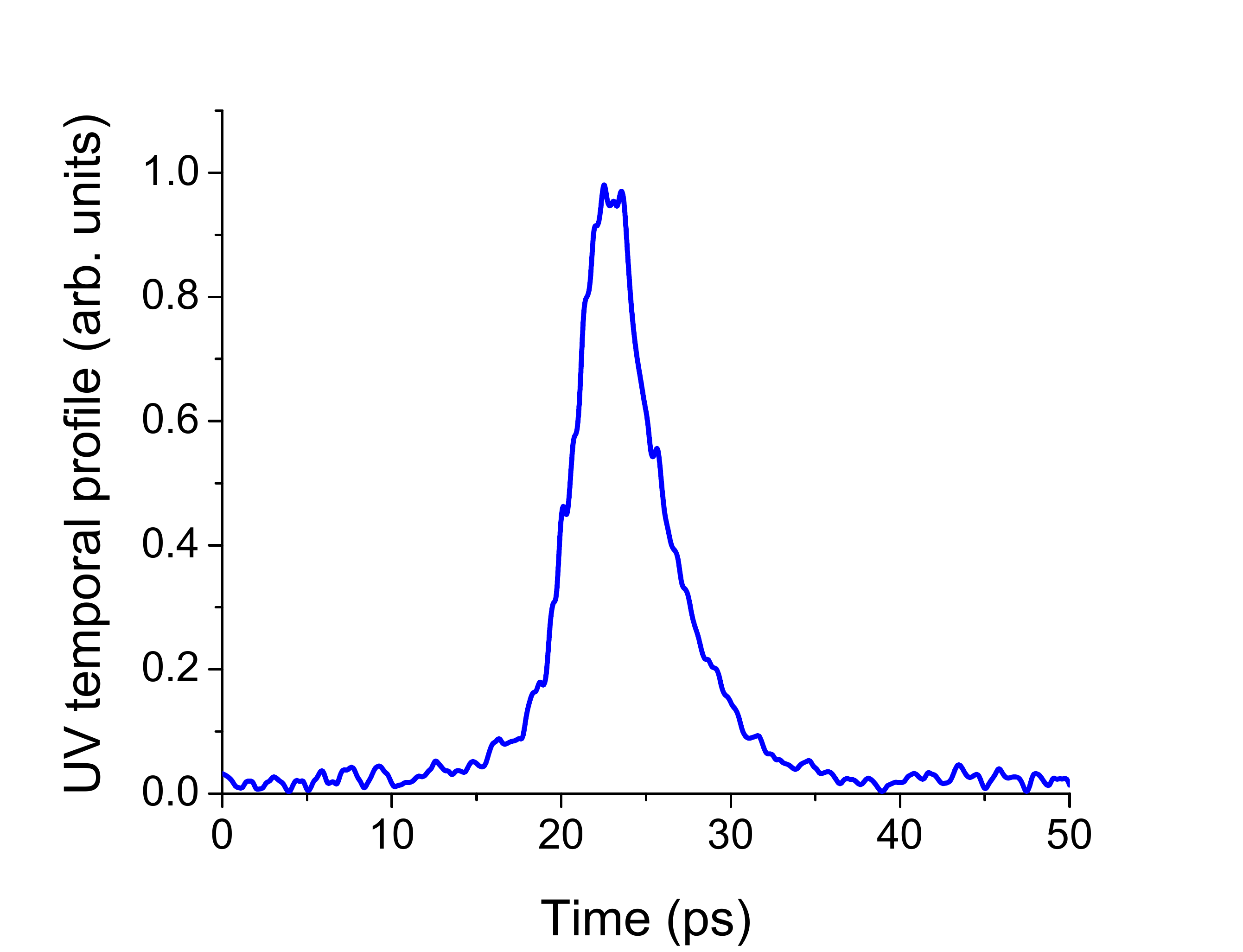}
  \caption{Longitudinal pulse profile of the Nd:YLF laser, measured
   with cross-correlation in the UV.}
  \label{fig:jaguarlong}
\end{figure}

In the following we give a short description of the two laser systems and their
performance for pulse shaping.
In the Nd:YLF laser system (Time-Bandwidth Products, Jaguar) the oscillator is
followed by a single regenerative amplifier stage. 
It delivers transform-limited 10~ps long pulses at a wavelength of 1047~nm with
up to 2~mJ pulse energy. 
These pulses are frequency quadrupled in two subsequent collinear type-I second
harmonic generators. 
For second-harmonic generation an anti-reflection (AR) coated BBO (beta barium 
borate) crystal 
($\theta$ = 23$^\circ$, AR/AR at 1048~nm/1048~nm + 524~nm, $d$ = 4~mm)
and a 2~mm thick uncoated BBO crystal ($\theta$ = 44$^\circ$) are used. 
The resulting 8~ps long, Gaussian-like pulses at 262~nm show a high energy 
stability (0.3\% rms at 260~nm over a 10 minute period) thanks to the direct 
diode pumping of the Nd:YLF crystal. 
Figure~\ref{fig:jaguarlong} shows a cross-correlation measurement of the 
Gaussian UV longitudinal profile.
The system can be operated either in a single-pulse or a double-pulse mode.
For the latter a pulse replica is created in the UV using 
polarizing beam splitters. 
The amplitude and the delay of the two pulses can be adjusted to produce
electrons in two buckets of the same rf pulse 28~ns apart.

\begin{figure}[hbt]  
  \includegraphics*[width=1\linewidth]{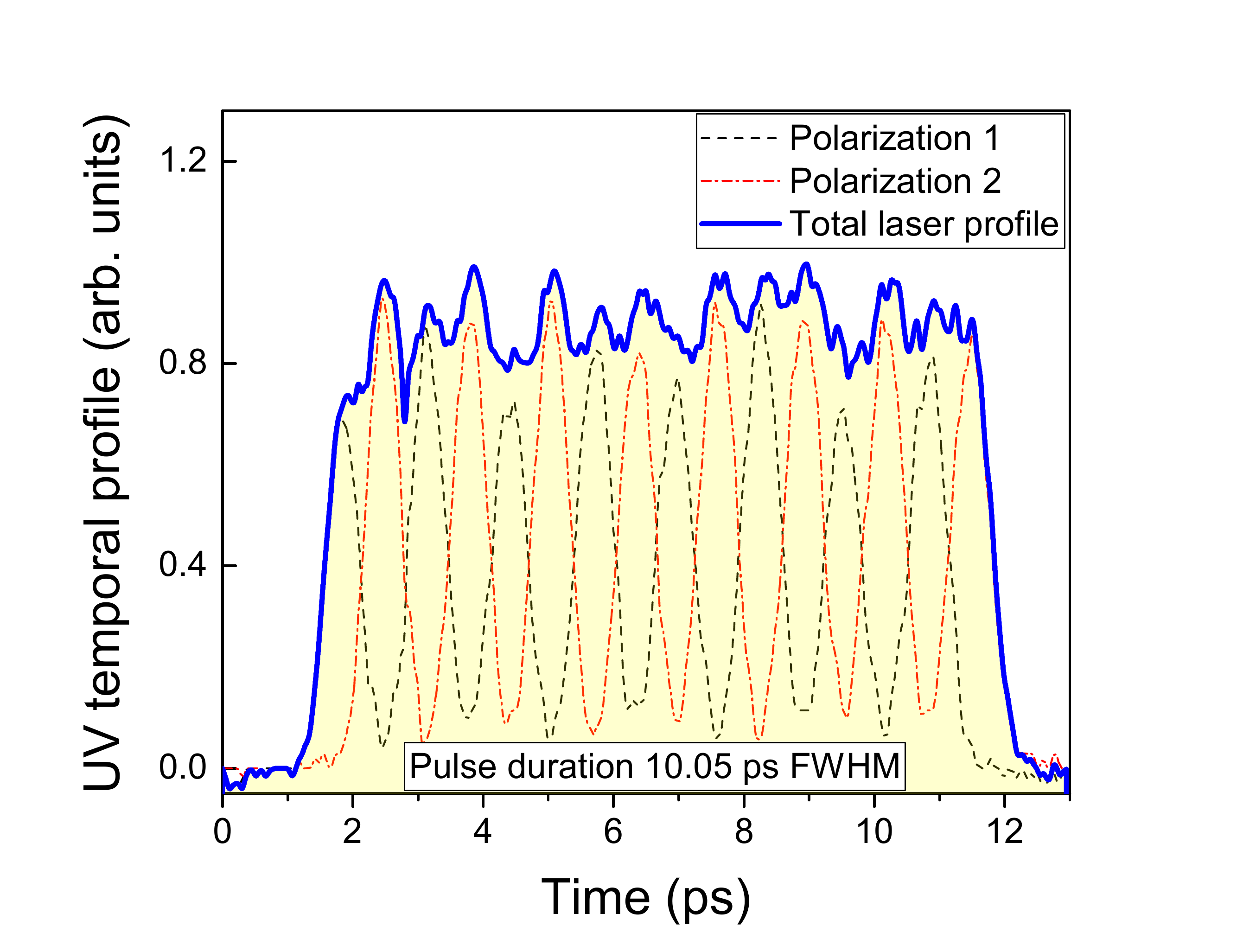}
  \caption{Longitudinal pulse profile of the Ti:sapphire laser, measured
   with cross-correlation in the UV.
   The profile is obtained by pulse stacking with four crystals, resulting in
   16 pulse replicas. 
   The cross-correlations for s- and p-polarized pulses are done in 
   sequence (black and red curves) and summed up numerically (blue curve).}
  \label{fig:pulsarlong}
\end{figure}

The Ti:sapphire amplifier system (Amplitude Technologies, Pulsar) is based on 
broadband chirped-pulse amplification. 
The laser is equipped with advanced pulse shaping technology based on 
acousto-optic programmable modulators (Fastlite, Dazzler and Mazzler) that 
allows the production of broadband ($\Delta\lambda$ = 100~nm) or alternatively 
narrow-band pulses tunable within a range of 770--830~nm~\cite{Tri11}. 
To achieve the required pulse stability six pump lasers (Quantel Laser, Centurion) 
are multiplexed in the final amplifier stage. 
This results in an improved pulse stability of 0.4\% rms in the near IR, which 
is a factor of $\sqrt{6}\approx$ 2.5 better compared to a single pump laser
(but entails higher maintenance cost). 
The near-IR output is frequency converted into the UV range by collinearly
mixing the fundamental and second harmonic pulses in a type-I BBO 
($\theta$ = 44.3$^\circ$, $\phi$ = 0$^\circ$, $d$ = 0.5~mm, AR/AR coated). 
The 50~fs short UV pulses with an energy stability of 1.1\% rms are then 
stretched to 500~fs by employing a 10~cm long CaF$_2$ block after temporal 
pulse shaping. 
Among the various explored temporal shaping schemes including 
UV Dazzler~\cite{Tri11a}, chirp mixed UV generation~\cite{Vic12} and 4-f pulse
shaping, the technique of pulse stacking turned out to be the most efficient 
to produce flattop-like pulses in the picosecond range. 
Three to five ($n$) birefringent crystals of increasing thickness 
($L = 2^{(n-1)} \times$ 400 \textmu m, $n = 1,\ldots,5$) are used to produce 
2$^n$ replicas equally delayed in time and forming the 3 to 10~ps long UV pulse
(Fig.~\ref{fig:pulsarlong}). 
The inherent amplitude modulations along the temporal pulse shape could be 
reduced to less than 20\% (peak-to-peak). 

\begin{figure*}[t]    
  \includegraphics*[width=0.8\linewidth]{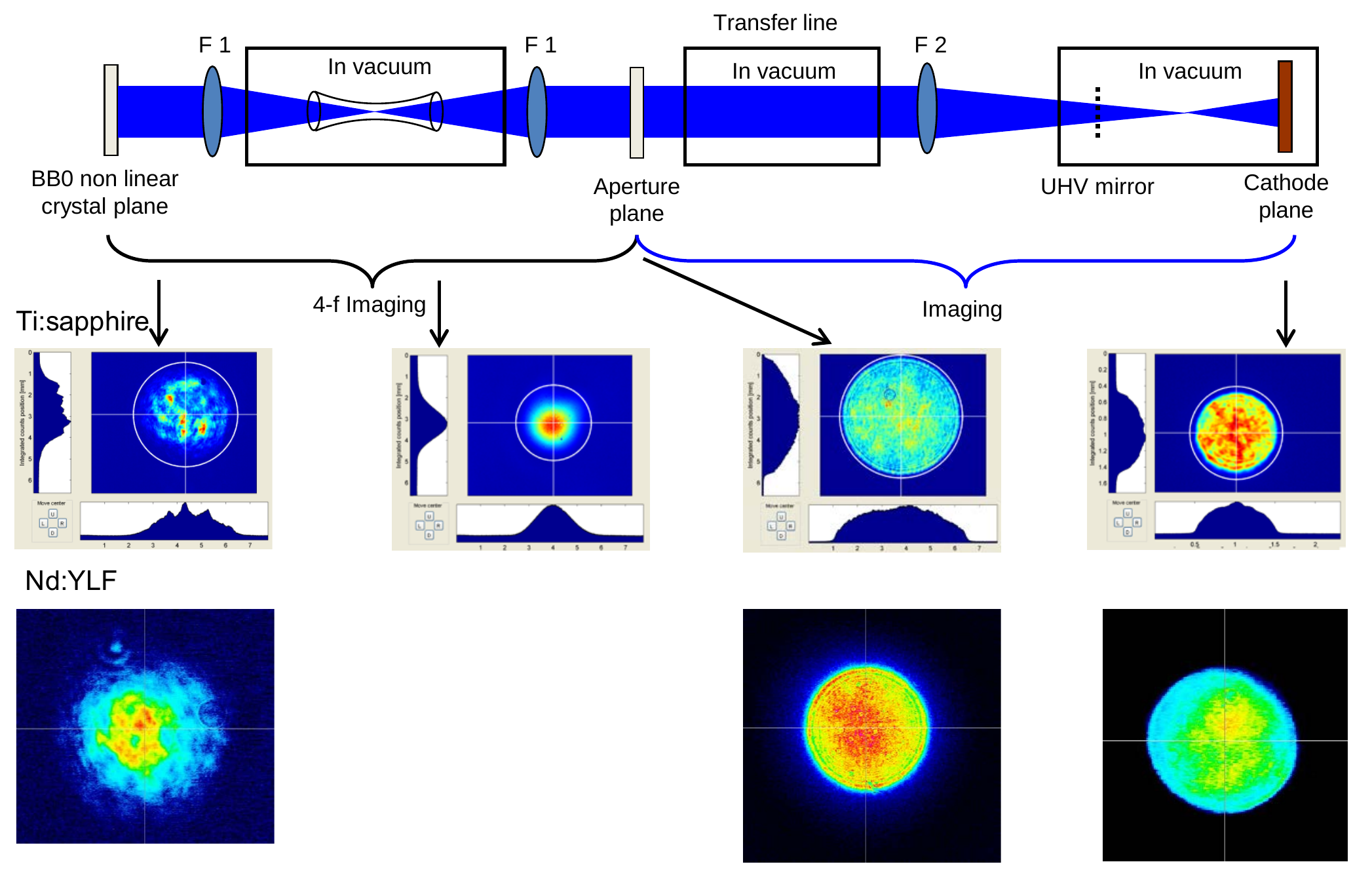}
  \caption{Beam transport and transverse beam profiles along the beam line for
    the Ti:sapphire and Nd:YLF gun laser systems.}
  \label{fig:lasertrans}
\end{figure*}

A homogeneous transverse laser beam profile on the cathode is essential for the
production of a low-emittance electron beam. 
Figure~\ref{fig:lasertrans} schematically illustrates the beam transport and 
imaging sequence for the gun lasers in use at the SITF.
The UV transverse beam profile is cleared from high-frequency spatial 
components by means of an in-house designed biconical glass capillary acting as
spatial Fourier filter.
The filter is located in air for the Nd:YLF system and in vacuum for the 
Ti:sapphire system. 
Prior to beam transport the Gaussian-like UV beam intensity profile is 
truncated by a hard aperture of variable diameter.
This approach has turned out to be more robust than other explored shaping 
schemes such as microlens arrays or the aspheric beam shaper. 
Beam transportation towards the cathode is performed in a high-vacuum 
transfer line (10$^{-7}$~mbar) sealed with UV-transparent UV-fused silica windows
(AR coated at 260~nm). 
A single imaging lens projects the truncated profile onto the copper cathode in
the rf gun with a demagnification ratio of 2.5:1. 

The imaging lens is mounted on a motorized stage and allows adjustment 
depending on the wavelength chosen for operation. Part of the beam is reflected
off from an AR coated wedge before entering the electron beam line. 
The first reflection is sent onto a K6 UV scintillator (Metrolux), placed at a 
plane at a distance equivalent to that to the cathode, acting as a virtual 
cathode. 
The visible light is then imaged onto a camera. 
The image is used to monitor beam profile, position and size. 
A double-mirror piezo system in conjunction with the image analysis allows for 
pointing feedback stabilization. 
The second reflection from the wedge is used to monitor the laser energy 
(LaserProbe, RjP-465). 
Motorized stages allow inserting pyroelectric detectors (Gentec, QE25) into the
direct beam for cross-calibration. 
A He-Ne laser is co-propagated with the UV beam as a reference for alignment 
onto the cathode.
Frequently used diagnostics are integrated into the injector control system and
remotely accessible to operators. 
This includes a remotely controlled cross-correlation of the stacked pulse, 
online diagnostic of the UV pulse energy, the virtual beam profile at the 
cathode position and beam pointing.

For SwissFEL a different technology based on a Yb:CaF$_2$ amplifier was
chosen for the gun laser. 
This selection was motivated by the fact that it combines many advantages of 
both laser systems described above. 
It provides, for example, pulses with a high energy stability (due to direct 
pumping) and directly emits transform-limited 500~fs pulses, which can be used 
for pulse stacking without need for additional dispersion. 
Furthermore, it features a significantly reduced complexity compared to a 
Ti:sapphire laser, with favorable consequences for system uptime and 
maintenance expenses. 

\subsection{\label{sec:cath}Photocathodes}

\begin{figure}[b]    
  \includegraphics*[width=\linewidth]{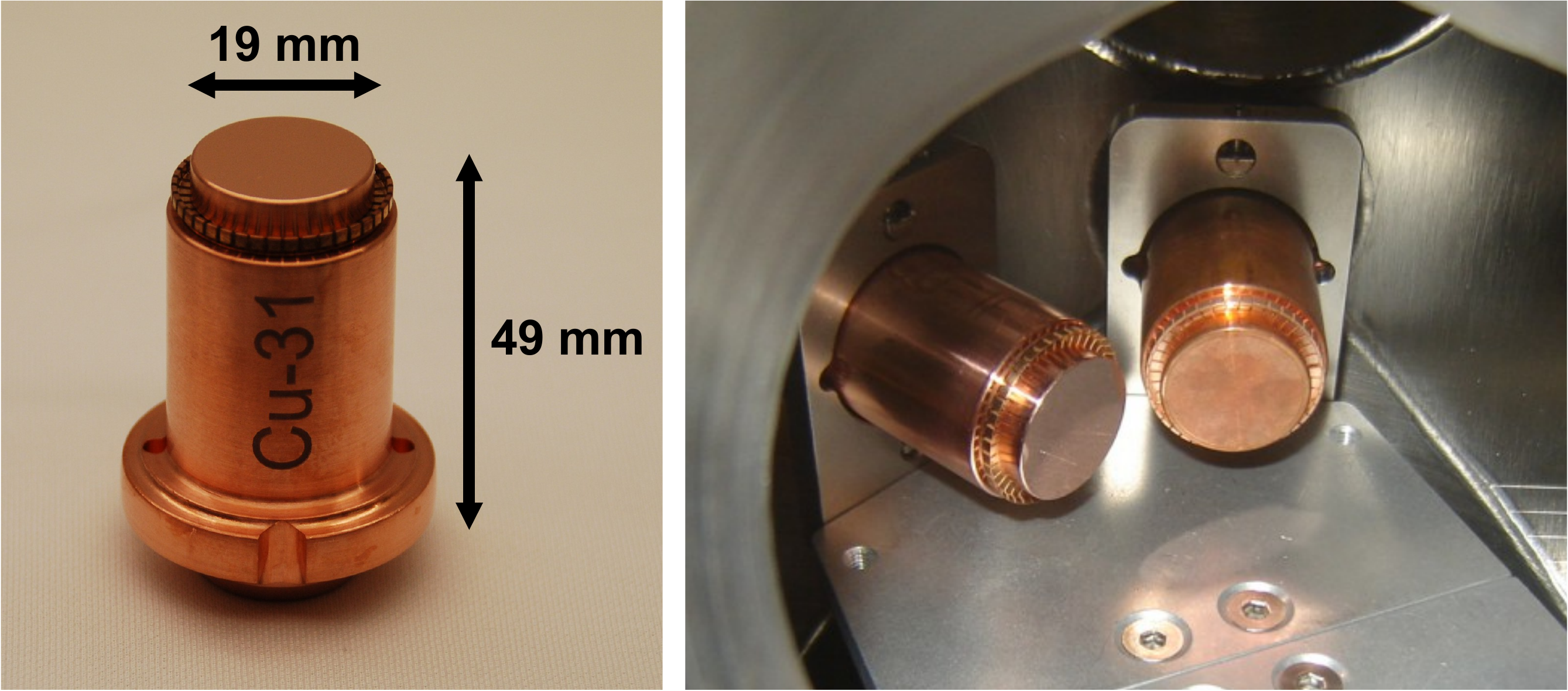}
  \caption{SwissFEL gun cathode plug (copper, no coating), with dimensions 
  indicated (left). 
  Two cathode plugs mounted in the carousel holder of the load-lock chamber
  (right, see also Fig.~\ref{fig:cath2}).
  The rf contacts can be recognized around the circular cathode surfaces.}
  \label{fig:cath1}
\end{figure}

\begin{figure}[t]  
  \includegraphics*[width=1\linewidth]{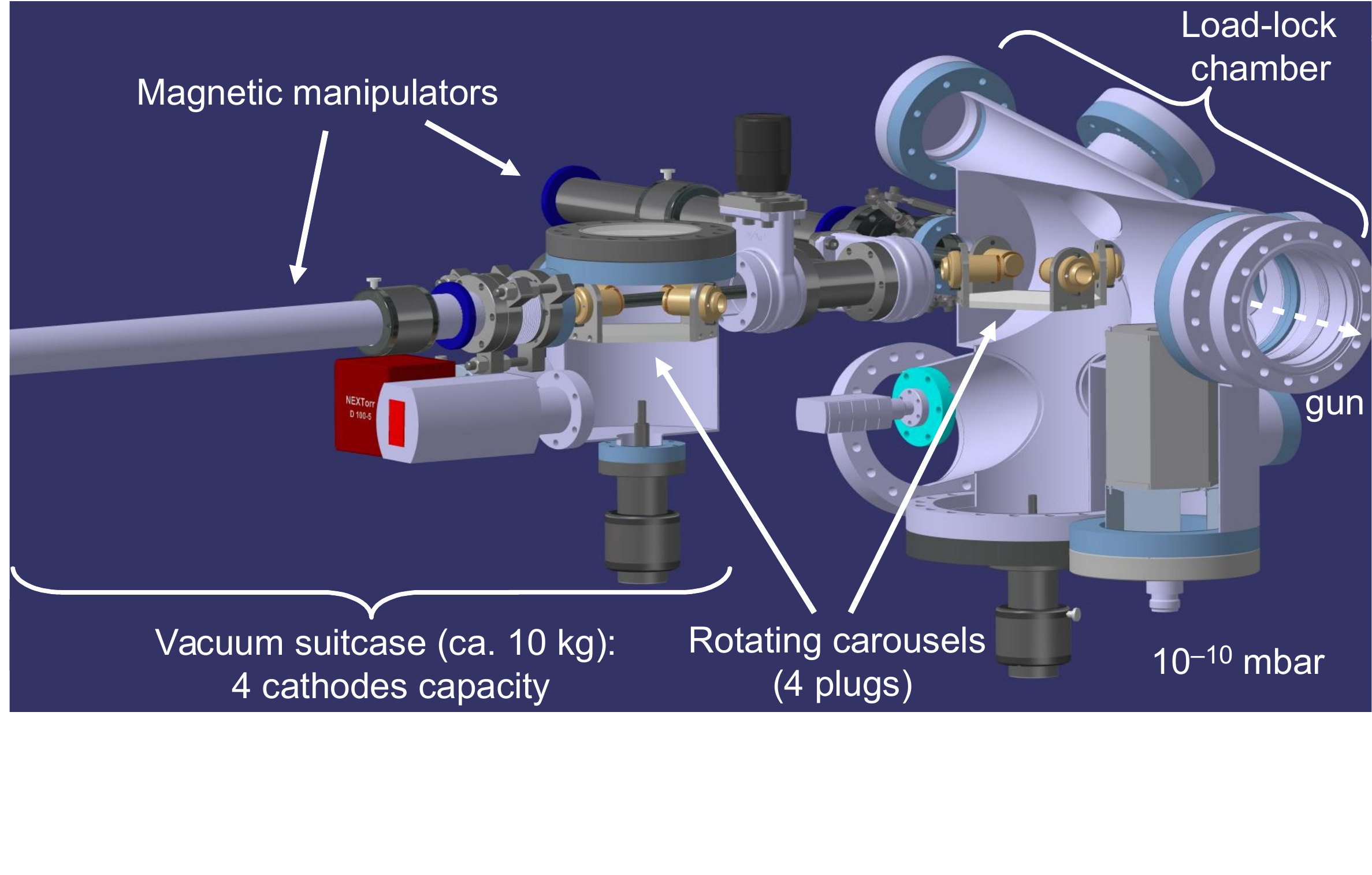}
  \caption{3D drawing of the vacuum suitcase connected to the load-lock chamber
  showing the cathode transfer principle and the two storage carousels.}
  \label{fig:cath2}
\end{figure}

The backplanes of both the CTF and the SwissFEL gun feature circular openings
where cathode plugs can be inserted. 
The cathode plugs are based on earlier designs developed at CERN for the CLIC 
test facility (see, e.g., Refs.~\cite{Sub96,Aul96}). 
An array of rf spring contacts arranged circularly around the cathode surface 
ensures electrical contact to the gun body (see Fig.~\ref{fig:cath1}). 
Thanks to a load-lock system and a vacuum suitcase (Fig.~\ref{fig:cath2} and
Ref.~\cite{Gan13}), cathodes can be cleaned and, if desired, coated with 
cesium telluride (Cs$_2$Te) in the lab and then loaded behind the gun while 
staying under ultra-high vacuum (less than 10$^{-9}$ mbar pressure). 
The load-lock chamber facilitates the cathode exchange considerably since no 
venting of the gun is necessary and, more importantly, it enables the use of 
semi-conductor cathodes, which cannot be transported through air. 
The exchange of a cathode then only takes about half a day, including the rf
conditioning of the new cathode.
Up to four cathodes can be stored under vacuum in the load-lock chamber.

\begin{figure}[b]    
  \includegraphics*[width=1\linewidth]{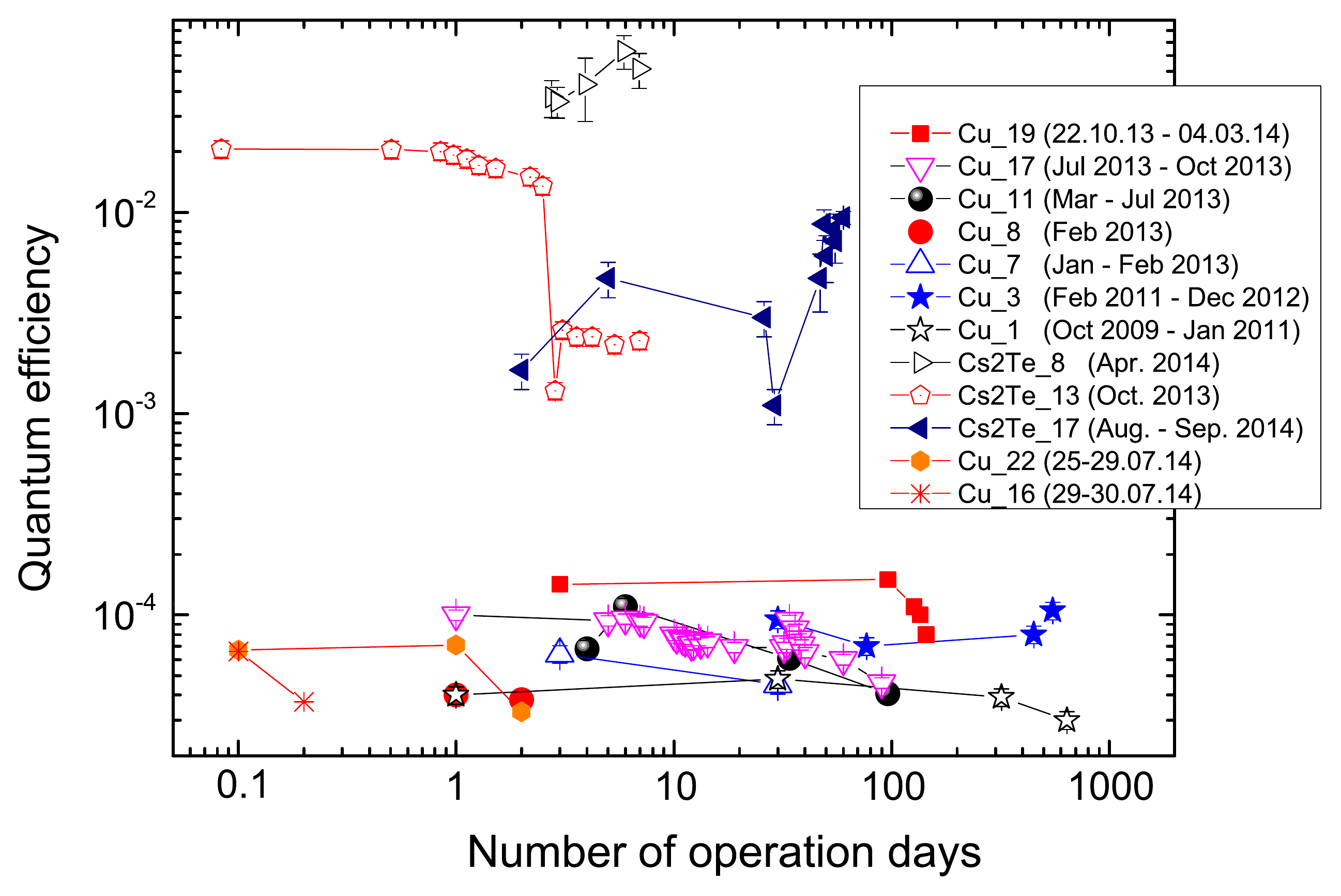}
  \caption{Quantum efficiency evolution of the cathodes used at the SITF 
  between October 2009 and September 2014.}
  \label{fig:cath3}
\end{figure}

The fundamental parameters of a cathode are the local work function and the 
electric field enhancement parameter (see also Sec.~\ref{sec:bd-source}). 
These two parameters are strongly related to the material type, the surface 
roughness and the surface contamination. 
Surface contaminants can be accumulated both during cathode preparation in air, 
as well as during operation in the gun, so that the initial QE can easily vary 
by a factor 10 from cathode to cathode (see Fig.~\ref{fig:cath3}). 
The QE also fluctuates with time because of the contaminants' activity on the 
surface (migration, bombardment, desorption etc.). 
To improve the QE stability and reproducibility of copper cathodes, we have 
established a preparation procedure~\cite{Gan13}, the most important step of
which is the annealing procedure at 250$^\circ$C over 10 hours to desorb the 
most detrimental contaminants like water. 
This procedure was implemented together with the load-lock installation in July
2013 and ensured initial QEs above 10$^{-4}$ for newly installed copper 
cathodes. 

The same treatment of the copper surface was applied for the preparation
of cesium telluride cathodes, before evaporating 15~nm of Te and 
25~nm of Cs onto the surface~\cite{Bos14}. 
Indeed the load-lock chamber offers the possibility to compare copper and 
cesium telluride cathodes under the same gun and laser conditions, whereby QE 
and intrinsic emittance are the most relevant parameters to assess cathode 
quality. 
A high QE gives more laser energy reserve, which may be invested in laser 
shaping (emittance improvement), while a low intrinsic emittance directly 
increases the beam brightness. 
It is well known that cesium telluride provides a QE much higher than copper, 
and it is generally assumed that this advantage comes at the price of a higher 
intrinsic emittance.
Measurements with the SwissFEL gun, described in more detail in 
Sec.~\ref{sec:bd-source}, showed a factor of 100 gain in QE at an increase in 
intrinsic emittance of only about 20\%~\cite{Pra15}.
What is more, the slower response time of cesium telluride relative to copper 
also helps to wash out the laser pulse time profile ripples in the case of 
flat-top laser pulses. 

In conclusion, since cesium telluride exhibits the more favorable properties
for long-term operation, if laser energy is limited,
it has been selected as the baseline option for SwissFEL.

\subsection{\label{sec:rf}Radiofrequency systems}

\begin{figure}[hbt]  
   \centering
   \includegraphics*[width=1\linewidth]{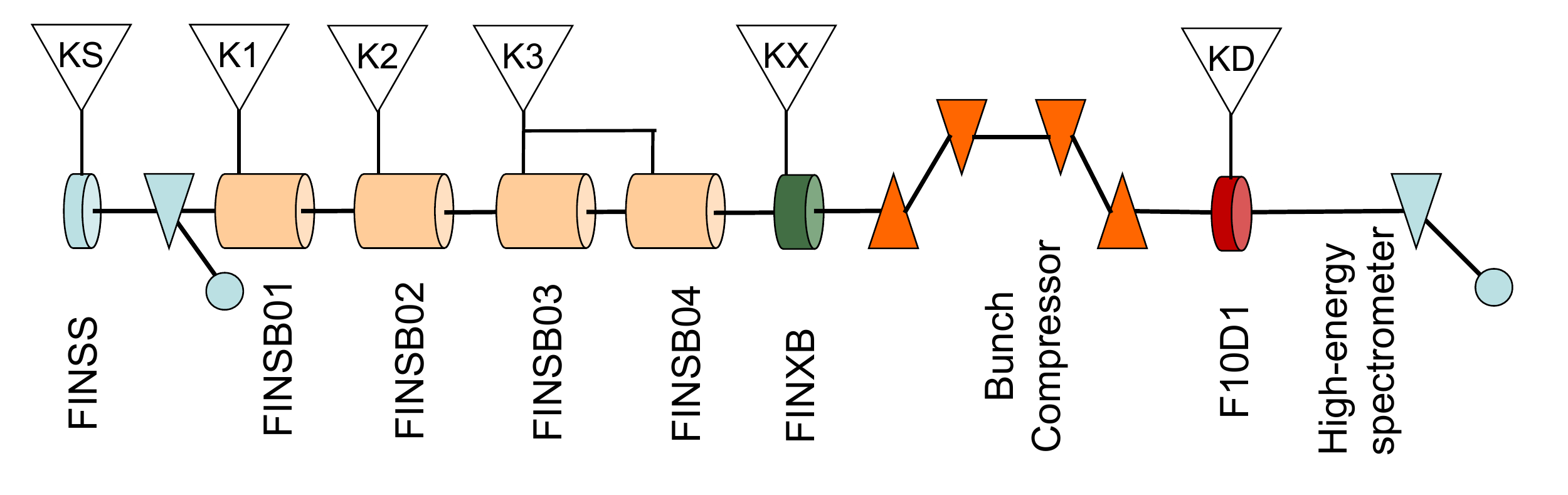}
   \caption{Schematic layout of the SITF rf system, showing the accelerating
     structures as cylinders with the corresponding power sources (klystrons).}
   \label{fig:rf-fig1}
\end{figure}

Figure~\ref{fig:rf-fig1} shows a schematic layout of the injector rf system, 
featuring the S-band photocathode rf gun (labeled FINSS), the S-band 
booster (FINSB), as well as the X-band linearizing system (FINXB) and the 
transverse deflector (F10D1) used for diagnostic purposes.
Each rf plant consists of a high-voltage modulator and a klystron driving the
connected accelerating structure (or structures).
The S-band booster consists of four accelerating structures powered by three rf
plants:
the first two structures (FINSB01 and FINSB02), usually run at the on-crest
phase, are driven by separate rf plants each, whereas a single rf plant powers
the last two structures (FINSB03 and FINSB04), which are operated off-crest to
generate the energy chirp whenever the beam is to be compressed in the ensuing
magnetic chicane. 

In the following we describe each of the rf subsystems, placing particular 
emphasis on the photocathode rf gun, for which a new design was developed and 
tested at the SITF.

\subsubsection{\label{sec:rf-gun}Photocathode rf gun}

\begin{figure}[bt]  
   \centering
   \includegraphics*[width=0.7\linewidth]{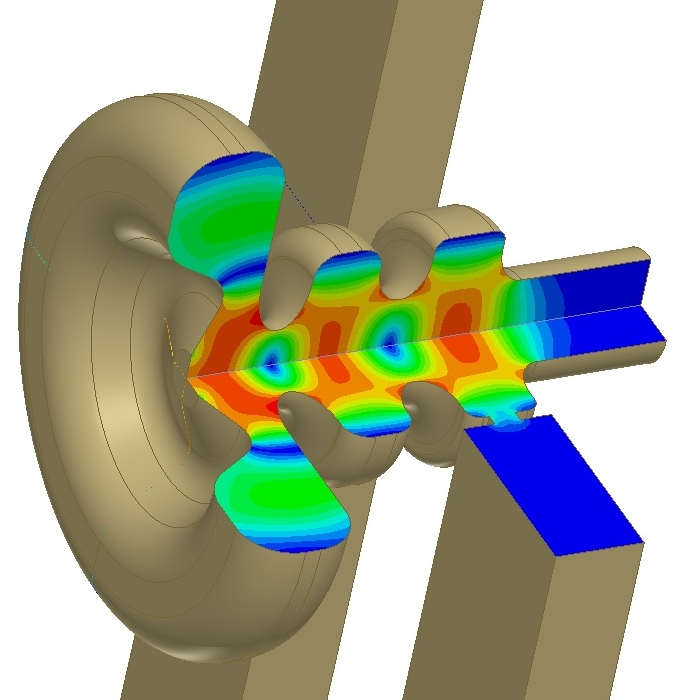}
   \caption{Geometry of the CTF gun No.~V, with magnitude of the electric field
    in the horizontal and vertical planes.}
   \label{fig:rf-fig2}
\end{figure}

The first commissioning phases of the SITF commissioning relied on the CTF gun
No.~V, a 2.6-cell gun originally developed for high-current, multi-bunch 
operation at the CLIC test facility~\cite{Bos95,Bos96}. 
The general geometry of this gun, as used in numerical simulations, is depicted
in Fig.~\ref{fig:rf-fig2}, along with electric field magnitudes in the 
horizontal and vertical planes.
The unique feature compared to other designs is the large diameter of the first
half cell, where the TM$_{02}$ resonance is used as the main accelerating mode. 
This resonance is particularly well suited for compensating the beam loading 
associated with extremely high beam currents. 
At the modest currents required for SwissFEL, however, this feature is of no
relevance. 

The gun structure is rotationally symmetric except for the tuners, the pickups 
and the coupling irises located in the last cell. 
The field asymmetries introduced in the accelerating mode have a diluting 
effect on the beam emittance, which is, however, well below the initial thermal 
emittance from the photocathode~\cite{Deh09}.

Operating at an rf power of 21~MW and with a pulse length of typically 
2~\textmu s, the CTF gun provides an on-axis peak electric field of 100~MV/m,
with a maximum field of 85~MV/m at the cathode. 
The extracted electrons reach a nominal momentum of 7.1~MeV/$c$ at the gun 
exit. 
The repetition rate was generally kept at 10~Hz because the overall cooling 
channels are not compatible with operation at 100~Hz.
Indeed, for repetition rates higher than 20~Hz a significant overheating of the
tuners was observed.

To meet the operational requirements of SwissFEL, in particular long lifetime
and high reliability at a repetition rate of 100~Hz, a new rf photocathode
gun was designed and manufactured in house~\cite{Rag12}. 
The geometry of the new SwissFEL gun is inspired by the 2.5-cell PHIN 
gun~\cite{Los06,Rou07} and adopts some mechanical design aspects of the LCLS 
gun~\cite{Xia05,Dow08}. 
The SwissFEL gun design, shown in a 3D view in Fig.~\ref{fig:rf-fig3}, 
is based on 2.6 cells operating with a near-perfect 
rotationally symmetric $\pi$-mode at S-band frequency, where each cell is 
equipped with a dedicated pickup.
The middle cell is coupled to two rectangular waveguides, symmetrically arranged
to cancel the dipolar component of the field. 
To reduce pulsed surface heating, the z-coupling scheme developed for the 
LCLS gun~\cite{Dow08} is adopted.
The racetrack interior shape of the coupling cell is optimized to keep the 
quadrupolar field components at a minimum~\cite{Hai97}.

\begin{figure}[bt]  
   \includegraphics*[width=1\linewidth]{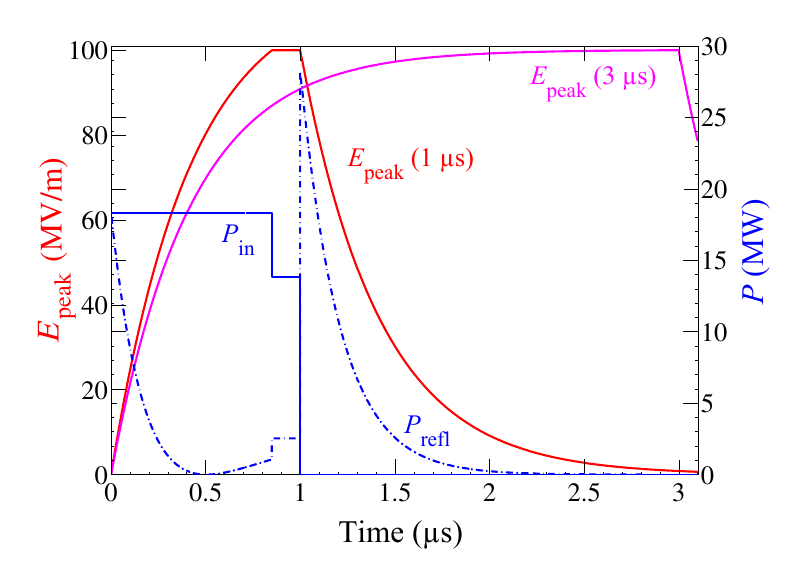}
   \caption{Amplitude modulation scheme for the SwissFEL gun: 
   input rf power (solid blue line), reflected power (dash-dotted blue line)
   and resulting peak gradient (red line) as a function of time for the 
   nominal modulated rf pulse of 1~\textmu s total length.
   For comparison, the peak gradient for an unmodulated 3~\textmu s pulse 
   is also shown (magenta line)~\cite{Rag12}.}
   \label{fig:ampmod}
\end{figure}

\begin{figure}[hbt]  
   \includegraphics*[width=0.8\linewidth]{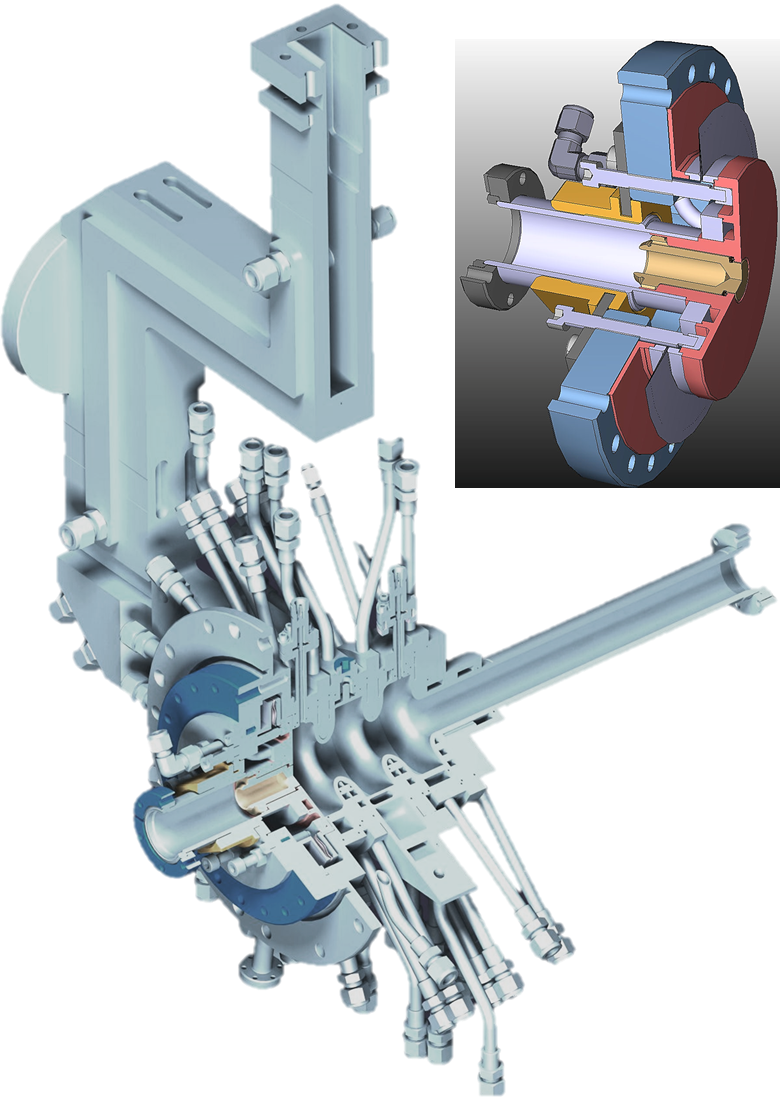}
   \caption{Cutaway 3D views of the SwissFEL gun. Details of the backplane
     and the cathode plug are visible in the upper right figure.}
   \label{fig:rf-fig3}
\end{figure}

Similar to the LCLS gun, our design opts for a large coupling coefficient 
($\beta \approx 2$) to decrease the filling time of the cavity. 
As a consequence, the rf pulse needed to reach the accelerating gradient can
be shortened, resulting in a lower heat load and hence less mechanical stress 
on the gun body. 
The shorter rf pulse has the additional advantages of reducing both the amount 
of dark current originating from the gun and the probability of electric arcs.
The length of the rf pulse can be further reduced through the application of 
a suitable amplitude modulation scheme of the rf input power~\cite{Sch02a}.
In such a scheme, the input power level during the filling of the cavity is
higher than that required to reach the desired accelerating gradient, but is 
then quickly reduced to a lower level once this gradient has been attained. 
The lower power level is sufficient to maintain the nominal accelerating 
gradient in the cavity and is provided for as long as the field is needed.
In the case of the SwissFEL gun, the initial power level of about 18~MW is
maintained for 850~ns and then reduced to 14~MW for the following 150~ns,
yielding a flat-top pulse compatible with the two-bunch operation with 
28~ns bunch separation foreseen for SwissFEL (see Fig.~\ref{fig:ampmod}). 

Nominally, the gun provides an on-axis peak electric field of 100~MV/m.
In contrast to the CTF gun, the location of the on-axis peak field coincides 
with the cathode.
Therefore the electrons, which are extracted at a similar laser injection
phase, experience a higher gradient.
The mean electron momentum at the gun exit is again 7.1~MeV/$c$ as in the CTF 
setup.

The cooling layout of the SwissFEL gun allows operation at repetition rates up
to 400~Hz.
At the nominal repetition rate of 100~Hz and a pulse duration of 1~\textmu s 
the required input power is 18~MW while the dissipated power averages to
900~W.

The parts constituting the first SwissFEL gun were premachined and brazed 
at PSI~\cite{Ell14}, whereas the final machining, involving ultra-precision
turning and milling, was performed at VDL Groep within the specified
tolerances on the surface roughness of $R_a$ $\leq$ 25~nm.
The finished gun was installed in the SITF in April 2014. 

As described in Sec.~\ref{sec:cath} the gun features a circular opening
in the backplane allowing the straightforward exchange of a cathode plug 
under vacuum through a load-lock system.
The upper right of Fig.~\ref{fig:rf-fig3} shows a cutaway view of the 
cathode plug inserted in the backplane. 

\begin{figure}[bt]  
   \centering
   \includegraphics*[width=1\linewidth]{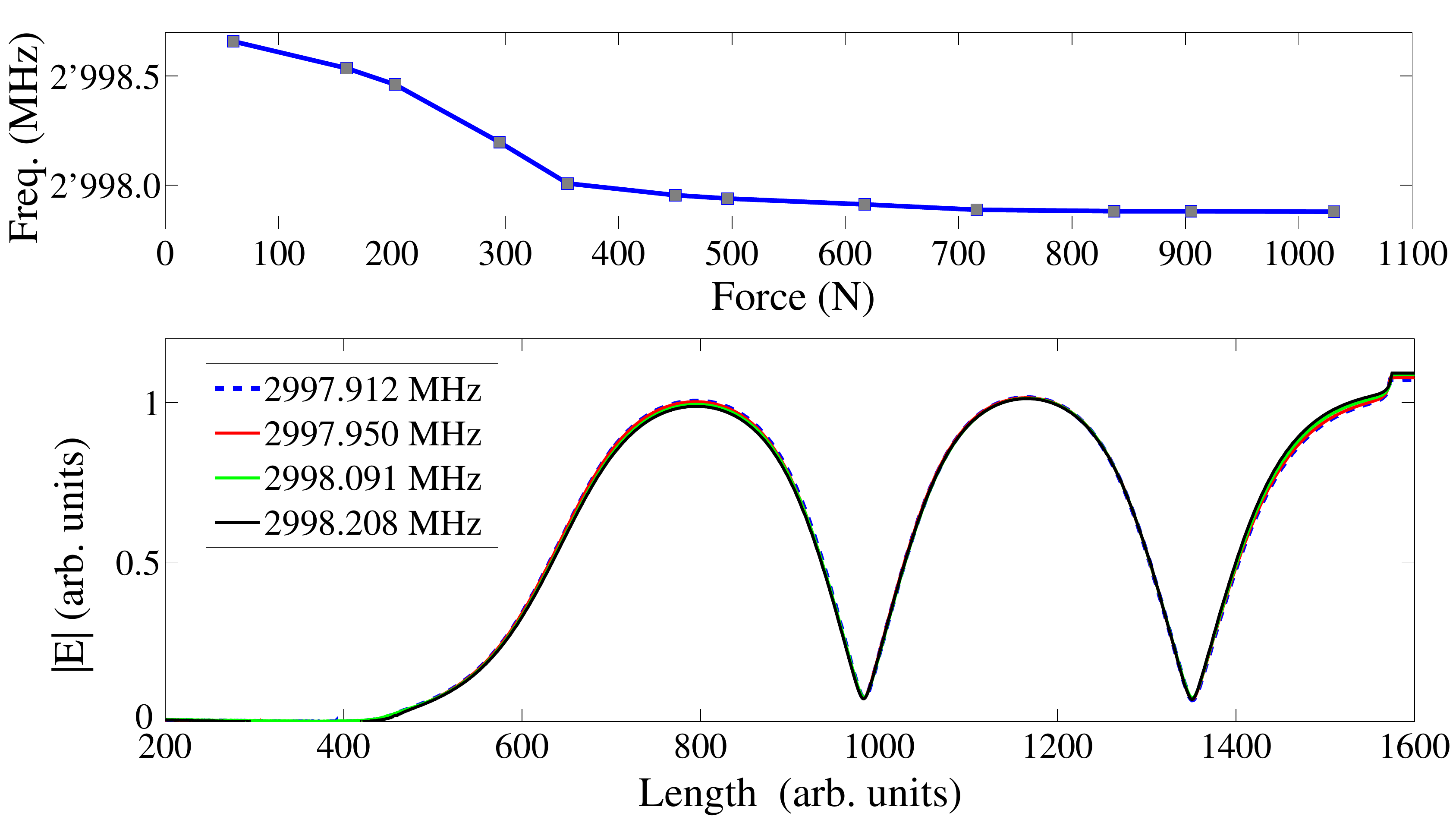}
   \caption{Resonant frequency of the $\pi$-mode as a function of the force 
     applied to the cathode plug (top).
     On-axis electric field profiles for different frequencies, obtained by 
     varying the position of the cathode plug (bottom).}
   \label{fig:rf-fig4}
\end{figure}

\begin{figure}[bt]  
   \centering
   \includegraphics*[width=1\linewidth]{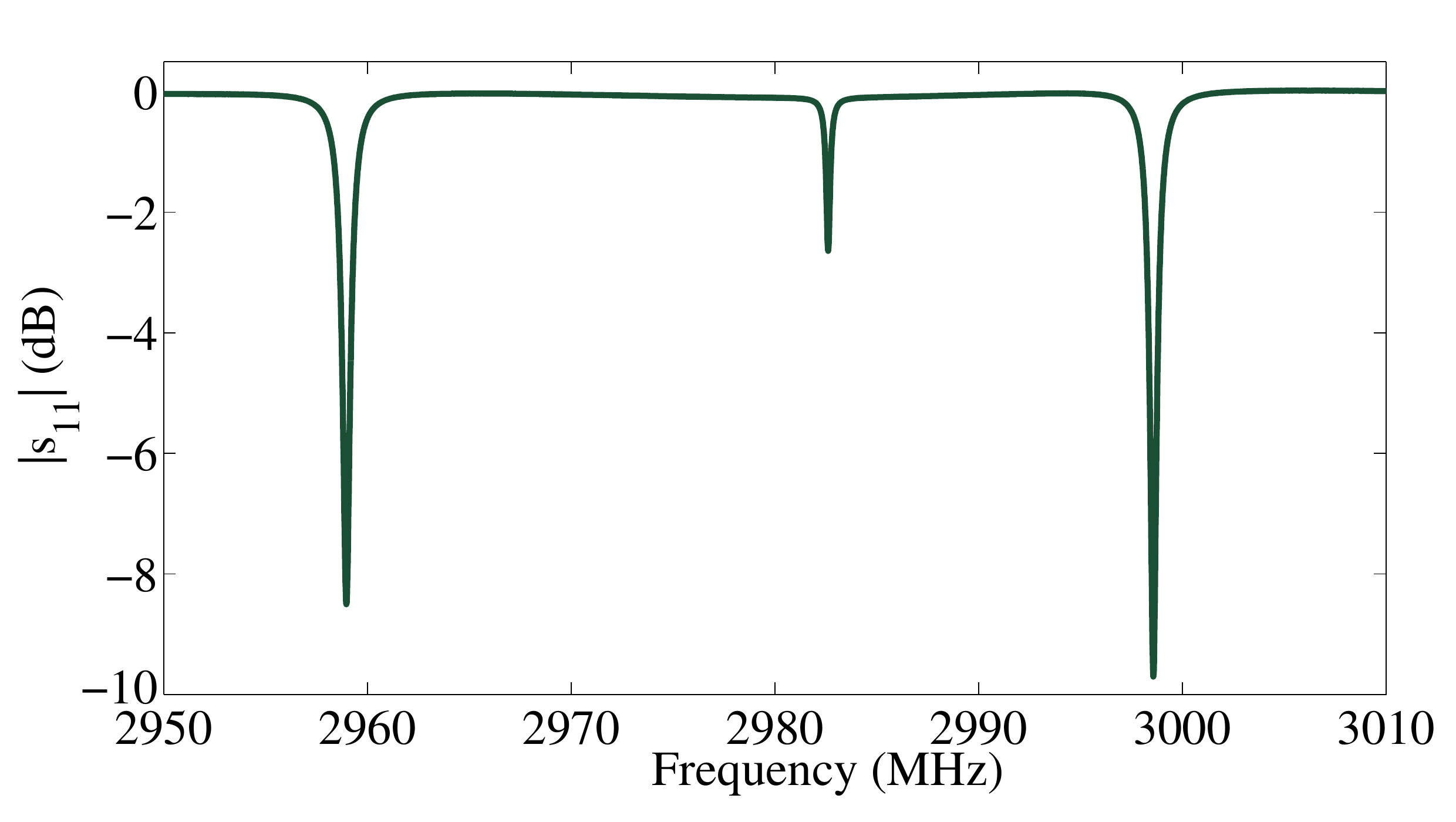}
   \caption{Measured reflection coefficient ($|s_{11}|$) as a function 
     of frequency revealing the frequency spectrum of the SwissFEL gun. 
     The operating frequency in vacuum is 2\,997.912~MHz at 53$^\circ$C. 
     The frequency separation to the adjacent mode is 16.2~MHz.}
   \label{fig:rf-fig5}
\end{figure}

Table~\ref{tab:rfgun} lists the most important parameters of the gun, 
comparing measurement results to values obtained with rf simulations performed 
with the ANSYS High Frequency Structure Simulator (HFSS). 
Measurements at room temperature in the clean room showed that the resonant 
frequency depends on the force with which the cathode plug is pushed into the 
gun.
Therefore the frequency spectrum and field balance were directly verified
with the bead-pull method at the SITF before the start of rf conditioning and 
with the load-lock chamber installed.
To measure the frequency and field balance as a function of the force applied
to the cathode plug, the plug manipulator was equipped with a calibrated 
mechanical pressure gauge.
Figure~\ref{fig:rf-fig4} (top) illustrates the results of the measurements for 
different forces. 
The rf gun is tuned at the nominal SITF S-band frequency of 2\,997.912~MHz with
a balanced field when a force of more than 600~N is applied at a working 
temperature of 53.0$^\circ$C. 
The lower part of Fig.~\ref{fig:rf-fig4} shows on-axis electric field profiles
for different positions of the cathode plug as obtained from bead-pull 
measurements with a ceramic bead of 3~mm diameter. 
The cathode plug used for these measurements has a central hole to accommodate 
the bead-pull wire. 
At the nominal frequency the field ratios (balances) are $E_2/E_c$ = 1.02 and 
$E_1/E_c = 1.01$, where $E_c$ is the field on the cathode and $E_2$ and $E_1$ 
are the fields in the middle and last cell, respectively. 
The frequency spectrum at operational conditions is shown in 
Fig.~\ref{fig:rf-fig5}. 
The frequency separation from the adjacent mode is 16.2~MHz, in excellent 
agreement with the simulation result.

\begin{table}[hbt]
  \caption{Simulated and measured SwissFEL gun rf parameters.
    The time constant refers to the rise time of the electromagnetic field 
    inside the cavity.}
   \begin{ruledtabular}
     \begin{tabular}{lcc}
       Parameter                & HFSS      & Measurement   \\      
       \colrule                                
       $\pi$-mode freq.\ (MHz)  & 2\,997.912  & 2\,997.912      \\
       $\beta$-coupling         &  1.98     & 2.02          \\
       Quality factor $Q_0$     &  13630    & 13690$\pm$100 \\
       Time constant (ns)       & 485       & 481           \\
       Mode separation (MHz)    & 16.36     & 16.20         \\
       Field balance (\%)       & $>$98     & $>$98         \\
       Operating temperature ($^\circ$C) & 57.7 & 53.0       \\      
     \end{tabular}
   \end{ruledtabular}
   \label{tab:rfgun}
\end{table}

The frequency and field balance measurements indicate that no additional tuning 
on the rf gun geometry is necessary, once the adequate force is applied on the 
cathode plug.
Furthermore, the operating temperature is only 4.7$^\circ$C away from the design
value (Table~\ref{tab:rfgun}), which corresponds to a difference in frequency 
of about 220~kHz.
These results bear testimony to the excellent mechanical precision with which
the gun parts were manufactured.

\begin{figure}[tb]  
  \centering
  \includegraphics*[width=1\linewidth]{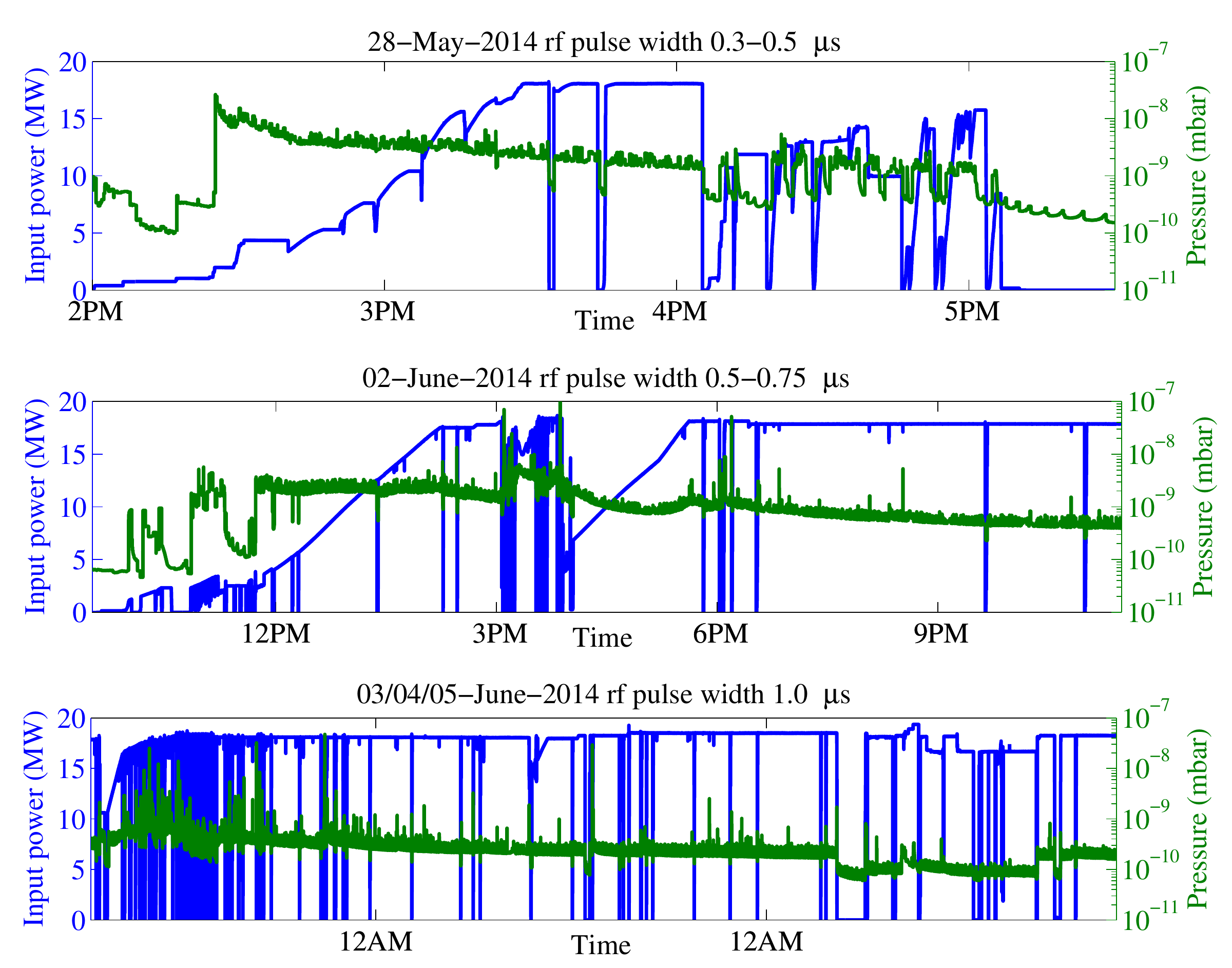}  
  \caption{Progress in rf conditioning the SwissFEL gun towards high power.
    The pressure gauge is mounted on the rf waveguide at the input of the
    rf gun.}
  \label{fig:rf-fig6}
\end{figure}

The SwissFEL gun was tested at full power in the SITF bunker using the rf 
distribution system of the injector equipped with a ScandiNova K2
solid-state modulator and a Thales TH 2100L (45~MW) 100~Hz S-band klystron. 
The klystron output is connected to the gun via a 5~kW average-power
(35~MW peak-power) rf circulator (AFT microwave).
An active temperature control system (Regloplas 90S) stabilizes the gun body 
temperature.
Typical curves illustrating the rf conditioning progress are shown in 
Fig.~\ref{fig:rf-fig6}. 
The rf conditioning was performed at a pulse repetition rate of 100~Hz, by 
gradually increasing the rf power with the help of an automatic conditioning 
tool while increasing the pulse width from 0.3 to 1.0~\textmu s. 
After the conditioning period the vacuum level with rf power was measured to be
around 10$^{-10}$~mbar. 
The nominal rf power of 18~MW for an rf pulse width of 1~\textmu s was already 
attained in the first week of operation.

\begin{figure}[bt]
\centering
   \includegraphics*[width=1\linewidth]{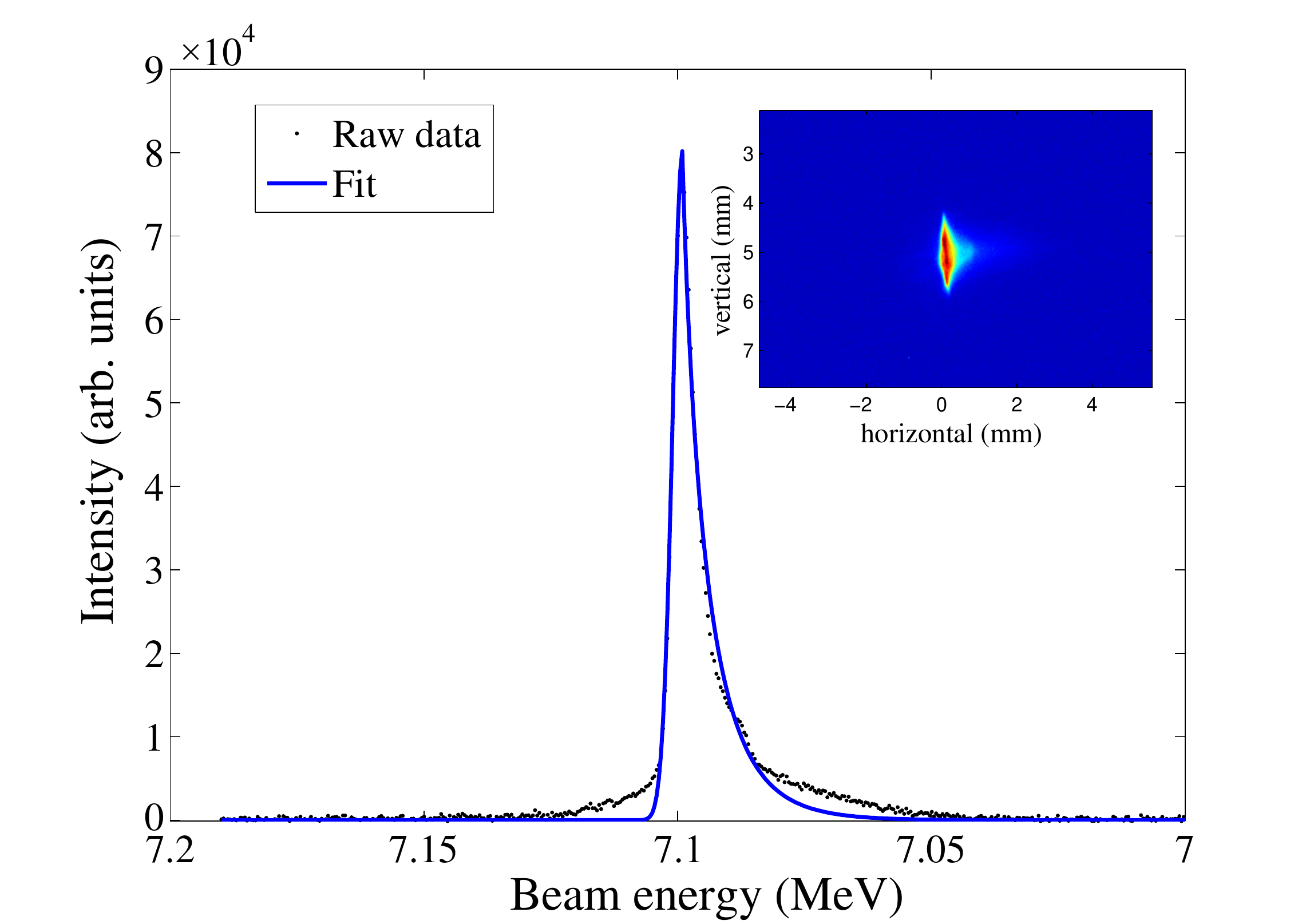}
   \caption{Example of a non-Gaussian horizontal beam profile in the 
    spectrometer at 7.1~MeV (black points), with a modified super-Gaussian 
    function fitted to the data (blue line).
    The inset shows the original screen image.}
   \label{fig:rf-fig7}
\end{figure}

\begin{figure}[tb]
   \centering
   \includegraphics*[width=1\linewidth]{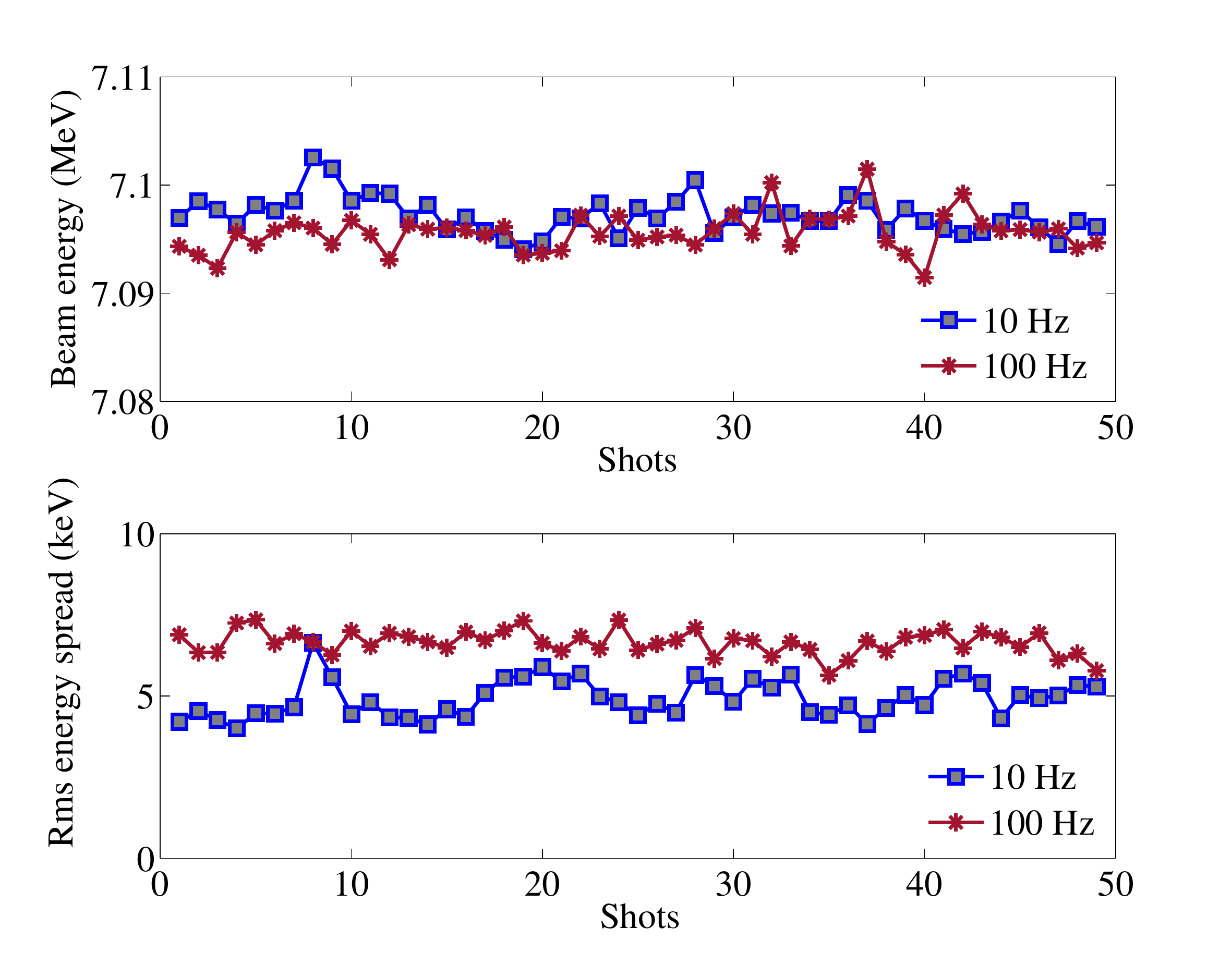}
   \caption{Bunch mean energy jitter (top) and bunch rms energy spread 
   (bottom) at the gun exit determined from 49 nonconsecutive shots over 30~s 
   with a laser pulse repetition rate of 10~Hz for rf running at 10~Hz (blue
   squares) and 100~Hz (red asterisks).} 
   \label{fig:rf-fig8}
\end{figure}

The momentum and momentum spread of the electron beam can be measured with
the low-energy spectrometer arm located after the rf gun and before the first 
traveling-wave structure. 
To limit the contribution of the natural beam size to the dispersion observed
in the spectrometer, the beam is focused horizontally by a small quadrupole 
magnet in front of the bending magnet.
The rms energy resolution at the spectrometer is 4~keV plus a systematic error 
arising from uncertainties of the dipole magnetic field of less than 0.1\%. 
Figure~\ref{fig:rf-fig7} shows an example of non-Gaussian horizontal profile 
at the spectrometer YAG:Ce scintillating screen (Sec.~\ref{sec:scrmon}) with a 
modified super-Gaussian function (see Ref.~\cite{Pen13} for details) fitted to 
it. 
The horizontal axis is converted to MeV according to the dipole dispersion of 
0.387~m. 
Figure~\ref{fig:rf-fig8} shows the jitters in bunch mean energy and energy 
spread at the gun exit determined from 49 nonconsecutive shots over 30~s with 
a laser pulse repetition rate of 10~Hz for rf running at 10~Hz and 100~Hz. 
The rms relative mean energy jitters are 0.023\% and 0.020\% for 100~Hz and 
10~Hz, respectively. 
The rms jitter of the energy spread is 0.5~keV for both repetition rates.

During the high-power tests the emitted dark current was also measured, an 
important quantity in view of the long-term operation of SwissFEL. 
Earlier measurements performed on the CTF gun resulted in alarmingly high dark 
current values, triggering systematic studies on possible collimation 
strategies~\cite{Bet13}.

The average charge arising from the photocathode and the rf gun structure
within one rf pulse was measured as a function of the peak cathode field for 
different pulse durations by use of an insertable coaxial Faraday cup 
positioned at the gun exit (see Sec.~\ref{sec:charge}). 
In these measurements, the gun solenoid is adjusted in each case to maximize 
charge collection of the Faraday cup.
At nominal operating conditions with a peak electric field on the cathode of 
100~MV/m and a total rf pulse length of 1~\textmu s, the integrated dark 
current collected by the Faraday cup is approximately 50~pC for the SwissFEL 
gun~\cite{Cra14}. 
For comparison, the integrated dark current observed from the CTF gun amounts 
to about 1.4~nC under nominal conditions (85~MV/m on the cathode, 2~\textmu s
rf pulse length)~\cite{Bet13}. 

\begin{figure}[tb]
   \centering
   \includegraphics*[width=1\linewidth]{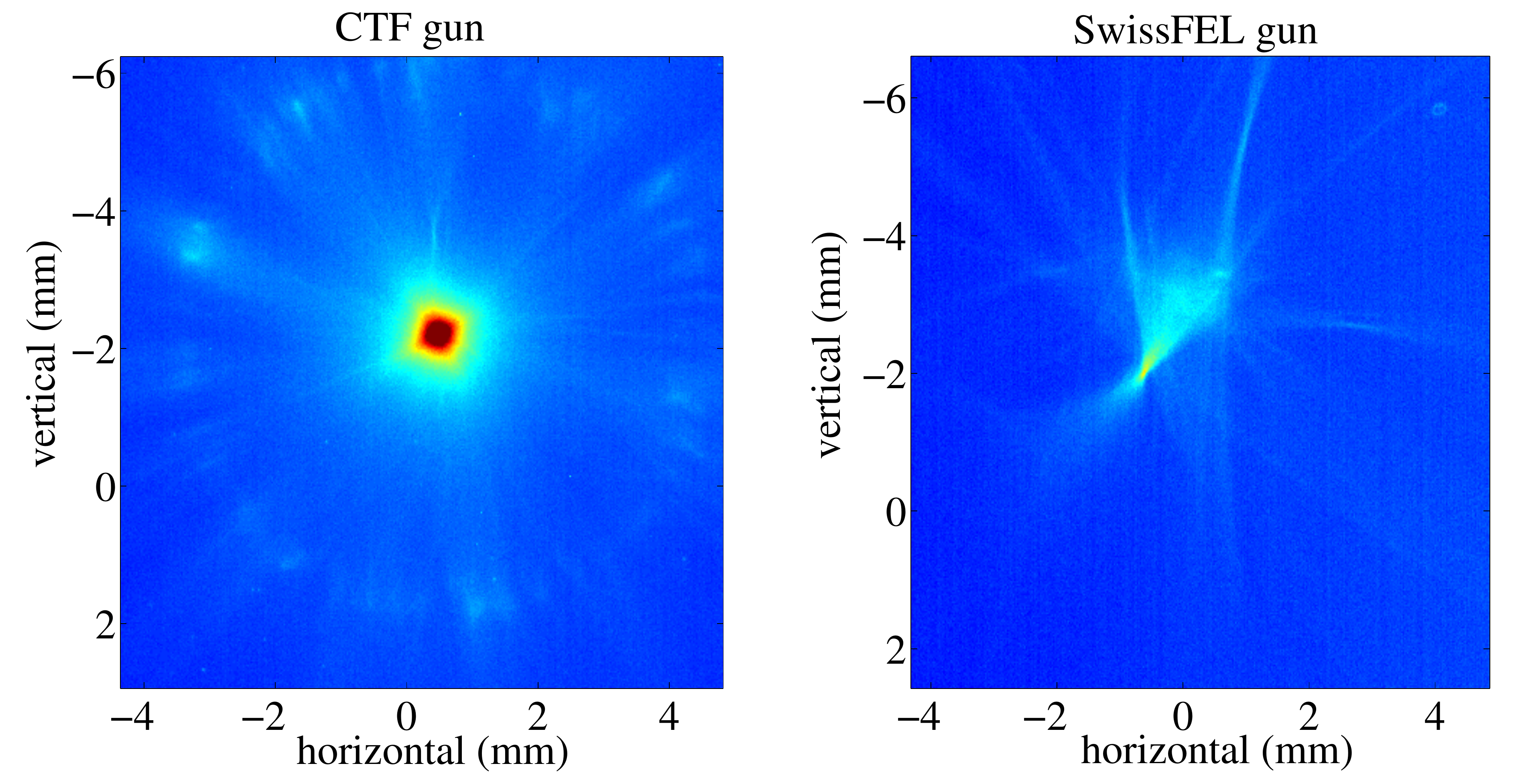}
   \caption{Example images of dark current at nominal conditions on a YAG:Ce 
   screen near the gun exit for the CTF (left) and SwissFEL (right) guns.}
   \label{fig:rf-fig9}
\end{figure}

The dark charge was also imaged using a 200~\textmu m thick YAG:Ce 
scintillating screen placed in the same housing as the Faraday cup. 
Figure~\ref{fig:rf-fig9} shows a comparison between images of the dark charges
observed in the CTF and SwissFEL guns. 
In both cases the electrons are focused with the gun solenoid on the same 
view screen. 

\begin{figure}[tb]
   \includegraphics*[width=1\linewidth]{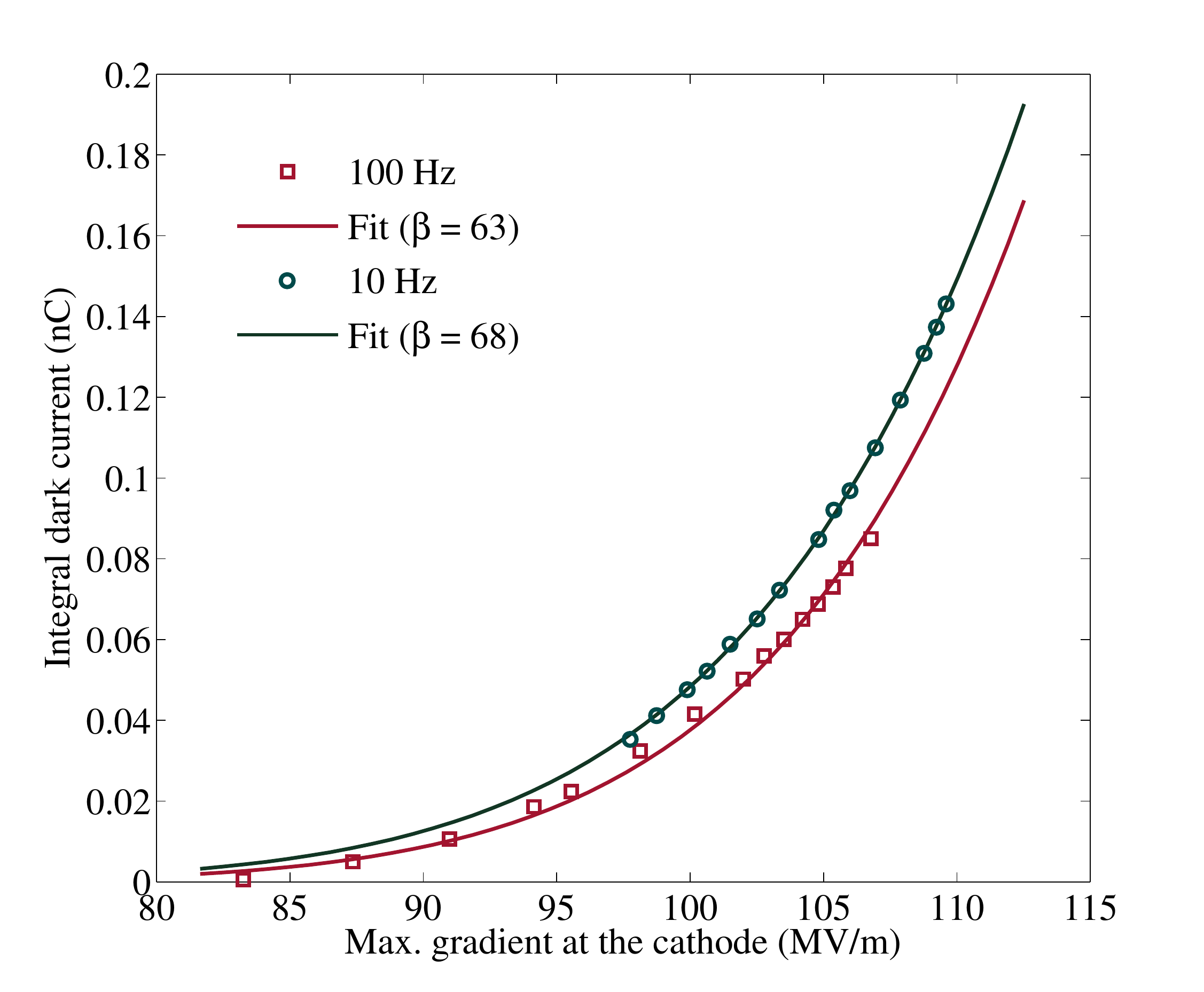}
   \caption{Integrated dark current per rf pulse as a function of the peak 
   electric field at the cathode at 10~Hz and 100~Hz repetition rate 
   (SwissFEL gun).}
   \label{fig:rf-fig10}
\end{figure}

The increase of the integrated dark current with the cathode field, as measured
with the SwissFEL gun for a diamond-milled polycrystalline copper cathode, is 
plotted in Fig.~\ref{fig:rf-fig10} for repetition rates of 10~Hz and 100~Hz. 
The emission of electrons in a period of the rf field is essentially generated 
by field emission, which is governed by the well-known Fowler-Nordheim 
equation~\cite{Fow28}. 
Using this equation with a copper work function of 4.65~eV, and taking into 
account the transient behavior due to the filling time (field rise time) of the 
gun, we estimate the field enhancement factor $\beta$ to lie in the range 
63--68.

\subsubsection{\label{sec:rf-booster}S-band booster}

The S-band booster comprises four constant-gradient traveling-wave 
accelerating structures with 2$\pi$/3 phase advance~\cite{Rag12a}. 
The 4.15~m long structures are separated by short drifts, in which 
corrector magnets, BPM and screen monitor are accommodated, and consist of 
122 cells each: 120 regular cells and two coupler cells. 
The latter are of the double-feed type with a racetrack cross-section to
cancel dipolar and minimize quadrupolar field components at lower beam 
energies.
The structures were designed at PSI and manufactured at RI Research 
Instruments.
Table~\ref{tab:rf-sband1} summarizes a few important parameters of the 
S-band structure design.

\begin{table}[hbt]
   \caption{Main parameters of the S-band booster constant-gradient structures.
   Where available and of relevance measured values are given in brackets.}
   \begin{ruledtabular}
     \begin{tabular}{lr}
       Frequency (MHz)       & 2\,997.912 \\
       Overall length (mm)   & 4150     \\         
       Phase advance         & 2$\pi$/3 \\         
       Attenuation (Np)      & 0.69 (0.68--0.70)    \\  
       Filling time (ns)     & 955 (960--970)      \\    
       Gradient/power ((MV/m)/$\sqrt{\text{MW}}$) & 3.22 (3.07--3.10)\\
       Maximum pulse repetition rate (Hz) &  100     \\  
       Operating temperature ($^\circ$C) & 40.0 \\  
     \end{tabular}
   \end{ruledtabular}
   \label{tab:rf-sband1}
\end{table}

The four booster structures are driven by three rf plants (see 
Fig.~\ref{fig:rf-fig1}), each consisting of a ScandiNova K2
solid-state modulator and a Thales TH 2100L (45~MW) 100~Hz S-band klystron.
After some initial difficulties (see Sec.~\ref{sec:operation}) the S-band 
booster has fulfilled its requirements to provide acceleration and stable 
energy chirp for the magnetic compression with reasonable reliability.
Table~\ref{tab:rf-sband2} lists the energy gains supplied to the electron 
beam by all the S-band rf plants (including the SwissFEL gun) at nominal 
operation conditions when accelerating the beam at the respective on-crest 
phases.
Also listed are the corresponding klystron operational parameters.
\begin{table}[t]
   \caption{Energy gains provided by the S-band rf plants and corresponding
    klystron operational parameters.}
   \begin{ruledtabular}
     \begin{tabular}{lcccc}
       rf plant       & Energy         & High    & Anodic  & Output   \\      
                      & gain           & voltage & current & power    \\
                      & (MeV)          & (kV)    & (A)     & (MW)     \\
       \colrule
       FINSS (gun)    & 7.1            &  235    & 225     & 17.0     \\
       FINSB01        & 56.3           &  235    & 262     & 20.8     \\
       FINSB02        & 75.6           &  285    & 312     & 37.5     \\
       FINSB03/04     & 2$\times$60.8  &  302    & 327     & 48.5     \\
     \end{tabular} 
   \end{ruledtabular}
   \label{tab:rf-sband2}
\end{table}

\begin{figure}[hbt]
  \includegraphics*[width=1\linewidth]{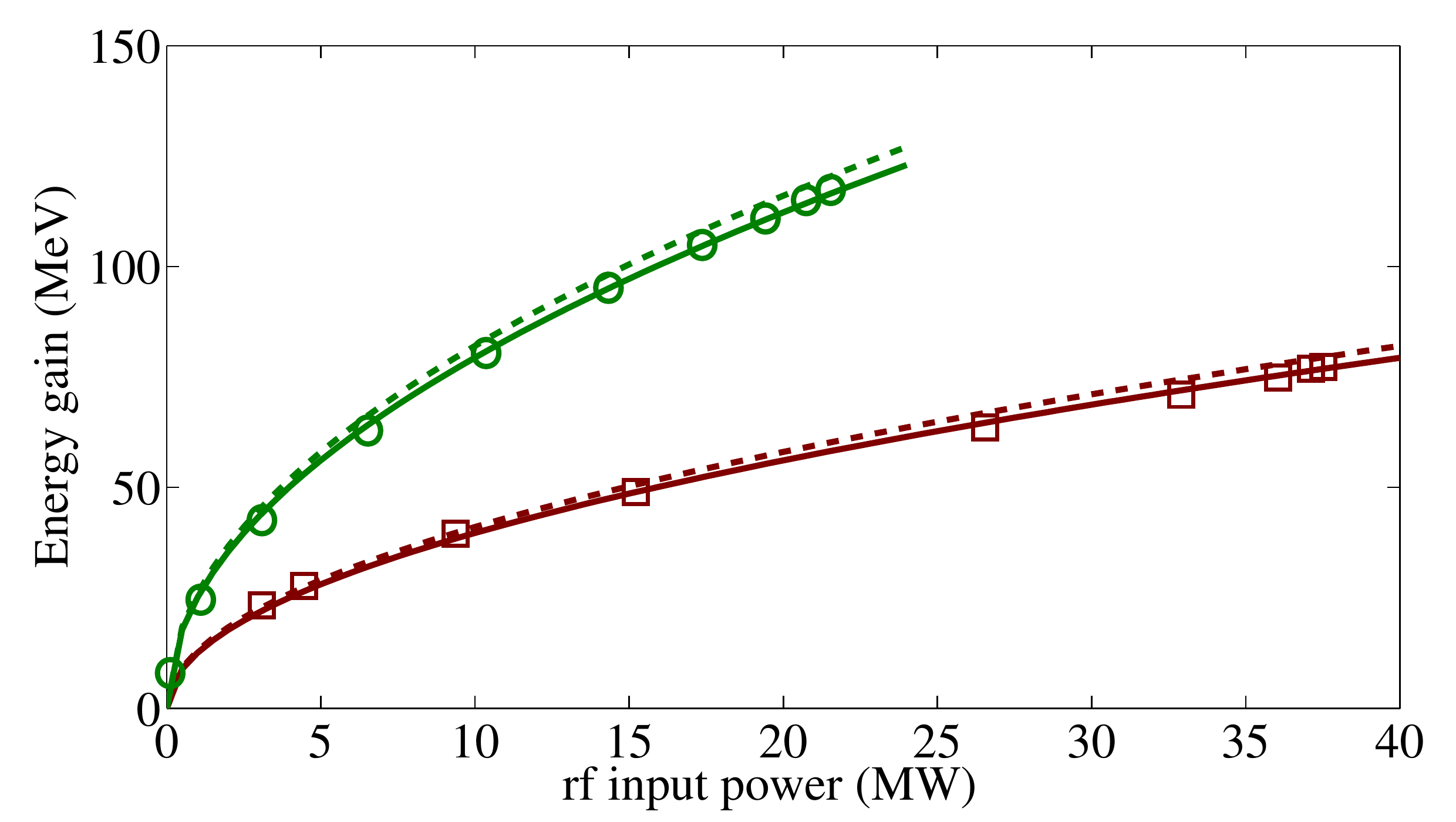} 
  \caption{Measured beam energy gains per accelerating structure as a function 
    of the rf input power for the FINSB02 (red squares, one structure) and 
    FINSB03/04 (green circles, two structures) rf stations.
    The solid lines of the corresponding colors represent square-root fits to 
    the data, the dashed lines show the respective design expectations. }
 \label{fig:rf-fig11}
\end{figure}

To verify the design gradient-to-power ratios given in 
Table~\ref{tab:rf-sband1} we measured the energy gains as a function of the rf 
input power for the FINSB02 and FINB03/04 rf plants. 
The results are shown in Fig.~\ref{fig:rf-fig11}.
The gradient-to-power ratios of the structures were determined by fitting 
square-root functions to the data as 
3.07~(MV/m)/$\sqrt{\text{MW}}$ and 
3.10~(MV/m)/$\sqrt{\text{MW}}$ 
for FINSB02 and FINSB03/04, respectively.
This corresponds to an integrated accelerating voltage of about 
12.5~MV/$\sqrt{\text{MW}}$ for a single structure.

\subsubsection{\label{sec:rf-xband}X-band system}

\begin{figure}[bt]
   \includegraphics*[width=1\linewidth]{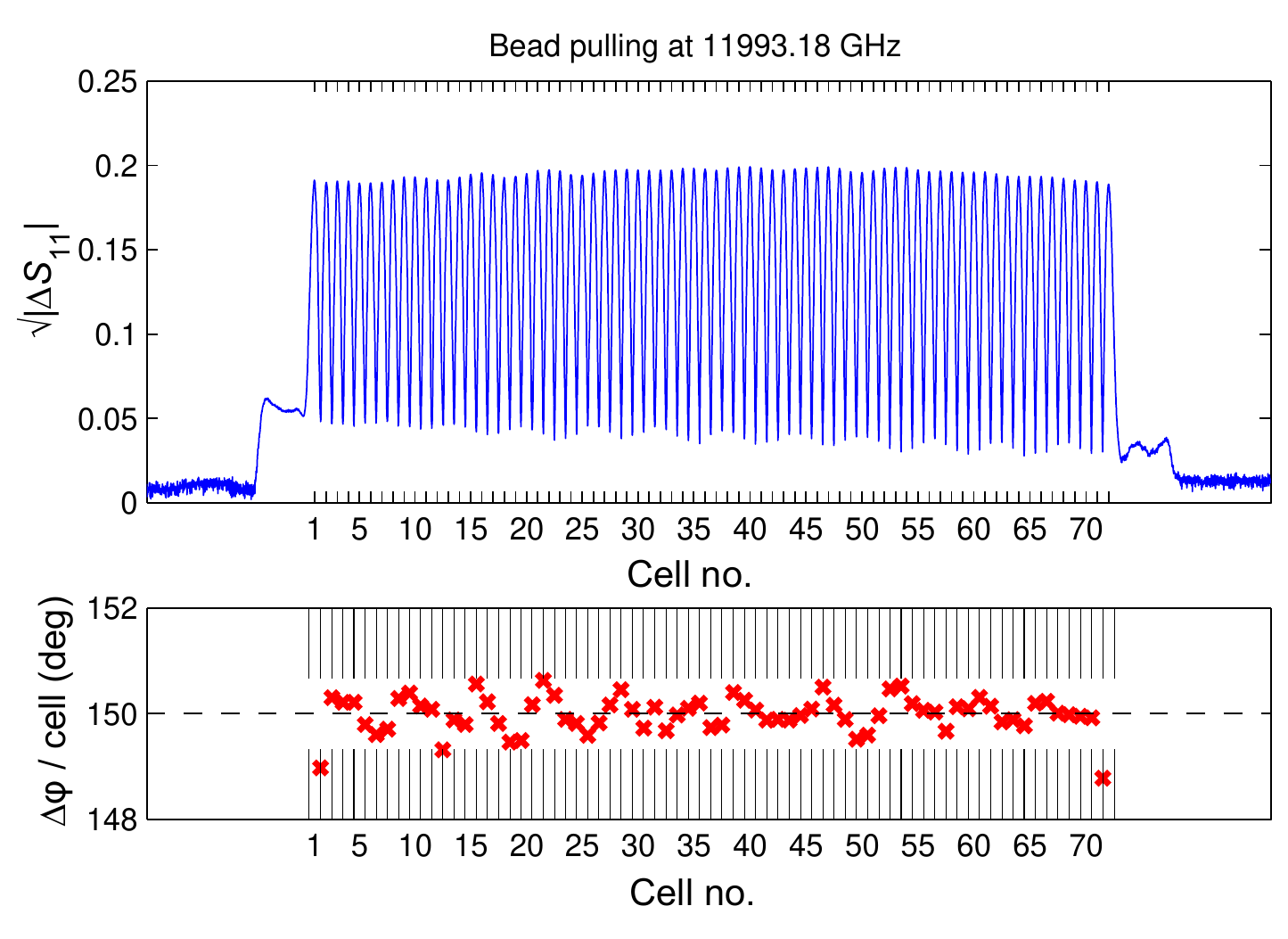}  
   \caption{Field amplitude (top) and phase advance (bottom) measurements of
     the X-band structure using the bead-pull method, after tuning. 
     Measurements and tuning were carried out at CERN~\cite{Shi13}.}
   \label{fig:rf-xband1}
\end{figure}

The linearization of the longitudinal phase space needed for a
uniform compression of the bunch is provided by a 12-GHz structure 
(X-band, fourth harmonic of S-band) installed upstream of the bunch compressor. 
The development of the structure was initiated by PSI and CERN~\cite{Deh09a},
and subsequently carried out by a collaboration including also Sincrotrone
Trieste~\cite{Deh11,Deh12}.
The constant-gradient structure consists of 72 cells with a total active 
length of 750~mm, a phase advance of 5$\pi$/6 and an average iris diameter of 
9.1~mm. 
Table~\ref{tab:rf-xband} lists the main rf parameters of the structure. 
Figure~\ref{fig:rf-xband1} shows the results of the field amplitude and the 
phase advance measurements using the bead-pull method~\cite{Shi13}. 

\begin{table}[t]
  \caption{Main parameters of the X-band constant-gradient structure.
    Where available and of relevance measured values are given in brackets.}
  \begin{ruledtabular}
    \begin{tabular}{lr}
      Frequency (MHz)       & 11\,991.648 \\
      Overall length (mm)   &  1017       \\         
      Active length (mm)    &   750       \\         
      Phase advance         &  5$\pi$/6   \\         
      Attenuation (Np)      & 0.50 (0.56) \\
      Filling time (ns)     & 100 (94)    \\
      Maximum pulse repetition rate (Hz) & 100   \\  
      Gradient/power ((MV/m)/$\sqrt{\text{MW}}$) & 7.55 (7.15) \\
      Operating temperature ($^\circ$C) & 43.5 \\  
    \end{tabular}
  \end{ruledtabular}
  \label{tab:rf-xband}
\end{table}

The geometry adopted, similar to the one of the NLC type H75~\cite{Li02}, 
represents a compromise between a high shunt impedance, to be achieved with 
smaller iris diameters, and a low transverse kick (reduced wakes), calling for 
larger iris diameters. 
The input and output coupler design is based on the mode launcher design
(see Ref.~\cite{Nan04}), which converts a rectangular TE$_{10}$ waveguide mode 
into the circular TM$_{01}$ waveguide mode of the X-band structure. 
The advantage of this design is a reduction of the surface electric field due 
to the rounded and thickened matching cell irises and the coupling of the rf 
power through the broad wall of the feeding waveguide.

\begin{figure}[bt]
   \includegraphics*[width=0.9\linewidth]{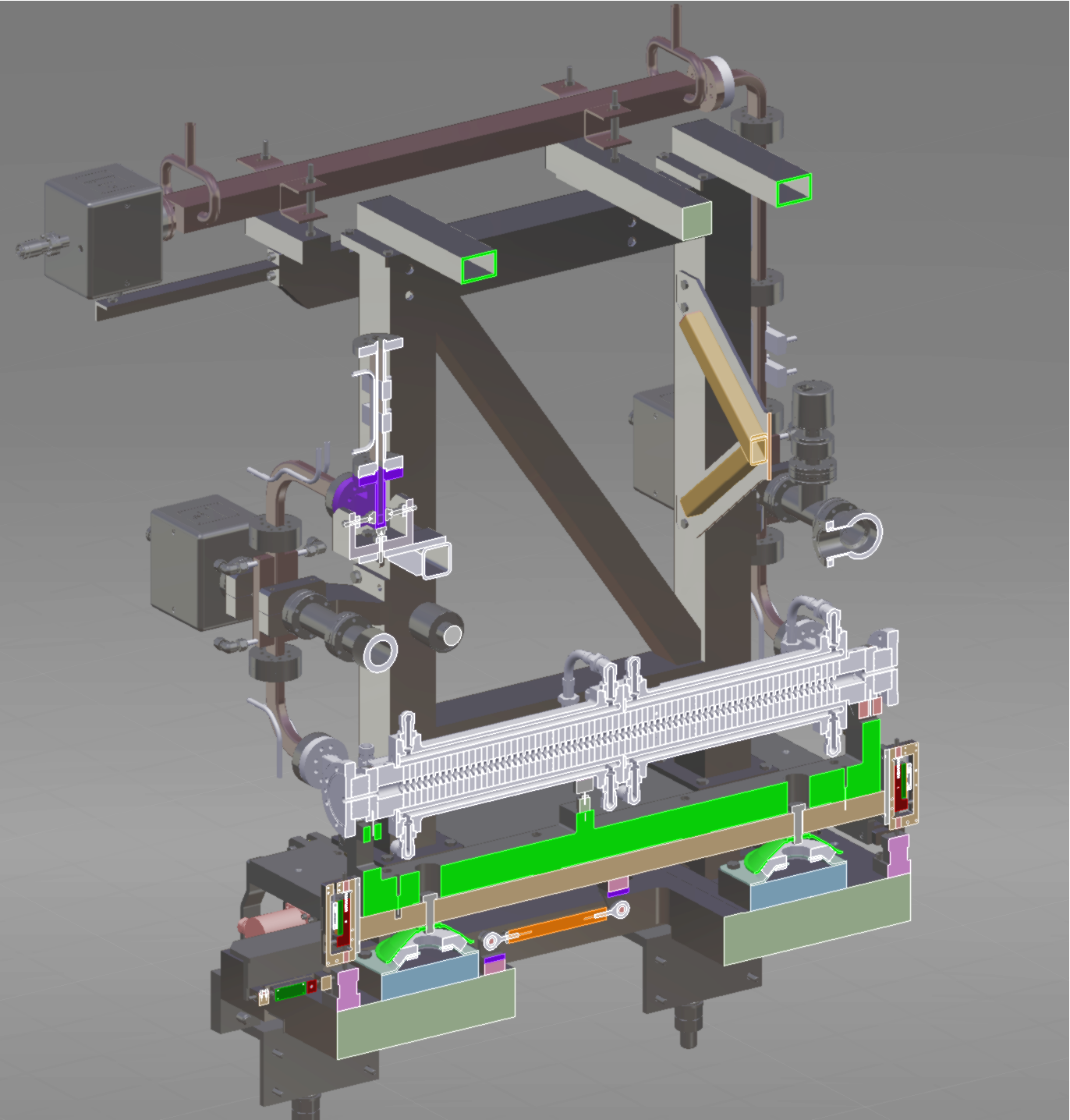}
   \caption{Cutaway drawing of the X-band structure including waveguide system 
     and motorized support.}
   \label{fig:rf-xband2}
\end{figure}

The structure integrates two alignment monitors for accurate beam steering and 
trajectory correction (see Sec.~\ref{sec:wfm}). 
To facilitate beam-based alignment the cavity is placed on a rigid motorized 
support that can be re-positioned remotely by a maximum of 2~mm with a 
resolution less than 2~\textmu m and a reproducibility better than 
$\pm$5~\textmu m. 
Figure~\ref{fig:rf-xband2} shows a longitudinal cut of the X-band structure 
with waveguide connection and motorized support.

The associated rf plant is equipped with a ScandiNova K2-3X solid state 
modulator and a 50~MW and 100~Hz X-band klystron. 
The klystron is a SLAC XL5 klystron~\cite{Spr10}, which is a scaled version of 
the XL4 tube~\cite{Car97}, developed in the framework of a cooperation 
agreement between SLAC, CERN, PSI and Sincrotrone Trieste. 

\begin{figure}[bt]
   \includegraphics*[width=1\linewidth]{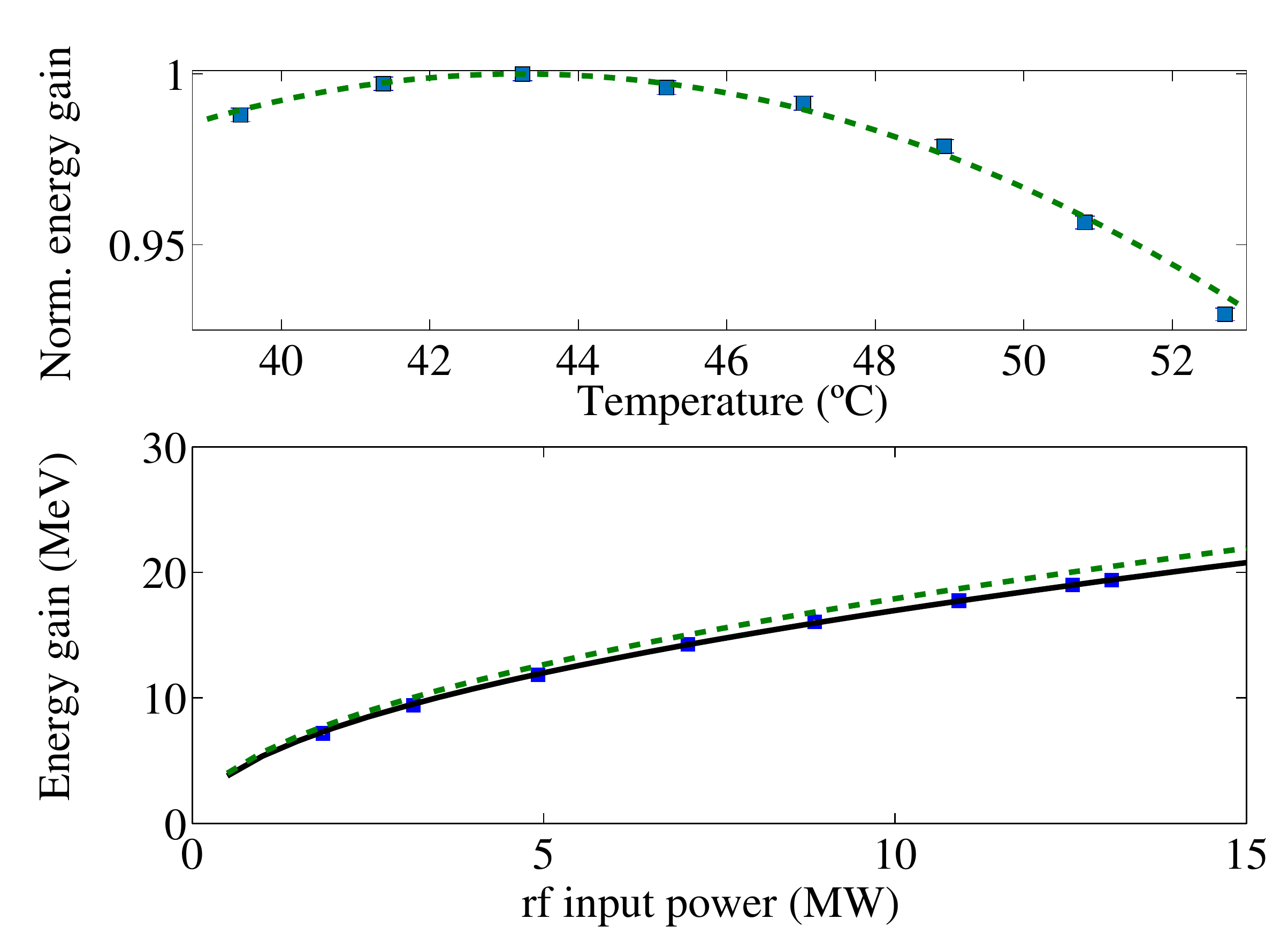}
   \caption{Measured beam energy gains (blue squares) for the X-band structure 
     as a function of operating temperature (top) and rf input power (bottom).
     The dashed lines correspond to the design expectation, the solid line is
     the square-root fit used to extract the structure's gradient-to-power 
     ratio.}
   \label{fig:rf-xband3}
\end{figure}

The ideal operating temperature was found by measuring the beam energy gain as 
a function of structure temperature.
As shown in Fig.~\ref{fig:rf-xband3} (top), maximum acceleration is obtained
at an operating temperature of 43.5$^\circ$C.
Similar to the S-band structures, we determined the gradient-to-power ratio
by recording the beam energy gain as a function of the rf input power
(Fig.~\ref{fig:rf-xband3}).
In the case of the X-band structure, the square-root fit to the measurements 
yields a ratio of 7.15~(MV/m)/$\sqrt{\text{MW}}$.

\subsubsection{\label{sec:rf-tds}S-band transverse deflecting cavity}

In the diagnostics section following the bunch compressor a transverse 
deflecting rf cavity~\cite{Loe65} is used to streak the electron bunches in 
the vertical direction for diagnostics purposes.
The induced correlation between longitudinal and vertical coordinate along the
bunch allows measurements of bunch length and current profile with regular
transverse profile monitors~\cite{Akr01,Akr02}.

The deflector is a standing-wave structure consisting of five cells operating
in the $\pi$-mode. 
It was manufactured by INFN Frascati, according to a design that evolved from 
an earlier one developed for the SPARC project~\cite{Ale06} by scaling it 
to the SITF operating frequency.
A summary of the most important parameters is given in Table~\ref{tab:rf-defl}.
The rf power is provided by a Thales TH 2157 (7.5~MW) 100~Hz S-band klystron
with a ScandiNova K1-2S solid state modulator.
The klystron output is connected to an 3.75~kW average-power (7.5~MW peak-power)
rf circulator (AFT microwave).
While installed at the SITF, the rf deflector was only operated at 10~Hz 
repetition rate. 
It is, however, planned to use the same cavity at the SwissFEL facility
at a rate of 100~Hz. 
Details on the calibration and the performance of the transverse deflector will
be given in Sec.~\ref{sec:long}.

\begin{table}[tb]
  \caption{Main parameters of the S-band deflecting cavity.
    Where available and of relevance measured values are given in brackets.
    The time constant refers to the rise time of the electromagnetic field 
    inside the cavity.}
  \begin{ruledtabular}
    \begin{tabular}{lr}
      Frequency (MHz)                                   &       2\,997.912  \\
      Overall length (mm)                               &              441  \\
      Phase advance                                     &            $\pi$  \\
      Quality factor                                    & 15\,600 (15\,500) \\
      Time constant (ns)                                &         800 (855) \\
      Transverse shunt impedance (M$\Omega$)            &               2.5 \\
      Deflecting voltage/power (MV/$\sqrt{\text{MW}}$)  &       2.24 (2.20) \\
      Maximum pulse repetition rate (Hz)                &          100 (10) \\
      Operating temperature ($^\circ$C)                  &              40.0 \\
    \end{tabular}
  \end{ruledtabular}
  \label{tab:rf-defl}
\end{table}

The initial injector design had provisions for a single-cell transverse 
deflecting cavity for the longitudinal analysis of the electron bunch emerging
from the rf gun~\cite{Fal09}.
Eventually the plan for such a low-energy rf deflector was not realized at the 
SITF due to substantial beam optics problems associated with the deflection of
relatively long bunches at very low beam energies.

\subsection{\label{sec:llrf}Low-level rf system}

The primary goal of the implementation of a low-level rf (llrf) system at the 
SITF consisted in testing suitable concepts and operational issues for the 
SwissFEL facility rather than the validation of the final SwissFEL llrf hard- 
and software.

\subsubsection{\label{sec:llrf-overview}System overview} 

Each of the five S-band and one X-band rf stations at the SITF
is controlled by a digital llrf system. 
The pulsed S-band rf signals from pickups and directional couplers are 
down-converted to an intermediate frequency (IF) of 46.842~MHz 
(64th sub-harmonic of the S-band frequency 2\,997.912 MHz). 
An additional down- and up-conversion stage is used for the X-band system, 
which translates the 11\,991.648~MHz signals to S-band frequencies and vice 
versa. 
Therefore the same llrf system as for the S-band systems can be used for the 
X-band system. 
Local phase-locked oscillator (PLO) units at each rf station generate the 
required S-band rf frequency and additionally 9 and 12~GHz signals at the 
X-band rf station. 
All PLO units are locked to the master oscillator by means of a 
temperature stabilized coaxial reference distribution system.

The common IF signals are digitized by 16-bit analog-to-digital (A/D) 
converters running at a sampling rate of 124.913~MHz (the 24th sub-harmonic 
of the S-band frequency). 
A so-called non-IQ algorithm~\cite{Sch08} implemented in FPGA down-converts 
the digitized signals to baseband, providing I/Q values of the rf signals. 
These waveforms cover a time window of about 8.2~\textmu s. Amplitude and phase 
waveforms are calculated in firmware and transferred to the EPICS control 
system input/output controller (see Sec.~\ref{sec:controls}). 
The drive signal to the amplifier chain can be controlled by a baseband vector 
modulator with its I/Q inputs connected to a 16-bit digital-to-analog (D/A) 
converter with the same output rate as the A/D converters. 
This allows arbitrary amplitude and phase waveform shaping of the drive signal 
within a bandwidth of slightly more than 20~MHz. 

\begin{figure}[tb]
  \includegraphics*[width=1\linewidth]{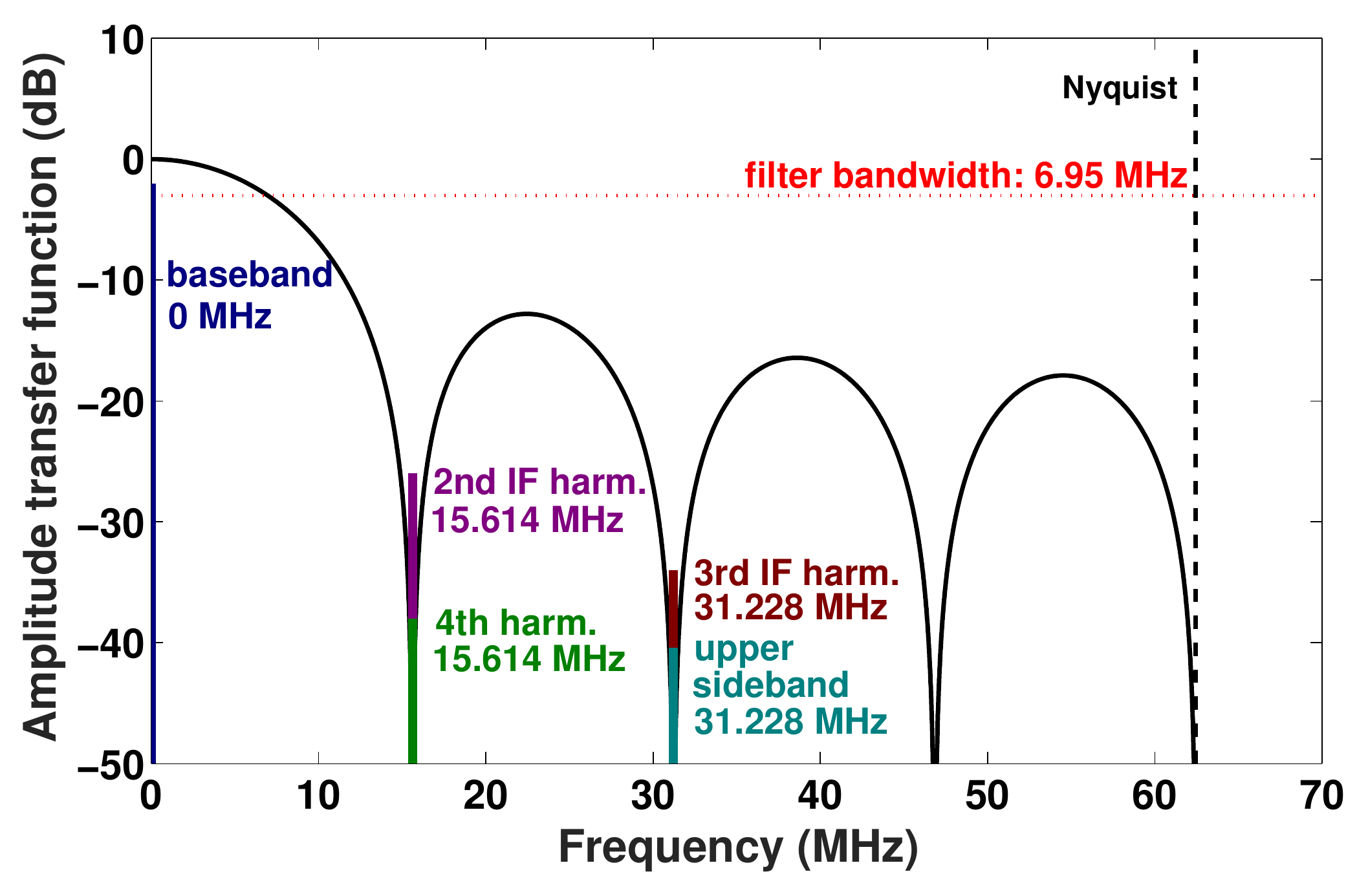}
  \caption{\label{fig:nonIQ_filter}Amplitude transfer function of the non-IQ 
    digital filter of the llrf system.}
\end{figure}

The implemented non-IQ algorithm with its IF to sampling frequency ratio of 
$f_{\text{IF}}/f_s = 3/8$ offers the advantage of its simplicity with 
respect to the digital implementation in firmware, and the benefit that the 
digitally generated upper sideband in the down-conversion process at 
$2 f_{\text{IF}}$ as well as harmonics up to 6th order are automatically
suppressed by the non-IQ filter characteristics. 
The amplitude transfer function of the non-IQ algorithm is shown in 
Fig.~\ref{fig:nonIQ_filter}.
The detection bandwidth of the complete llrf system is mainly dominated by 
this digital filter and covers about 7~MHz ($-$3~dB bandwidth). 
It turns out that the choice of IF and sampling frequencies with the 
implemented down-conversion algorithm is a good compromise between detection 
bandwidth and achievable resolution under the constraint of the performance of
A/D converters at this time. 

\begin{figure}[b]
  \includegraphics*[width=1\linewidth]{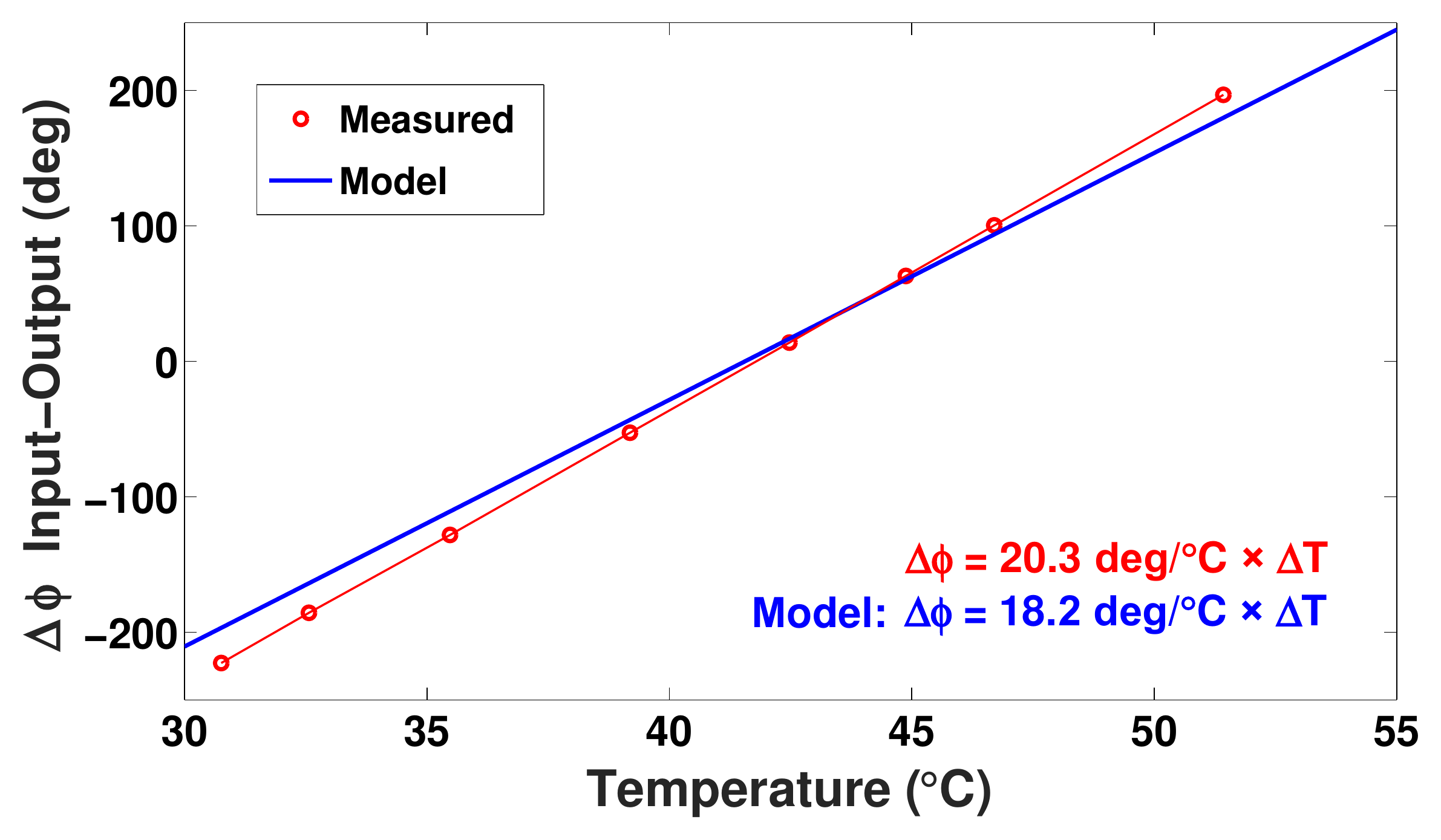}
\caption{\label{fig:phase_slip}Phase slippage between output and input of the 
  S-band traveling wave structures due to temperature variations of the 
  structure.
  The ideal operating temperature, where output and input phases are equal, is 
  found to be 41.7~$^\circ$C in this example. 
}
\end{figure}

The proof-of-concept system is realized with FPGA evaluation boards 
(with Xilinx Virtex-5 FPGA), custom crates and A/D as well as D/A converter 
modules. 
The maximum repetition rate of this setup is limited to 10~Hz whereas 
a 100~Hz processing is required for SwissFEL. 
Amplitude and phase averages of all acquired rf waveforms are calculated 
(typically within a window of about 1~\textmu s for the S-band structures and 
100~ns for the X-band structure), which are then used in pulse-to-pulse 
feedback loops implemented at the EPICS level to compensate for slow drifts in 
the rf system. 
Due to the very short rf pulse lengths no intra-pulse feedback has been 
realized. 
The pulse-to-pulse stability is therefore predominantly depending on the 
stability of the drive chain.
The temperature stability of the traveling-wave structures is maintained to 
better than $\pm0.1\,^{\circ}\mathrm{C}$. 
Nevertheless, any temperature variation within this range results in a change 
of rf phase velocity, which then translates into a phase slippage between rf 
wave and electron beam. 
The model for the installed S-band structures is compared to experimental 
results in Fig.~\ref{fig:phase_slip}.
The effective electron beam accelerating phase is very close to the average 
of in- and output phase of the rf wave. 
Therefore, all rf based feedbacks are implemented as vector-sum feedbacks to 
keep constant the phase of the vector sum of those signals.

The llrf system is located in temperature stabilized huts, which provide the 
same constant temperature as inside the accelerator tunnel to a tolerance of
$\pm$0.1$^\circ$C. 
The huts are located directly adjacent to the linac tunnel resulting in maximum
cable lengths of 10~m. 
The phase-drift-compensated 3/8-inch Andrew Heliax rf cables have expected 
drifts of less than 0.1~ps (about 0.1$^\circ$ S-band phase) in this entirely 
temperature stabilized environment.

\subsubsection{Operational experience, performance and measurements}

The operation of the llrf system has turned out to be very reliable. 
Besides the standard control tasks the main duties of the llrf system are to 
mitigate rf amplitude and phase drifts below the required rf stability 
tolerances and monitor the pulse-to-pulse jitter of the drive chain, which 
entirely depends on the stability of the individual components such as rf 
reference signal, vector modulator, preamplifier and modulator. 
Table~\ref{tab:LLRF_performance} gives an overview of the required rf stability
tolerances and the performance of the llrf measurement system for the different
frequency bands.

The integrated amplitude and phase of an accelerating structure seen by the 
beam is closely related to the bandwidth of the structure and thus to the 
filling time. 
Therefore, the pulse-to-pulse jitter, reported in 
Table~\ref{tab:LLRF_performance}, is calculated within the time window of 
interest, which is approximately the filling time of the structure.
The phase jitter performance of the X-band system is mainly limited by the 
preamplifier stability. 
In addition, the X-band klystron cannot be operated in saturation due to 
missing drive power which then limits the amplitude stability of the forward 
power.

Due to the fact that all llrf local oscillator (LO) and clock frequencies are 
derived from a single rf reference signal at the location of the rf station, 
the reference signal can be used to determine drifts of the llrf measurement 
system. 
It turns out that the main drift originates from the drift of the LO 
frequency, which can easily be compensated by the so-called 
``reference tracking'' method. 
The reference signal is split and fed into one of the rf receiver channels. 
Any observed long-term drift on this channel is subtracted from all other 
measurement channels thus minimizing the effect of common drifts. 
Measurements showed that the remaining channel-to-channel drift amounts to 
less than 0.1$^{\circ}$ per day.

The typically achieved stability figures for the injector accelerating 
structures over one hour of operation are given in 
Table~\ref{tab:RF_stability}. 
In most cases the obtained stabilities fall short of the SwissFEL requirements 
for the pulse-to-pulse stability of the drive chain. 
Upgrades of preamplifiers and high-voltage modulators will be required when 
reusing them at the SwissFEL facility.

\begin{table}[hbt]
  \centering
  \caption{\label{tab:LLRF_performance}
    Pulse-to-pulse jitter resolutions of the llrf measurement system for the 
    two frequency bands, compared to SwissFEL tolerances (rms values).
    The averaging window size for the jitter measurements is 900~ns (100~ns) 
    for S-band (X-band).} 
  \begin{ruledtabular}
    \begin{tabular}{llcc}
      Frequency  &  Parameter  & SwissFEL rf  & Measurement  \\
      band       &             & tolerance    & resolution   \\
      \colrule
      S-band     & amplitude (relative)   & $1.8 \times 10^{-4}$ & $4.4 \times 10^{-5}$  \\
                 & phase              & $0.018^{\circ}$       & $0.0034^{\circ}$       \\
      \colrule
      X-band     & amplitude (relative)  & $1.8 \times 10^{-4}$ & $1.8 \times 10^{-4}$   \\
                 & phase             & $0.072^{\circ}$       & $0.019^{\circ}$         \\
    \end{tabular}
  \end{ruledtabular}
\end{table}

\begin{table}[hbt]
  \caption{\label{tab:RF_stability}
    Overview of typical rf stability figures (rms values) over one hour of 
    operation for all rf stations operated at the SITF.}
  \begin{ruledtabular}
    \begin{tabular}{lcc}
      rf station & Amplitude  & Phase  \\
      \colrule
      \multicolumn{3}{l}{\textit{S-band:}}                     \\
      FINSS (rf gun) & $1.3 \times 10^{-4}$ & $0.020^{\circ}$     \\
      FINSB01        & $1.4 \times 10^{-4}$ & $0.030^{\circ}$     \\ 
      FINSB02        & $1.2 \times 10^{-4}$ & $0.027^{\circ}$     \\
      FINSB03/04     & $2.1 \times 10^{-4}$ & $0.032^{\circ}$     \\
      \colrule
      \multicolumn{3}{l}{\textit{X-band:}}                     \\
      FINXB\footnotemark[1] & $2.1 \times 10^{-3}$ & $0.17^{\circ}$      \\
    \end{tabular}
    \footnotetext[1]{Limited by klystron preamplifier stability.}
  \end{ruledtabular}
\end{table}

\begin{figure*}[tb]
  \includegraphics*[width=0.9\linewidth]{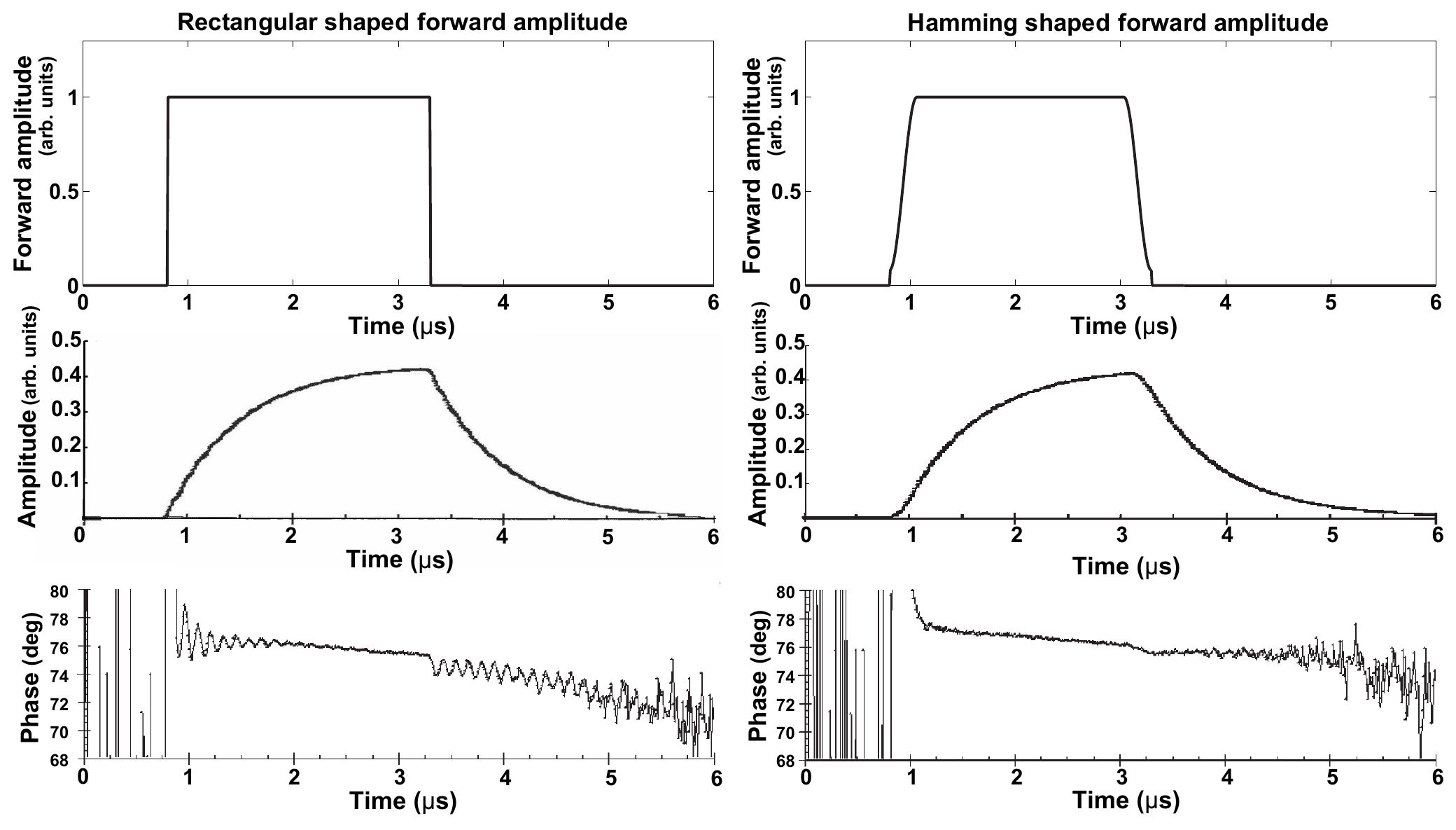}
  \caption{\label{fig:finss_gun}Gun rf amplitude and phase with ``rectangular'' 
    (left plots) and with ``Hamming'' (right plots) rf pulse shape excitation. 
    The rectangular shaped forward pulse also excites the nearest undesired 
    mode 8.1~MHz away from the $\pi$-mode.}
\end{figure*}

The possibility to drive the output of the vector modulator with arbitrary 
waveforms offers the flexibility to shape the rf pulse as desired. 
We use this, for instance, to alleviate the problem of exciting undesired 
passband modes of the rf gun. 
During the initial operation of the first rf gun (CTF gun, see 
Sec.~\ref{sec:rf-gun}) a rectangular shaped pulse of the forward rf power 
excited the adjacent passband mode, only 8.1~MHz away from the fundamental 
$\pi$-mode. 
Introducing a Hamming shaped window at the raising and falling edges of the 
forward amplitude avoids this undesired excitation 
(see Fig.~\ref{fig:finss_gun}).

\subsubsection{Outlook for the SwissFEL low-level rf system}

Based on the experience gained with the operation of the prototype llrf system 
at the SITF several improvements are planned for the SwissFEL series systems. 
The digital processing platform will be changed to a decentralized system. 
Several standard VME64x PowerPC boards with commercial off-the-shell A/D and 
D/A FPGA mezzanine cards (FMC standard) will provide a well defined future 
hardware upgrade path and will remove bottlenecks in the data acquisition chain
to comply with operation at the 100~Hz pulse repetition frequency.
Furthermore, the operation of a large number of almost identical rf stations
calls for automated approaches for startup procedures, operating point 
determinations, failure analyses and recovery actions~\cite{Gen14}.
A first prototype of the new SwissFEL llrf system has been installed and tested
at the SwissFEL C-band rf test stand at PSI~\cite{Hau14}.

\subsection{\label{sec:bc}Bunch compression chicane}

Bunch compression is achieved with a four-bend magnetic chicane, where the two 
central dipole magnets are installed on a motorized movable girder to allow for 
different deflection angles.
The bunch compressor support unit consists of a system of girders made of 
granite, a material with excellent thermal stability.
The bending magnets and the beam-line elements are mechanically linked by 
movably mounted steel girders.
This mounting concept, called ``train link,'' provides an exact mechanical 
reference for the beam transport through the dipole magnets as well as for the 
adjustment of diagnostics components.
The construction design of the bunch compressor unit was optimized for stability
against static and dynamic stress, taking into account the structure's natural
frequencies and possible external excitations such as earthquakes.

\begin{figure}[tb]
   \centering
   \includegraphics*[width=1\linewidth]{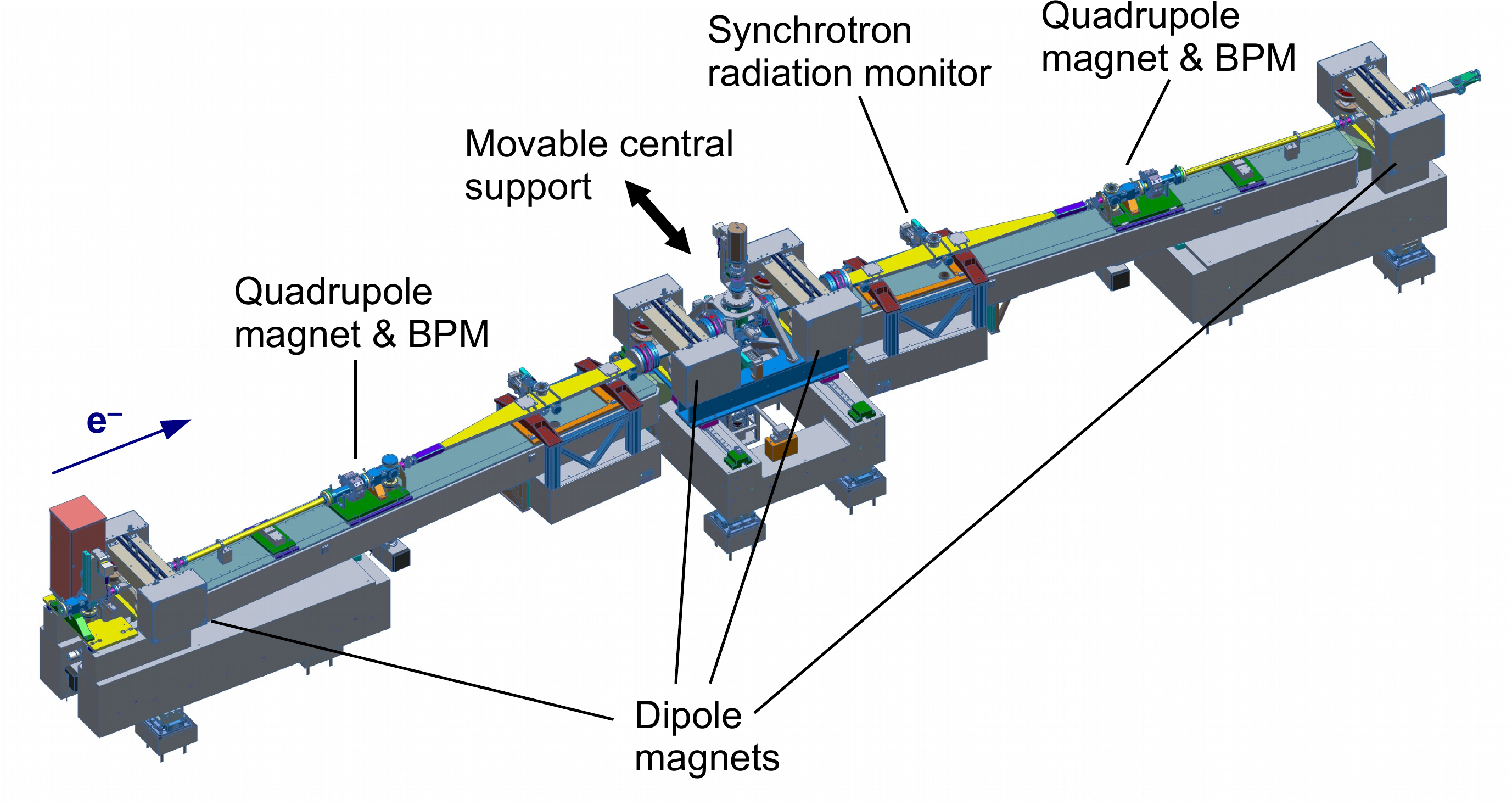}
   \caption{3D drawing of the bunch compression chicane showing the initial
    configuration featuring dipole magnets with return-yoke windings. 
    The thick black arrow indicates the lateral movement of the central support
    for the two inner dipole magnets, which also houses a screen monitor and 
    insertable collimation masks.}    
   \label{fig:bc-drawing}
\end{figure}

The four dipole magnets are powered in series to ensure an equal field in each 
dipole.
Small, individually powered corrector dipoles are integrated in the main 
dipoles with the original purpose of compensating possible field imperfections.
Since these imperfections turned out to be small, the corrector dipoles are 
mainly used to compensate the geomagnetic field, see Sec.~\ref{sec:bc-geomag}.
Two small quadrupoles attached to the outer arms are used to correct for 
residual dispersion.
In March 2013, additional skew quads were added to enable studies related to
beam-tilt correction based on dispersion~\cite{Gue15}.
The transverse profile of the dispersed beam can be observed by means of a 
movable YAG:Ce screen in the central chamber of the chicane, which is imaged by
a CCD camera with a low-resolution lens.
More detailed information on the beam can be obtained, in a nondestructive 
manner, from the synchrotron radiation monitor described in Sec.~\ref{sec:srm}.
Two horizontally movable beam scrapers, also installed in the central chamber, 
can be used to collimate the dispersed beam and thus select a certain energy 
range.
Figure~\ref{fig:bc-drawing} shows a 3D drawing of the compression chicane, 
Table~\ref{tab:bc-pars} lists the most relevant parameters.

\begin{table}[hbt]
  \centering
  \caption{Main bunch compressor parameters.} 
  \begin{ruledtabular}
    \begin{tabular}{lrr}
       Parameter & Nominal & Maximal \\ 
       \colrule
           $R_{56}$                    & --46.8 mm    & --68.9 mm    \\
           Displacement                & 0.333 m     &  0.404 m     \\
	   Deflection angle            & 4.07$^\circ$ &  5.0$^\circ$  \\
           Bend magnetic field         & 0.18 T      &  0.22 T      \\
       \colrule
           Total length                &   &  11.16 m  \\ 
	   Inner drift length          &   & 0.75 m  \\
           Outer drift length          &   & 4.39 m  \\
           Dipole length (effective)   &   & 0.25 (0.303) m \\
    \end{tabular}
  \end{ruledtabular}
  \label{tab:bc-pars}
\end{table}

\begin{figure}[tb]
   \centering
   \includegraphics*[width=0.8\linewidth]{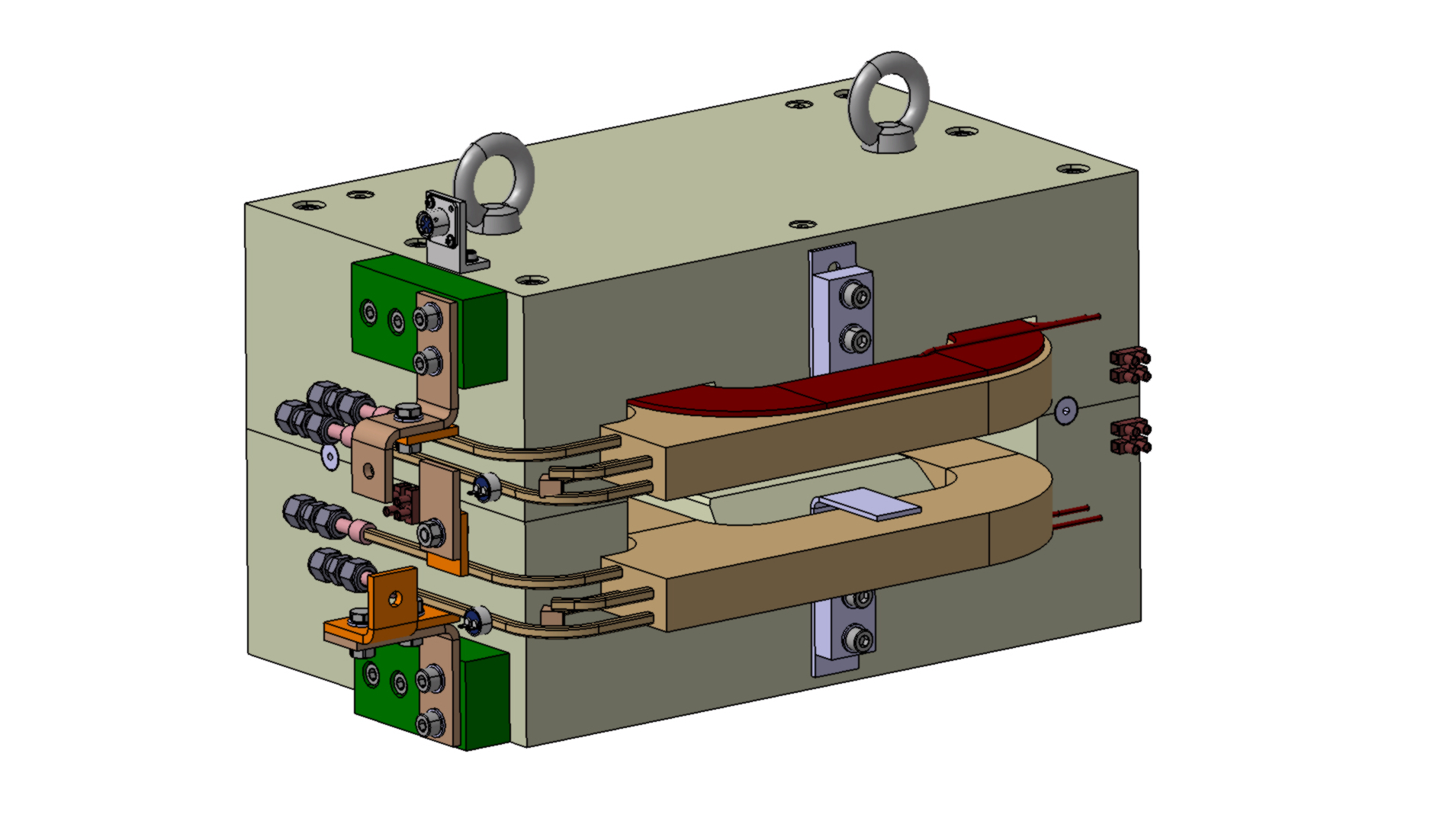}
   \includegraphics*[width=1\linewidth]{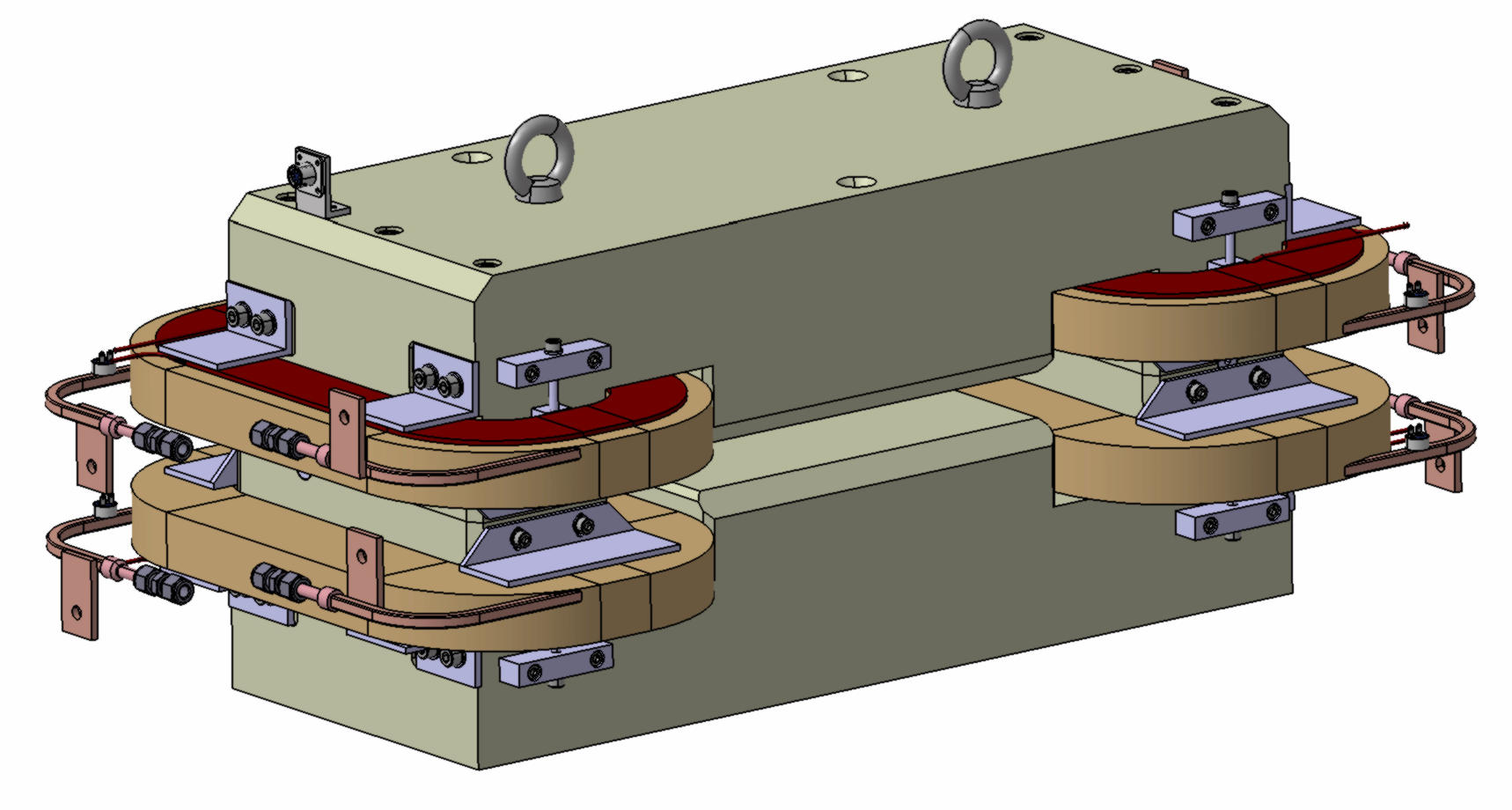}
   \caption{The two dipole magnet designs tested in the SITF bunch compressor.
   Top: original design with pole windings.
   Bottom: second design with larger pole width, later discarded because of
   the long-range stray fields resulting from the return-yoke windings.}
   \label{fig:bc-dipoles}
\end{figure}

The large dispersion in the central region of the bunch compressor gives rise
to a transverse beam size of several millimeters and orbit excursions within
the dipole fields of similar magnitude. 
Therefore a wide transverse good-field region in the main dipole magnets of the
chicane is essential.
The original design of the bunch compressor dipole magnets, shown in 
Fig.~\ref{fig:bc-dipoles} (top), is a traditional H-type magnet with coil 
windings around the upper and lower poles, providing a good-field region of 
60~mm after shimming.
(Here, the good-field region is defined according to the requirement 
$\int (\Delta B / B) ds < 10^{-4}$.)
Following heightened concerns about a possible emittance growth due to 
nonlinear dispersion and nonlinear focusing effects in the dipole fields
based on experience gained at LCLS~\cite{Wel08}, a second design with a
larger pole width was proposed and realized (Fig.~\ref{fig:bc-dipoles}, bottom).
By arranging the coils in two pairs, each around the return yokes of the H-type
magnet, it was possible to extend the size of the transverse good-field region  
to 110~mm~\cite{Neg12}.

The initial bunch compression studies at the SITF were carried out with dipole
magnets of this second type, providing the larger good-field region.
Unfortunately, however, the beam measurements revealed that the magnets with 
the return-yoke windings exhibit a long-range stray field, perturbing the beam 
orbit in locations as far as as 2~m before the bunch compressor entrance.
While bunch compression studies at the SITF were still possible with suitable
orbit corrections, the impairment by the stray fields was deemed unacceptable
for the future operation of SwissFEL.
As an obvious alternative to examine, the originally foreseen dipole magnets 
with coil windings around the poles were installed in a brief shutdown in 
July 2014 (see Sec.~\ref{sec:phaseIV}).
Subsequent measurements performed with these dipole magnets exposed no problems
related to the magnet design, either from stray fields or due to the limited
good-field region.

Based on the experience gained at the SITF, the SwissFEL bunch compressors will
be equipped with dipole magnets of the first design, i.e., with coil windings
around the poles.
Furthermore, as a consequence of the observed susceptibility of the transverse 
beam emittance to coherent synchrotron radiation effects (see 
Sec.~\ref{sec:bd-comp} for details), the compression chicanes at SwissFEL will 
be somewhat longer than that at the SITF (13.75~m and 16.95~m for the first and 
second chicane, respectively, as compared to 
11.16~m at the SITF).

\subsection{\label{sec:ediag}Electron beam diagnostics}

Here we briefly review the main diagnostics systems used for the routine
operation and beam characterization tasks at the SITF~\cite{Isc14}, namely 
charge monitors (Sec.~\ref{sec:charge}), beam position monitors based on 
resonant striplines (Sec.~\ref{sec:bpm}), screen monitors 
(Sec.~\ref{sec:scrmon}) and the combination of transverse deflecting rf cavity 
and screen monitor for longitudinal diagnostics (Sec.~\ref{sec:long}).
Additional diagnostics systems not used for routine operation but nevertheless
profiting from the available electron beam for test purposes will be 
described in Sec.~\ref{sec:diag-dev}.

\subsubsection{\label{sec:charge}Charge monitors}

The determination of the bunch charge along the beam line is essential to ensure
full transmission and for the correct interpretation of all beam physics
measurements.
At the SITF we mainly rely on the strip-line BPMs, to be described in 
Sec.~\ref{sec:bpm}, for the monitoring of the bunch charge.
While the BPMs provide a rather precise relative measurement of the bunch 
charge, they cannot give an absolute charge reading and need to be calibrated
with an independent charge measurement.
For the absolute charge measurement, a wall current monitor in the gun section 
is used.
This monitor measures the voltage across 3$\Omega$ resistors around the
circumference of a ceramic gap in the beam pipe, which is proportional to the
wall current flowing across the gap.
At our typical bunch charges this voltage is very small, leading to a readout
noise equivalent to about 10 pC.
With its bandwidth of several GHz the wall current monitor is capable of 
distinguishing the photoelectron beam from the dark current.

In the same section, an insertable coaxial Faraday cup is also available for 
charge measurements. 
It is primarily used for dark current measurements (see Sec.~\ref{sec:rf-gun}).
Both the wall current monitor and the Faraday cup are based on designs already
applied at the Swiss Light Source (SLS)~\cite{Dac00}.
No dedicated read-out system has been devised for these two monitors---they are
simply read out with oscilloscopes.

At SwissFEL cavity BPMs (see Sec.~\ref{sec:cbpm}) will be used primarily to 
monitor variations in bunch charge, along with integrating current transformers
(ICT) for absolute charge measurements.
A few ICTs, both of the standard kind and prototypes, have been installed and 
tested at the SITF to assess their performance in view of the SwissFEL 
requirements.
These tests are described in Sec.~\ref{sec:ict}.

\subsubsection{\label{sec:bpm}Resonant stripline beam position monitors}

The significantly different requirements and time scales of the SITF and 
SwissFEL projects led to the development of two separate beam position monitor
(BPM) systems for the two machines.
To cover the need for a robust and reliable BPM system operational from day
one, a system of 19 resonant-stripline BPMs based on proven technology and  
existing implementations at PSI was built for the SITF~\cite{Kei10}.
The use of button BPMs was discarded due to their inferior resolution at low
bunch charges~\cite{Tre13}.
At the same time, the development of an advanced cavity BPM system was started,
aiming to fulfill the much more stringent requirements of the SwissFEL
facility arising from the varying beam pipe diameter, the presence of an 
undulator section and other factors~\cite{Kei13}.
For this development, a dedicated test area was made available at the SITF, 
located in the downstream part of the facility.
Here we describe the resonant-stripline BPM system.
The tests of the cavity BPM system performed at the SITF are summarized in 
Sec.~\ref{sec:cbpm}.

\begin{figure}[hbt]
   \centering
   \begin{minipage}{.5\linewidth}
     \includegraphics*[width=1\linewidth]{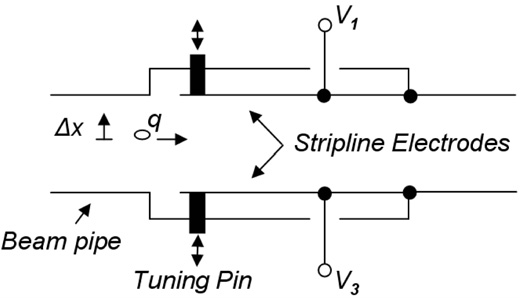}
   \end{minipage}
   \hspace{.05\linewidth}
   \begin{minipage}{.4\linewidth}
     \includegraphics*[width=1\linewidth]{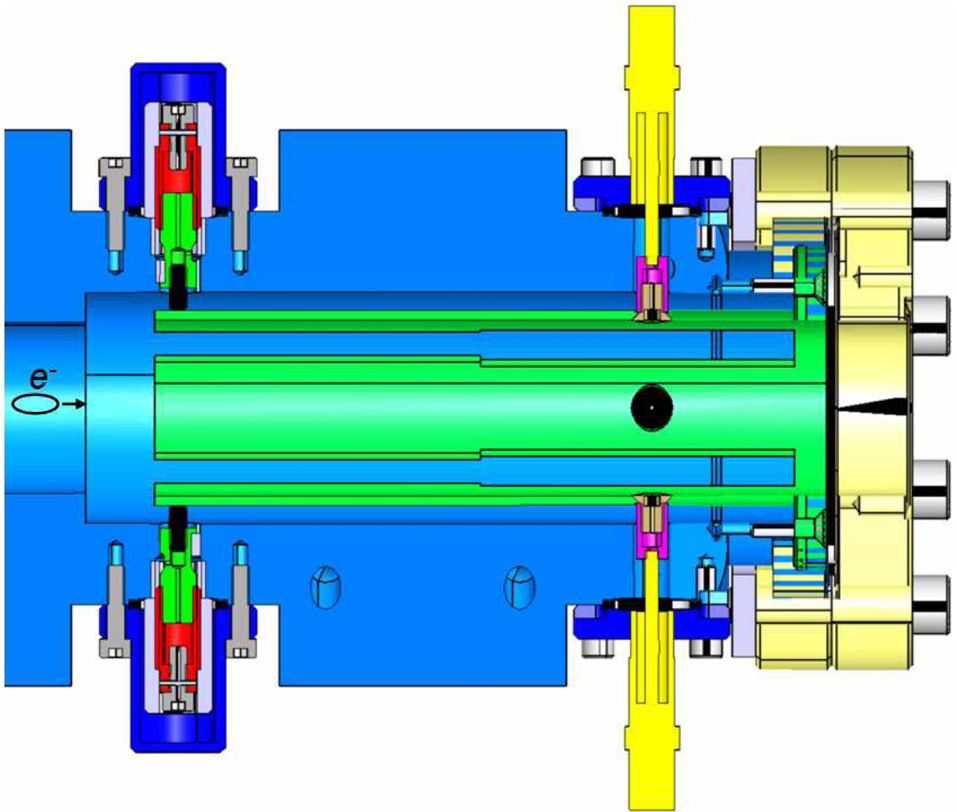}
   \end{minipage}
   \caption{Resonant stripline BPM pickup schematics (left) and mechanical
   drawing (right)~\cite{Cit09}.}
   \label{fig:bpm-pickup}
\end{figure}

\begin{figure}[tb]
   \centering
   \includegraphics*[width=0.8\linewidth]{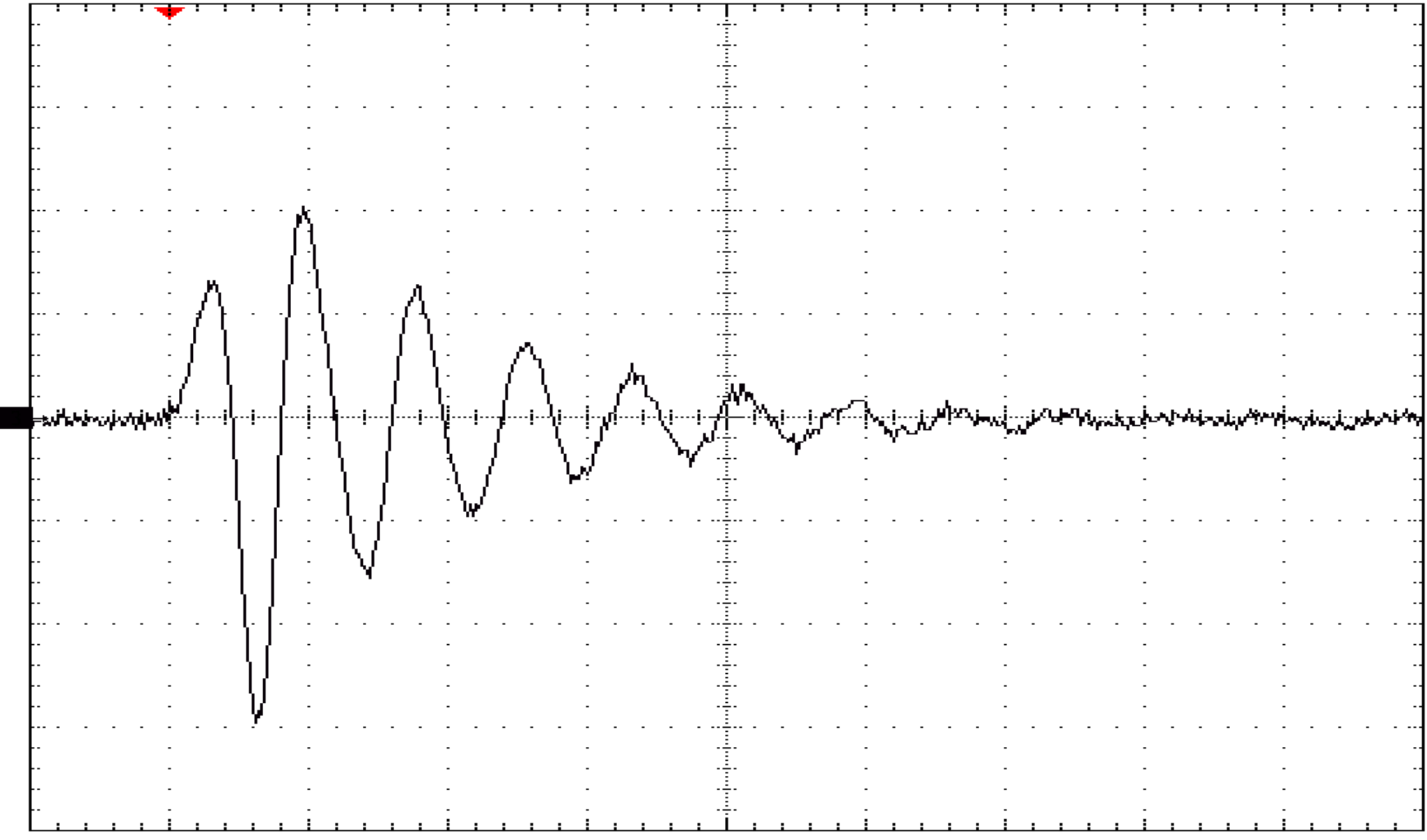}
   \caption{Beam-induced resonant stripline BPM pickup signal voltage 
    (vertical axis, 1~V/div.) versus time (horizontal axis, 2.5~ns/div.), 
    sampled with a 20~GSample/s oscilloscope (screenshot). Only the 500~MHz 
    fundamental mode is visible, higher harmonics are suppressed with a 
    low-pass filter.}
   \label{fig:bpm-signal}
\end{figure}

The pickups (see Fig.~\ref{fig:bpm-pickup}) contain four strips mounted 
parallel to the beam direction, each with an open and a shorted end 
($\lambda$/4 resonator configuration).
The inner diameter of the pickup corresponds to the beam pipe dimension of
38~mm.
A passing electron bunch excites a decaying sine signal at 500~MHz on each of
the strips (Fig.~\ref{fig:bpm-signal}).
The design of the pickups is based on earlier monitors in use at the linac and 
transfer lines of the SLS at PSI, but was optimized for the use 
at the SITF, for instance by adding mechanical resonance frequency 
tuners~\cite{Cit09}. 
Moreover, profile monitors (not shown in Fig.~\ref{fig:bpm-pickup}) were 
integrated into the BPM enclosures to save costs and to enable straightforward
cross calibration between view screen and BPMs. 
The choice of 500~MHz as fundamental resonance frequency of the resonant 
stripline pickup created synergies with present and future BPM systems at the
SLS, which operate at the same frequency~\cite{Kop12}.

\begin{figure}[tb]
   \centering
   \includegraphics*[width=1\linewidth]{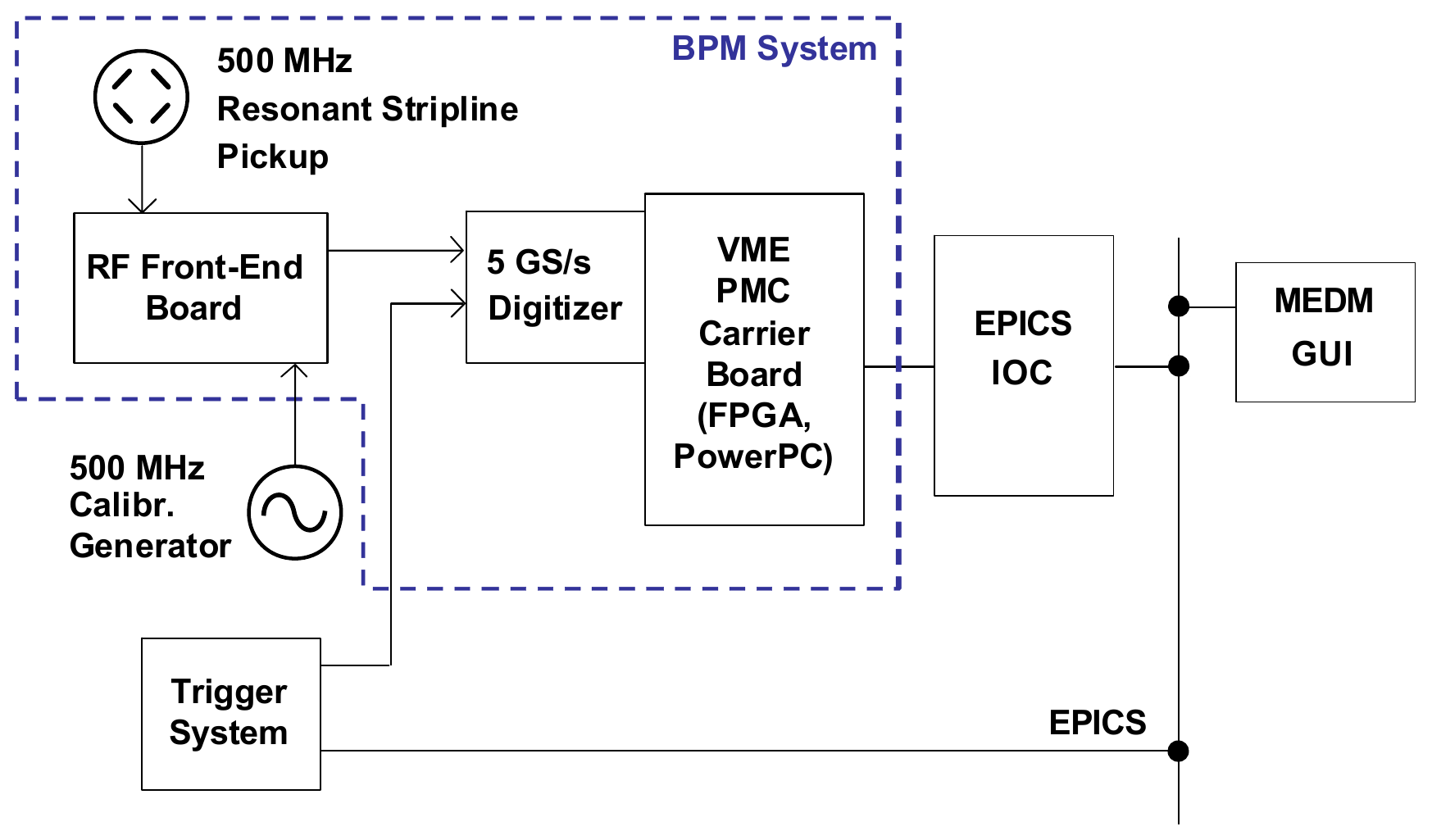}
   \caption{BPM system block schematics, including the linkage to the external
   trigger system and the EPICS control system (see Sec.~\ref{sec:controls})~\cite{Kei10}.}
   \label{fig:bpm-elec}
\end{figure}

The four stripline signals are coupled out via antennas near the shorted ends
of the strips before being filtered and amplified in an rf front-end (rffe)
board.
The four rffe output signals are connected to a digitizer board, consisting of 
a mezzanine module for analog-to-digital conversion, which is plugged onto an 
FPGA carrier board for digital signal processing and calculation of beam 
position and charge (Fig.~\ref{fig:bpm-elec}). 
The FPGA carrier board design was already used for other systems at PSI, e.g., 
for proton accelerator BPM systems~\cite{Kei06}, or for the MEG particle 
physics experiment at PSI~\cite{Ada13}. 
In the latter application, the mezzanine modules of the FPGA board feature an 
analog sampling chip called DRS4 (domino ring sampler version 4)~\cite{Rit10},
which was developed at PSI for the cost-effective digitization of several 
thousand detector signals at rates of up to 5~GSample/s.
Since the chip is well suited for sampling 500~MHz resonant stripline pickup
signals, we use it for our BPM readout electronics, accommodated in newly 
developed mezzanine modules optimized for the BPM system.
The analog input stage of this mezzanine module has bandpass filters and 
variable gain amplifiers for each of the four input channels, leading to an 
overall BPM system gain range of 63~dB.

The modest cost of the DRS4 chip allows it to be used in a redundant way.
For instance we only use four of the nine available input channels on each 
chip (one per stripline).
Using only every second channel reduces crosstalk, which is most pronounced 
between neighboring channels.
Moreover the four stripline signals are read out independently by two chips.
This not only improves the noise at higher bunch charge by averaging the 
readings of both chips, but also improves the mean time between failures of 
the system, enabling the detection of rare performance degradations or failures
of the DRS4 chips (2.5\% failure rate in 4 years of SITF operation).

Each of the two DRS4 chips on the mezzanine records the four BPM signals with 
5~GSample/s and stores the signal waveforms internally in an array of 1024 
sample-and-hold (S/H) capacitors per channel. 
The sampling (i.e., the domino wave) is stopped for the readout of the 
waveforms stored in the S/H cells.
For each channel, a dedicated multiplexer inside the chip reads out the voltage
of each S/H cell one by one. 
The signal is then digitized using a commercially available multi-channel ADC 
working at 33~MSample/s, connected to the multiplexer output of the DRS4 
chip.

The resonant-stripline BPMs require an external trigger from the timing system
to ensure the measurement of the BPM signal at the right time.
During the initial commissioning phase, which involved frequent changes in the 
gun laser or synchronization setup, the actual bunch arrival time could vary
with respect to the external trigger by several nanoseconds on short 
timescales.
Since the BPMs automatically detect the peak of the envelope waveforms, they 
can reliably measure beam position and charge even in the 
presence of arrival time variations exceeding 100~ns.
In the event of even larger shifts of the BPM trigger relative to the arrival 
time (e.g., due to major modifications of the gun laser or the timing system), 
the beam signal ends up outside of the $\approx$200~ns sampling window of the 
BPMs. 
In this case, a manual readjustment of the BPM or gun trigger delay based on an 
independent signal, e.g., from the wall current monitor, or from the 
oscilloscope readout of a BPM pickup, is needed.

The design principle of the DRS4 chip causes the input signals to be sampled at
nonequidistant points in time.
This so-called time nonlinearity is systematic and mostly reproducible, 
therefore it can be calibrated in the lab.
To this end we use a novel method in which 500 waveforms of a 500~MHz 
continuous-wave (CW) sine signal are acquired at different phases, where a fit 
procedure determines the time differences between the waveform sampling 
points~\cite{Kei10}.
Further lab calibration is needed for the individual voltage gains and offsets
for each S/H cell of the DRS4 chip.
During beam operation the FPGA then uses the calibration data to apply the
appropriate time, gain and offset corrections to the measured BPM signals in
real time.

\begin{figure}[tb]
   \centering
   \includegraphics*[width=1\linewidth]{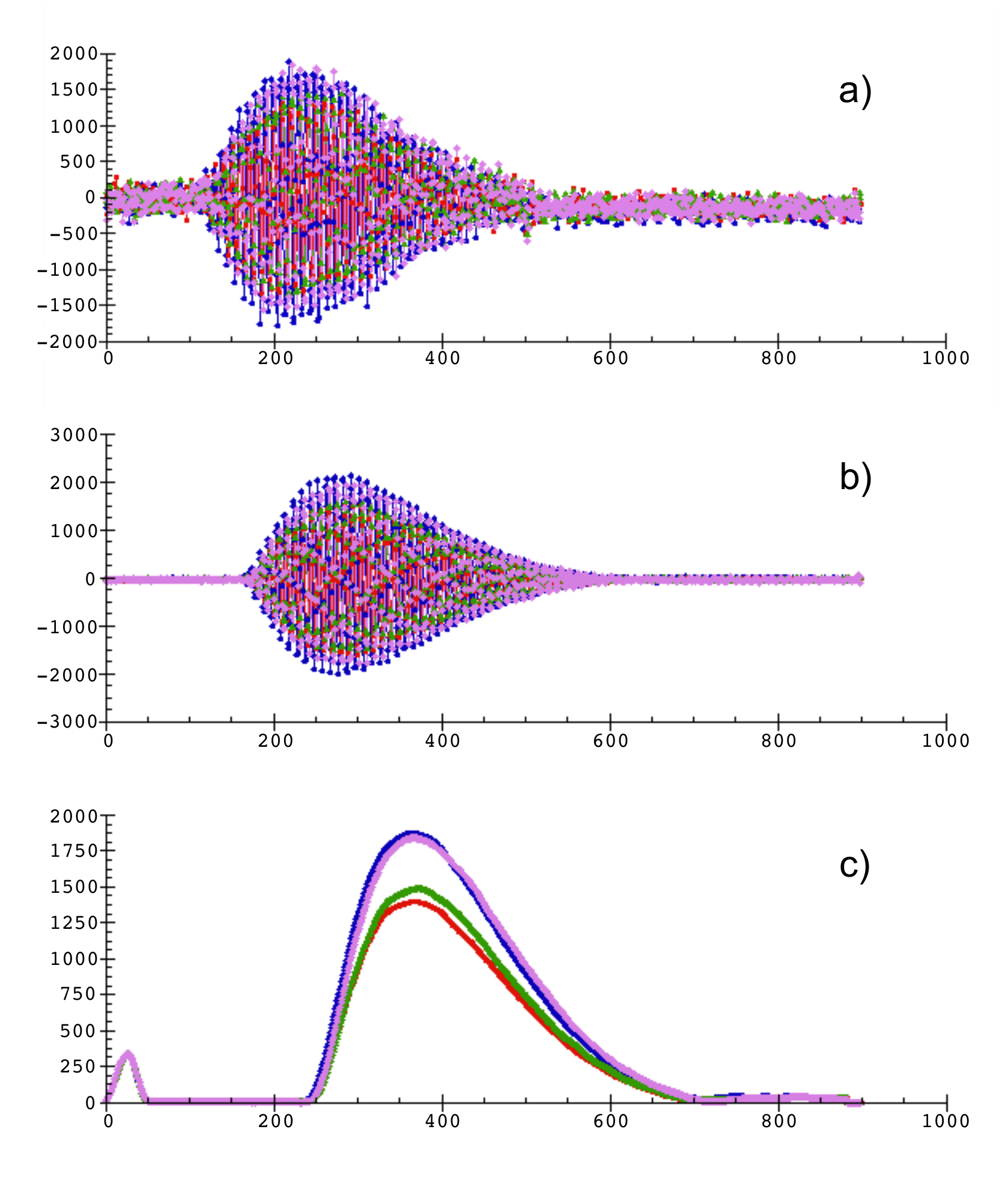}
   \caption{Raw BPM electronics waveforms (ADC raw value versus sample
   number), sampled by the DRS4 chip at 5~GSample/s (a).
   The same waveforms after offset gain and time-nonlinearity correction,
   performed by the FPGA for each sample individually (b).
   Envelopes of the four BPM electronics signals as calculated by 
   the FPGA (signal amplitude versus sample number) (c).
   The additional delay is due to digital filtering.}
   \label{fig:bpm-wave}
\end{figure}

The digital signal processing algorithm on the FPGA board is implemented in a 
mixture of VHDL (performed in the FPGA fabric) and C code (executed on a 
PowerPC 405 processor embedded in the FPGA). 
The FPGA on the carrier board calculates the beam position and charge from the 
sampled waveforms (shown in Fig.~\ref{fig:bpm-wave}a) in several steps. 
First, the gain and offset of each sample are corrected using the calibration
look-up table. 
Then the waveform, consisting of nonequidistant samples, is converted to a 
waveform with equidistant samples by use of a cubic spline interpolation 
algorithm.

The resulting waveform (shown in Fig.~\ref{fig:bpm-wave}b) is then 
bandpass-filtered with a 500~MHz  finite impulse response (FIR) filter to 
suppress noise and higher harmonics. 
For the filtered waveform, the FPGA reconstructs a 90$^\circ$ phase shifted 
signal (Q) from the original signal (I) via Hilbert transform, performs a 
Cartesian-to-polar conversion of the resulting I/Q vectors, and applies a 
low-pass filter to obtain the envelopes of the rffe output signal waveforms 
(shown in Fig.~\ref{fig:bpm-wave}c). 
For the first 50 samples of these waveforms, the pipeline of the FIR filter in 
the FPGA is not yet fully populated with valid samples, resulting in the 
artificial signal peak around sample No.~25.
For this reason only samples with index greater than 50 are used for further
processing. 

The signal amplitude of each envelope is then determined from the envelope 
waveform by integrating over an adjustable number of samples.
A number of about 200 samples was empirically found to result in the best
resolution.
The position of the integration window is dynamically adjusted by the FPGA 
during operation:
The FPGA detects the peak of the envelope and starts the integration a certain 
number of samples before.

Finally, the horizontal and vertical beam positions, $x$ and $y$, are 
calculated from the signal amplitudes (i.e., the integrals) A, B, C and D of 
the four channels according to the approximate formulae
\begin{align*}
  x &= k \frac{(B+C)-(A+D)}{A+B+C+D} , \\
  y &= k \frac{(A+B)-(C+D)}{A+B+C+D} ,
\end{align*}
with the constant $k$ = 16.2~mm. 

Similarly, the bunch charge $Q$ is obtained as
\begin{equation*}
  Q = c_\text{abs} c_\text{gain} (A+B+C+D),
\end{equation*}
where $c_\text{gain}$ is a scaling factor that depends on the (variable) gain of
the electronics, and $c_\text{abs}$ is a constant scaling factor obtained 
individually for each BPM by cross-calibration with a dedicated, absolutely 
calibrated charge monitor---at the SITF normally the wall current monitor (see
Sec.~\ref{sec:charge}).

Position and charge resolutions at bunch charges above about 5~pC are dominated
by random sampling jitter of the DRS4 chip and thus independent of the bunch 
charge~\cite{Kei10}. 
To reduce the impact of this jitter on the position resolution of the BPMs, 
the relatively short pickup signals are stretched in the rffe by applying a 
narrow-band 500~MHz filter realized in stripline technology on add-on printed 
circuit boards (PCBs). 
By choosing a filter quality factor about four times higher than the loaded 
quality factor of the pickups, the number of usable ADC samples for the 
position calculation is also increased by a factor of four, thereby reducing 
position and charge noise by a factor of about two. 
A further reduction in sampling noise is achieved by the parallel sampling of 
all signals with two DRS4 chips mentioned above. 

The position and charge resolutions are determined by correlating 
the data of several BPMs. 
For charges above 5~pC, we measured a charge-independent position resolution 
of typically 7~\textmu m (rms). 
For very low bunch charge, the resolution is dominated by thermal rffe noise 
and thus scales inversely with the charge, with a product of charge and 
resolution of about 30~\textmu m pC. 
This is nearly seven times better than current state-of-the-art button BPM 
systems operating at similar beam pipe diameters~\cite{Tre13}. 
The relative charge resolution is generally 0.13\% at higher bunch charges 
(e.g., 130~fC resolution at 100~pC charge).
At very low bunch charges the resolution is again dominated by thermal rffe
noise leading to an absolute charge resolution between 3 and 4~fC.
Since the charge resolution of the BPMs is far superior to that of any 
dedicated charge monitor available at the SITF, beam operation mainly relied 
on the BPM system for charge measurements, using the dedicated charge monitors
only for gun studies or cross-calibration of the BPM charge readings.

\subsubsection{\label{sec:scrmon}Screen monitors}

Among the three types of transverse profile monitors installed in the SITF,
wire scanners, synchrotron radiation monitors, and screen monitors, only the
last were routinely used for beam optics measurements.
The commissioning of the wire scanner and synchrotron radiation monitor 
systems performed at the SITF primarily served as development work for SwissFEL
and will be described in Sec.~\ref{sec:trpromon}.

View screens provide direct two-dimensional information on the beam profile and
play an important role, both in the commissioning and the fine tuning of an 
electron linac: they are simple and efficient beam finders at the start-up of
the machine and later precise and fast tools to characterize the phase space
of the beam via optics dependent measurements, including the longitudinal
phase space in case a transverse deflecting structure and a dispersive section
are available.
 
\begin{figure}[hbt]
   \includegraphics*[width=0.9\linewidth]{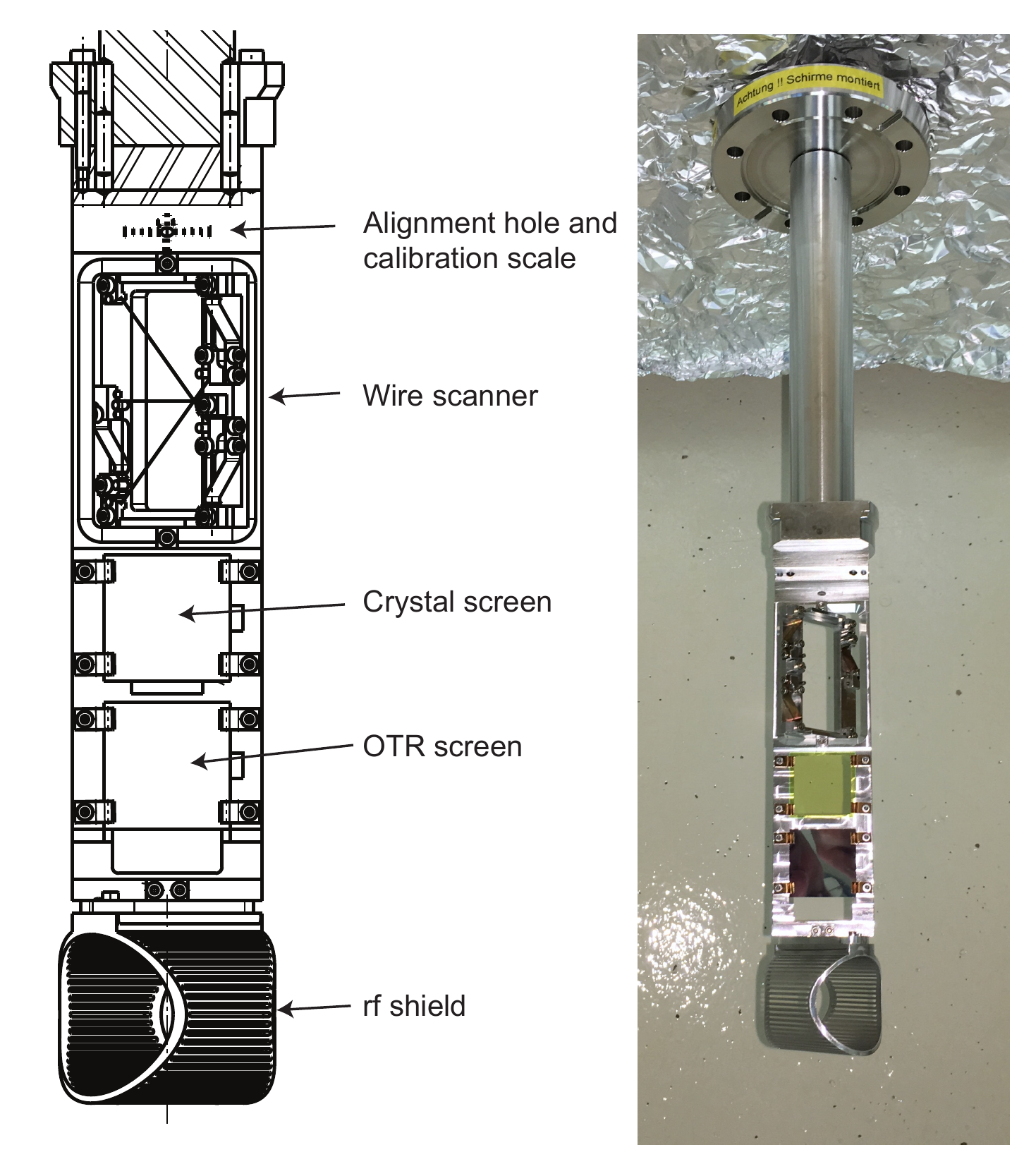}
   \caption{Technical drawing and photograph of the ultra-high-vacuum 
   feedthrough insertion accommodating alignment targets, three scanning wires,
   two screens (crystal scintillator and OTR), and an rf shield.}
   \label{fig:scr-ins}
\end{figure}

The screen monitors of the SITF are combined scintillator and OTR screen 
monitors including wires for transverse profile scanning in a single 
device~\cite{Isc10}. 
One OTR screen, consisting of an aluminum-coated silicon mirror, one 
scintillator, one alignment target, and three  25~\textmu m tungsten wires 
(one horizontal and two diagonal) are mounted on a linear ultra-high-vacuum 
feedthrough and can be inserted into the beam by means of a stepper motor
(see Fig.~\ref{fig:scr-ins}). 
The scintillator consists either of cerium doped yttrium aluminum garnet 
(YAG:Ce) or cerium doped lutetium aluminum garnet (LuAG:Ce), the latter 
offering a twice as large light output at the expense of higher beam losses.
Crystal thickness varies between 20~\textmu m and 1~mm depending on the 
specific detection efficiency and optical resolution needs at a given location.

In the low-energy sections of the machine, only scintillator screens are 
installed since OTR screens would not yield sufficient signal.
These screens are installed perpendicular to the electron beam and viewed 
through mirrors at 45$^\circ$ incidence angle, placed directly behind the
scintillators.
At higher energies scintillator and OTR screens are mounted at an angle of
45$^\circ$ with respect to the electron beam. 
Backward transition radiation is thus emitted perpendicular to the electron 
beam direction. 
For convenience, the scintillating crystal can be moved to the same location 
as the OTR screen, such that a re-alignment and refocusing of the 
camera is not required. 
Two cameras with different magnifications are installed, one for overview 
images $(M = -1/5.3)$, and one with a larger magnification $(M = -1/1.1)$ of 
the center of the screen.

Image acquisition was initially performed by FireWire (IEEE 1394 bus standard) 
cameras, with most of the FireWire network installed in the accelerator tunnel.
This setup unfortunately resulted in frequent system crashes requiring manual 
reboots of the cameras.
Moreover, the FireWire solution would not work at the 100~Hz repetition rate
envisaged for SwissFEL.
Therefore, new camera types based on the GigE Vision and Camera Link standards
have been tested at the SITF and are now planned for use at SwissFEL.

A total of 26 of such monitor stations were installed in the SITF in the 
beginning and used extensively for the initial setup of the beam optics as well
as for early emittance optimizations. 
Despite the solid performance of these profile monitors, certain limitations
suggested a redesign for SwissFEL.
First, OTR screens are not suited for the observation of micro-bunched beams
due to the occurrence of coherent optical transition radiation (COTR), as 
evidenced by measurements performed at LCLS~\cite{Loo08}, FLASH~\cite{Wes09},
and, more recently, SACLA~\cite{Tan11} (see also Ref.~\cite{Lum14}).
Since the orientation of the scintillating crystals matches that of the OTR
screens, the scintillators would not be exempt from COTR either.
Second, the horizontal resolution of the scintillators is limited by the 
crystal thickness, as even an arbitrarily thin beam would mark a line in 
the volume of the scintillating crystal. 
Given that electron beams at SwissFEL are expected to be a few tens of 
micrometers in diameter, and that scintillator crystals below 100~\textmu m 
thickness are difficult to produce and handle (in addition to providing only
weak signal at low charge), this limitation would dominate the resolution in 
the central regions of the screens.
To overcome these short-comings, the design of the transverse profile monitors
was revised.
We describe the new design and its use at the SITF in Sec.~\ref{sec:profimg}.

\subsubsection{\label{sec:long}Longitudinal diagnostics}

The primary technique applied at the SITF for longitudinal beam diagnostics,
i.e., bunch length measurements, current profiles, time-resolved (``slice'')
emittance evaluations, involves the use of the transverse deflecting rf 
cavity and a transverse profile monitor, typically a view screen, located at an
appropriate distance.
Technical details on the rf deflector can be found in Sec.~\ref{sec:rf-tds}, 
while the screens are described in Secs.~\ref{sec:scrmon} and \ref{sec:profimg}.
More advanced, and in particular nondestructive longitudinal diagnostics 
schemes based on coherent bunch radiation or electro-optic methods were the 
subject of vigorous research and development efforts at the SITF and will be 
described in Sec.~\ref{sec:diag-dev}.

During normal beam operation, the deflecting rf cavity is powered on, but the
timing of its rf pulses is shifted in such a way that the electron bunches 
traverse the cavity completely unaffected. 
Only when a longitudinal measurement is required, the timing of the deflector
pulse is adjusted to coincide with the passage of the electron bunches.
The rf phase is tuned to synchronize the bunch centroid with the zero crossing
of the transverse deflecting rf field.
Thus, depending on the applied rf power, the bunch is more or less streaked in 
the vertical plane without deflection of its centroid.
The resulting correlation between the electrons' time coordinates and their
vertical positions gives access to longitudinal measurements via normal 
screens~\cite{Akr01,Akr02}.

\begin{figure}[tb]
   \includegraphics*[width=1\linewidth]{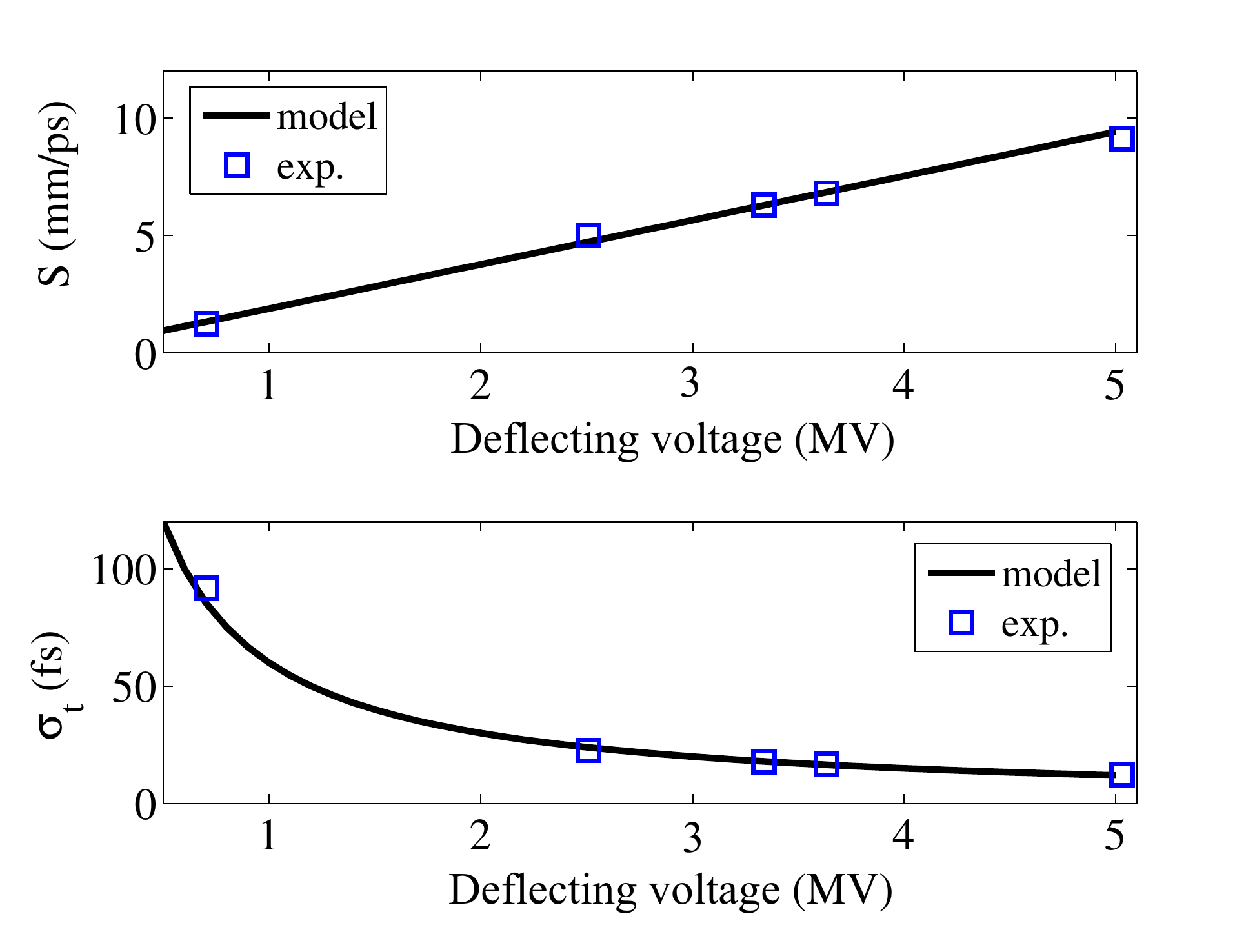}
   \caption{Deflector calibration factor (top) and time resolution (bottom) as 
   a function of the integrated deflecting voltage, measured in each case at
   the zero crossing of the rf phase, for a beam energy of 200~MeV and compared
   to model expectation~\cite{Cra13}.}
   \label{fig:rf-tds}
\end{figure}

A calibration of the setup is easily achieved by relating a change in rf 
phase (in this case S-band frequency) to the corresponding displacement of the
beam centroid on the observing screen. 
In Fig.~\ref{fig:rf-tds} we show a comparison between measured and analytically
expected calibration factor (conversion of mm on screen to ps along the bunch)
and time resolution as a function of the integrated deflecting voltage.
The analytical model agrees well with experimental data.
At the highest deflection voltage of 5 ~MV the time resolution is approximately
12~fs~\cite{Cra13}.

Other methods to streak the beam for time-resolved measurements exist and
have been investigated at the SITF.
Instead of an actively applied rf field, the beam induced wakefields in an
appropriately shaped dielectric structure may be utilized to generate the 
desired correlation between the longitudinal and a transverse coordinate.
For a report on corresponding beam experiments performed at the SITF we refer 
to Ref.~\cite{Bet16}.
The same type of correlation can be achieved by means of dispersion combined
with energy chirping.
We will briefly expand on this technique during the discussion of time-resolved
measurements, see Sec.~\ref{sec:bd-long}.

\subsection{\label{sec:controls}Control system}

The SITF control system~\cite{Dac11,Che13} is implemented in a way that allows 
incremental expansions and upgrades as the project proceeds.
With this approach new requirements could be accommodated and new hardware
could be incorporated in the course of the different project phases.
In general the system is based on well proven technologies and concepts
used at existing PSI facilities, in particular the Swiss Light 
Source (SLS)~\cite{Hun01}.
Following this working example it is built around the Experimental Physics 
and Industrial Control System (EPICS) tool kit~\cite{Dal94}. 
The hardware and software platforms, boot and archive services, configuration 
management and logistics already developed for the other PSI accelerators are 
thereby used as far as possible.
The linear geometry of the SITF, however, required several adaptations of the
existing solutions and led to new designs and concepts for SwissFEL.
In the following we describe our approach, with particular emphasis on new 
developments for SwissFEL.

\subsubsection{EPICS}

The use of EPICS not only provides benefits from in-house know-how and the
experience gained with other PSI facilities, but also, through its 
collaborative nature, allows us to tap into the broad pool of expertise
accumulated at other laboratories.
The SITF control system is based on EPICS version 3.14.8.

The control-room consoles run EPICS client applications, which connect over 
an Ethernet network to distributed EPICS servers called input/output 
controllers (IOCs).
The IOCs have a direct connection to the hardware devices to be controlled.
In addition, central services like the archiver run on dedicated servers 
located in a server room.   
On all these computers customized programs required for controlling and 
characterizing the machine can be implemented and integrated into EPICS.

The most relevant parts of the EPICS toolkit used at the SITF are:
\begin{itemize}
\item the network-based client-server model implementing Channel
  Access as network protocol;
\item the distributed record database that contains information on
  current machine parameters;
\item EPICS extensions such as Alarm Handler, Archiver and MEDM (Motif Editor 
  and Display Manager, a Motif based graphical user interface) well-known to 
  control-room operators;
\item available Channel Access interfaces to common programming
  languages (e.g., C, C++, Java, Tcl/Tk or MATLAB).
\end{itemize}

In the course of SITF operation some of the tools in use approached
the end of their lifecycles or became difficult to maintain and will
be replaced for SwissFEL.  
For instance the graphical user interface toolkit MEDM will be replaced 
by caQtDM~\cite{Mez13}, or the scripting language Tcl/Tk substituted by Python.

Many high-level applications used at the SITF for processing and visualizing 
information obtained from EPICS channels are written in MATLAB using the
MOCHA interface~\cite{Chr13}, which is based on an in-house C++ Channel Access 
interface library called CAFE~\cite{Chr11}.
While this approach enables the quick development of tailor-made measurement
programs well suited for a test facility, the resulting applications are
typically difficult to maintain.
For this reason most of the tools needed for the standard operation of SwissFEL
are likely to be re-implemented in Python, again using the CAFE 
interface~\cite{Chr15}.

\subsubsection{Network structure}

The network layout for the SITF was designed from the start with the SwissFEL 
requirements in mind.
For the SwissFEL facility the required number of network ports is too large to
fit into a standard class C Ethernet subnet, which is limited to 254 addresses.
The SITF was therefore intended to be a testbed for a setup integrating
several such networks in one facility, although in this case one network would
have been enough to cover all addresses. 
In comparison to a class B network, the advantages of a class C network are
its increased security, scalability and easier error isolation, among other
benefits.

The initial SITF network consisted of three class C subnetworks, where the 
connections between the subnets were controlled by Channel Access 
Gateways~\cite{Eva05} to minimize network traffic.
Unfortunately, however, this solution did not perform as expected. 
The Channel Access Gateways turned out to be bottlenecks that would hamper
even normal IOC operation due to bandwidth limitations.
The problem was most severe for long waveforms transmitted at high rates.
Owing to the many operational disturbances the Channel Access Gateways were
removed in early 2013.
From then on, directed broadcasts were used to reach EPICS records in the 
different subnetworks.

\subsubsection{Input/output controllers}

To profit from existing know-how and to economize on stock keeping the 
hardware platforms used as IOCs at the SITF were, as far as possible, selected
from the range of platforms already employed at PSI.
The following IOC hardware solutions are installed at the SITF:

\begin{itemize}
\item VME64x running with the VxWorks operating system (Wind River), mainly
  using the Motorola MVME5100 card as CPU. 
  A rich set of modules, e.g., for digital and analogue I/O, are supported.
\item Motor drivers developed at PSI, using the MAXv-8000 VME controller 
  (Pro-Dex).
\item Digital power supply controllers developed at PSI.
\item Programmable logic controllers (PLCs), used for I/O and applications
  that require high reliability but only moderate speed.
  The PLCs are based on Siemens S7 series controllers.  
  Typical applications are vacuum interlock and other component protection
  systems.
\item So-called microIOCs (Cosylab) were successfully tested for the gun laser
  system.
  They will, however, not be used further for SwissFEL, since the corresponding
  product line has been discontinued.
\end{itemize}

To optimize the management of spare components, the VME hardware for the SITF 
is housed in 21-slot VME crates of the same type as used for the SLS.
The linear, relatively sparse arrangement of the SITF devices, however,
resulted in many idle slots.
For this reason, smaller 7-slot VME crates have been designed and developed
in view of SwissFEL. 
The new crates in addition provide an easy-to-use crate monitoring system,
which is interfaced to EPICS. 
Both old and new crates are manufactured by Trenew Electronic AG.

Ethernet is used as a fieldbus substitute for devices that do not require
tight timing synchronization.
These devices are read out through so-called Soft IOCs running on central
servers.

Special requirements by the diagnostics and llrf systems suggested the use
of FPGA-based solutions (see Secs.~\ref{sec:bpm} and \ref{sec:llrf-overview},
respectively).
These led to the in-house development of new boards for SwissFEL: 
the GPAC (Generic PSI ADC Carrier) for diagnostics~\cite{Kei12}
and, in collaboration with IOxOS Technologies, the IFC 1210 for llrf and
as a future VME CPU board, running embedded Linux as operating 
system~\cite{Kor13}. 

\subsubsection{Configuration management}

Software distribution and installation for the SITF follow the procedures
established for other accelerators at PSI~\cite{Kre13}.
The adopted system facilitates the reproducible deployment of software to the
IOCs using a repository based on the Concurrent Versions System (CVS) and a 
relational database that contains all essential information such as operating
system or EPICS versions.
The IOC configuration files and the necessary drivers are deployed to a central
location, from where the IOCs load them during boot time.

In comparison to, e.g., the SLS, the SITF uses a broad range of IOC hardware,
which necessitated a few adaptations of the distribution and installation system
to deal with multiple versions of drivers and configurations for several
hardware platforms and operating systems.
In view of SwissFEL, the integration of Windows-based systems (mainly camera
drivers) turns out to be of particular importance.

\subsubsection{\label{sec:timevsys}Timing and event system}

At a pulsed machine such as the SITF the synchronization of different components
and control devices at the pulse-by-pulse level is of crucial importance.
The synchronization of the various SITF elements is ensured by an MRF timing
and event system (Micro-Research Finland Oy).
An external reference clock of 124.913~MHz 
(the 12th sub-harmonic of the local reference frequency of 1\,498.956~MHz) 
and the 50~Hz mains supply synchronize an event generator.
Event information is broadcast in encoded form via optical fibers to the event 
receivers.  
The data is decoded by the receivers and appropriate output signals for 
triggering external devices or interrupts are generated. 
Further details can be found in Ref.~\cite{Kal03}.

For the characterization of individual bunches measurements of beam 
parameters need to be synchronized with the electron bunch and tagged with a
unique bunch identification.
To this end the timing system supplies a monotonically increasing bunch number
consisting of 52 bits.
This allows for $2^{52}$ $\approx$ $4.5 \times 10^{15}$ bunch numbers, more than
enough for the lifetime of the facility.

The bunch-synchronized recording of beam data across multiple IOCs was made 
possible through the development of a so-called beam-synchronous data 
acquisition (BS-DAQ) mechanism~\cite{Kal11} using the same event system.
In this scheme the IOCs hold the data to be recorded synchronously in local
buffers, which are read out for analysis at the end of a predefined acquisition 
period.
A BS-DAQ run can be configured dynamically by the user via the specification
of parameters such as the number and spacing of acquisitions and then be 
initiated and controlled through a graphical user interface.
The BS-DAQ application supports the acquisition of waveforms and simultaneous
operation by several users.

At SwissFEL the presence of more data sources, the higher event rates and the
peculiar mix of rf frequencies~\cite{Hun14} 
call for an upgrade of the timing and event system.
The upgrade includes an efficiency enhancement of the BS-DAQ system through the
implementation of the buffering on a dedicated fast file system.

\subsection{\label{sec:timsynch}Timing and synchronization}

The tight tolerances on rf phase stability at FELs represent a major challenge 
for the distribution of the rf frequencies along the accelerator.
At the SITF, two distribution systems were installed in parallel, an electrical
system as the baseline and an optical system using fiber-optic links as a test 
bed for the SwissFEL implementation.

\subsubsection{\label{sec:refgen}Reference generation and distribution}

The timing reference generation and distribution at the SITF is based on an 
electrical master oscillator working at a fixed repetition rate of 
214.136\,571\,4~MHz, 
an rf amplifier, a coaxial cable distribution and local phase locked 
oscillators with output frequencies of 
1\,498.956~MHz, 
2\,997.912~MHz, and 
11\,991.648~MHz~\cite{Hun09}. 
While this approach offered the fastest way to start with a precise timing 
signal for the main components of the injector facility, such as the gun 
lasers, the optical master oscillator (OMO), the timing event generation system
and the llrf stations, the system is not optimized for drift performance,
needs compensation for attenuation and has a limited bandwidth.
Since the electrical distribution of the reference signal was soon found to 
meet the requirements for operating the SITF entirely, not much effort went 
into stabilizing its cables (e.g., temperature stabilization). 
Instead further work focused on the development and test of the optical system
planned for the SwissFEL facility.

For the final system at SwissFEL the precise timing reference will be based on 
a purely optical distribution, with an OMO at a repetition rate of 142.8~MHz 
serving as reference source. 
This oscillator will be synchronized to an rf master oscillator to obtain the 
best jitter performance in the spectral range from 10~Hz to 10~MHz. 
The optical distribution along the 740~m long facility will be provided, 
depending on the performance needs, via highly stabilized CW or pulsed links or
via not specifically stabilized links.
Where needed, a cable based subdistribution will be installed close to the
client location for transmitting rf frequencies over short distances.
At the SITF, a synchronized OMO at 214.137~MHz was operated for testing and 
development purposes and for developing and operating the bunch arrival-time 
monitors (see Sec.~\ref{sec:bam}).

\subsubsection{Stabilized optical links}

One of the key advantages of stabilized optical links, CW or pulsed, in view 
of their application at FELs, is their excellent drift performance.
In a CW link the optical carrier is modulated with the rf reference signal from
an external master oscillator in a Mach--Zehnder electro-optical modulator. 
In a pulsed optical link the short pulses from an OMO are transmitted and 
cross-correlated with the partly back-reflected pulses from the remote 
location. 
At the remote location the optical pulses can be used for applications such as
arrival-time monitoring of the electron bunches, optical synchronization of
lasers via two-color optical cross-correlation, or extraction of required rf 
frequencies from the optical spectrum. 
In both types of stabilized optical links, phase and amplitude stabilization 
loops are present to further minimize drifts.

We designed, built and characterized two prototypes of pulsed optical links,
one based on cross-correlation and one with temperature-stabilized receivers.
The latter type uses both phase and amplitude stabilization loops and was
used for the characterization of the bunch arrival-time monitor.
In the light of the increasing commercial availability of pulsed optical links,
we decided to not focus on building a series of complete systems ourselves.

\begin{figure}[tb]  
  \includegraphics*[width=1\linewidth]{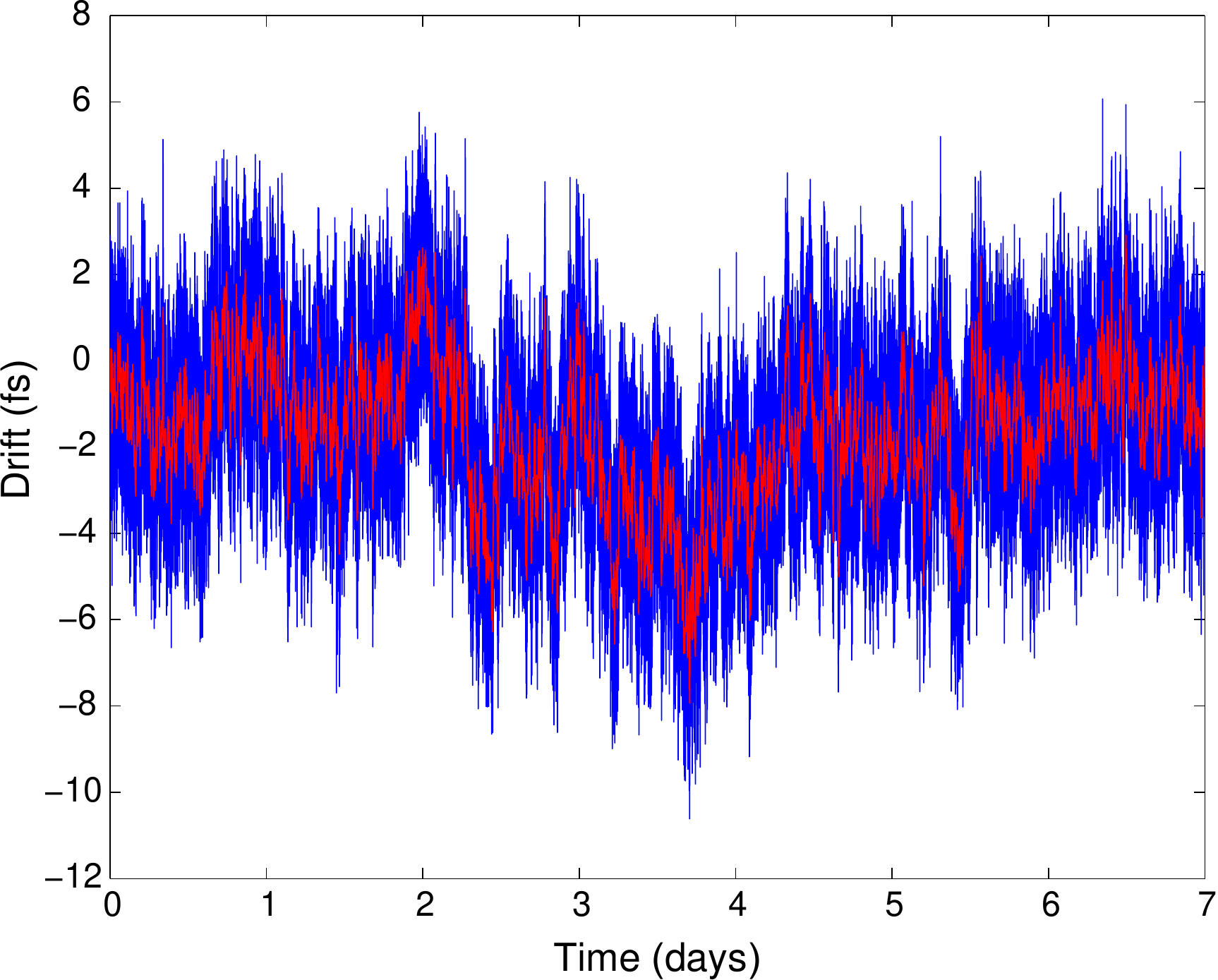}
  \caption{Long-term drift measurement of a 500~m long CW optical link:
   raw data at 1~Hz (blue curve), and filtered with a time window of five
   minutes (red curve). See text for details.}
  \label{fig:timsync1}
\end{figure}

We developed and tested CW optical links in-house, in collaboration with 
Instrumentation Technologies~\cite{Ore14}.  
The CW link system consists of a transmitter and a receiver, which are 
connected via a pair of optical fibers. 
The transmitter is fed with an rf frequency (2\,997.912~MHz in this case), 
which is also the output of the receiver. 
In a linac, the system will transport the rf frequency over long distances, 
with transmitter and receiver far apart. 
For the system's characterization, however, they need to be close, because 
input and output have to be compared with a very sensitive 
temperature-stabilized phase detector using short rf cables.
For the long-term stability and drift measurement presented in 
Fig.~\ref{fig:timsync1}, the 500~m long fiber link was looped back from the
transmitter to the nearby receiver.
The data were taken at 1~Hz over seven days (blue curve in 
Fig.~\ref{fig:timsync1}). 
The observed drift values depend on the details of the analysis.
For the determination of peak-to-peak values, a well suited method is based
on Savitzky-Golay filtering, in our case applied with a third-degree polynomial
and a frame size of five minutes, adequate for a measuring time of 
seven days (red curve in Fig.~\ref{fig:timsync1}). 
This results in a peak-to-peak drift of 10.9~fs. 
Shortening the time window to 30 seconds increases the drift to 13.3~fs.
The rms drift of the raw data amounts to 2.6~fs for the entire seven days.

For those cases where drift performance is not an issue, e.g.\ the BPM system,
not specifically stabilized optical links were developed for cost reasons. 
The working principle of these links is also based on rf modulation of an 
optical carrier.

\subsubsection{Laser synchronization}

The synchronization of the laser with respect to the gun rf field is of crucial
importance for the operation of a photoinjector.
Also in this case, the synchronization can be achieved by an electrical
or an optical system.

At the SITF we developed an electrical synchronization unit for the 
synchronization of the Nd:YLF gun laser (see Sec.~\ref{sec:laser}, for which 
the commercial solution did not fulfill the stringent requirements of 
photoinjector operation.
In addition we successfully tested the optical synchronization by locking the 
Ti:sapphire oscillator with a 1560~nm mode-locked erbium-doped fiber laser 
(EDFL) over a period of 60 minutes. 

For SwissFEL, the principal gun laser (Yb:CaF$_2$) will be initially stabilized 
with an electrical synchronization unit and later on additionally by optical 
cross-correlation with the pulses from the OMO, which will consist of an 
erbium-ytterbium glass (Er:Yb:glass) laser. 

\paragraph{Electrical synchronization of the Nd:YLF gun laser}

\begin{figure}[b]  
  \includegraphics*[width=1\linewidth]{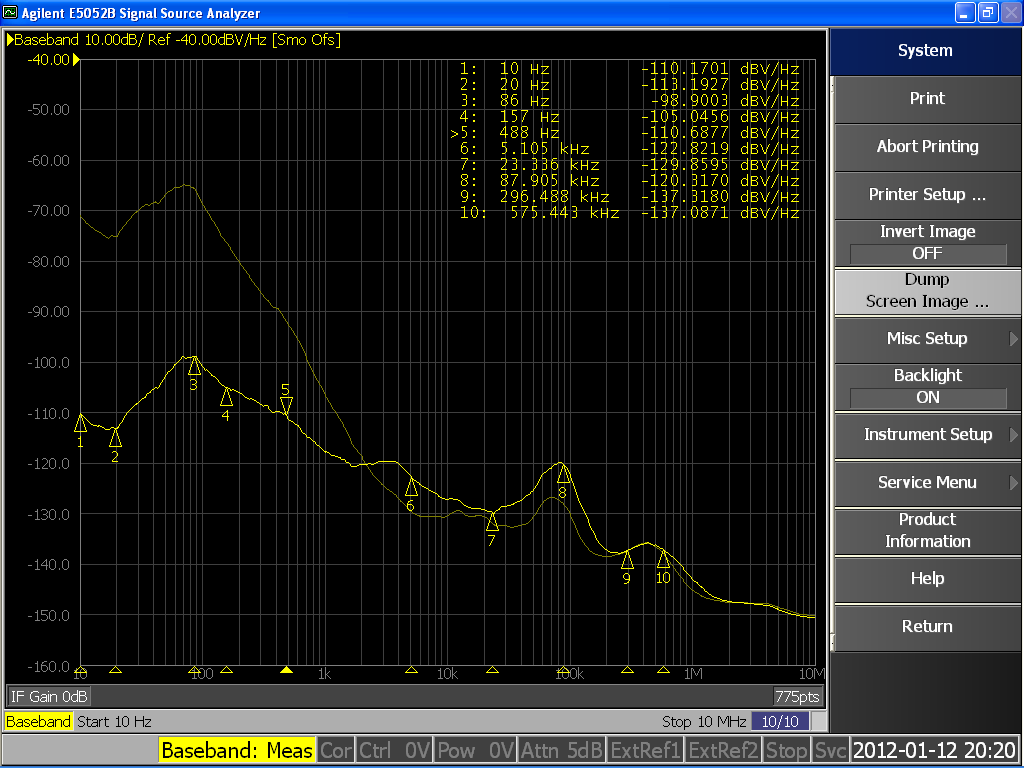}
  \caption{Signal source analyzer screenshot illustrating the improvement of 
    the jitter of the Nd:YLF laser:
    Using the in-house built synchronization unit with gain optimized within 
    the controller bandwidth, the laser exhibits a jitter of 45~fs in the 
    spectral range between 10~Hz and 10~MHz (bright yellow curve with arrows).
    For comparison, the original synchronization box results in a jitter of
    about 600~fs in the same spectral range (pale yellow curve).}
  \label{fig:timsync2}
\end{figure}

To improve the synchronization of the Nd:YLF gun laser, we implemented our own
electrical synchronization unit (phase-locked loop, PLL).
It reduced the integrated jitter in the range between 10~Hz and 10~MHz by an 
order of magnitude with respect to the original laser synchronization box, 
down to about 45~fs (see Fig.~\ref{fig:timsync2}).
A better value cannot be achieved since the noise floor of the laser itself 
in the higher frequency range cannot be reduced due the limited bandwidth of
the PLLs of up to several kHz.
Further improvements can only be realized during the assembly of the laser.

The described synchronization unit has been integrated into the EPICS control 
system, allowing for automated locking and tuning via software running on a 
middle-layer server.
The tuning is necessary to prevent the fine-tuning actuator from leaving its
locking range due to changes in the environmental conditions.
This delocalized approach for controlling the laser synchronization proved to
be very reliable and stable, allowing the laser to stay synchronized with 
respect to the reference rf for weeks without interruption.

The operation with our synchronization unit results in low jitter values at
the oscillator output~\cite{Div14}.
Drift measurements of the laser oscillator output including the PLL yield
values around 200~fs over 12 hours, depending on the environmental changes
and the amplitude-to-phase conversion of the photoreceiver and the phase 
detector. 
Since the PLL is not optimized for drift, a further development is under way 
making use of cross-correlator based optical synchronization.
This effort and a proof-of-principle demonstration carried out at the SITF
are described in the next subsection.

\paragraph{Optical synchronization of the Ti:sapphire laser}

Ultimately, the operation of the SwissFEL user facility will require the
synchronization of several laser systems (gun, seed, pump-probe lasers) at a 
level below 10~fs over hours.
Such a degree of stability requires the stabilization of the entire chain from 
the oscillator to the end point, e.g., the photocathode or the experimental 
probe. 
An important milestone towards achieving this ultimate goal is the optical 
locking of two mode-locked oscillators.  
At the SITF we successfully demonstrated the stability of an optical PLL by 
using the Ti:sapphire laser oscillator at 800~nm as a reference to synchronize
an EDFL at 1560~nm~\cite{Ars11}.
(The Ti:sapphire oscillator was chosen as reference simply to minimize 
interference with beam operation.)
The Ti:sapphire pulses were transported through an approximately 5~m long 
free-space transfer line to optically synchronize the EDFL via two-color 
balanced optical cross-correlation in a 0.5~mm thick 
phase-matched BBO crystal. 
Commercial electronics were used for the locking of the EDFL, which at the 
time had not been optimized.
The results should therefore be regarded as a proof of principle rather than a
demonstration of ultimate performance. 

\begin{figure}[tb]  
  \includegraphics*[width=1\linewidth]{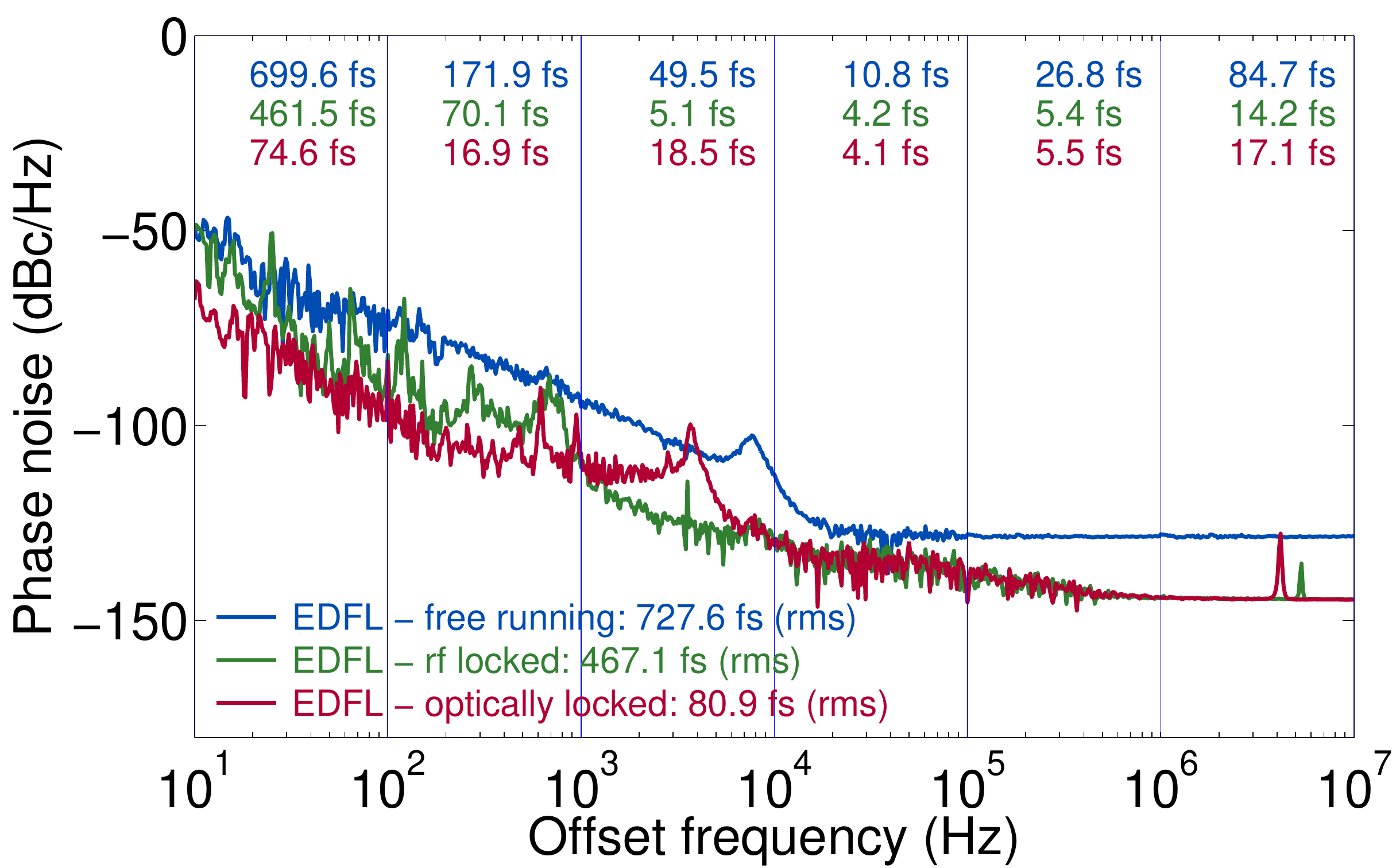} 
  \caption{Phase noise stability comparison between free running, rf-locked, and 
  optically locked EDFL~\cite{Ars11}.
  The values given in the upper part of the figure refer to rms jitters
  measured in the respective frequency ranges.
  The values in the lower part of the figure are jitters measured over the 
  entire frequency range (10~Hz to 10~MHz).}  
  \label{fig:optsync1}
\end{figure}

A comparison of the phase noise stability of the free running, the rf-locked, 
and the optically locked EDFL is shown in Fig.~\ref{fig:optsync1}. 
The values refer to the absolute jitter including the contribution of the 
reference.
It can be seen that the optical lock gives the lowest noise floor.
Normally, the free running and the rf-locked EDFL feature broad noise peaks 
around 7.5~kHz originating from their controller units. 
In the case of the rf-locked EDFL, the peak is suppressed by the rf PLL
(green trace in Fig.~\ref{fig:optsync1}).
For the optical synchronization, a peak at 5.4~kHz becomes dominant.
By tuning the PI gains, this peak can be reduced considerably, albeit at the 
expense of the stability of the optical lock.
With a jitter variation between 80.9~fs and 217.8~fs (rms) in the range between 
10~Hz and 10~MHz on different days and for different PI gain settings, the 
optical synchronization provides better stability than the rf one, for which 
jitter values between 318~fs and 5.6~ps (rms) in the same range have been 
measured. 
The wide span of measured jitter values in the case of an rf synchronization 
comes from the multiple noise spikes in the range between 10~Hz and 1~kHz, which
are always present in the phase noise spectra in this setup.

The measured jitter of the optically locked EDFL includes contributions from
the phase noise of the PLO, with which the Ti:sapphire is synchronized, and
from the transfer line. 
In the range between 10~Hz and 1~kHz the optically locked EDFL jitters between 
76.5~fs and 118.5~fs (rms), as measured on different days and under different 
environmental conditions. 
When the PLO contribution of 75.7~fs (rms) in this frequency range is 
subtracted, the remaining jitter contribution for the optical loop and the 
transfer line is between 11.0~fs and 91.2~fs (rms).

\begin{figure}[tb]  
  \includegraphics*[width=1\linewidth]{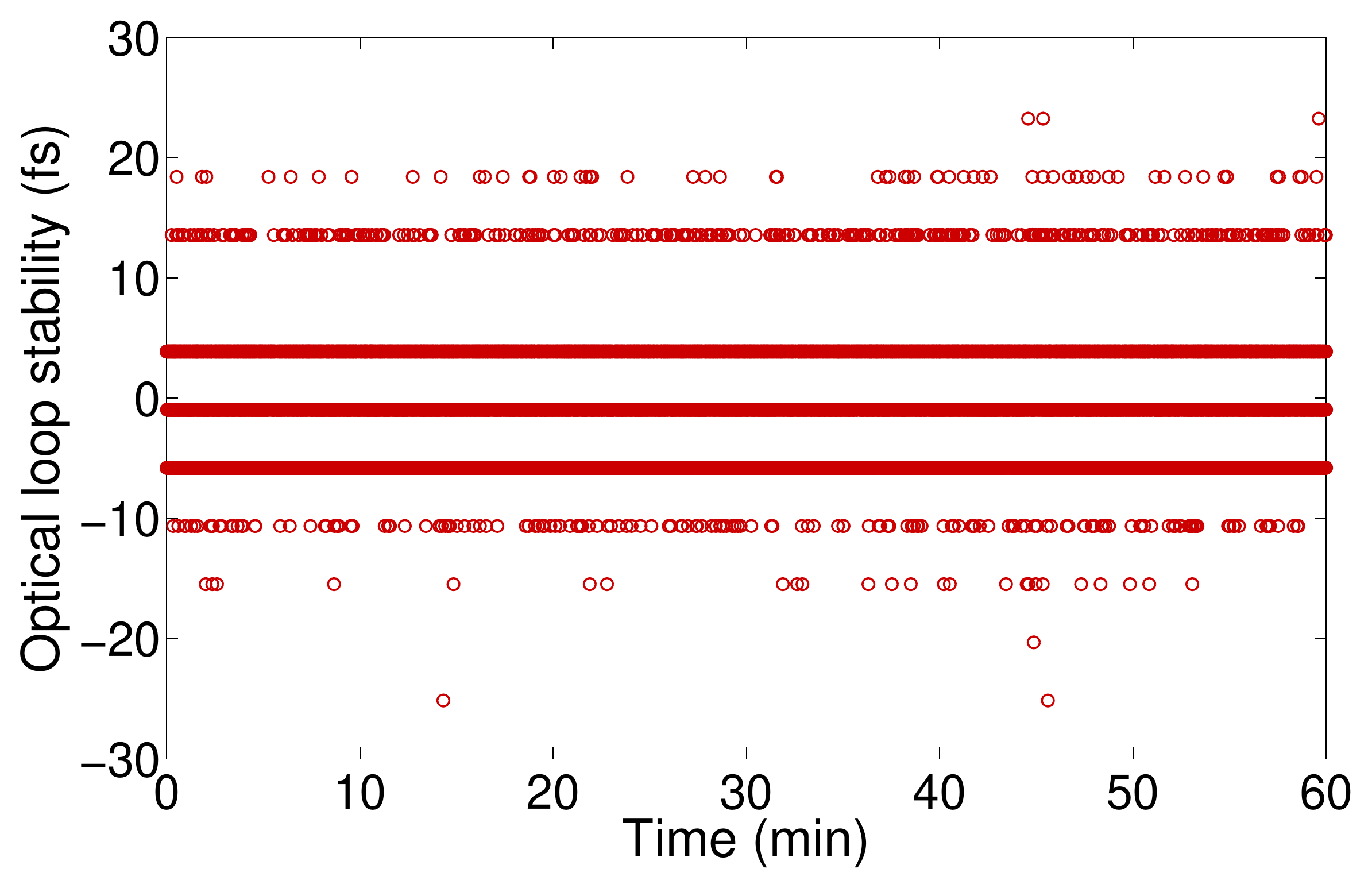} 
  \caption{Long-term in-loop optical lock stability of the EDFL 
  OMO~\cite{Ars11}.
  The integrated timing jitter over 60 minutes is 4.6~fs (rms) in the
  range between 1~Hz and 100~kHz.
  The clustering of the stability values at distinct values is a consequence of 
  the limited resolution of the sampling clock.}
  \label{fig:optsync2}
\end{figure}

\begin{figure}[tb]  
  \includegraphics*[width=1\linewidth]{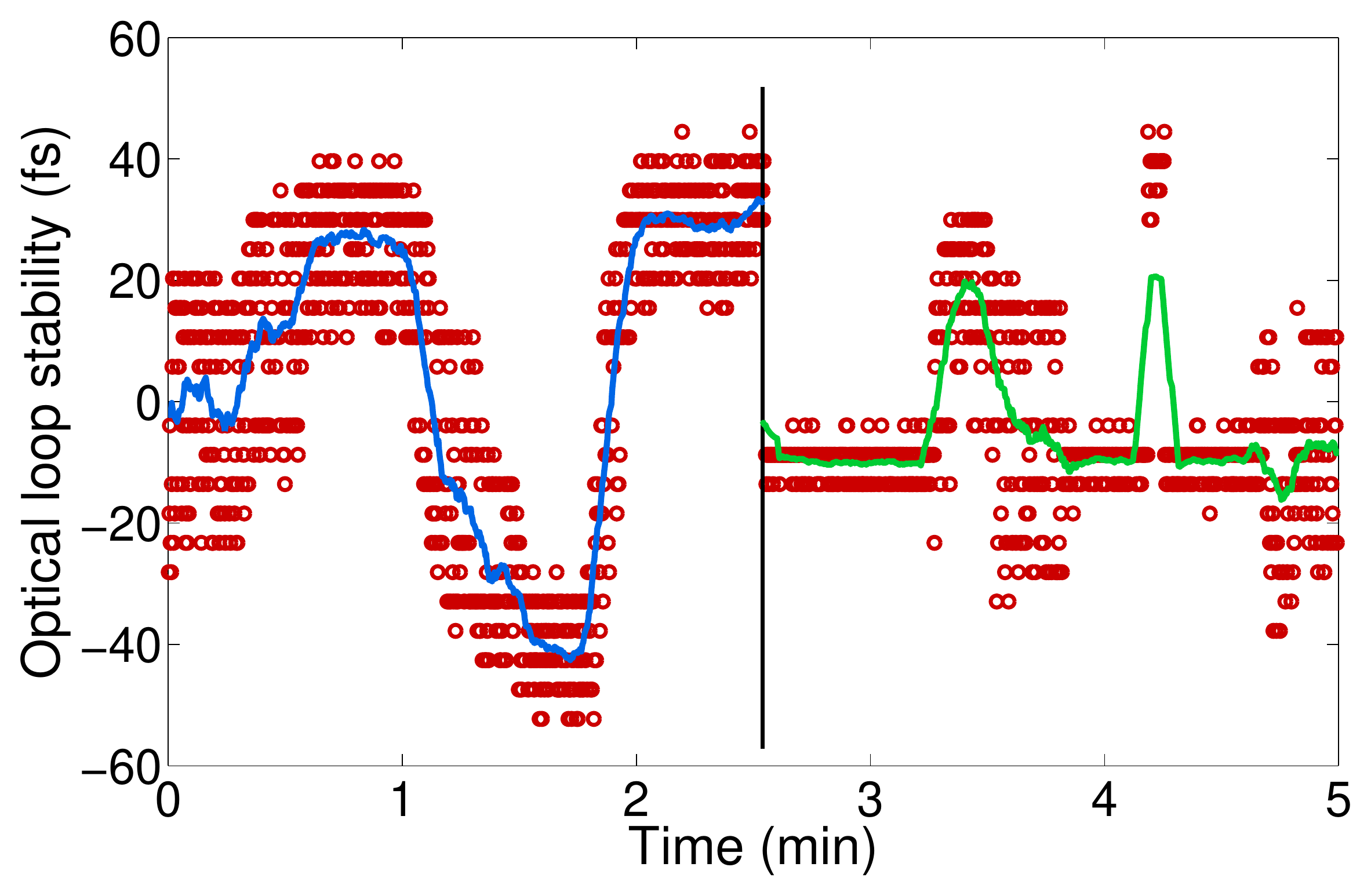}
  \caption{Test of the stability of the optical phase detector by addition of
  a manually varied voltage offset to the error signal~\cite{Ars11}.
  The initial curve shows stable operation following the induced changes, 
  until the loop breaks after 2.5 minutes (indicated by the vertical line).
  The optical lock is recovered once the perturbation is removed and the optical
  gains are readjusted (curve to the right of the vertical line).
  The blue and green curves represent sliding averages over 20 consecutive
  measurements.
  The integrated timing jitter is 11~fs (rms) in the range between 1~Hz and 
  100~kHz. 
  (As in Fig.~\ref{fig:optsync2}, the resolution of the sampling clock leads to 
  a clustering of the stability values at distinct levels.)}
  \label{fig:optsync3}
\end{figure}

The long-term in-loop optical lock stability of the EDFL is shown in 
Fig~\ref{fig:optsync2}.
The integrated timing jitter over 60 minutes is 4.6~fs (rms). 
The ADC sampling rate is 10 samples per second with a bandwidth of 100~kHz. 
Since these are in-loop measurements, they do not incorporate the drifts. 
Without a zero-crossing feedback on the optical S-curve, the drift will 
eventually break the optical loop. 
To test its stability under the influence of an external perturbation, 
e.g.\ a drift, an offset voltage is added to the optical error signal and 
varied manually, see Fig.~\ref{fig:optsync3} (left part). 
The induced changes are followed by the optical PLL within a range of 95~fs 
(peak-to-peak) before the loop breaks. 
Figure~\ref{fig:optsync3} (right part) illustrates how the optical lock is 
swiftly re-established after the removal of the external perturbation and a
readjustment of the optical gains.

\subsubsection{Superperiod synchronization}

We refer to the greatest common divisor of all rf frequencies synchronized to a
facility's master oscillator the as ``superperiod'' of that facility.
At the SITF the superperiod frequency in use is 2.974~MHz.
It is distributed via the event timing system, which is synchronized with the 
master oscillator.
The superperiod lock provides an absolute zero reference for all synchronized 
devices thereby ensuring reproducible timings throughout the machine. 
The gun laser, for example, with its repetition rate of 83.275~MHz, has 18 
possibilities to lock to the local reference frequency of 1\,498.956~MHz. 
Additional synchronization to the superperiod frequency ensures the same timing
of the laser and thus for other components along the machine. 

The electronics providing the superperiod synchronization for the Nd:YLF laser, 
as well as for the SITF OMO, was built in-house. 
It consists of an rf vector modulator, which shifts the phase of the reference 
timing until coincidence with the superperiod signal is achieved. 
The process is automatically steered by a controller board accessible via an 
RS-232 interface and integrated into the EPICS control system.
The superperiod synchronization has performed reliably throughout the operation 
time of the facility and will be used in a similar way during the initial phase 
of SwissFEL.

\subsection{\label{sec:drm}Dose rate monitors}

Dose rate monitors will be required at SwissFEL to prevent
radiation-induced demagnetization of the in-vacuum undulators. 
At the SITF, the SwissFEL dose rate monitoring system was tested during the
phase of the undulator experiment.

In its final state the system will consist of 48 integrating dosimeters based
on radiation-sensing field-effect transistors (RADFET) to be placed on each
undulator flange, close to the beam pipe, and a few other locations spread
throughout the SwissFEL accelerator.
The RADFETs are of the type RFT-300-CC10 (REM Oxford Ltd) and are read out by 
the four-channel dosimetry system DOSFET-L02, developed for
FERMI at Elettra~\cite{Fro13}.

\begin{figure}[tb]    
  \includegraphics*[width=1\linewidth]{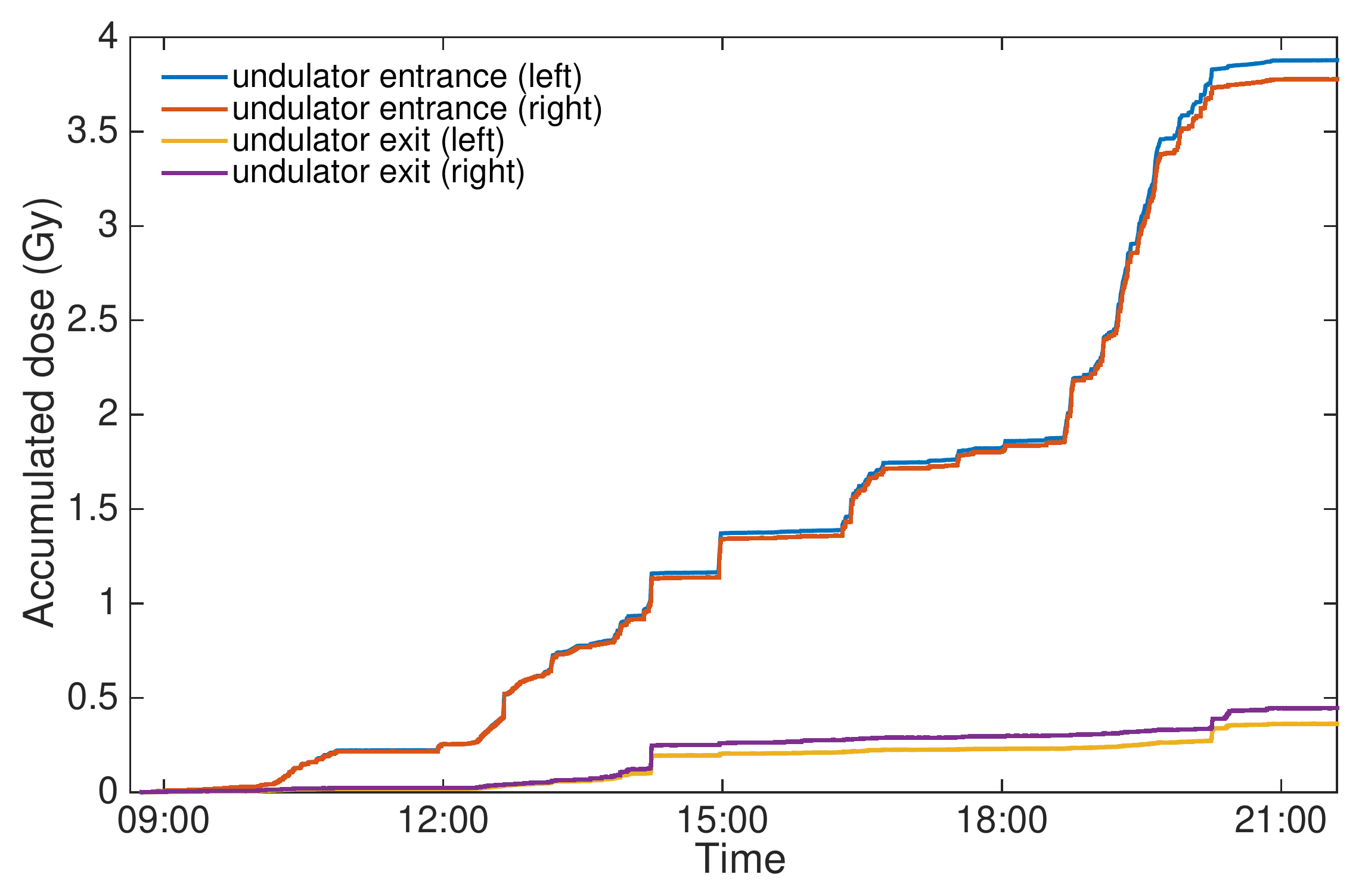}
  \includegraphics*[width=1\linewidth]{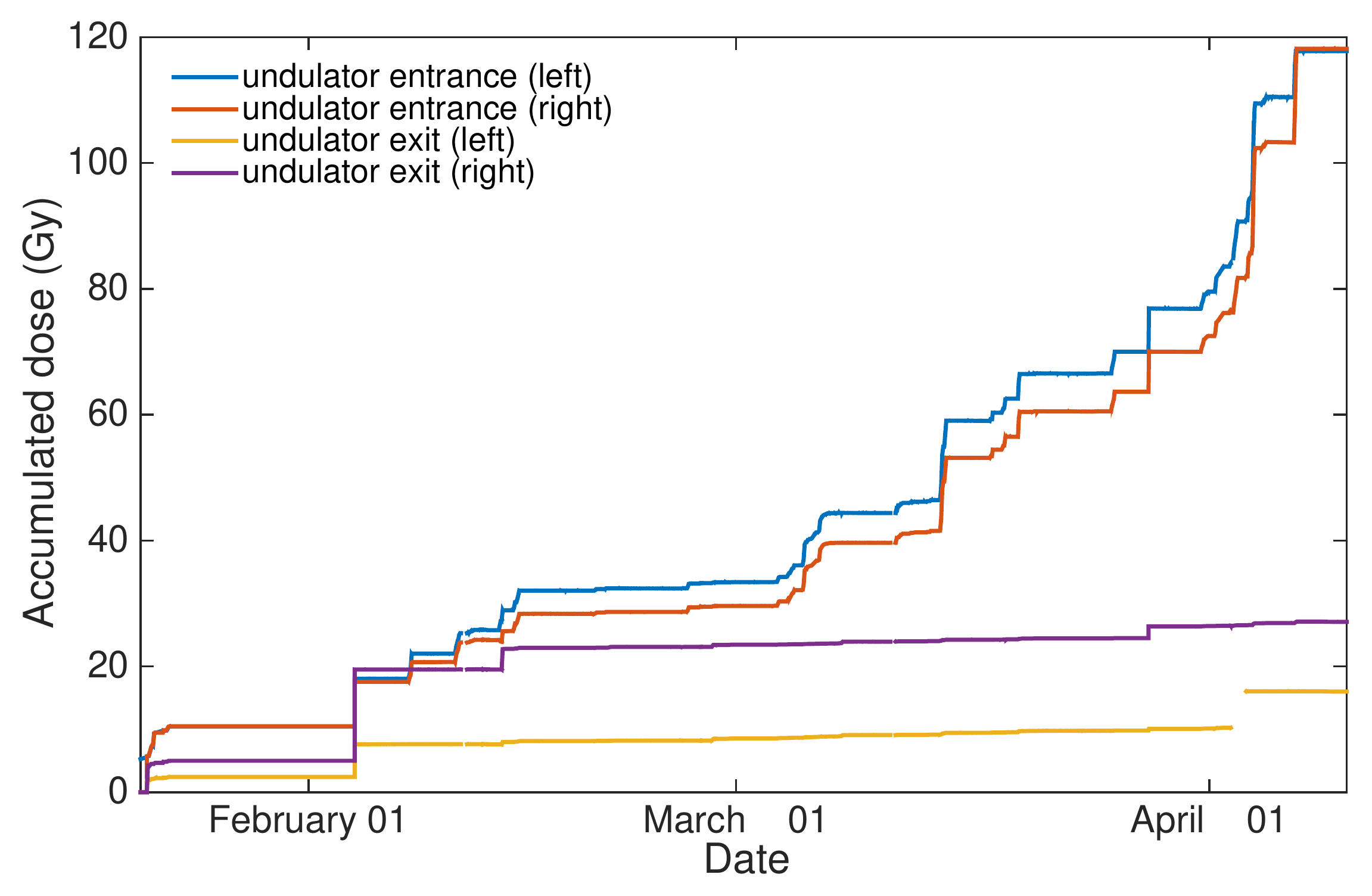}
  \caption{Top: Snap-shot of the first day of RADFETs monitoring radiation 
  exposure of the undulator.
  The four curves correspond to the four RADFETs installed at the undulator
  entrance and exit. 
  Significant doses (of the order of a Gray) can be accumulated in a few 
  minutes, demonstrating the need for a fast interlock.
  When the machine settings are properly adjusted the remaining losses are
  moderate.
  Bottom: RADFETs monitor data for the entire period of the undulator 
  experiment.}
  \label{fig:drm}
\end{figure}

The test at the SITF involved four RADFETs, operated in 25 V bias mode and
each acquiring one reading every 20 seconds.
The application of a positive bias significantly increases the response at the 
expense of limiting the total measurable dose~\cite{Fro13a}. 
In this mode, integrated doses of up to 118~Gy could be measured during the
approximately 2.5 months of undulator testing.
The RADFET readings were cross-checked against the accumulated dose 
measured with Gafchromic XR-RV3 film dosimeters positioned behind the 
sensors and found to be in good agreement.
Figure~\ref{fig:drm} shows the dose rate monitored during the first day (top) 
and the entire duration (bottom) of the undulator experiment.

\section{\label{sec:diag-dev}Diagnostics developments}

In this section we describe diagnostics systems, which were not primarily 
used for the operation of the SITF, but were utilizing the SITF for tests and
developments towards a possible implementation in SwissFEL~\cite{Isc13}.
An exception is the novel transverse profile imager (Sec.~\ref{sec:profimg}), 
which was developed and tested at the SITF, and immediately adopted for routine 
emittance measurements.

\subsection{\label{sec:ict}Integrating current transformers}

The reference bunch charge measurement at SwissFEL will be provided by 
integrating current transformers (ICT)~\cite{Uns89,Uns90}.
At the SITF, two types of ICTs have been tested (but were not used for
routine operations), Bergoz Instrumentation's standard ICT with BCM-IHR readout 
electronics~\cite{Bergoz}, and Bergoz Instrumentation's Turbo-ICT-2 with 
BCM-RF2 readout electronics~\cite{Art12,Art13}.

The standard Bergoz ICT is a calibrated device that measures the total charge 
in an integration window of 5~\textmu s. 
Thus it cannot resolve the two bunches envisaged at SwissFEL. 
Furthermore it is not immune to dark current, spurious noise of rf, kickers, 
etc., which introduce additional noise.

The Turbo-ICT-2 is an upgrade of the Bergoz Turbo-ICT featuring absolute
charge calibration to within 4\% and two-bunch resolving capability with
negligible beam position and bunch length dependence. 
Thanks to its fast readout of the beam induced charge at higher bandwidth it
is insensitive to dark current.
It also provides usable readings for arrival time jitter less than 1~ns. 

\begin{figure}[hbt]  
  \includegraphics*[width=1\linewidth]{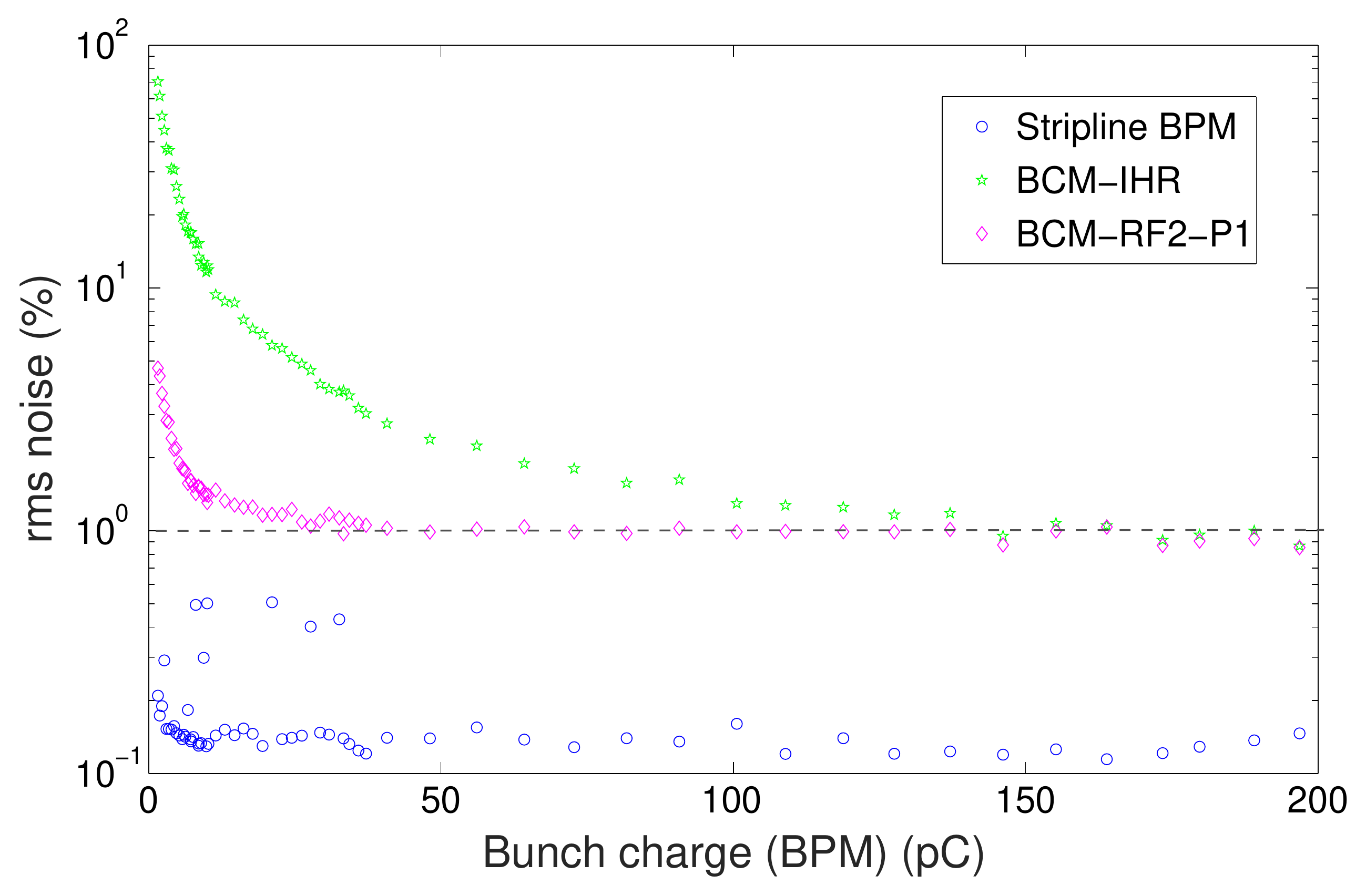}
  \caption{Noise (rms) measurements of the standard ICT (green stars), 
  Turbo-ICT-2 (first pulse shown, magenta diamonds) and a stripline BPM 
  located close to the ICTs (blue circles), as a function of bunch charge 
  (as measured with a BPM).}
  \label{fig:ict}
\end{figure}

We carried out a systematic comparison of the performances of the charge 
monitors using an uncompressed beam (but using the bunch compression
chicane to eliminate dark current) with bunch charge varying between 1 and 
200~pC.
The rms noise of the charge monitors was determined using the relative charge 
noise of one stripline BPM with respect to another, and by comparing one 
stripline BPM to the two ICT monitors. 
As shown in Fig.~\ref{fig:ict}, the standard ICT has a noise level of about 
10\% at 10~pC, while the Turbo-ICT-2 is at 1.2\% throughout the  nominal bunch 
charge range between 10 and 200~pC, thereby fulfilling the SwissFEL 
requirements.

The Turbo-ICT-2 was also successfully tested in two-bunch mode for a charge 
range between 1 and 50~pC per single bunch.
The measurements showed that the performance for both bunches are within 1\%
over the explored charge range.

Considering the requirements for SwissFEL, the Turbo-ICT-2 with BCM-RF2 
readout will be used to provide charge measurements and data for beam 
development and operation. 
These devices can also be used for BPM cross-calibration throughout the 
duration of machine operation.
A further advantage of the Turbo-ICT-2 is its on-device calibration pulser, 
which allows for performance self-checks over long-term beam exposure.

\subsection{\label{sec:cbpm}Cavity beam position monitors}

While the daily operation of the SITF relied on the resonant stripline beam 
position monitors (BPMs) described in Sec.~\ref{sec:bpm}, the development of a
more advanced cavity BPM system fulfilling the higher demands for SwissFEL
was pursued in parallel~\cite{Kei13}.
This development included numerous beam tests carried out at the SITF in a 
dedicated test section near the beam dump of the high-energy spectrometer.

The injector, linac and beam transfer lines of the SwissFEL accelerator will
be equipped with stainless steel (AISI 316N) cavity BPM pickups with apertures 
of 38~mm and 16~mm.
These pickups have a nominal frequency of 3.2844~GHz, which is well below the 
beam pipe cutoff frequency for the 38~mm cavity BPM, and a loaded quality 
factor $Q_L$ $\approx$ 40 for the working modes. 
In two-bunch operation, at the nominal bunch separation of 28~ns, the signal
of the first bunch has decayed to 0.07\% of its initial amplitude by the time
the second bunch arrives, resulting in very low crosstalk of position and 
charge readings between the two bunches.

For the SwissFEL undulators, where the BPMs have an aperture of 8~mm, 
we chose a cavity BPM design with a higher $Q_L$ of about 1000 and a working 
frequency of 4.9266~GHz, since these BPMs only need to measure single bunches 
at 100~Hz repetition rate. 
The higher $Q_L$ enhances the range-to-resolution ratio for position 
measurements and simplifies the electronics design described below, while the 
higher frequency improves the resolution at low charge. 
In contrast to the 38~mm and 16~mm cavity BPMs, the ones for the 8~mm aperture 
with their much higher quality factor require copper resonators to reduce 
resistive losses at the resonator walls to an acceptable level.
These resonators are brazed into an outer hull of stainless steel that includes
the vacuum flanges.
The brazing is done before the machining of the resonators, which is 
cost-effective and avoids known issues arising from alternative designs
based on thin copper coating of stainless steel resonators.
 
All three SwissFEL cavity BPM pickup types have two resonators: 
The reference resonator measures the bunch charge via the TM$_{010}$ mode, while
the position resonator measures the product of position and charge via the 
TM$_{110}$ mode.
The position resonator is equipped with mode-selecting waveguide couplers to 
reject undesired modes. 
Both resonators have the same frequency and loaded quality factor. 
Combined with a symmetric design of position and charge signal channels of the 
BPM electronics, temperature-induced drifts of the beam position readings 
(where the drift depends on the signal frequency) cancel out to first order
when calculating the beam position from the ratio of position and reference 
cavity signal amplitudes. 
In addition, critical parts of the cavity BPM electronics are actively 
temperature stabilized to minimize drift effects in the presence of ambient 
temperature changes.

The performance requirements for the injector, linac and transfer line BPMs 
could also have been met by resonant or nonresonant stripline BPMs. 
The choice of cavity BPMs for the undulators was clear, however, as these can 
most easily fulfill the required sub-micron resolution and drift limit for 
undulator beams that are typically near the center of the beam pipe.
In addition, as it turned out, the cavity BPM pickups developed for SwissFEL 
cost less than the resonant stripline BPM pickups made for the SITF while 
providing sub-micron resolution and drift at higher bunch charges~\cite{Kei13}.
The decision to use only cavity BPMs for the entire SwissFEL accelerator was 
therefore a straightforward one.

While sub-micron resolution and drift are mandatory for the undulator BPMs to 
ensure the alignment and stabilization of the beam trajectory and to optimize 
electron-photon overlap, it is also highly beneficial for the larger apertures
in the accelerator sections upstream of the undulator.
The better performance will enable, for instance, high-resolution measurements
of the beam energy jitter in dispersive sections, where the BPMs will be used 
as beam energy monitors with an expected relative resolution of 10$^{-5}$.
It will also allow for the optimization of the FEL photon intensity based on 
correlations of beam trajectory or energy variations with FEL pulse energy 
during normal FEL user operation. 

\begin{figure}[bt]  
  \includegraphics*[width=1\linewidth]{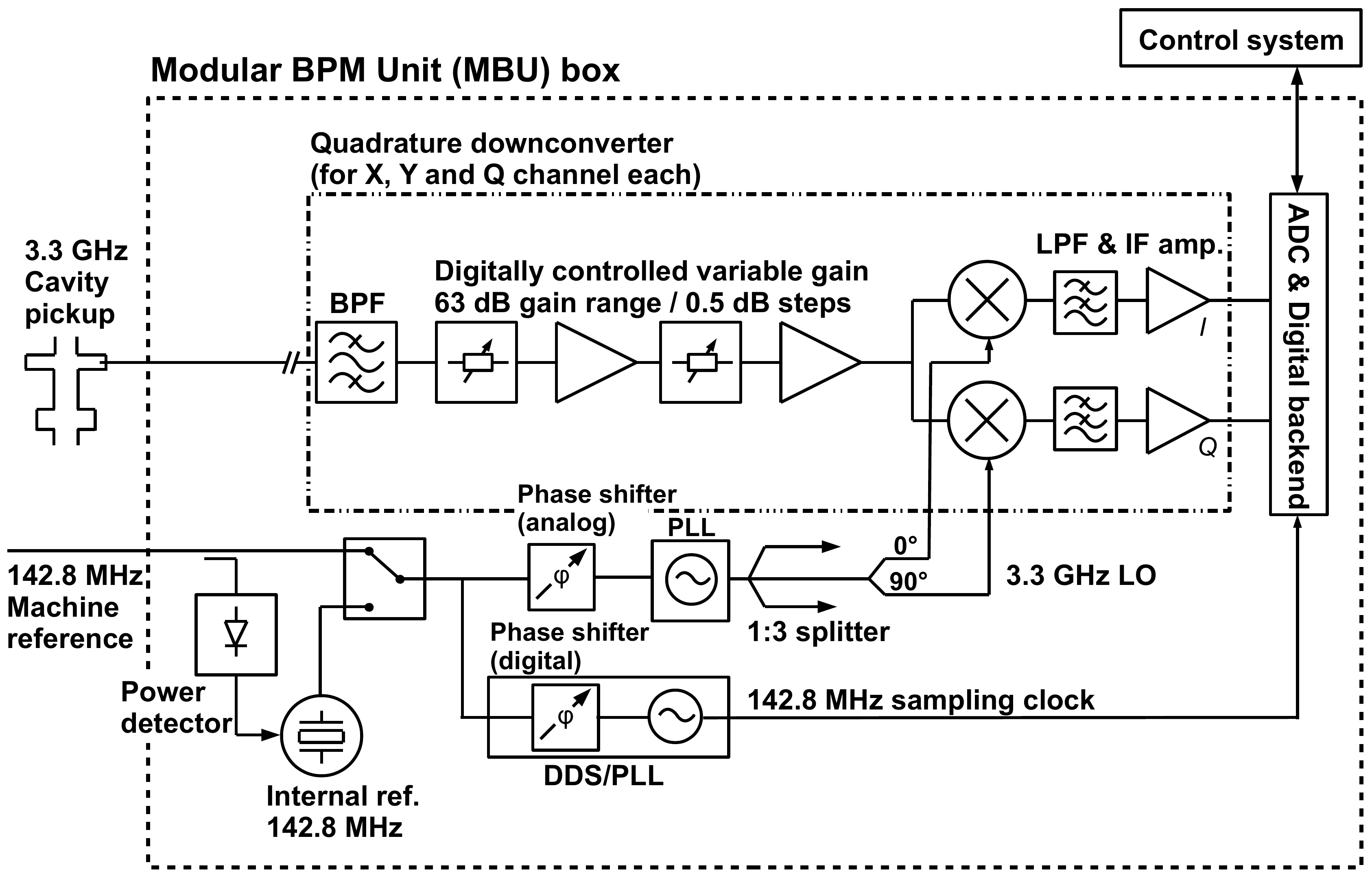}
  \caption{Simplified schematic of low-Q cavity BPM rf front-end electronics,
  showing only one of its three input channels (one reference and two position
  signal channels)~\cite{Kei13}.}
  \label{fig:cbpm_rffe}
\end{figure}

Beam tests performed at the SITF involved prototypes of all three cavity BPM 
pickup types and a preseries version of the readout electronics for the 38~mm 
and 16~mm pickups.
Figure~\ref{fig:cbpm_rffe} shows a simplified schematic of this electronics~\cite{Sta14}.
For the 16~mm cavity BPM, a position resolution of $<$0.8~\textmu m was 
measured for a bunch charge of 135~pC (with 0.35~mm beam offset and 
$>$2~mm peak-to-peak range).
The position resolution is determined by correlating the position readings of 
adjacent BPMs~\cite{Kei13}. 

A first prototype of the 8~mm cavity BPM electronics was designed and 
successfully tested, reaching, e.g., 310~nm rms position noise down to 
10~pC charge and within a 1.3~mm measurement range.

The design is similar to the low-Q electronics for the larger apertures, but 
has only one mixer per channel, where the three pickup signals are mixed to an 
intermediate frequency (IF) of about 130~MHz rather than to baseband, thus 
reducing the number of required ADC channels per BPM from six to three. 
Beam position and charge are then determined digitally by the FPGA from the 
undersampled IF waveform. 
A prototype version of the 8~mm cavity BPM pickup was tested successfully with
beam at the SITF~\cite{Sta14a}.
After the decommissioning of the SITF, the final BPM system is currently 
being tested with beam at the injection linac of the SLS.

In addition to high-resolution position measurements, all cavity BPMs will also
be used for beam charge measurements. 
The signal scaling factor will be precalibrated in the lab, with an expected 
scaling factor error of less than 10\%.
A more accurate beam-based calibration will be performed in situ comparing the
readings to those of dedicated charge monitors such as integrating current
transformers (see Sec.~\ref{sec:ict}).
During beam tests at the SITF with the low-Q cavity BPMs, we typically measured
a relative charge resolution of 0.07\% (rms) for high bunch charges, where the 
absolute resolution is proportional to the charge, and an absolute charge 
resolution of 8~fC (rms) at very low charges, where the absolute resolution
is charge independent.
The measured values exceed the SwissFEL specifications of 0.1\% and 30~fC, 
respectively.

\subsection{\label{sec:wfm}X-band wakefield monitors}

The small aperture of the X-band structure may give rise to strong short-range 
transverse wakefields induced by bunches passing through with a transverse
offset.
At the relatively moderate beam energies of the SITF the induced wakefields
may significantly deteriorate the beam emittance and must therefore be reduced
to a minimum by a proper alignment of the structure.
Since the mechanical axis of the structure does not necessarily coincide with
the electrical axis producing the smallest wakefields, a beam-based alignment
is essential.
Besides the evaluation of the beam emittance downstream of the structure as a
function of its position and orientation, a direct measurement of the 
wakefields via their coupling to the higher-order modes (HOM) of the structure
may be used as guidance for the structure alignment.

\begin{figure}[t]  
  \includegraphics*[width=0.8\linewidth]{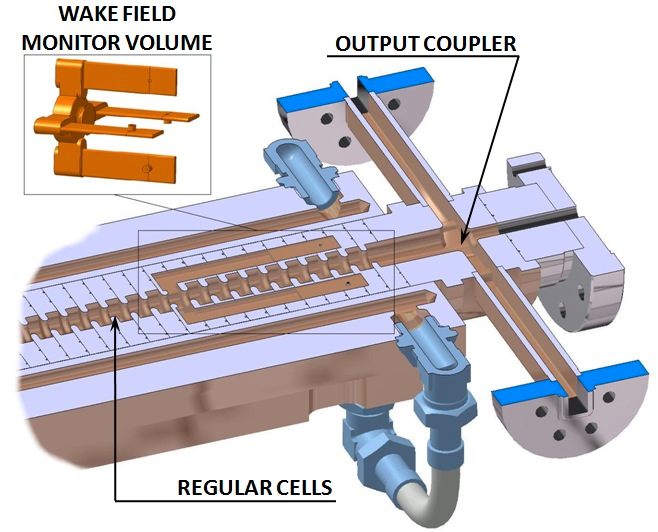}
  \caption{Cutoff view of the downstream end of the X-band structure showing 
  the location and geometry of the wake field monitor coupling cell relative
  to adjacent regular cells, high-power output coupler and cooling 
  circuit~\cite{Deh11}.}
  \label{fig:wfm1}
\end{figure}

The X-band structure installed in the SITF is equipped with two HOM couplers,
one upstream and one downstream of the structure, designed to be used as
wakefield monitors~\cite{Deh09a,Deh13}.
The couplers are special cells coupling to the transverse higher order
modes in the structure (see Fig.~\ref{fig:wfm1}).
The main reason for using two couplers at either end of the structure lies in 
the constant-gradient design of the X-band structure, in which the smooth 
variation of the cell dimensions along the structure compensates for internal 
losses and keeps the gradient of the fundamental mode constant.
This leads to a spread of the synchronous frequency of the 
position-dependent dipole modes in the 15 to 16~GHz range, i.e., an 
offset beam will excite lower dipole-mode frequencies upstream than downstream.
These modes are typically trapped, their fields extending only over parts of 
the structure.
Two couplers are required to cover the full range of modes.

\begin{figure}[b]  
  \includegraphics*[width=1\linewidth]{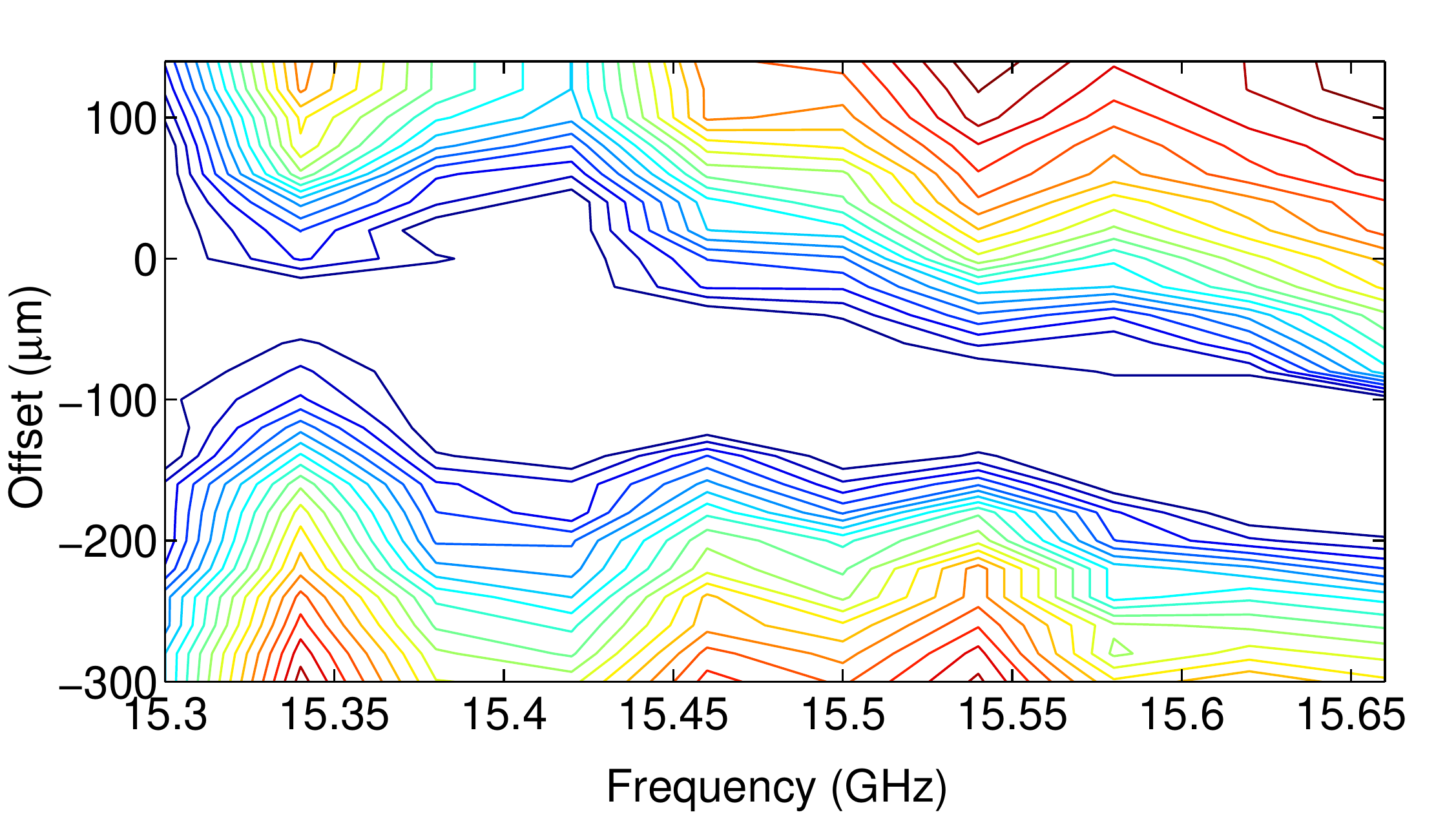}
  \caption{Signal output measured by the upstream wake-field monitor at 
  different frequencies for a varying structure offset.
  The structure is tilted on purpose by an angle of 0.5~mrad.}
  \label{fig:wfm_tilt}
\end{figure}

The correlation between signal frequency and cell number along the structure
can be exploited to diagnose higher-order misalignments. 
For instance, if the structure is tilted or bent, the offset between the cells 
and the beam varies along the length of the structure, resulting in a distinct 
spectral signature. 
An example of such a measurement is shown in Fig.~\ref{fig:wfm_tilt}, where the
structure was intentionally tilted and the spectrum of the upstream pickup was 
measured for various offsets.
The signal amplitude near 15.3~GHz, corresponding to the upstream end of the
structure, shows a minimum at roughly --40~\textmu m, whereas the center of
the structure, probed at a frequency of 15.67~GHz, has an offset of 
--140~\textmu m. 
Tracing the minimum signal versus the frequency clearly reveals the tilt of 
the structure.

The clean measurement of the HOM spectrum at frequencies between 15 and 16~GHz
in a radiation prone environment represents a technical challenge.
In the framework of the EuCARD-2 project~\cite{EuCARD2}, a front end based on 
electro-optical techniques, promising significant advantages in bandwidth, 
sensitivity and radiation hardness, is currently under development at 
PSI~\cite{Deh14}.

In the SwissFEL injector, two X-band structures will be used to linearize the
beam's longitudinal phase space at an energy of 320~MeV.
Both structures will be mounted on motorized support structures and equipped 
with wake field monitors.
While the associated readout electronics will not be part of the initial 
project phase, their implementation is envisaged in an early upgrade of the 
system.

\subsection{\label{sec:trpromon}Transverse profile monitors}

Most of the SITF beam operation relied on screens 
(Secs.~\ref{sec:scrmon} and \ref{sec:profimg}) for beam profile measurements.
In addition to these invasive monitors, we tested and operated minimally 
invasive and completely noninvasive transverse profile monitors at the SITF:
wire scanners and synchrotron radiation monitors, respectively.
We describe these systems, and the experience gained with them, in the 
following two subsections.

\subsubsection{\label{sec:wsc}Wire scanners}

At SwissFEL, wire scanners are intended to complement the screens in various
aspects: 
First, they provide a quasi-nondestructive monitoring of the transverse
profile of the electron beam during FEL operations.
Second, they can resolve the 28~ns time structure of the two bunches at 100~Hz.
And third, they offer an alternative profile measurement in case the imaging
performance of the screens is degraded by coherent transition radiation or 
other effects.

In a wire scanner measurement, a precise reconstruction of the single-axis 
projection of the beam transverse profile requires the beam-synchronous 
acquisition (see Sec.~\ref{sec:timevsys} and Ref.~\cite{Kal11})
of the readout of a beam-loss monitor picking up the electromagnetic shower 
produced by the wire scanning the beam and the encoder readout of the wire 
position, together with the charge and position measurements of adjacent 
beam-position monitors. 

\begin{figure}[bt]  
  \includegraphics*[width=1\linewidth]{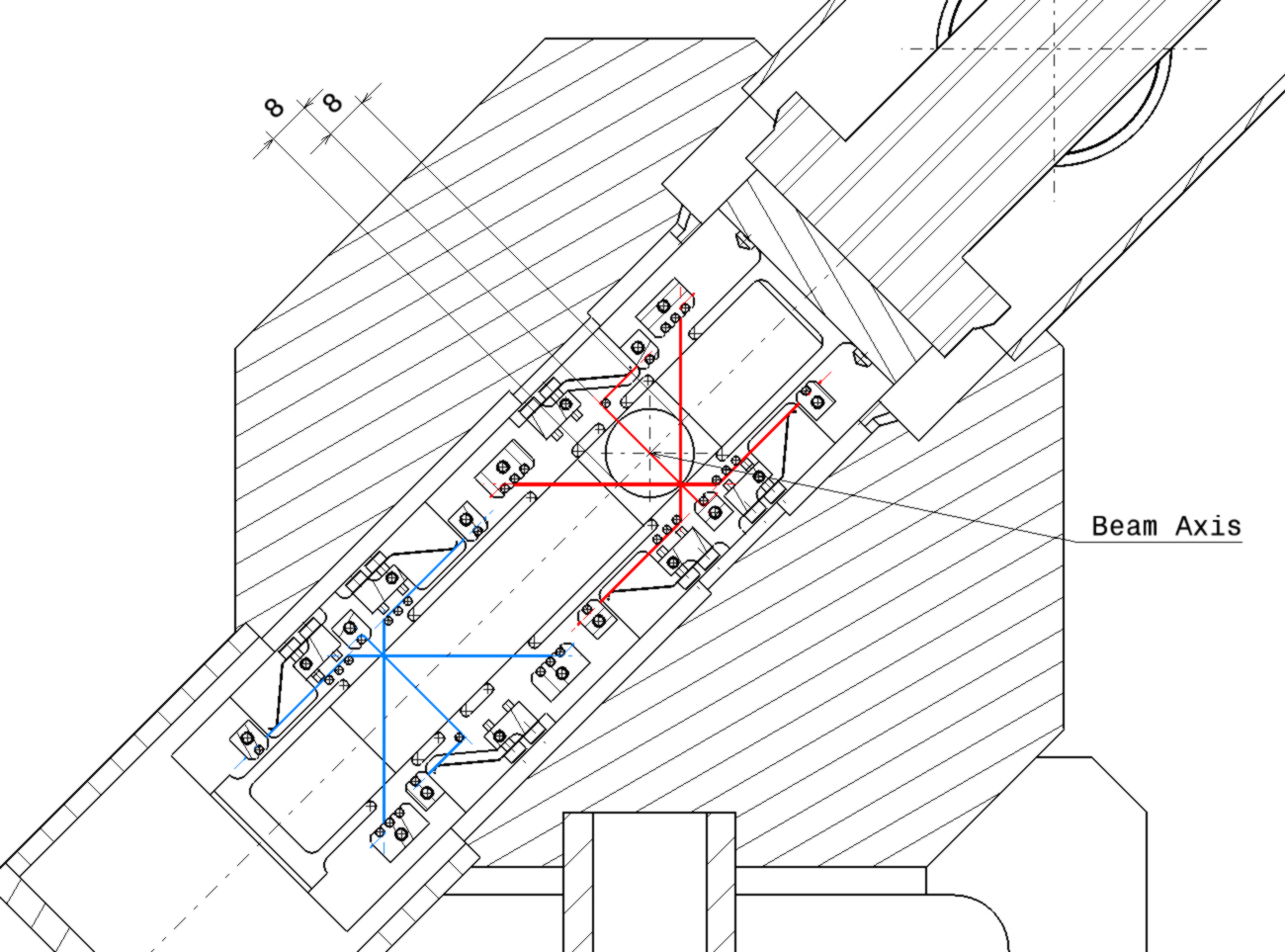}
  \caption{Front-view of the wire scanner vacuum chamber and the wire-fork. 
  The wire-fork is equipped with 3 different pin-slots where 5~\textmu m 
  tungsten wires can be stretched (indicated as red and blue lines). 
  The distance between the ``wire vertex'' and the vacuum-chamber center can be 
  set to 8, 5.5 and 3~mm corresponding to the 3 different pin slots.}
  \label{fig:wsc}
\end{figure}

Initial tests of the wire scanner system at the SITF were performed with the 
wires integrated in the screen monitor devices (Sec.~\ref{sec:scrmon}), where
the charge losses were recorded with BPMs.
Later studies focused on the final SwissFEL design, for which a prototype 
installation was integrated into the SITF in July 2014.
In this design the wire induced radiation losses are measured by means of 
scintillator fibers (polymethyl methacrylate, PMMA) wound around the beam pipe 
and directly connected to a photomultiplier~\cite{Orl14}. 
Figure~\ref{fig:wsc} shows a drawing of the wire fork as it was installed in 
the SITF beam pipe.

To finalize the design of the SwissFEL wire scanners, a prototype was 
experimentally characterized on a test bench.
The tests included the verification of the stability of the in-vacuum 
feed-through of the wire-fork under a stepper-motor driver and the measurement
of the induced vibrations of the wire stretched on the fork.
The measured wire vibration was about 1~\textmu m, less than the expected 
scanning resolution of the wire for the scanning velocities foreseen at
SwissFEL~\cite{Orl14}.

During beam tests at the SITF, several issues concerning the functionality and
reliability of the overall system were addressed, such as the beam-synchronous
readout of the encoders and the loss monitor, the optimization of the 
loss monitor signal versus the distance between wire and loss monitor, and the 
loss monitor sensitivity at low bunch charges scanned by very thin wires
(5~\textmu m diameter).

Recent experimental tests of a prototype of the SwissFEL wire scanner at
the FERMI@Elettra FEL facility demonstrated the viability of the SwissFEL wire 
scanner design also at the GeV energy scale~\cite{Pen15}.

\subsubsection{\label{sec:srm}Synchrotron-radiation monitor}

The deflection of the electron beam in magnetic chicanes offers the opportunity
to monitor the beam in a nondestructive way by observing the synchrotron
radiation emitted in one of the dipole magnets of the chicane.
At the SITF, a prototype setup of such a synchrotron radiation monitor (SRM) 
using the light emitted by the electron beam at the entrance of the third 
dipole magnet of the bunch compressor was commissioned and operated.
If the chicane dispersion is known, the horizontal projection of the 
synchrotron light spot allows the determination of the energy spread of the
beam induced by the off-crest rf operation.

The setup consists of a two-mirror periscope imaging the synchrotron light
from the third dipole magnet to a camera and lens system.
Within a range of $\pm$1$^\circ$  around the nominal bunch compressor bending
angle (4.07$^\circ$) the beam profile of the electron beam can be monitored
continuously and nondestructively over the entire range of compression 
factors~\cite{Orl12,Orl13}. 
The sCMOS camera (pco.edge) provides 100 frames per second and, in conjunction 
with a 300~mm lens, can resolve the beam profile with a projected pixel size of 
0.04~mm.
An image analysis algorithm running directly on the camera server can process
the images at a rate faster than 100~Hz, providing beam position and size 
information to the control system for possible use in feedback systems.

\begin{figure}[tb]  
  \includegraphics*[width=0.9\linewidth]{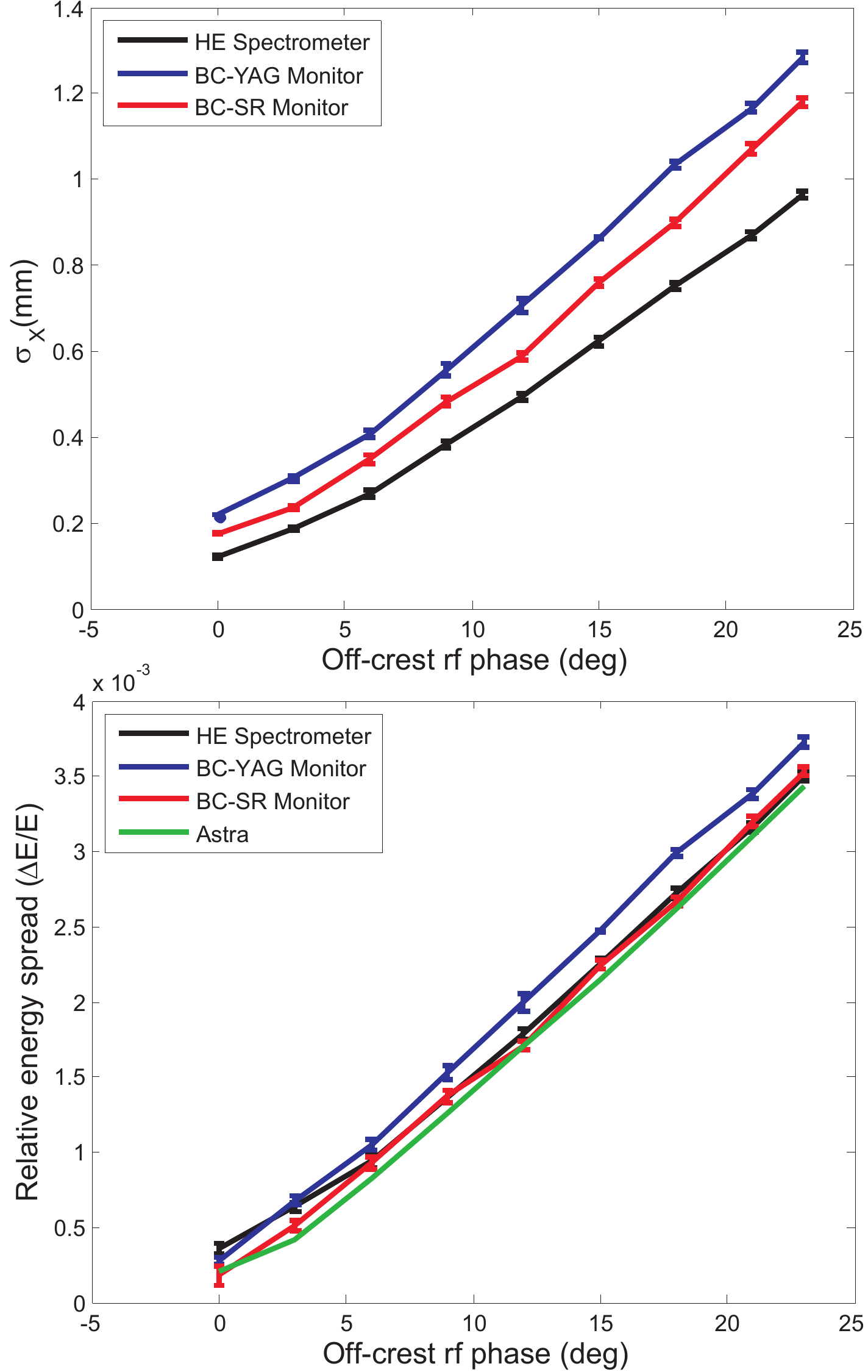}
  \caption{Top: Horizontal beam size versus off-crest rf phase as measured 
    with the high-energy spectrometer, the YAG screen in the bunch compressor 
    (BC-YAG), and the synchrotron radiation monitor in the bunch compressor 
    (BC-SR) for a beam energy of 230~MeV and a bunch charge of 12~pC.
    Bottom: Relative energy spreads derived from the measurements, compared
    to the result of a numerical simulation based on Astra~\cite{ASTRA}.}
  \label{fig:srm}
\end{figure}

Measurements of the beam energy spread performed with compressed beams
confirmed that the SRM sensitivity is sufficient to monitor the light spot 
from a 10~pC bunch over the entire compression range, see Fig.~\ref{fig:srm}. 

\subsubsection{\label{sec:profimg}Novel transverse profile imager}

The limitations of the original transverse profile monitors installed at the
SITF mentioned in Sec.~\ref{sec:scrmon} called for a redesign of the screen
geometry.
The new design takes into account both the Snell-Descartes law of refraction 
and the Scheimpflug imaging condition~\cite{Isc15}. 
In this geometry, (coherent) transition radiation is directed away from the 
camera, and the scintillating line inside the volume of the crystal is imaged 
onto a single point, thereby achieving a beam size resolution much better
than the crystal thickness.
The Scheimpflug imaging ensures focusing over the entire field of view.

\begin{figure}[tb]
   \centering
   \includegraphics*[width=0.8\linewidth]{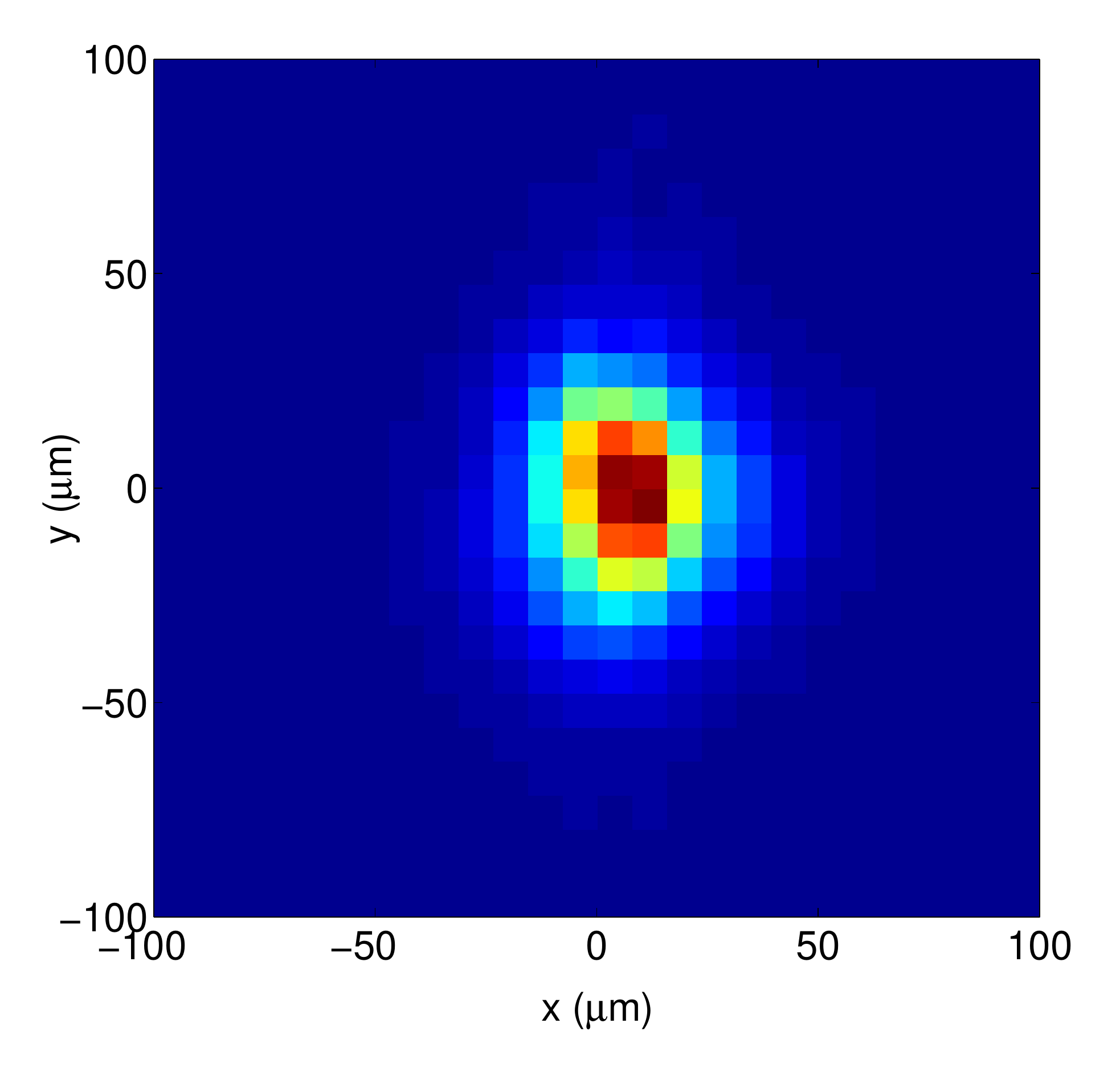}
   \caption{Example of a beam-size measurement for a bunch charge of 6~pC 
     performed with the novel transverse profile imager. 
     The horizontal and vertical rms beam sizes obtained from Gauss fits are 
     about 15 and 20~\textmu m, respectively.}
   \label{fig:smallbeam}
\end{figure}

A prototype of this novel screen monitor including a high-resolution readout
by a CMOS sensor (pco.edge 5.5 camera with global shutter) was installed and
tested in the SITF.
Cooled by a Peltier element, the sensor has sufficient sensitivity not only 
for the design operating range of 10 to 200~pC, but also for charges well below
1~pC as measured in thermal emittance studies (see Sec.~\ref{sec:bd-source}).
Figure~\ref{fig:smallbeam} shows an example measurement of a beam with 
transverse rms beam size of the order of 20~\textmu m, much smaller than the 
crystal thickness, in this case 100~\textmu m of YAG:Ce scintillator.

To verify the geometric suppression of COTR at higher beam energies (see 
Sec.~\ref{sec:scrmon}) the profile imaging setup was also tested at a location
after full compression in the LCLS linac, which is notorious for intense 
COTR~\cite{Loo08}.
The measurements, performed at a beam energy of 13.1~GeV and with bunches of
20~pC charge, show no significant variation of the light intensity as a function
of the bunch compression~\cite{Isc14a}.

At the SITF, the prototype screen with its outstanding resolution was quickly 
adopted as the standard screen for all beam optics and emittance measurements 
in the diagnostics section.
The imaging principle worked as expected, and only minor revisions to the 
mechanical design needed to be done for the SwissFEL series production.

\subsection{\label{sec:cdrmon}Coherent-diffraction-radiation based bunch
compression monitor}

\begin{figure}[htb]
   \centering
   \includegraphics*[width=0.8\linewidth]{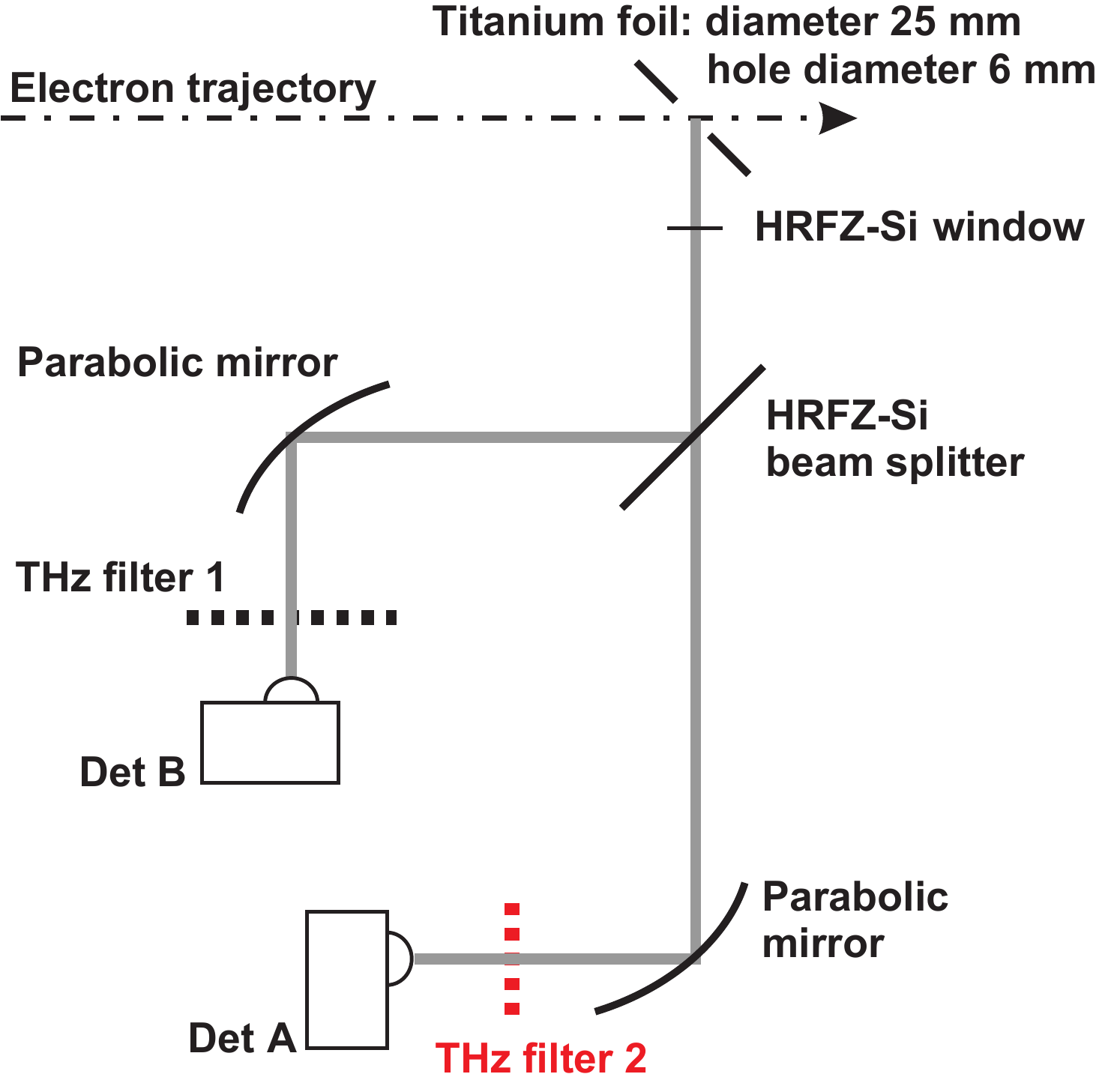}
   \caption{Schematic setup of the electron bunch length monitor based on 
     coherent diffraction radiation. A window of 
     high-resistivity float-zone (HRFZ) silicon is used to transmit the
     THz radiation.}
   \label{ff84_f1}
\end{figure}

The electron bunch length is one of the crucial beam parameters determining
the FEL performance and thus requires constant monitoring on a single-shot
basis.
At the SITF we performed preliminary tests of electron bunch compression 
monitors based on coherent edge and synchrotron radiation emitted at the last 
dipole magnet of the bunch compressor chicane, see Refs.~\cite{Isc11} and
\cite{Fre13} for brief overviews. 
In this section we focus on the more detailed studies and tests carried out 
with an electron compression monitor based on coherent diffraction radiation 
(CDR) generated by the electron bunch when passing through a hole of 6~mm 
diameter in a 1~\textmu m thick titanium foil (see Fig.~\ref{ff84_f1}), 
located a few meters after the bunch compressor.

The generated spectral density distribution radiated by a bunch is dominated by
the longitudinal form factor, defined by the Fourier transform of the 
longitudinal charge distribution~\cite{Ger11}. 
For rms bunch lengths of 250--500~fs, coherent radiation is expected up to the 
THz spectral region. 
By integrating the spectral intensity of the coherent radiation in a certain 
frequency range, variations in bunch compression lead to a corresponding
variation in the detected intensity. 

To increase the sensitivity of the measured signal with respect to small 
variations in the compression phase, so-called ``thick grid'' THz high-pass 
filters are used to select two spectral bands (from 0.26 to 2~THz and from 
0.6 to 2~THz). 
The focused coherent radiation is subsequently detected by fast Schottky 
diodes capable of resolving two consecutive bunches separated by 
28~ns.

\begin{figure*}[tb]
   \includegraphics*[width=0.6\linewidth]{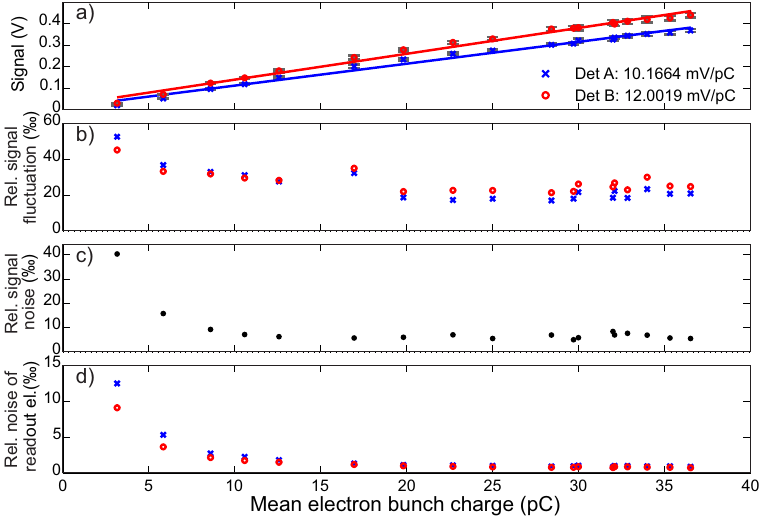}
   \caption{Detector signals for two identical THz filters (a), 
   relative signal fluctuation (b), 
   detector signal noise as calculated from the correlation from the two 
   detection channels (c), 
   and noise contribution from read-out electronics (d), 
   all shown as a function of the electron bunch charge.}
   \label{ff84_f2}
\end{figure*}

To assess the resolution of the setup, the system was characterized using two 
identical THz high-pass filters (both from 0.26 to 2~THz). 
In this way, correlated fluctuations can be attributed to changes in the 
electron beam, while uncorrelated fluctuations are assumed to be caused by 
detector or readout noise. 
To characterize the setup at different signal amplitudes, the integrated 
bunch charge was varied between 2 and 38~pC. 
Since for large signal levels, the Schottky diode signal is proportional to the
electric field strength, we expect a linear dependence on the bunch charge 
(i.e., the number of electrons) in the case of coherent radiation. 
This is confirmed by our measurements shown in Fig.~\ref{ff84_f2}a.
The slightly different frequency responses of the two detectors give rise
to the small discrepancy between the two slopes.

\begin{figure}[bt]
   \centering
   \includegraphics*[width=1\linewidth]{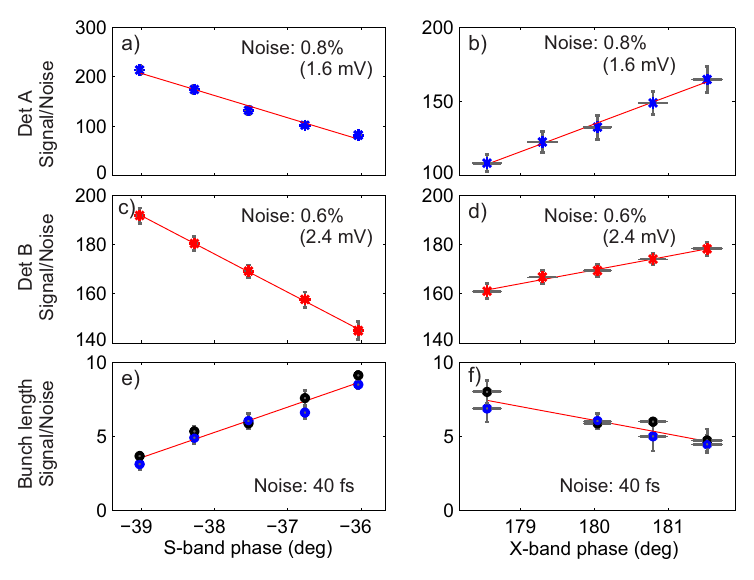}
   \caption{Detector signals as a function of S-band and X-band rf phases
    controlling the bunch compression, shown for two different spectral bands:
    the signals of detector A, plots a) and b), are shown after a high-pass
    filter at 0.6~THz, those of detector B, plots c) and d), after a high-pass
    filter at 0.26~THz.
    For comparison, plots e) and f) show corresponding bunch-length 
    measurements performed with the rf deflector.
    All signal values are displayed in relation to their corresponding noise 
    levels as noted explicitly in the figures.}
   \label{ff84_f3}
\end{figure}

Comparing the relative deviations (Fig.~\ref{ff84_f2}c) with the noise 
(Fig.~\ref{ff84_f2}b) of the detector including cabling and read-out indicates 
that the main contribution to the overall fluctuations shown in 
Fig.~\ref{ff84_f2}a is indeed due to the electron beam.
These results were subsequently used to estimate the significance of the 
measured signals with respect to phase changes of the accelerating rf fields 
as shown in Fig.~\ref{ff84_f3}.

The primary application of the compression monitors at SwissFEL will consist
in providing a diagnostic signal for the active stabilization of the electron 
bunch length via control of the relevant rf compression phases.
To assess the monitor's suitability for this task, a systematic study was 
performed, in which the THz signal was recorded as a function of the common 
phase of the last two S-band accelerating structures near the nominal 
compression phase as well as the phase of the X-band structure near the maximum
deceleration phase. 
The results are shown in Fig.~\ref{ff84_f3}.
For the relatively short electron bunch length of 260~fs, a high-pass filter
at 0.6~THz clearly enhances the sensitivity.
The inferred resolutions in S-band and X-band phase along with the stability
goals for SwissFEL are summarized in Table~\ref{ff84_tab1}.
A more complete description of the response of the bunch compression monitors,
as well as other diagnostics systems in use at the SITF, to rf phase and 
amplitude variations can be found in Ref.~\cite{Fre14}.

\begin{table}[hbt]
  \caption{Equivalent resolution in rf phase achieved by the CDR based 
  electron bunch length monitors at the SITF, in comparison with the 
  resolution goals for SwissFEL.}
  \begin{ruledtabular}
    \begin{tabular}{lcc}
       rf band & SITF measurement & SwissFEL requirement \\ 
       \colrule
           S-band & 0.022$^\circ$  & 0.018$^\circ$  \\
           X-band & 0.054$^\circ$  & 0.072$^\circ$  \\
    \end{tabular}
  \end{ruledtabular}
  \label{ff84_tab1}
\end{table}

For SwissFEL, electron bunch compression monitors based on CDR are planned to
be installed after the first and the second bunch compressor. 
Since the operation of the monitors requires the insertion of a foil with a 
hole into the beam pipe, they will not be entirely parasitic.
For this reason, alternative bunch length monitors based on coherent edge
and synchrotron radiation from the bunch compressors' last dipole magnets
will also be installed at SwissFEL.
It is expected that the intensity from coherent edge radiation is higher than
the CDR intensity.
These monitors, along with a further optimization of the high-pass THz filter, 
may allow us to achieve the rf phase resolution goal also in the case of the 
S-band rf phase.

\subsection{\label{sec:eomon}Electro-optic bunch length monitor}

\begin{figure}[bt]  
  \includegraphics*[width=0.9\linewidth]{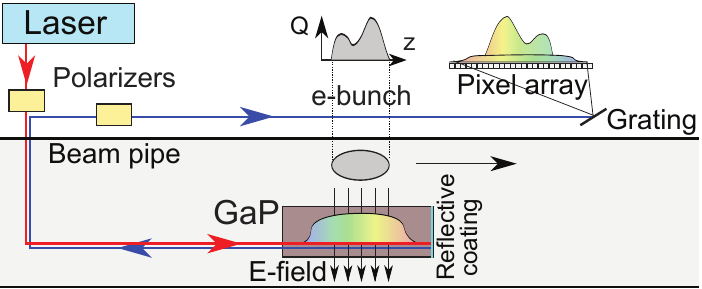}
  \caption{Schematic representation of the electro-optic bunch length
  monitor.}
  \label{fig:eomscheme}
\end{figure}

A bunch length monitor based on the electro-optic effect was developed for 
SwissFEL~\cite{Mue10,Mue11} and tested at the SITF~\cite{Iva14}. 
The measurement principle, known as electro-optic spectral decoding, is 
schematically shown in Fig.~\ref{fig:eomscheme}. 
A chirped laser pulse, i.e., with time-to-wavelength correlation, is 
linearly polarized and sent through a 2~mm thick GaP crystal installed in the
vacuum beam pipe.
The laser polarization is changed in the crystal via the electro-optic effect
caused by the electric field of the passing electron bunch.
A second polarizer after the crystal turns the polarization modulation into an
intensity modulation with time-to-wavelength correlation known from a direct
calibration procedure.
The bunch profile can then be reconstructed from the intensity pattern measured
by a grating spectrometer.
Since the technique is nondestructive it can be used for the online monitoring
of the bunch length on a pulse-to-pulse basis at rates up to 200~Hz, restricted
only by the spectrometer readout.

\begin{figure}[tb]  
  \includegraphics*[width=0.9\linewidth]{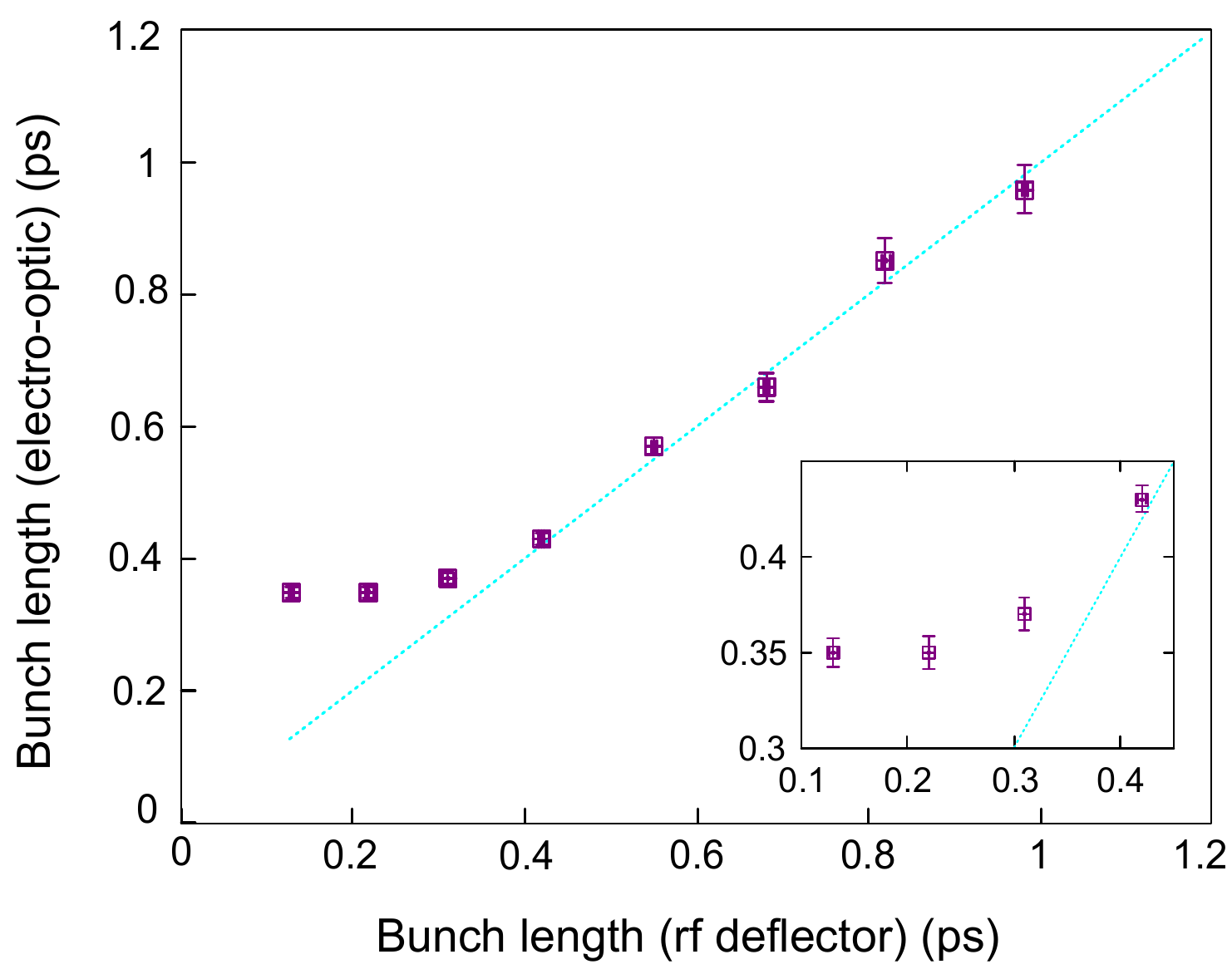}
  \caption{Comparison of bunch length measurements obtained with the 
  electro-optic monitor and the transverse deflecting rf cavity.}
  \label{fig:eomtdc}
\end{figure}

One of the studies conducted at the SITF was dedicated to the measurement of 
the resolution limit of the electro-optic monitor.
To this end, electron bunches with 200~MeV energy were compressed in the 
magnetic chicane and measured under the same conditions with both the 
electro-optic monitor and the transverse deflecting cavity, which has a 
resolution well below 100~fs (see Sec.~\ref{sec:long}).
A comparison of the bunch lengths obtained with the two methods is shown in
Fig.~\ref{fig:eomtdc} and indicates saturation of the electro-optic monitor
at about 350~fs.
The main resolution limiting factor for this method was found to be the 
frequency mixing effect (see Ref.~\cite{Mue11}). 
For the crystal installed at the SITF, the best resolution that can be expected
is around 200~fs.

For budget reasons, electro-optic monitors are not included in the current 
plans for SwissFEL.
It is possible, however, that such monitors may be added in a future upgrade
of the SwissFEL beam diagnostics system.

\subsection{\label{sec:bam}Bunch arrival-time monitor}

For the monitoring and optimization of the longitudinal linac stability,
the SwissFEL concept foresees the nondestructive measurement of the electron 
bunch arrival time at several locations with resolution better than 10~fs.
Two prototype setups of such bunch arrival-time monitors (BAM) have been 
installed and tested at the SITF, one upstream and, later, one downstream of 
the bunch compression chicane.
To meet the SwissFEL requirement of time resolution below 10~fs in the entire 
bunch charge range between 10 and 200 pC, these prototypes have undergone 
several iterations towards an optimal use of the component bandwidths and an
improved signal-to-noise ratio at low bunch charges.
In the following we describe the general design of the SwissFEL BAM system and
its implementation at the SITF (Sec.~\ref{sec:bam-design}), as well as the 
evaluation of rf pickups (Sec.~\ref{sec:bam-pickups}) for SwissFEL. 

At the SITF, the BAMs have been used to characterize bunch arrival-time 
drift and jitter~\cite{Ars13}, beam energy jitter~\cite{Ars14}, as well as the 
longitudinal dispersion (the $R_{56}$ parameter) of the bunch compression 
chicane.
In Sec.~\ref{sec:bam-drift} we give a brief account of the arrival-time drift 
and jitter measurement. 
The longitudinal dispersion measurement is described later in the section on 
bunch compressor characterization (\ref{sec:bc-char}).
For the beam energy jitter measurement we refer to Ref.~\cite{Ars14}.

\subsubsection{\label{sec:bam-design}Design and implementation}

The SwissFEL BAM concept is based on a Mach--Zehnder-type electro-optic 
modulator (EOM)~\cite{Loe10}, which is interfaced to an optical master 
oscillator (OMO, see also Sec.~\ref{sec:refgen}).
The OMO is a phase-locked-loop stabilized mode-locked laser oscillator
with a wavelength of 1560~nm and a repetition rate of 214.137~MHz.
It delivers reference pulses of a few 100~fs duration to the BAM stations via
single-mode optical fiber links, whose lengths are stabilized via optical
cross-correlation~\cite{Kim08a}.
This pulsed optical reference is highly stable against drift and jitter, at a
level below 10~fs.

An electron bunch passing through the beam pipe at the location of the BAM 
generates an S-shaped bipolar transient with a steep slope in an rf vacuum
pickup in the beam pipe.
The pickup signal is transmitted to the rf port of an EOM, which is DC
biased in quadrature (i.e., the optical signal is reduced to half its nominal 
amplitude at the output).
These EOMs are part of the electro-optic front-end, which is housed in 
an enclosing (the ``BAM box'') close to the beam pipe to avoid bandwidth losses 
in the rf cables leading to the EOM.
The box is temperature stabilized via its base plate at 28$^\circ$C to within 
0.01$^\circ$C and shielded against neutrons and X-rays.

Two EOMs are available in the front-end, which provides the flexibility to 
operate in high-sensitivity as well as high-dynamic-range configurations.
The second channel operates at the same zero crossing as the first but uses a 
separate pair of pickups.
Furthermore, the slew rate is filtered for that channel such that the rf slope 
is less inclined, resulting in a larger dynamic range of the BAM.
At the SITF, the second EOM channel was mainly used to simultaneously test 
different EOMs, rf pickups or other rf components such as attenuators or 
limiters.

Accommodating the end of the optical fiber link, the BAM box also houses two 
erbium-doped fiber amplifiers (EDFA) and a Faraday rotating mirror (FRM).
The first EDFA, which is part of the optical link and precedes the FRM, serves 
as an actuator for the link amplitude feedback loop.
After the FRM the second EDFA (out of loop) controls the power sent to the BAM 
photoreceiver (see below).

In the EOM the rf pickup signal is sampled at zero crossing by one of the 
reference (OMO) laser pulses.
The coincidence between a laser pulse and the zero crossing of the rf signal
is ensured by setting the laser timing accordingly.
Since the OMO laser is locked to the superperiod of the machine reference, this
zero-crossing overlap provides an absolute timing reference to any device that
is locked to the same superperiod (for instance the gun laser).
Any temporal offset due to a change in the electron time of flight leads to
an additional positive or negative voltage input at the EOM's rf port, thereby
increasing or decreasing the amplitude modulation of the one reference laser 
pulse which overlaps with the electron bunch.
Thus the bunch arrival time is encoded in the amplitude of a single reference 
laser pulse, while all the other pulses remain unaffected (i.e., premodulated 
to half amplitude). 

The optical pulses from the two BAM channels are received, together with a
sampling clock signal, in a data acquisition back-end located in the technical 
gallery outside the accelerator tunnel.
The sampling clock is generated by the same reference pulse train, split in the
BAM box before the EOM.
The pulses are conditioned in an in-house designed optical photoreceiver module
before being readout by a fast 12-bit analog-to-digital converter (ADC) with
500~MHz bandwidth.
The data acquisition chain is synchronous to the bunch rate.

The photoreceiver module uses high-bandwidth photodiodes followed by 
bandpass filters and high-dynamic-range trans\-impedance amplifiers.
The bandpass filters extract either the sine component of the sampling clock 
or quasi-Gaussian pulses, with 90~ps broad peaks and a few nanoseconds long 
intervals with a flat baseline in between, from the BAM signals. 
After the pulse form manipulation, the reference laser pulses are additionally 
amplified and matched to the ADC input. 
To ensure the optimal use of the ADC dynamic range in view of a better time
resolution, a DAC module (the PSI Analogue Carrier Board, PAC~\cite{Ozk15}), 
providing additional DC offset and amplitude control, should be used.
(This module only became available for the second BAM station downstream of the 
bunch compressor.)

The ADC does not sample a waveform, but only the peak or baseline points of a
large number of reference laser pulses, most of which precede the pulse that is
modulated by the electron bunch.
Therefore, a single-shot measurement determines not only the bunch arrival time,
but also the error limiting the resolution due to the laser amplitude jitter and
the influence of the subsequent opto-electronic chain.
In other words the BAM is a self-referenced system capable of measuring its own
resolution on a shot-to-shot basis.
The BAM resolution is essentially a function of the slew rate of the rf 
transient and the effective amplitude jitter of the reference laser pulses, as 
measured by the ADC.
For our system, the latter effect was found to dominate.
The measured effective laser amplitude fluctuations vary between 0.25\% and
0.3\%.

Since both the sampling clock and the signal are generated from the same 
optical pulse train, the ADC sampling position on the reference laser pulse 
remains unchanged under temporal shifts of the laser reference.
Such temporal shifts of the laser are brought about, for instance, during the 
calibration of the system, when the modulated laser pulse is scanned over the
rf transient to map the extent of the amplitude modulation to the laser delay. 
In a similar vein, a dedicated delay stage in the BAM box is applied to run a  
zero-crossing feedback: each time the bunch arrival time exceeds a predefined
limit determined by the slope of the rf transient, the reference laser pulse is
shifted back to the zero crossing, thereby keeping the BAM acquisition within
the dynamic range of the transient.

\subsubsection{\label{sec:bam-pickups}Evaluation of rf pickups}

Two types of rf pickups were evaluated in the first BAM station (upstream of 
the bunch compressor), a button and a ridge waveguide, both mounted on a 
common 38~mm vacuum chamber~\cite{Ars13}.

The ridge wave guide features a high coupling strength to support measurements
at low bunch charges and an intrinsic bandwidth of 16~GHz.
Measurements with this pickup type revealed a strongly nonuniform bunch charge
dependence of the BAM signal with maximum slew rates (resulting in the best
BAM resolution) around 100~pC bunch charge and dropping again at higher 
charges.
In view of the limited dynamic range with respect to bunch charge, this design
was considered impractical for SwissFEL.

The button pickup is designed as a 50~$\Omega$ coaxial line, starting with a  
cone shaped tip with 2.5~mm base and smoothly tapered over the 28~mm length to 
a coaxial glass-bead feedthrough and a 2.92~mm connector. 
Its nominal intrinsic bandwidth is 80~MHz.
Test results obtained with the button pickup show a uniform bunch charge
dependence.
The slew rates, however, are limited by the coupling strength at low bunch 
charges and by the feedthrough bandwidth of 20~GHz. 

Based on this experience, the initial plan for SwissFEL was to increase
the coupling strength by shortening the 50~$\Omega$ coaxial line to 6~mm and
by extending the bandwidth of the feedthrough to 40~GHz. 
The design diameter of the button remained at 2.5~mm and is a compromise 
between signal strength and bandwidth (40~GHz). 
The production of such vacuum feedthroughs, however, turned out to be a 
substantial technical challenge.
Two approaches were pursued, one based on commercial high-bandwidth 
bore-silicate glass beads and another wherein a kovar glass is directly 
sintered to the flange.
Unfortunately the corresponding prototypes turned out to be both costly and 
time-consuming in production and did not meet the design specifications.
We therefore decided to abandon this development and started testing two 
pickup prototypes developed by TU Darmstadt and DESY~\cite{Ang12,Ang15}.
The second BAM station at the SITF, positioned after the bunch compression
chicane is equipped with rf pickups of this type.
The arrival time resolutions obtained with this BAM station, using an rf pickup 
prototype adapted to the SwissFEL 16~mm vacuum chamber, are 35~fs for a bunch
charge of 10~pC, 20~fs for 20~pC and vary in the range from 13~fs down to 7~fs
for bunch charges between 30~pC and 200~pC.
For the SwissFEL BAM stations in the undulator section, where the beam pipe
diameter is only 8~mm, a further improvement in resolution at low bunch charges
by about a factor of two is expected.

\subsubsection{\label{sec:bam-drift}Arrival time drift and jitter measurement}

\begin{figure}[tb]  
  \includegraphics*[width=1\linewidth]{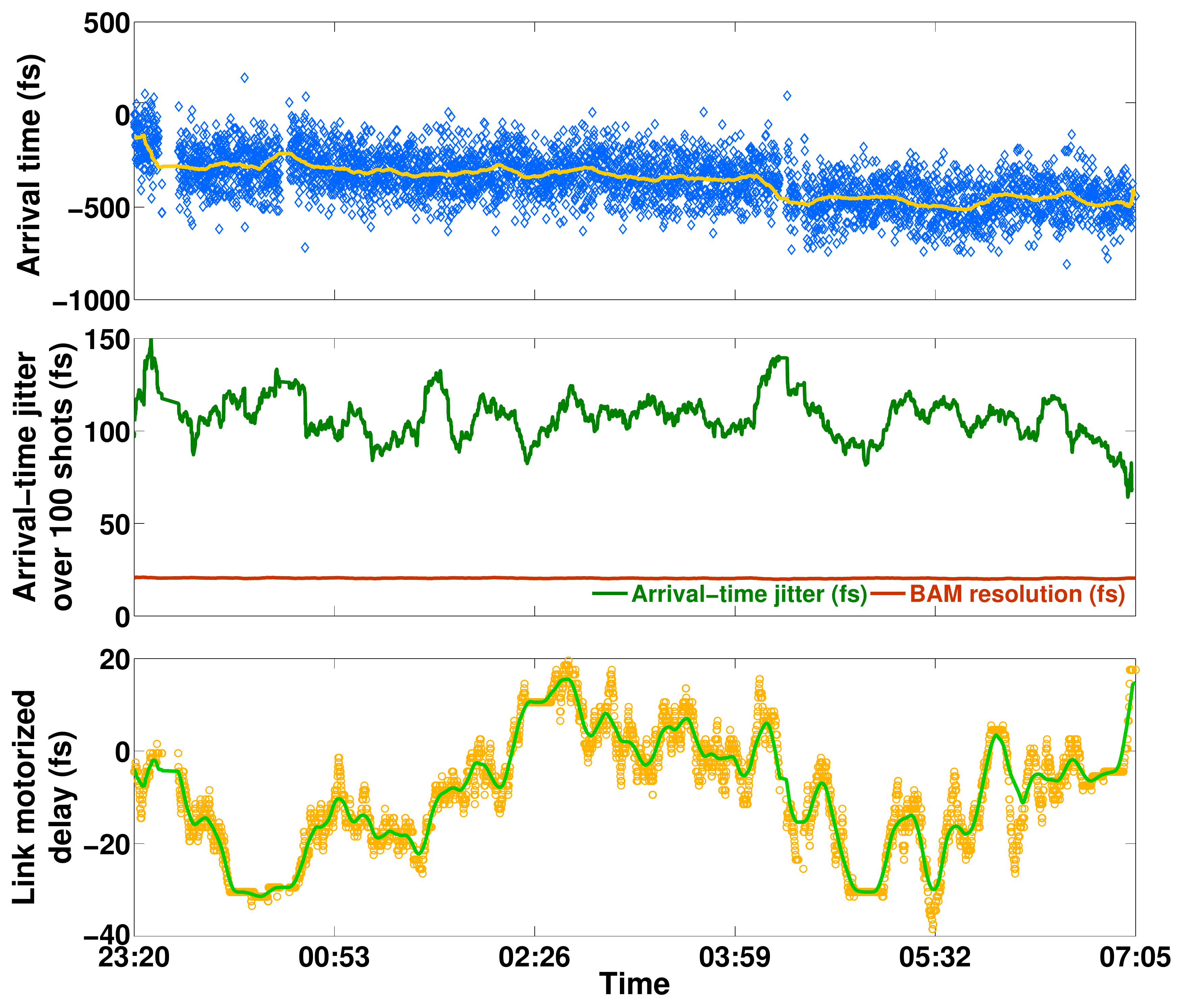}
  \caption{Long-term bunch arrival-time measurement over a period of eight 
    hours, using the BAM upstream of the bunch compressor at a bunch charge
    of 130~pC (top), arrival-time jitter over 100 shots  
    (middle) with instantaneous resolution (red curve) and link length
    compensation applied by the motorized delay line (bottom).} 
  \label{fig:BAMdrift}
\end{figure}

An example of a long-term bunch-arrival-time measurement performed with the
BAM located upstream of the bunch compressor over approximately eight hours is 
given in Fig.~\ref{fig:BAMdrift}.
For this measurement, the Nd:YLF based gun laser system was used, locked to
the 1\,498.956~MHz rf master oscillator and to the machine superperiod.
The average bunch charge during the measurement period was 130~pC.
The optical fiber link was stabilized with an accuracy of 21~fs.
During the run the link motor had to compensate up to 47~fs of fiber drift
(Fig.~\ref{fig:BAMdrift}, lower plot).
A feedback ensures that the BAM acquisition is always kept near the zero 
crossing of the BAM pickup slope.
The resolution is 20~fs, which was typical for our initial pickup design, where
the bandwidth was limited by the vacuum feedthroughs (20~GHz) and the 
electro-optic intensity modulators (12~GHz).
The bunch arrival-time drift, determined as a sliding average over 100 shots
(i.e., approximately 10~s), is 410~fs (peak-to-peak) over the 8 hour period.
During this period several rf interlocks occurred, leading to beam interruptions
visible as data gaps in the drift plot (Fig.~\ref{fig:BAMdrift}, upper plot).
After recovery of the rf the BAM acquisition was resumed automatically.
The middle plot of Fig.~\ref{fig:BAMdrift} shows the jitter, defined as the 
deviation of the instantaneous arrival time from the smoothed average described
above.
The jitter ranges between 70 and 150~fs, with an rms value of 110~fs.
Similar jitter values were obtained even before the commissioning of the 
optical fiber link.
A substantial fraction of the observed jitter arises from the gun laser 
amplifier and transfer lines after the stabilized laser oscillator.
A dedicated laser arrival-time monitor is currently under development for
SwissFEL with the goal of characterizing and mitigating this 
effect~\cite{Div15}.

\section{\label{sec:bd-procs}Beam and lattice characterization procedures}

In addition to advanced beam diagnostics, the detailed characterization of the 
electron beam generated at the SITF also relied on elaborate methods and 
procedures, which first needed to be established at our facility, and later
refined and adapted for routine use.
In this section, we describe these procedures, before presenting the main 
results in Sec.~\ref{sec:bd-results}.

\subsection{\label{sec:lattice}Lattice characterization and setup}

The detailed knowledge of the lattice acting on the beam is essential for the
proper interpretation of all beam physics measurements.
We present our efforts to characterize and adjust the beam orbit in all parts
of the machine, including the effects of the bunch-compressor and geomagnetic
dipole fields. 

\subsubsection{\label{sec:bba}Beam-based alignment}

All machine components were mechanically aligned during the installation phase
with a precision better than 100~\textmu m.
In addition, a beam-based alignment was performed for those beam-line components
that are relevant for beam optics measurements.

For the beam-based alignment of the rf gun elements, we essentially follow the
procedures described in Ref.~\cite{Kra05}.
In a first step, the laser position on the cathode of the electron gun is 
adjusted to the central axis of the gun rf field.
In the case of the CTF gun, this is determined by the center of the dark 
current halo, as it can be observed on one of the first view screens of the gun
section with gun solenoid and laser both turned off.
The position of the laser spot on the cathode is then adjusted until the 
photoelectron beam and the dark current halo are concentric.
During the procedure, all corrector dipole magnets between the gun and the
view screen are turned off since in general the energy of the dark current
electrons differs from that of the photoelectrons.
To improve the visibility of the photoelectron beam on the screen, the rf
phase is shifted towards the phase of maximum charge extraction, resulting
in a focusing of the photoelectron beam by the rf field, and the rf gradient
is reduced by 10--20\% to increase the contrast between the dark current halo
and the photoelectron beam.

The laser position found in first approximation by centering on the dark 
current halo may be further readjusted by verifying that the beam position on
the screen is insensitive to changes of the gun rf phase.
In the case of the SwissFEL gun (from April 2014 on), the amount of dark 
current produced was insufficient for the simple alignment procedure based on
the dark current halo and an rf phase scan was needed for the laser alignment.

Once the gun laser spot is fixed, the gun solenoid is aligned by minimizing 
the movement of the photoelectron beam, again observed on one of the first 
view screens of the gun section, in response to changes in the strength of the
solenoid field.
The gun solenoid is mounted on a motorized support with four degrees of 
freedom (the longitudinal position and the rotation angle around the beam 
axis are fixed).
For a large variation of the solenoid excitation current from 20 to 150~A, 
a variation of the beam position well below 1~mm, measured on a screen 47~cm
downstream of the solenoid exit, is achieved routinely.
The nominal current is around 120~A, corresponding to an integrated field of
about 0.006~Tm.

All beam-line elements following the gun section, with the exception of the 
X-band structure, are not motorized and cannot be aligned mechanically by 
remote control (the bunch compression chicane can only be moved in one 
direction).
Instead the beam reference orbit, i.e., the orbit to be approximated by a 
manual or automated orbit correction procedure (feedback) is adjusted for 
optimal beam propagation.
In practice this is achieved by adding (or subtracting) offsets to the BPM 
readout positions before using them for orbit correction.
The task of alignment then consists in finding and applying the appropriate 
BPM offsets in each section.

\begin{figure}[b]
  \includegraphics*[width=0.9\linewidth]{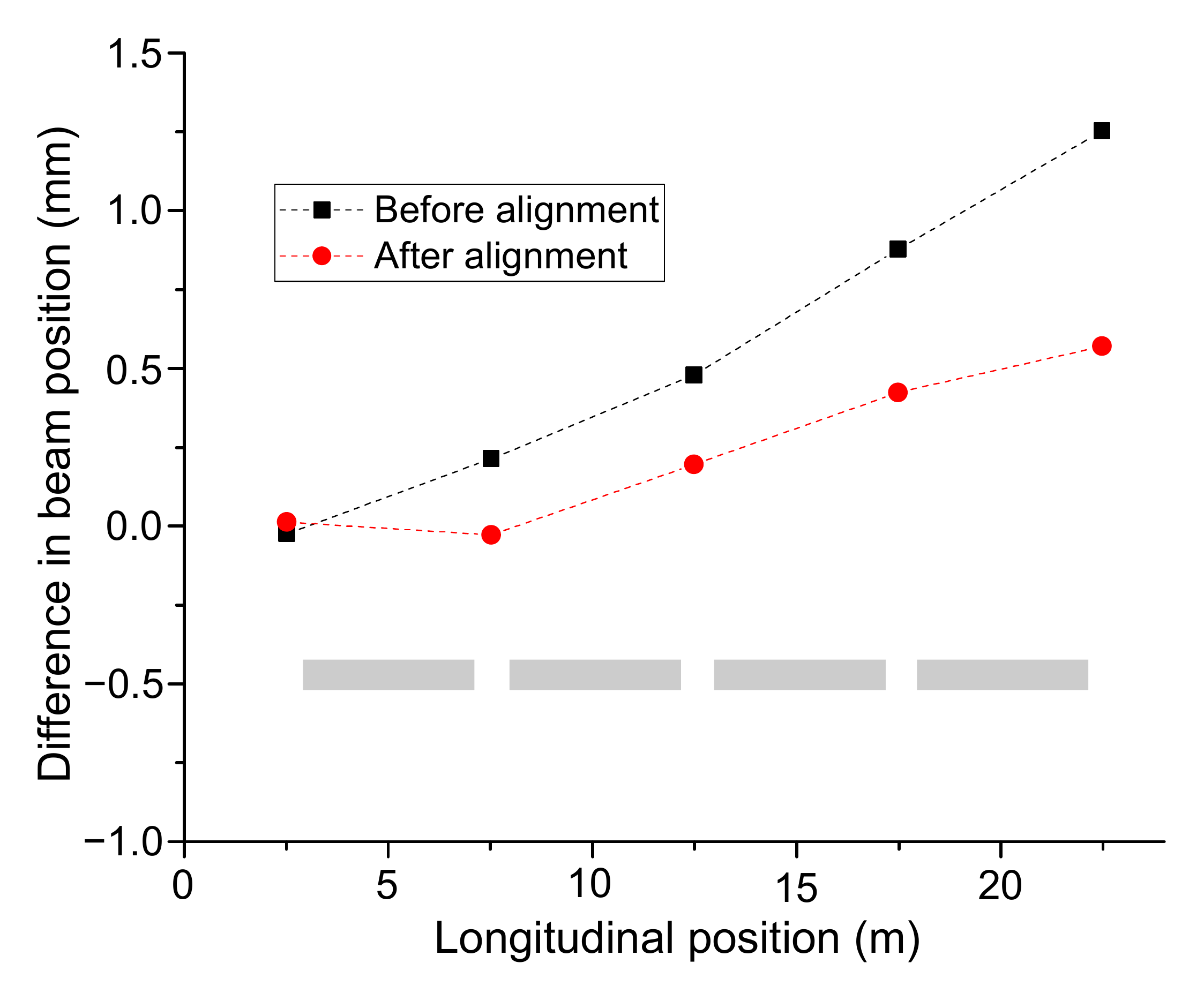}
    \caption{Measured beam position variation along the injector in the 
      horizontal plane in response to a change in rf gradient, before and 
      after the beam-based alignment of the first booster structure.
      The beam position at the second BPM immediately after the first S-band 
      structure is almost unchanged after the alignment, whereas further 
      downstream changes persist due to the still misaligned structures.
      The locations of the four booster structures are indicated as gray 
      boxes.}
    \label{fig:bba}
\end{figure}

In the case of the booster section, the reference orbit is adjusted to coincide 
with the central axis in each accelerating structure.
This coincidence is achieved when the beam trajectory remains constant after a 
change in rf gradient in the given structure. 
Figure~\ref{fig:bba} shows the difference of the beam trajectory before and
after the first booster section alignment in the horizontal plane.

Downstream of the booster the reference orbit is chosen to minimize spurious
transverse dispersion along the beam line (dispersion-free 
steering~\cite{Rau91}).
The amount of spurious dispersion after the booster is determined by 
observing the orbit deviation in response to a variation in the beam momentum.
The beam trajectory (position and angle) at the exit of the S-band section is 
fixed through an orbit feedback to compensate possible residual cavity kicks 
induced by the momentum change, which may still remain after the beam-based
alignment of the booster structures.
The variation in beam momentum is achieved by shifting the second S-band booster
structure from on-crest to off-crest acceleration.
We vary the phase and not the gradient to avoid possible temperature effects in
the structure following a change in gradient (rf power), which could matter 
over the timescale of the alignment procedure.

\begin{figure}[tb]
  \includegraphics*[width=0.9\linewidth]{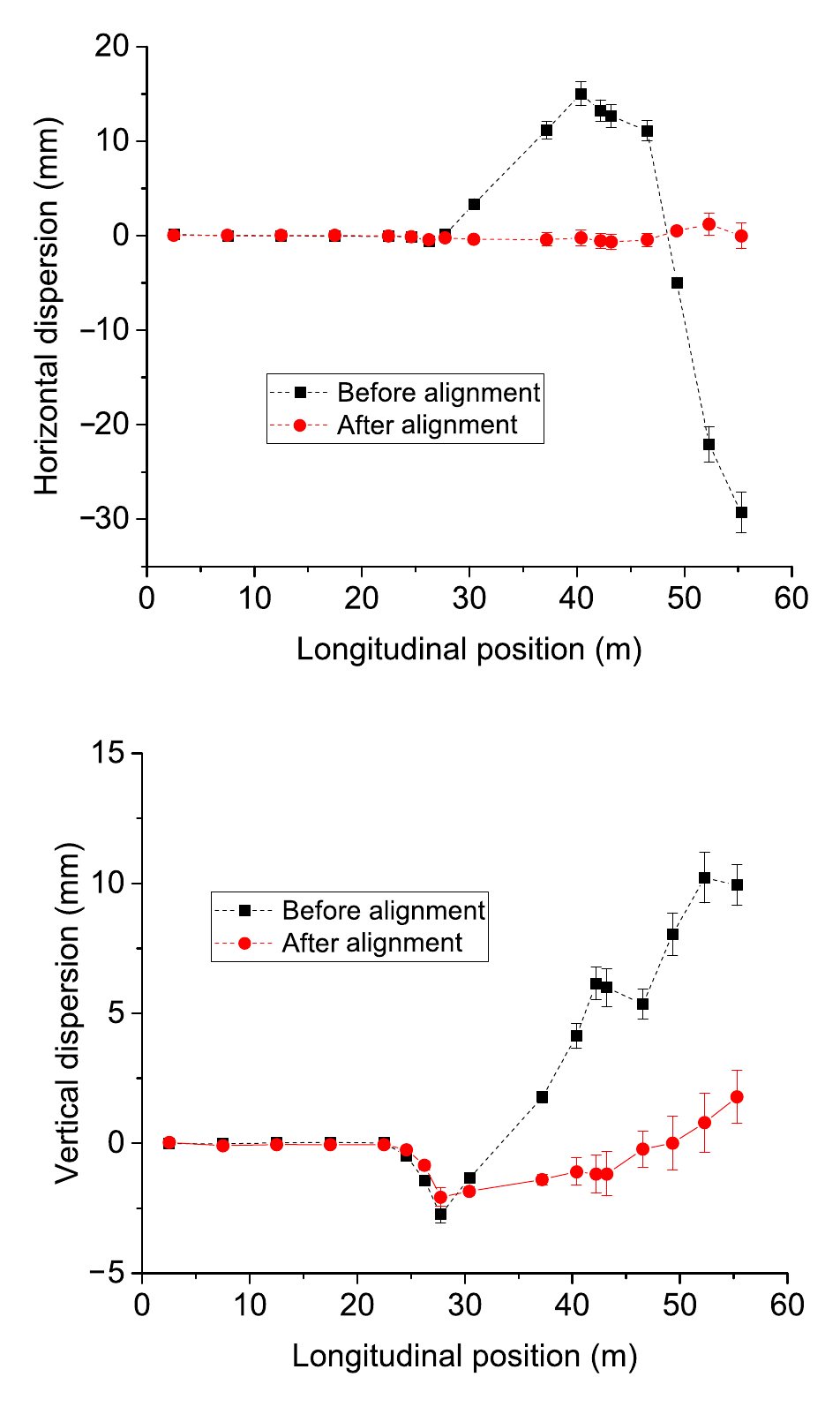}
    \caption{Measured spurious dispersion in the horizontal (upper panel) and
      vertical plane (lower panel) along the injector before and after the
      beam-based alignment (adjustment of the reference orbit).}
    \label{fig:disp}
\end{figure}

In Fig.~\ref{fig:disp} we show the measured spurious dispersion before and 
after adjustment of the reference orbit.
While the dispersion is reduced to a satisfactory level, the correction is only 
valid for the quadrupole setting at the time of the correction.
Since the matching quadrupoles in the X-band section are frequently adapted
depending on the beam shape at the exit of the S-band section, the BPM offsets
need to be updated whenever a significant change of the quadrupole settings has
occurred.

The X-band structure, positioned between the booster and the bunch compressor
to linearize the longitudinal phase space for optimal compression, can be 
aligned actively.
Similar to the gun solenoid, it is mounted on a remotely controlled movable 
platform with four degrees of freedom.
The position of the structure can be varied within a range of about $\pm$2~mm
in both transverse dimensions. 
The optimal position and orientation of the structure---relative to the 
reference orbit as determined by dispersion minimization---are found 
empirically by minimizing the measured emittance of the uncompressed beam.
In addition, the wake field monitors described in Sec.~\ref{sec:wfm} may be 
used to find the optimal orbit through the structure.
The two methods are found to agree well within errors.

\subsubsection{\label{sec:orbit}Orbit response}

The orbit response matrix relates changes in corrector magnet strengths to the
resulting changes in orbit measured at certain positions (the locations of the 
BPMs).
A measurement of the orbit response matrix can be used to calibrate quadrupole 
and corrector magnet strengths.
In contrast to circular accelerators, where a well determined rectangular 
matrix results in accurate values~\cite{Saf97}, the calibration in beam 
transport lines, such as linear accelerators, is compromised by the changing 
beam energy and by the limited information available for the first elements,
i.e., the matrix is only triangular due to the constraints of causality.
Nevertheless a measurement of the orbit response matrix still provides useful
information.

By comparing the measured orbit response matrix to that obtained from an 
ideal optics model, we gain an approximate verification of the quadrupole and 
corrector strengths, thereby confirming the field measurements performed before 
installation to within a few percent.
The intrinsic accuracy of the field measurements is, however, much higher, on 
the order of 0.1\%.
The orbit response measurement therefore primarily serves as a means to uncover
installation errors.

The response matrix is also applied for the correction of beam trajectory 
errors.
Either matrices derived from the optics model or from a response measurement
may be used, but the measured response matrix usually results in a better
performance, since it takes into account small empirical differences such as
BPM calibrations errors.

\subsubsection{\label{sec:bc-char}Bunch compressor characterization}

The magnetic chicane is characterized by both transverse and longitudinal
dispersions ($R_{56}$). 
The transverse dispersion is monitored by the two BPMs installed in either 
arm of the chicane (see next section).
The longitudinal dispersion, the most relevant characteristic of the bunch 
compressor and also called $R_{56}$ for the corresponding element of the 
transfer matrix, was verified by two independent methods.
In both methods, the change $\Delta t$ in arrival time in response to a small 
change $\Delta p$ in beam momentum is observed after the compression chicane.
Depending on the $R_{56}$ of the chicane, bunches with higher energy will 
arrive earlier than bunches with lower energy, since they pass the chicane
on a shorter path, according to
\begin{equation*}
 c \Delta t = R_{56} \frac{\Delta p}{p} ,
\end{equation*}
where $c$ is the speed of light.
The first, destructive method uses the temporal modulation of the transverse
beam position by the transverse deflecting cavity, measured by a BPM downstream
of the cavity, to derive $\Delta t$.
In the second, nondestructive method, the change in arrival time is directly 
recorded by the BAM described in Sec.~\ref{sec:bam}.

\begin{figure}[tb]    
  \includegraphics*[width=1\linewidth]{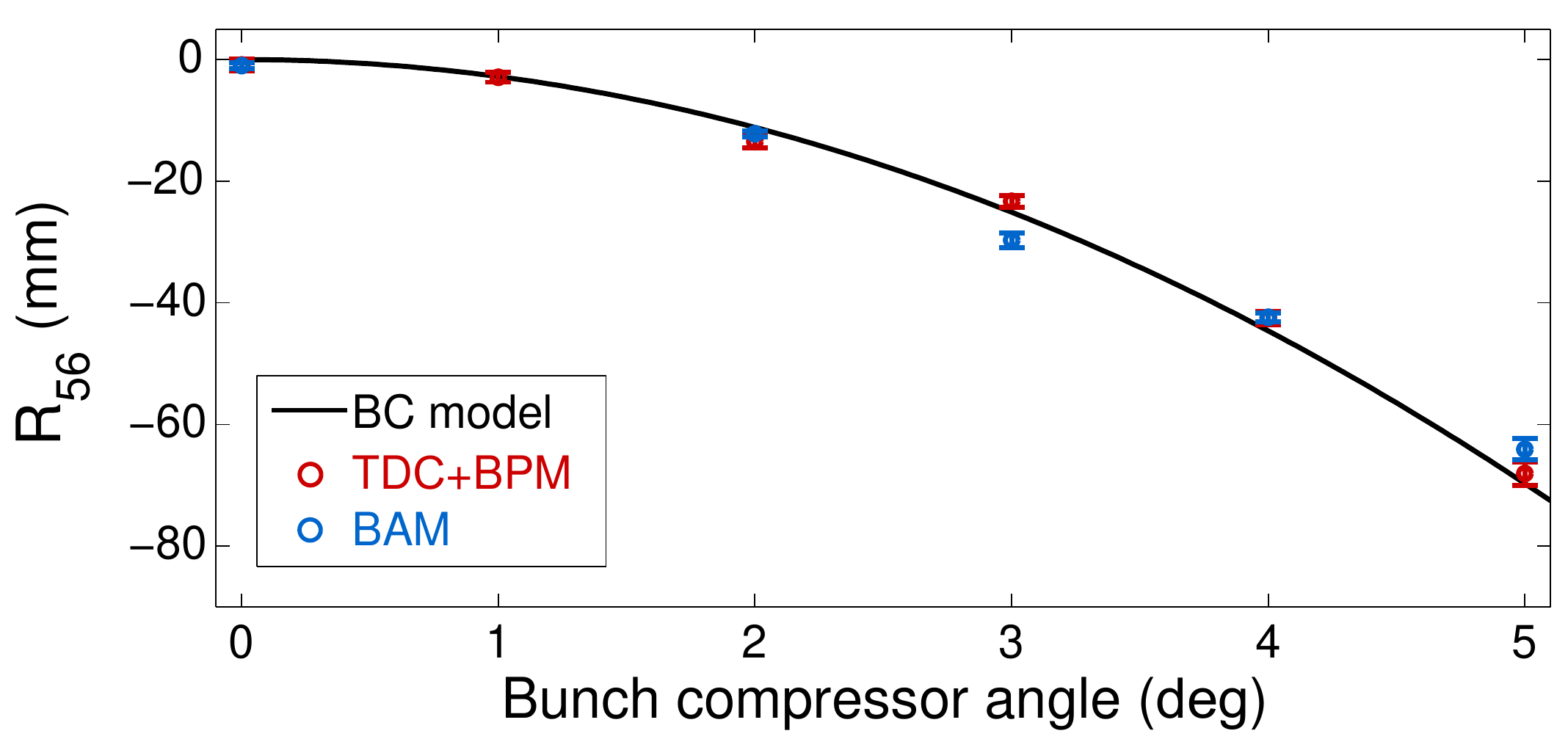}
  \caption{Comparison of measurements of the $R_{56}$ parameter using the 
    transverse deflecting cavity (TDC) and a downstream BPM (red points) and
    via bunch arrival-time using a BAM (blue points), for different bunch
    compressor angles.} 
  \label{fig:r56}
\end{figure}

Figure~\ref{fig:r56} shows the results of the two measurements, for various
bunch compressor angles between 0$^\circ$ and 5$^\circ$, compared to model 
expectations.
The measurement was performed at a bunch charge of 45~pC and a beam energy of
about 200~MeV, varied within a range of $\pm$1.5\% by adjusting the rf power
in the last two S-band structures.
All booster structures were operated on-crest for this measurement, i.e., the 
bunches were not compressed in the chicane.

\subsubsection{\label{sec:bc-geomag}Geomagnetic field compensation} 

The bunch compressor main dipoles are equipped with individually powered 
correction coils, designed to compensate possible differences in field 
integral between the four magnets (see Sec.~\ref{sec:bc}).
Since the main dipoles did not require much correction, we utilize the corrector
dipoles also for the compensation of the geomagnetic field, which is not 
negligible at our beam energy.
The SITF is almost north-south oriented, such that the effect is mostly in the 
horizontal plane.

\begin{figure}[b]    
  \includegraphics*[width=0.9\linewidth]{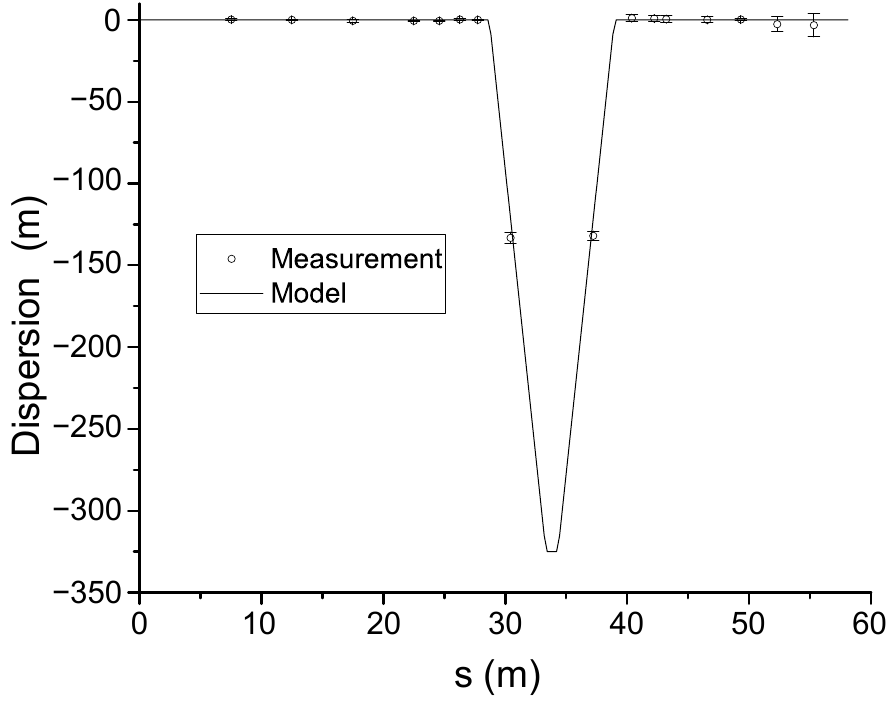}
    \caption{Measured and modeled dispersion along the injector, with the
      bunch compressor angle set to 4.0$^\circ$.}
    \label{fig:BCdisp}
\end{figure}

Measuring the geomagnetic field  with beam we find a vertical field component
of about 30~\textmu T. 
This results in a trajectory offset of about 0.5~mm at the BPMs situated in 
the bunch compressor arms. 
By setting the correction coil currents to values such that the additional 
integrated field corresponds to the integrated vertical geomagnetic field we
reduce the trajectory offset to almost zero.
To first order, the compensation is valid for any beam energy, since we directly
compensate one field by another.
The integrated geomagnetic field along the bunch compressor is about 
3 $\times$ 10$^{-4}$~Tm whereas the field differences between the magnets amount
to about 1.5 $\times$ 10$^{-4}$~Tm~\cite{Neg12}.
The dispersion chicane is almost perfectly closed once the correction is applied
as shown in Fig.~\ref{fig:BCdisp}.

\subsection{\label{sec:bd-trans}Transverse beam characterization}

One of the prime objectives of the SITF consisted in the demonstration of the
beam parameters needed for driving the SwissFEL linac and undulator lines to
reach the desired radiation power at short wavelengths.
A key parameter in this context is the transverse emittance of the electron
beam, as it has a strong influence on the FEL power that can be reached with a
given undulator length~\cite{Pra14}.
Consequently, a substantial fraction of the operational life of the SITF
was devoted to the characterization and optimization of the electron beam's 
transverse emittance.
In particular the emittance of individual longitudinal bunch slices in the
core of the bunch, the so-called slice emittance, should be kept as small as
possible to enable an optimal lasing process in the undulator section.
The projected emittance, i.e., the emittance of the whole bunch, should also be
reasonably small to prevent the deterioration of the FEL performance
due to transverse offsets between slices and optics mismatch along the bunch.
Moreover, if the projected emittance is of the same order as the slice 
emittance, the optics mismatch between the core and projected beams is reduced. 
Therefore, the procedure for the overall optics matching is simplified, as in 
this case it is sufficient to consider the projected parameters.

We measure both projected and slice emittance by means of beam-optical 
methods~\cite{Min03}, whereby the moments of the phase-space distributions are 
determined from beam-size measurements under varying phase advance between a 
reference and an observation point, the latter given by the measurement screen.
The measured rms beam sizes $\sigma_{x,i}$ ($i = 1,\ldots,N$) depend both on the 
respective beam transport matrices $R^i$ as well as on the initial beam moments 
$\left<x_0^2\right>$, 
$\left<x_0'^2\right>$, and 
$\left<x_0x_0'\right>$
at the reference point according to
\begin{equation}
\sigma_{x,i}^2 = {R^i_{11}}^2\left<x_0^2\right> +
             {R^i_{12}}^2\left<x_0'^2\right> +
             2R^i_{11}R^i_{12}\left<x_0x_0'\right>,  \label{eq:beamtrans}
\end{equation}
where $x$ and $x' = dx/ds$, referring to transverse coordinate and trajectory 
slope, respectively, are defined for the horizontal plane, but an equivalent 
formalism is valid for the vertical plane ($y$ and $y'$). 
The subscript $0$ denotes the reference point.
It follows that a sufficiently large set of beam size measurements $\sigma_i$ 
obtained under known transport conditions $R^i$ allows the reconstruction of 
the beam moments.
While in principle three such measurements would be sufficient to reconstruct
the beam moments, in practice we perform many more measurements and let a 
numerical least-square fit find the beam moments that best describe the
observations, to improve the accuracy of the reconstruction. 
From the beam moments, it is straightforward to derive the geometric rms 
emittance $\varepsilon_x$ as 
\begin{equation}
  \varepsilon_x  = \sqrt{
                   \left<x_0^2\right> 
                   \left<x_0'^2\right> - 
                   \left<x_0x_0'\right>^{2} } , 
\end{equation}
which may be normalized with respect to the beam energy according to
\begin{equation}
\varepsilon_{n,x} = \frac{p}{m_e c}\varepsilon_x ,
\end{equation}
with $p$ the beam momentum, $m_e$ the electron mass and $c$ the speed of 
light. 
Similarly, the Twiss parameters at the reference point are obtained as
$\alpha_{x,0} =  -\left<x_0x_0'\right> / \varepsilon_x$, 
$\beta_{x,0} =  \left<x_0^2\right> /\varepsilon_x$, and 
$\gamma_{x,0} =  \left<x_0'^2\right> / \varepsilon_x$.

We use two methods to vary the phase advance between the reference and the 
observation point.
In the first method, mainly used for the routine measurement of projected
emittance, a single quadrupole is used in the same scan to generate phase 
advance simultaneously and equivalently in the horizontal ($x$) and the 
vertical ($y$) plane.
This so-called symmetric single-quadrupole scan~\cite{Pra14a} requires a 
particular symmetric beam optics (identical in both planes) at the scanning 
quadrupole with $\beta$ equal to $\alpha$ times the distance $L$ between 
the quadrupole and the observing screen, i.e., 
$\beta_x=\beta_y=\beta_0 = \alpha_0 L$ with $\alpha_0=\alpha_x=\alpha_y$.
Under these conditions the beam waist occurs in both planes at the location of
the screen for zero gradient field in the quadrupole.
In the thin-lens approximation the minimum $\beta$ is given by 
$\beta_\text{min} = L^2/\beta_0$.
In our setup we choose $L$ = 10.83~m and $\beta_0$ = 15~m, therefore 
$\alpha_0$ = 1.385 and $\beta_\text{min}$ = 7.82~m.
Thanks to its symmetry this novel method has proven to be fast and intuitive
and thus has quickly become the standard method for routine emittance 
measurements at the SITF.

If greater freedom in the control over the measurement optics is required, as 
is the case in time-resolved measurements with a streaked beam  
(Sec.~\ref{sec:bd-long}), or the determination of the full set of transverse 
beam parameters including correlations between the two planes (4D beam matrix)
is desired~\cite{Pra14b}, then the use of several quadrupoles~\cite{Ten97} to 
control the phase advance independently in two planes, while fulfilling 
additional constraints, e.g., on the $\beta$-functions, is inevitable.
For this more elaborate method we use three to five quadrupoles downstream of 
the compression chicane to set up the different beam transport schemes required
for the beam optics measurements set out in Eq.~\ref{eq:beamtrans}.

In both methods, the beam originating from the booster should be matched to the 
optics required for the measurement to minimize the reconstruction errors.
This is done by setting the currents of a set of matching quadrupoles upstream
of the bunch compressor to appropriate values, normally obtained from a 
matching tool based on a suitable online machine model.
The quality of the optics matching is evaluated from a comparison between the 
measured Twiss parameters $\alpha$, $\beta$, and $\gamma$ and the corresponding
design values $\alpha_D$, $\beta_D$, and $\gamma_D$, usually quantified by the 
mismatch parameter~\cite{Min03}, defined as 
$\xi = (\beta_D\gamma - 2\alpha_D\alpha + \gamma_D\beta)/2$ and approaching
1 for a well matched beam.
Optics measurement and matching are iterated until a satisfactory result is
obtained. 
As a rule, we only consider emittance measurements to be valid if the mismatch 
parameter is below 1.1. 

For most of our emittance measurements we have used the novel profile monitor
described in Sec.~\ref{sec:profimg} to obtain the transverse beam sizes 
necessary for the evaluation based on Eq.~\ref{eq:beamtrans}.
We thereby approximate the rms width of a given beam profile in one dimension
by the rms width (standard deviation) of a Gaussian function fitted to the 
corresponding projection of the background-subtracted beam image. 
This fast and robust approach has proven to be fully adequate in the context
of the optimization of an FEL driver, where small contributions from far 
outlying tails in the beam profiles are not relevant.
More elaborate rms methods applied to appropriately noise-filtered beam 
images (see Refs.~\cite{Beu11, Beu10, Beu09}) were found to give no significant
advantages.

While the accurate absolute measurement of beam sizes is essential for the 
derivation of the beam emittance, it is not imperative for the determination
of the beam optics (Twiss) parameters.
Indeed, since the Twiss parameters $\alpha$, $\beta$, and $\gamma$ do not
explicitly depend on the absolute beam size but only on its variation along the 
beam line, it is possible to derive them from relative beam size measurements 
alone, in particular from the identification of beam waists.
A corresponding method was developed and successfully tested at the 
SITF~\cite{Aib14}.
It proved useful during the earliest beam commissioning stages when the profile 
monitors were still subject to calibration.

\subsection{\label{sec:bd-long}Longitudinal beam characterization and 
time-resolved measurements}

The experimental verification of short bunch lengths as well as the measurement
of key beam parameters (optics, emittance, energy and energy spread), as a 
function of the longitudinal position along the bunch, are of fundamental 
importance for the development of accelerators applied as FEL drivers.
At the SITF our main method for bunch length and time-resolved measurements
was based on the well established rf streaking technique~\cite{Akr01,Akr02} 
using a transverse rf deflector (see Sec.~\ref{sec:rf-tds}) operating at the 
same rf frequency as the booster linac and at the zero-crossing phase of the 
field to achieve a strong correlation between the longitudinal (time) and 
vertical coordinates along the bunch.
This correlation then allows for the measurement of the current profile along
the bunch as well as the horizontal emittance of individual slices using 
regular transverse profile monitors (see Secs.~\ref{sec:scrmon} and 
\ref{sec:trpromon}).

Similarly to the transverse beam size, we typically determine the rms bunch 
length from a Gauss fit to the background subtracted vertical profile of the 
streaked beam, although other methods for estimating the width of the 
distribution, e.g., based on the full-width at half-maximum or rms values,
may also be applied. 
The relation between vertical distance on the screen and time is derived from 
a comparison of different beam images, obtained at slightly different rf 
phases of the S-band deflecting cavity, for which 1$^\circ$ phase change amounts 
to a time difference of 927~fs.

\begin{figure*}[tb]    
  \includegraphics*[width=1\linewidth]{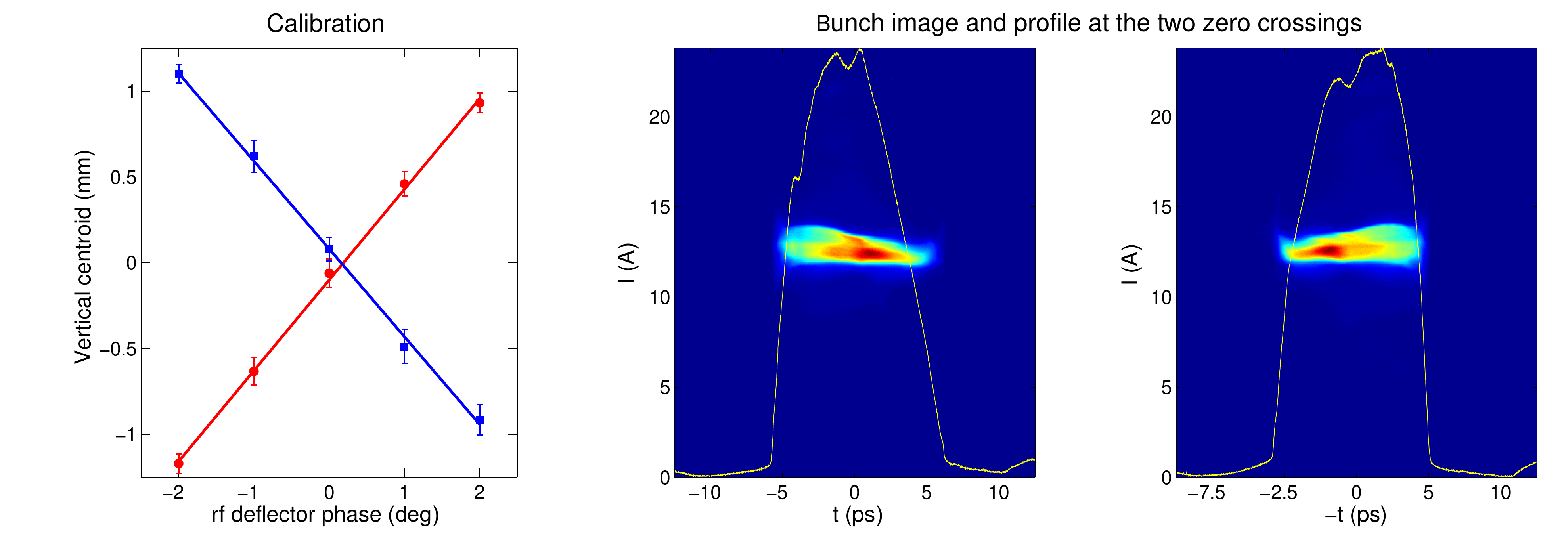}
  \caption{Example of a bunch length measurement with uncompressed beam, 
    showing the calibration measurement (left plot) and the current profiles
    overlaid on the images of the streaked beam, as observed at the two
    zero crossings of the rf deflector (right plots).}
   \label{fig:bunchlength}
\end{figure*}

The unstreaked beam can feature a correlation between the vertical and time 
coordinates, which can be caused, for instance, by wakefields in the 
accelerating structures. 
The reconstructed bunch length would be biased for an initially tilted bunch if 
the measurement is only performed at one zero crossing of the deflecting rf
field.  
We overcome this problem by performing the measurement at both zero crossings 
and comparing the two measurements following the procedure described in 
Ref.~\cite{Akr02}. 
In addition to the unbiased bunch length, this method also provides an estimate
of the incoming beam tilt. 
Figure~\ref{fig:bunchlength} shows an example of a bunch length measurement for
an uncompressed beam.

The combination of vertical streaking and horizontal dispersion, in our case
in the high-energy spectrometer, allows for a complete mapping of the 
longitudinal phase space (time and energy).
The visualization of the longitudinal phase space in real time constitutes
a valuable diagnostic when setting up the compression of the beam
(see Sec.~\ref{sec:bd-comp}).

For the measurement of the optical parameters and emittance of individual
slices along the bunch a rectangular region of interest is drawn around the
streaked beam and subdivided into horizontal bands corresponding to the bunch
slices.
Typically 10 slices per rms bunch length are chosen, but this number may vary
depending on the beam characteristics or measurement purpose.
During optics scans, it is important to associate measurements of the same
physical slice along the bunch with one another, regardless of fluctuations
in beam arrival time or deflector rf phase.
This requirement is satisfied by a slicing algorithm relying on beam-based 
reference points, see Ref.~\cite{Pra14} for details.

Once the beam is resolved into individual slices, the same optics scanning 
procedure for the determination of the optical parameters and the beam 
emittance described in the previous section may be applied.
In this case, the quadrupole settings used to provide the necessary phase 
advance for the emittance measurement, while controlling the beam size on the
observation screen, are further adjusted to keep the longitudinal (slice) 
resolution approximately constant during the scan.
As in the case for time projected measurements, we again aim for mismatch 
parameters---now determined individually for each slice---smaller than 1.1, at
least in the relevant central part of the bunch.

Instead of using an rf deflector, the correlation between time and 
transverse position may also be obtained by a combination of transverse 
dispersion and energy chirping~\cite{Qiu96,Dow03}.
This alternative method was further developed and generalized at the 
SITF~\cite{Pra14c}.
It has the distinctive advantages of enabling the measurement of slice 
parameters in both the vertical and the horizontal planes and anywhere along
the machine where dispersion can be generated.
The method has the limitations that it requires absolute calibration and an
energy chirp along the bunch, which precludes its use in conjunction with a
spectrometer to map the bunch's longitudinal phase space in general.

\section{\label{sec:bd-results}Beam physics results}

As one of the principal objectives of the SITF consisted in the demonstration 
of a high-brightness beam capable of driving a SASE FEL, the beam development
program focused on minimizing the transverse beam emittance at the source and 
on preserving the small emittance under beam transport, acceleration, and, in 
particular, compression.

The optimum matching of electron and photon beam resulting in transversely
coherent FEL radiation in the undulator line imposes the condition
$\varepsilon_n \lesssim \gamma\lambda/4\pi$~\cite{Kim86}
on the normalized transverse emittance, $\varepsilon_n$, where $\gamma$ is the
electron beam's Lorentz factor and $\lambda$ the FEL radiation wavelength.
Therefore the final beam energy required to drive an FEL at a given radiation 
wavelength is dictated by the transverse emittance that can be reached in the
rf photoinjector and preserved during beam transport and compression.
Beyond the beam energy requirement, a smaller emittance also results in higher
radiation power and allows FEL saturation to be reached with a shorter undulator
beam line~\cite{Pra14}.

In this section we describe our various efforts to optimize the beam emittance
originating from the the source (Sec.~\ref{sec:bd-source}), as well as further 
down the accelerator beam line, both for uncompressed 
(Sec.~\ref{sec:bd-uncomp}) and compressed beams (Sec.~\ref{sec:bd-comp}),
and summarize the beam parameters we were able to achieve with the SITF.

\subsection{\label{sec:bd-source}Source characterization}

The electron emission process at the gun cathode sets a fundamental lower 
limit on the achievable beam emittance and thus has a strong influence on the
final emittance of rf photoinjectors.
For instance, simulations performed for SwissFEL indicate that for an optimized
machine, around 70\% of the final emittance is due to the emittance related
to the cathode, also called intrinsic emittance.
At the SITF much effort went into characterizing the intrinsic emittance
of cathodes and finding ways to minimize it.

The intrinsic emittance of the cathode is related to the initial kinetic energy
or effective temperature of the electrons emitted from the cathode.
For laser-illuminated photocathodes, made of metal or semiconductor, 
the intrinsic emittance may be expressed as~\cite{Dow09, Flo97, Dow10}
\begin{equation}
\varepsilon_\text{int} = \sigma_l\sqrt{\frac{2E_{K}}{3m_ec^2}},
\label{eq_th}
\end{equation}
where $ \sigma_l$ is the rms laser beam size,  $E_K$ the average kinetic 
energy of the photo-emitted electrons, and $m_ec^2$ the electron's rest mass
energy. 
The ratio $\varepsilon_\text{int} / \sigma_l$ is independent of the laser beam
size and may be called the specific intrinsic emittance.
The electrons' average kinetic energy $E_K$ depends on the laser photon energy
$\phi_l$, material properties and the applied electric field.
It takes on a different form for metals and semiconductors.
For metal cathodes we have
\begin{equation}
2E_{K} = \phi_l-\phi_w+\phi_\text{Sch}, 
\end{equation}
with the work function $\phi_w$, whereas for semiconductors we have
\begin{equation}
2E_{K} = \phi_l-E_g-E_a+\phi_\text{Sch},
\end{equation}
where $E_g$ and $E_a$ refer to gap energy and electron affinity, respectively.
In both cases the kinetic energy is enhanced by the reduction of the potential
barrier due to the applied electric field, the so-called Schottky effect.
This contribution, denoted as $\phi_\text{Sch}$, is given by~\cite{Yus04}  
\begin{equation}
\phi_\text{Sch} =  \sqrt{\frac{e^3}{4\pi\varepsilon_0}\beta E_c},
\end{equation}
where $e$ is the electron charge, $\varepsilon_0$ the vacuum 
permittivity, $\beta$ the local field enhancement factor accounting for cathode
surface effects, and $E_c$ the rf field on the cathode applied during the 
extraction of the electrons. 

Apart from the cathode surface quality (described by $\beta$),
the intrinsic emittance of a photoinjector is thus influenced by three main 
parameters under our control:
\begin{itemize}
\item
the photon energy, given by the wavelength of the gun laser, affecting
directly the energy of the photo-emitted electrons,
\item 
the applied field on the cathode, determining the size of the Schottky effect,
and
\item 
the cathode material, specifying the work function in the case of a metal or the
gap energy and the electron affinity in the case of a semiconductor.
\end{itemize}
At the SITF we have performed experimental studies on all three relationships:
we confirmed, by using the wavelength-tunable Ti:sapphire laser system described
in Sec.~\ref{sec:laser}, that the intrinsic emittance decreases with laser 
wavelength~\cite{Div15a}; 
we observed the reduction of intrinsic emittance as a function of the applied
rf gradient~\cite{Pra15a};
and we compared the intrinsic emittances and general operational performance of 
metal (copper) and semiconductor (cesium-telluride) cathodes~\cite{Pra15}.

In all these studies we extract the specific intrinsic emittance
$\varepsilon_\text{int}/\sigma_l$ from measurements of the emittance at the final
booster energy as a function of the laser beam size on the cathode, controlled 
by adjusting the size of a circular laser aperture.
Special care must be taken to minimize or subtract possible contributions to
the emittance arising after the emission at the cathode, e.g., from space-charge
and rf-field effects.

To eliminate space-charge effects, the measurements are performed at very low 
bunch charges, well below the space-charge limit as confirmed experimentally.
During a laser-beam-size scan, the bunch charge is adjusted to keep the surface
charge density constant.
Possible chromatic and dispersive effects along the bunch are minimized by 
keeping the electron energy approximately constant along the bunch by means of
on-crest acceleration in the booster.
Furthermore, to avoid a possible impact on the projected emittance from the gun
rf field variation experienced by the bunch along its longitudinal extension, 
we only consider the slice emittance at the core of the bunch.
This core slice emittance is obtained by dividing the bunch into typically ten
slices per rms bunch length and averaging the measured slice emittance over an 
appropriate number of slices around the longitudinal center of the bunch.

\begin{figure}[tb]    
  \includegraphics*[width=1\linewidth]{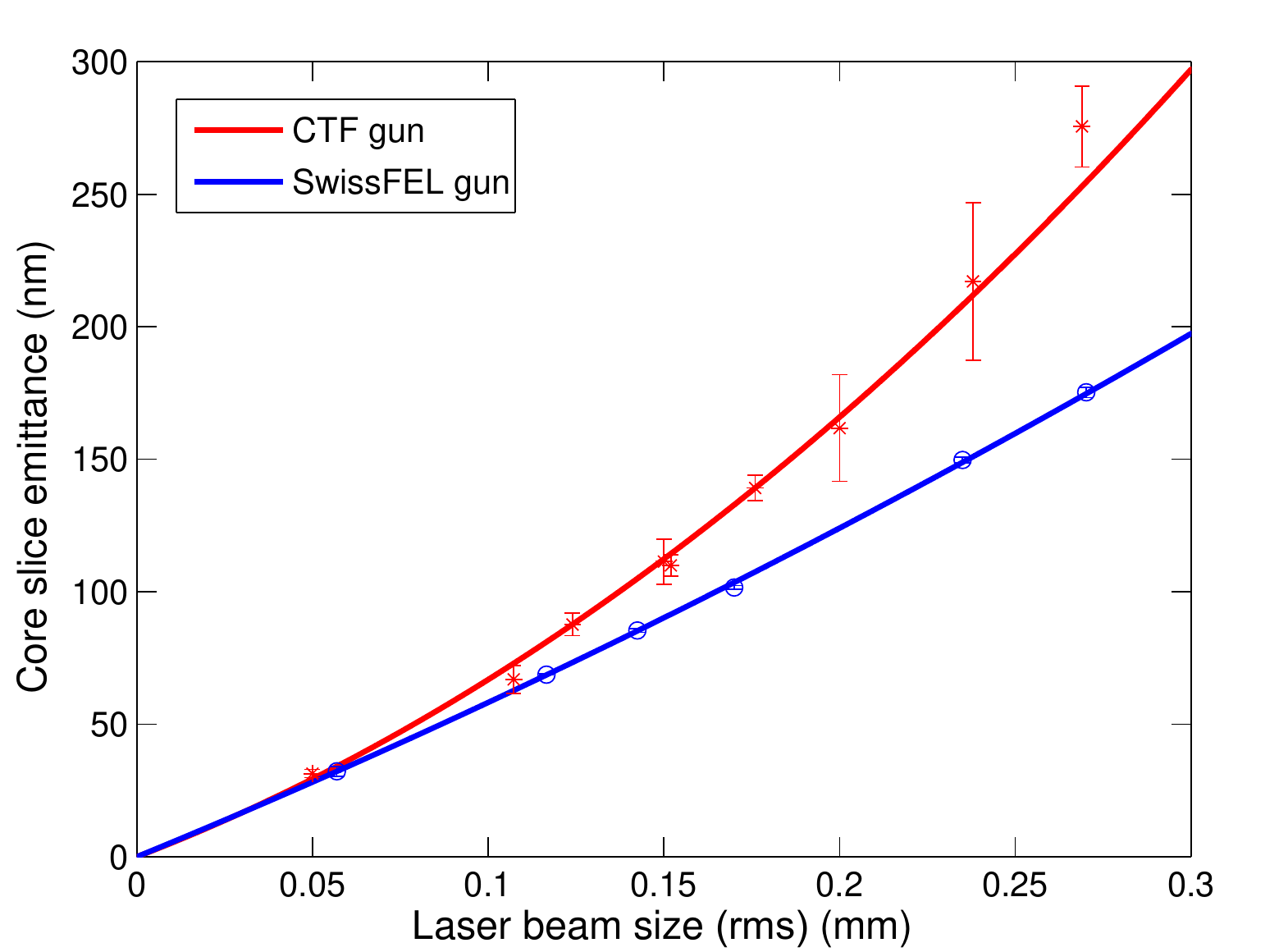}
  \caption{Two examples of aperture scans showing core slice emittance measured
   for various laser spot sizes with the CTF gun (red asterisks) and the 
   SwissFEL gun (blue circles).
   The dependence observed for the CTF gun features a distinct quadratic 
   component (see text for details).
   In both cases the electrons were extracted from a copper cathode.}
   \label{fig:apert-scan}
\end{figure}

Most of our intrinsic emittance measurements were performed with the CTF gun
(see Sec.~\ref{sec:rf-gun}), which features a symmetric rf feed design to
avoid dipolar kicks, but has no compensation of quadrupolar fields acting on
the electron beam.
These quadrupolar fields lead to a quadratic rise of the emittance as a 
function of the laser spot size.
We take this additional component into account by fitting a quadratic function
to the measurements, thereby obtaining the specific intrinsic emittance  as the
linear coefficient of the fit function.
Additional measurements performed with the new SwissFEL gun show that the 
quadratic effect is less pronounced in this case.
In Fig.~\ref{fig:apert-scan} we compare two examples of aperture scans performed
with the two rf guns.

The most relevant results of our systematic emittance studies with copper
cathodes are that the intrinsic emittance of such cathodes can indeed be 
reduced by about 20\% when tuning the laser wavelength from 262~nm to 
276~nm~\cite{Div15a} and that a similar reduction can be achieved by lowering 
the rf gradient at the extraction phase from 49.9~MV/m to 
16.4~MV/m~\cite{Pra15a}.
The wavelength tuning measurements confirm earlier studies~\cite{Hau10} 
carried out with a combined diode rf electron gun~\cite{Gan10}.

These observed reductions apply to individual cathodes measured under the same 
operating conditions within a relatively short time.
Measuring various cathodes at different times with respect to their 
installations we found that variations in intrinsic emittance between different
cathodes, originating, e.g., from differences in surface quality, as well as 
degradation effects are also in the 20\% range.
Therefore such effects can easily outweigh any benefits gained from 
laser-wavelength or rf-gradient tunings, which, however, come at the expense of
significant losses in quantum efficiency.

Similarly, our comparison between copper and cesium telluride cathodes resulted 
in emittance values for the two types of cathodes that lie in the same 
range~\cite{Pra15}:
for both cathode materials we obtained values between about 0.4~\textmu m/mm 
and 0.6~\textmu m/mm for the specific intrinsic emittance.
Earlier measurements performed on copper cathodes in the context of the
development of the LCLS at SLAC generally gave higher 
values~\cite{Gra01,Sch04,Din09,Qia12}, whereas a previous PSI experiment
based on a combined diode rf electron gun resulted in a specific intrinsic 
emittance of around 0.5~\textmu m/mm for this material~\cite{Hau10,Gan10}.
In the case of cesium telluride, experiments at the photoinjector test facility
at DESY, Zeuthen (PITZ), also indicated a higher specific intrinsic 
emittance~\cite{Ste10}, whereas estimates based on the observed electron 
emission spectrum~\cite{Ser04} agree with our measurements.

Some of the laser-beam-size scans involved emittance measurements at extremely
low bunch charges.
For instance, the measurement with the smallest laser aperture at the lowest
rf gradient was done on a bunch with only 30~fC charge.
The beam slices, for which individual emittances were determined in this 
measurement, contain around 1~fC of charge; in other words they are made up of
a few thousand electrons only.
The measured normalized core slice emittance in this case was slightly below 
25~nm~\cite{Pra15a}.

Given our experimental results on intrinsic emittance, the options of tuning
the wavelength of the gun laser or the rf gradient of the photoinjector
gun will not be pursued further at SwissFEL.
Moreover the baseline cathode material for the SwissFEL injector has been 
changed from copper to cesium telluride, since it provides a quantum efficiency 
about two orders of magnitude higher at comparable intrinsic emittance (see 
also Sec.~\ref{sec:cath}).
The copper option will be maintained nevertheless, for the case of major 
problems with the production or operation of the cesium telluride cathodes.

\subsection{\label{sec:bd-uncomp}Optimization of uncompressed beam}

Apart from the mitigation of emittance dilution effects, the optimization of 
an rf photoinjector involves trade-offs between the contributions to the beam 
emittance related to the intrinsic emittance at the cathode, space charge and 
rf fields.
To minimize the emittance at the SITF, we first performed a numerical 
optimization followed by a full empirical optimization of the actual machine.

The numerical optimization is based on simulations using the Astra 
code~\cite{ASTRA} coupled to a dedicated optimizer developed for 
SwissFEL~\cite{Bet15}.
The optimizer minimizes both the emittance and the mismatch along the central 
part of the bunch (75\% of the total length) at the exit of the second 
booster structure, via the construction of an appropriate figure of merit.
The main parameters to be optimized are the distance between the gun and the
first booster structure, the transverse laser spot size and the gun solenoid
focusing strength.
In the optimization procedure, as applied to the SITF design, the gun gradient 
is kept at its nominal value (see Sec.~\ref{sec:rf-gun}) and also the bunch 
length at the cathode (determined by the laser pulse length) is fixed to the 
original design values (FWHM) of 9.9~ps for the 200~pC bunch and 3.7~ps for the 
10~pC bunch~\cite{Kim08,SITF}.
One of the most important consequences of the numerical optimization was an 
increase of the distance between the gun and the first booster structure by 
about 20~cm~\cite{Bet15} with respect to the original design, which was based 
on the Ferrario working point~\cite{Fer00}.

The laser spot size determined by the numerical optimization for the 200~pC case
serves as a starting point for the empirical optimization.
The 10~pC bunch is simply reduced in all dimensions to have the same 
three-dimensional charge density.
Based on the simulation results we select the laser aperture that gives a
laser beam size as close as possible to the best value.
Later we verify empirically the optimality of the aperture size.
The longitudinal laser profile is an approximated flat-top profile obtained
by pulse stacking of 32 replicas (Fig.~\ref{fig:pulsarlong} shows the profile 
for 16 replicas).
Further manual optimization steps, after the careful beam-based alignment
of the laser beam on the cathode and the gun solenoid to the gun rf axis, as 
described in Sec.~\ref{sec:bba}, include:
\begin{itemize}
\item
the fine adjustment of the transverse laser profile towards maximum azimuthal
symmetry and homogeneity (see Sec.~\ref{sec:laser});

\item 
the setup of the longitudinal laser profile via pulse stacking (see again 
Sec.~\ref{sec:laser}) to approximate a flat-top distribution, which ensures
that most of the longitudinal beam slices have the same emittance and optics,
thus resulting in a small projected emittance (in contrast to, e.g., a 
Gaussian distribution);

\item
the setting of the gun rf phase and gradient to obtain the smallest beam energy
spread around the 7.1~MeV design energy as observed with the gun spectrometer;

\item 
the empirical optimization of the gun solenoid excitation current to achieve
minimum emittance through invariant envelope matching;

\item 
the systematic correction of cross-plane coupling terms with two corrector
quadrupole magnets (normal and skew) integrated into the gun solenoid and four 
solenoids around the first two S-band accelerating structures~\cite{Pra14b};

\item 
the centering of the orbit in the S-band booster structures, as described in 
Sec.~\ref{sec:bba}, to minimize wakefield contributions to the emittance;

\item 
the empirical correction of spurious dispersion downstream of the booster
affecting the emittance measurement via the generation of local orbit bumps.

\end{itemize}

\begin{figure}[bt]    
  \includegraphics*[width=1\linewidth]{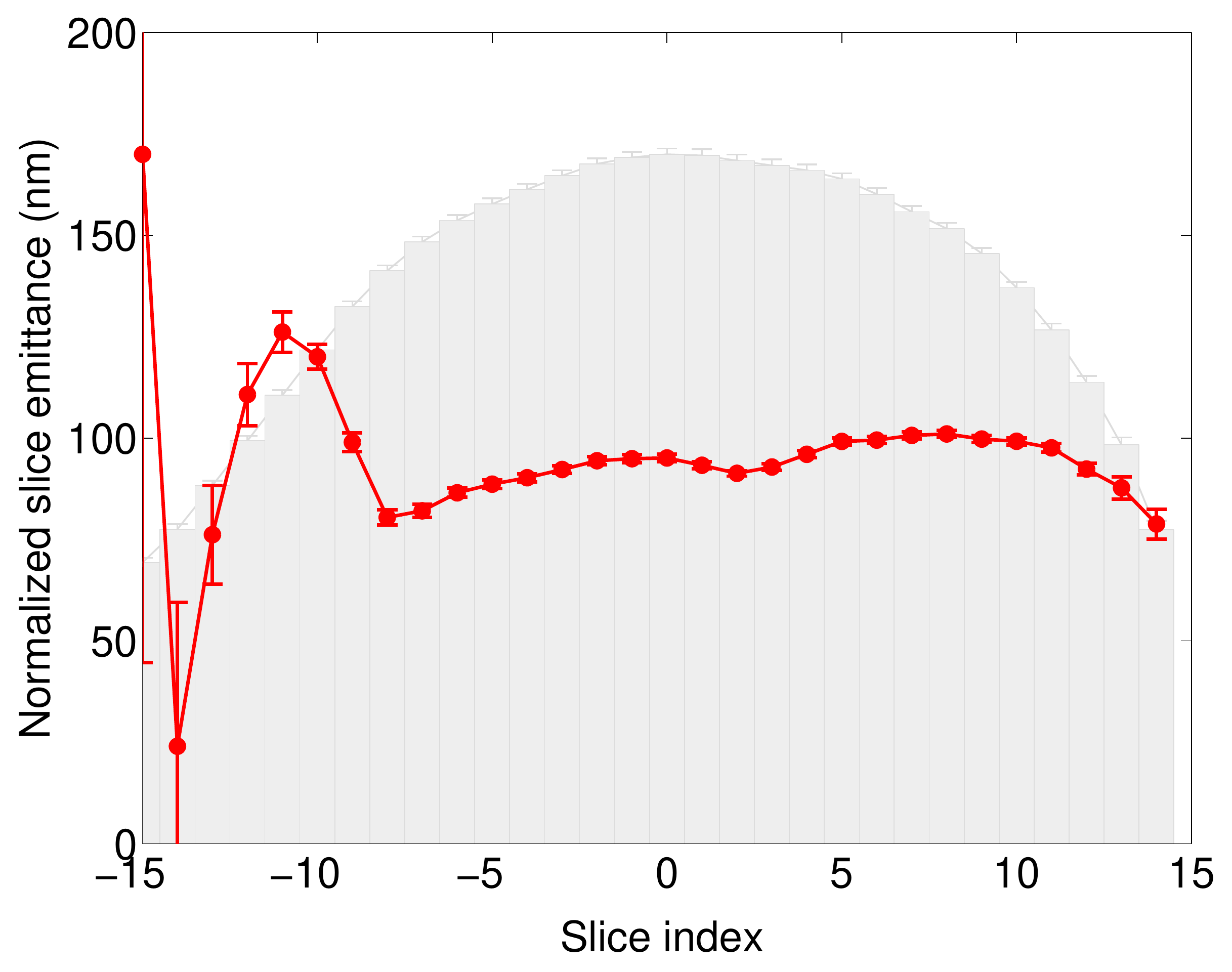}
  \caption{Normalized slice emittance measured with the SwissFEL gun for a 
   bunch charge of 20~pC (red). 
   The longitudinal charge profile is shown as gray bars. }
  \label{fig:20pC}
\end{figure}

\begin{figure}[tb]    
  \includegraphics*[width=1\linewidth]{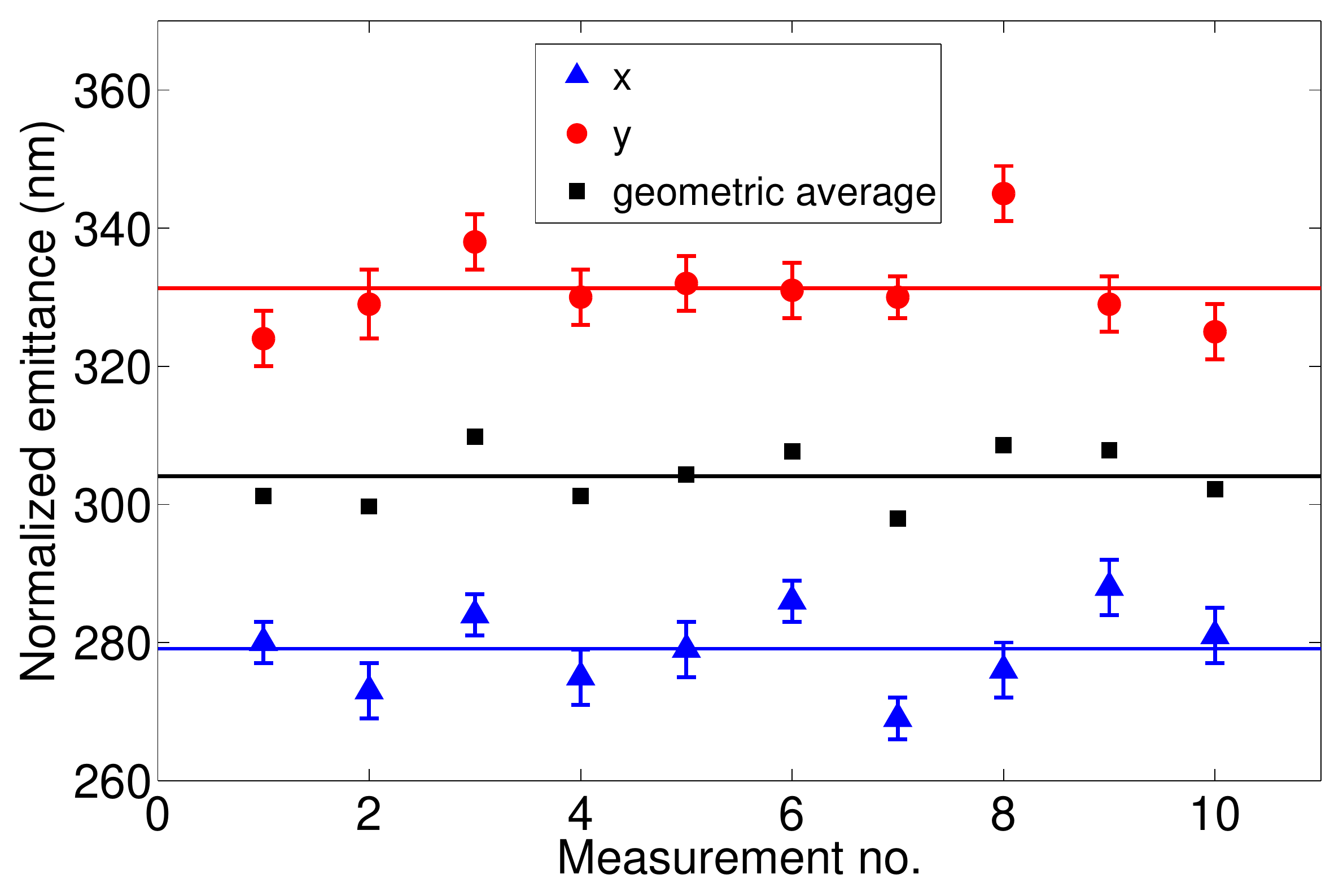}
    \caption{Normalized horizontal (blue) and vertical (red) projected 
      emittance obtained for a bunch charge of 200~pC from ten consecutive 
      measurements with the CTF gun, illustrating the stability of machine and 
      diagnostics chain.
      The geometric mean of the horizontal and vertical results is shown in 
      black.
      The continuous horizontal lines represent the averages over the ten
      measurements.}
    \label{fig:stability}
\end{figure}

After applying this optimization procedure we measured, for the 200~pC bunch
charge case, a normalized projected emittance of about 300~nm and a normalized
slice emittance of about 200~nm.
For a bunch charge of 10~pC, the optimized values were 140~nm and 100~nm for
projected and slice emittance, respectively.
These results, first obtained with the CTF gun, have been confirmed with the 
SwissFEL gun, and in some cases even improved upon.
For instance, during the last operation run in 2014 a slice emittance of around
100~nm was also achieved for a bunch charge of 20~pC.
This measurement is presented in Fig.~\ref{fig:20pC}; measurements at 10 and 
200~pC can be found in Ref.~\cite{Pra14}.
The obtained experimental results for both slice and projected emittance proved
to be stable under repeated measurements, as shown in Fig.~\ref{fig:stability},
and reproducible from scratch as long as the laser quality remains the same.
For a more detailed account of our emittance optimization procedure, as well as
a discussion of measurement resolutions and errors, we refer to 
Ref.~\cite{Pra14}. 

In the optimization procedure described above, the laser pulse length defining
the bunch length is treated as a given, originating from the design of the 
machine~\cite{Kim08,SITF}.
In the last operating months of the SITF, however, we performed a systematic 
study of the effect of the laser pulse length on the emittance both for 
uncompressed and compressed beams.
The study was carried through with the SwissFEL gun using a cesium telluride 
cathode.
Optimizing the beam separately for the design case with a laser pulse length
of 9.9~ps and for two additional cases with laser pulse lengths of 5~ps and 
2.5~ps (FWHM), we found no improvement either in emittance or in brightness 
when going to shorter laser pulse lengths.

\subsection{\label{sec:bd-comp}Bunch compression and emittance preservation}

Bunch compression utilizing path length differences in a dispersive section,
such as a magnetic chicane, has become a well established technique (see, e.g.,
Refs.~\cite{Doh05,Emm13}).
The SwissFEL design foresees two such bunch compressors to achieve final 
rms bunch lengths between 2 and 25~fs, depending on the bunch charge.
The first stage of this compression scheme was tested at the SITF with the
magnetic chicane described in Sec.~\ref{sec:bc}.
Indeed the demonstration of bunch compression with acceptable degradation of 
beam quality represented one of the main goals of the SITF experimental 
program.

\subsubsection{Compression procedure}

Our compression setup procedure is based on an analytical derivation of the
rf parameters needed to arrive at a certain bunch length for a given
configuration ($R_{56}$) of the bunch compressor.
In this procedure the first two booster structures are used for acceleration
only (operating at their on-crest phases), whereas the last two booster
structures are used for acceleration and energy chirping (operating at an
appropriate common off-crest phase).
The higher-harmonic X-band structure is operated at the on-crest phase providing
maximum deceleration to linearize the longitudinal phase space, including the
compensation of second-order terms induced by the bunch compressor.
This structure is not normally used to further modify the energy chirp.

For the case of the SITF, with one bunch compression chicane and without 
strong longitudinal wake-field interactions from C-band linacs, it is possible
to derive an analytic description of the compression setup.
(For SwissFEL with two bunch compressors and a C-band linac, such a 
straightforward approach is no longer applicable and a semi-analytical model
following Ref.~\cite{Zag11} will be adopted.)

In practice, we start the compression setup by determining the incident energy 
before the chirping structures, which is necessary in order to compute the 
appropriate compression parameters for a given target bunch length.
Our derivation of the compression parameters also assumes zero incoming
energy chirp.
We therefore, in a second step, remove a potential incoming energy chirp 
originating from the gun by observing the streaked beam in the final
spectro\-meter in response to small variations in the phase of the second 
booster structure.

To facilitate the ensuing optics matching of the compressed beam, the 
uncompressed beam at the final energy is matched. 
We then generate the required energy chirp by going to the predetermined 
off-crest phase, while simultaneously compensating for the energy 
loss by increasing the amplitude in the last two structures accordingly.
The amplitude of the X-band structure, operating at the on-crest decelerating 
phase, is then ramped up until its induced energy loss matches the value 
expected from the compression scheme, as confirmed directly with the 
spectrometer.
The associated energy loss is again compensated by a corresponding adjustment
in the amplitude of the last two booster structures.
Finally, the optics of the compressed beam are rematched, and the 
compression factor is verified with a bunch-length measurement. 

\begin{figure}[tb]    
  \includegraphics*[width=1\linewidth]{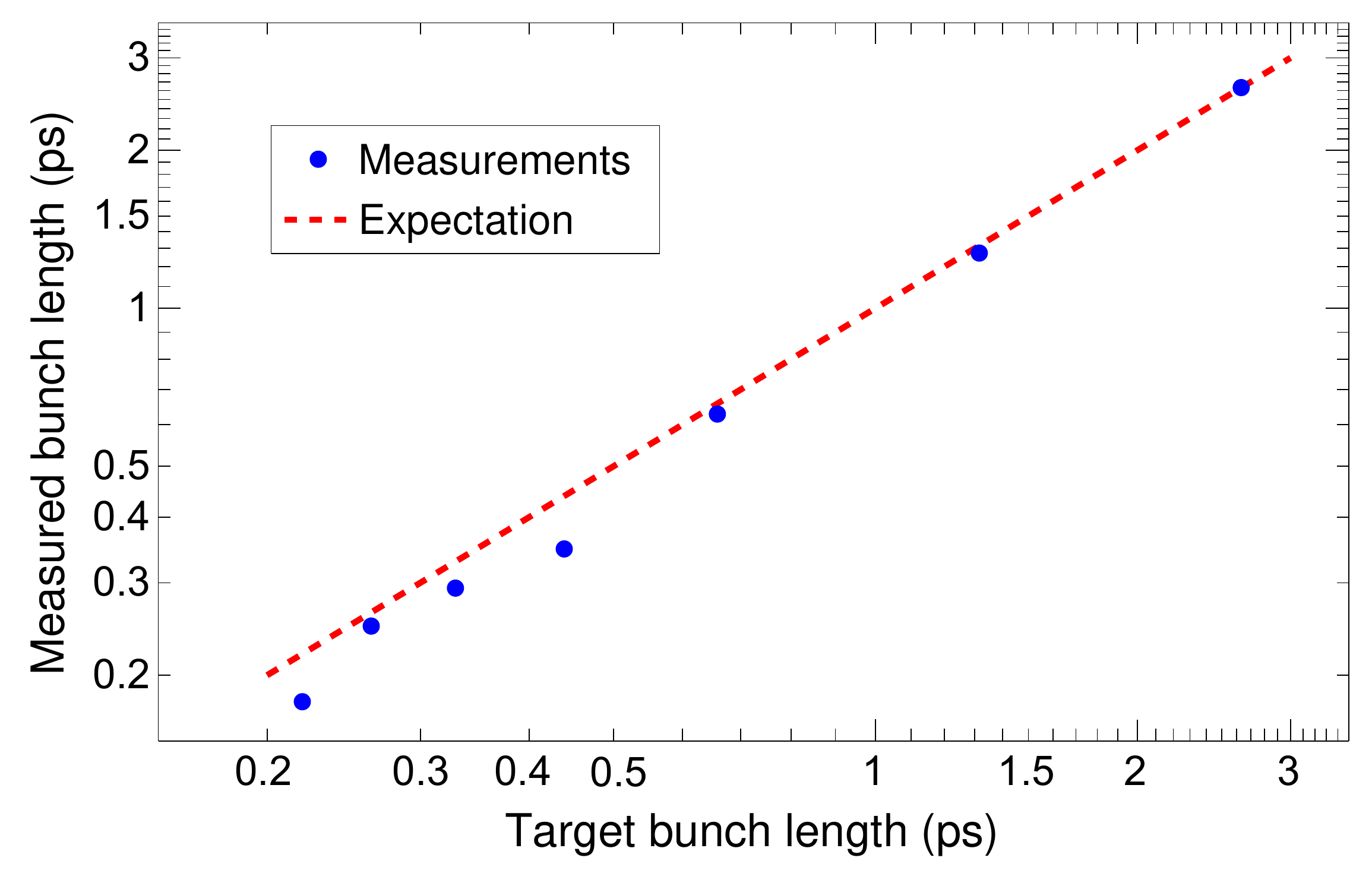}
    \caption{Measured rms bunch length after applying the compression procedure,
      compared to the respective target bunch length.
      (Statistical errors, obtained from repeated measurements, are smaller than
      the marker circles.)}
    \label{fig:compression}
\end{figure}

Following this procedure we can reliably establish compression factors up to
about 12 with a target accuracy of 10--20\%, as shown in 
Fig.~\ref{fig:compression}.
Small adjustments of the compression phase are then sufficient for the further
fine tuning of the bunch length.

\subsubsection{Phase space fragmentation of the compressed beam}

\begin{figure}[tb]    
  \includegraphics*[width=1\linewidth]{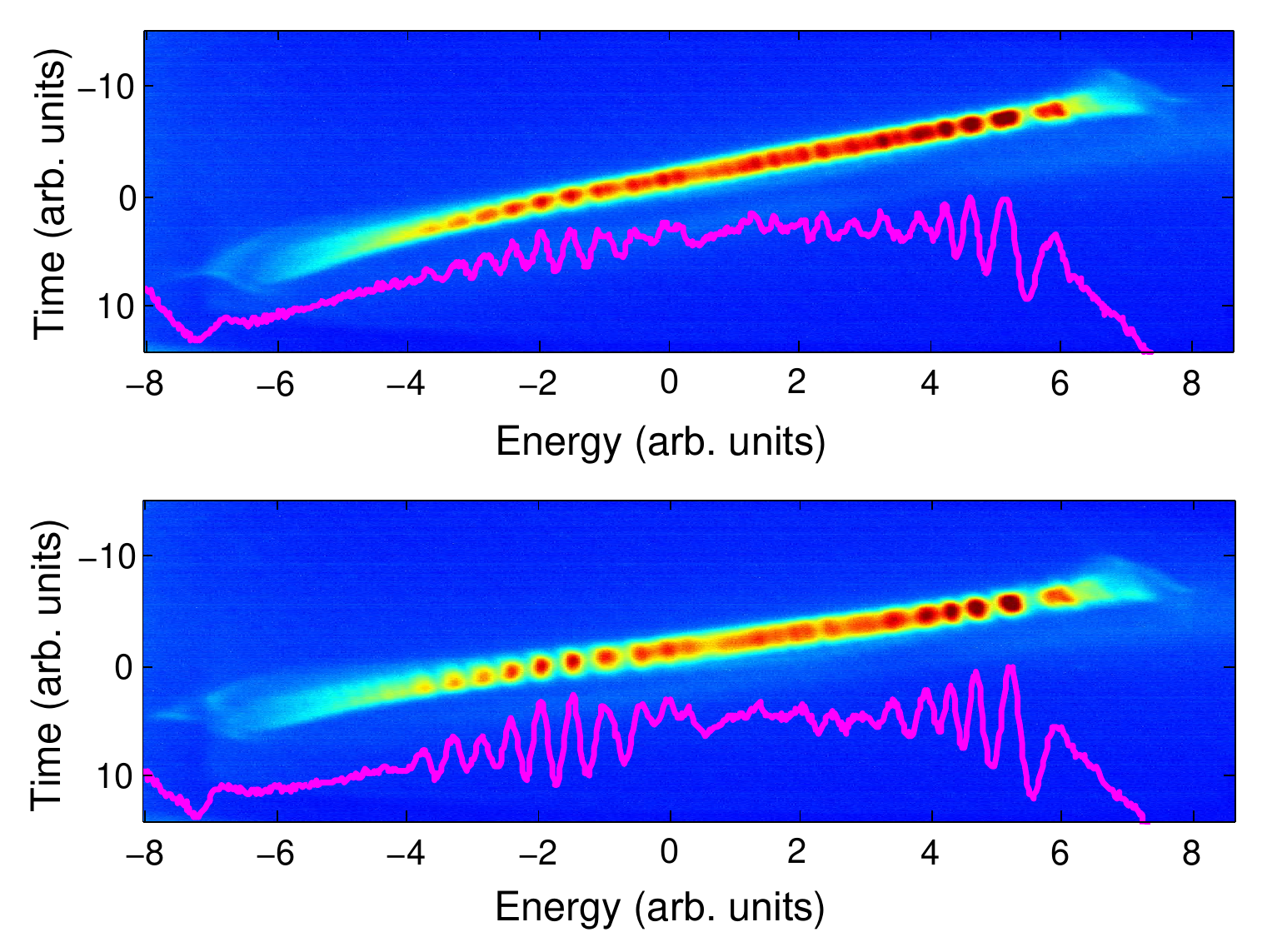}
    \caption{Examples of phase space fragmentation of compressed electron 
      beams generated from a copper cathode with a flat-top like laser profile 
      obtained with pulse stacking (see Fig.~\ref{fig:pulsarlong}).
      The figures show particle densities of the vertically streaked beam as 
      recorded on a screen in the dispersive section of the high-energy 
      spectrometer for compression factors of 6 (top) and 12 (bottom).
      The magenta lines are projections of the distributions onto the 
      horizontal (energy) axis, emphasizing the fragmentation effect.}
    \label{fig:phasespace}
\end{figure}

Preliminary compression experiments, using the flattop-like Ti:sapphire laser 
pulses built up from pulse stacking (see Fig.~\ref{fig:pulsarlong})
to generate electrons from a copper cathode, resulted in a strong fragmentation 
of the electron phase space into pronounced energy bands.
Similar distortions of the longitudinal phase space after compression have 
been observed at other facilities~\cite{Gra01a,Sha04,Hue01}.
The phase space fragmentation has an adverse impact on beam emittance and 
energy spread and compromises the emittance measurement itself.
It also aggravates the risk of further beam degradation with respect to FEL 
performance through the microbunching instability~\cite{Hua05}.
Figure~\ref{fig:phasespace} shows two examples of fragmented phase space 
distributions, measured in the dispersive section of the high-energy 
spectrometer while simultaneously streaking the beam with the rf deflector.
The measurements presented here were obtained with a laser profile similar to 
the one shown in Fig.~\ref{fig:pulsarlong}.

To avoid complications from this effect without changing the photocathode 
material, the optimization of the compressed beam, as described in the following
section, was continued with the Nd:YLF laser, which delivers relatively 
smooth Gaussian pulses (see Fig.~\ref{fig:jaguarlong}).

\begin{figure}[hbt]    
  \includegraphics*[width=1\linewidth]{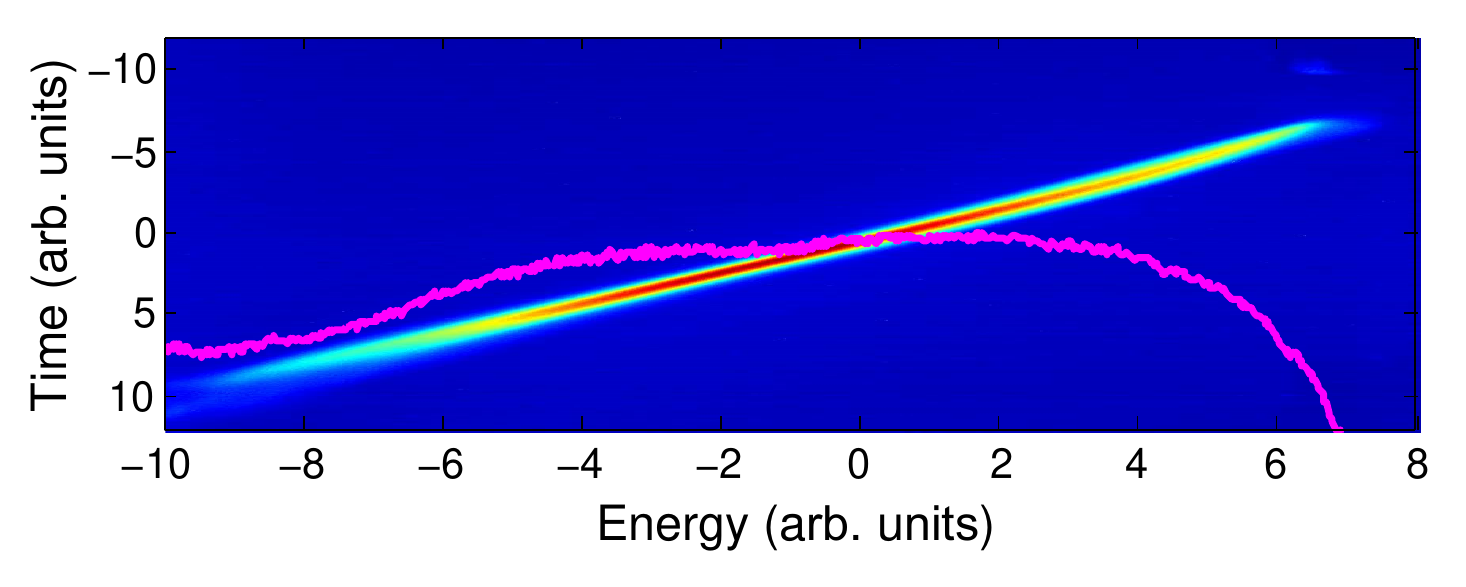}
    \caption{Example of a compressed beam with smooth phase space distribution 
      as obtained with a cesium telluride cathode using the same pulse-stacked
      laser profile as in Fig.~\ref{fig:phasespace}.
      The applied compression factor is 6, corresponding to the upper panel of 
      Fig.~\ref{fig:phasespace}.}
    \label{fig:phasespace2}
\end{figure}

The problem of phase space fragmentation arising from the longitudinal charge 
modulation induced by the pulse-stacked laser profile can also be mitigated by 
choosing a photocathode material with a response time longer than the 
characteristic temporal substructure of the laser pulse.
At the SITF we have verified that a cesium telluride cathode provides enough
time spread to smooth out significantly the phase space of electron bunches 
generated with a pulse-stacked laser profile, even after strong compression,
see Fig.~\ref{fig:phasespace2}.
This observation provided strong additional support for the choice of 
cesium telluride as baseline photocathode material for SwissFEL.
Further suppression of the microbunching instability at SwissFEL will be 
achieved with a laser heater system~\cite{Ped14} based on 
conceptual~\cite{Hua04} and experimental~\cite{Hua10} work carried out for the 
LCLS project at SLAC.

\subsubsection{Optimization of compressed beam}

The first round of emittance measurements on compressed beams, performed with 
a smooth longitudinal laser profile and at moderate compression factors between
5 and 10, indicated a significant increase in the core slice emittance 
resulting from the compression, similar to earlier observations made at 
FLASH~\cite{Roh06} and LCLS~\cite{Zho15}.
Systematic studies showed that the observed emittance growth is independent of 
the beam energy, but does increase with the dipole bending angle in the bunch
compressor (at constant compression factor)~\cite{Bet16a}.
Deliberately degrading the emittance of the incoming beam we also found that 
the emittance increase mainly occurs for beams with low initial emittance, 
whereas an already large emittance is hardly affected by the compression.
Furthermore, a strong dependence of the emittance growth on the beam optics 
along the bunch compressor was observed.
For instance, a change of one unit in the horizontal $\alpha$-function at the 
entrance of the bunch compressor can lead to a relative increase of the core 
slice emittance of the compressed beam of up to 50\%.

\begin{figure*}[tb]    
  \includegraphics*[width=0.7\linewidth]{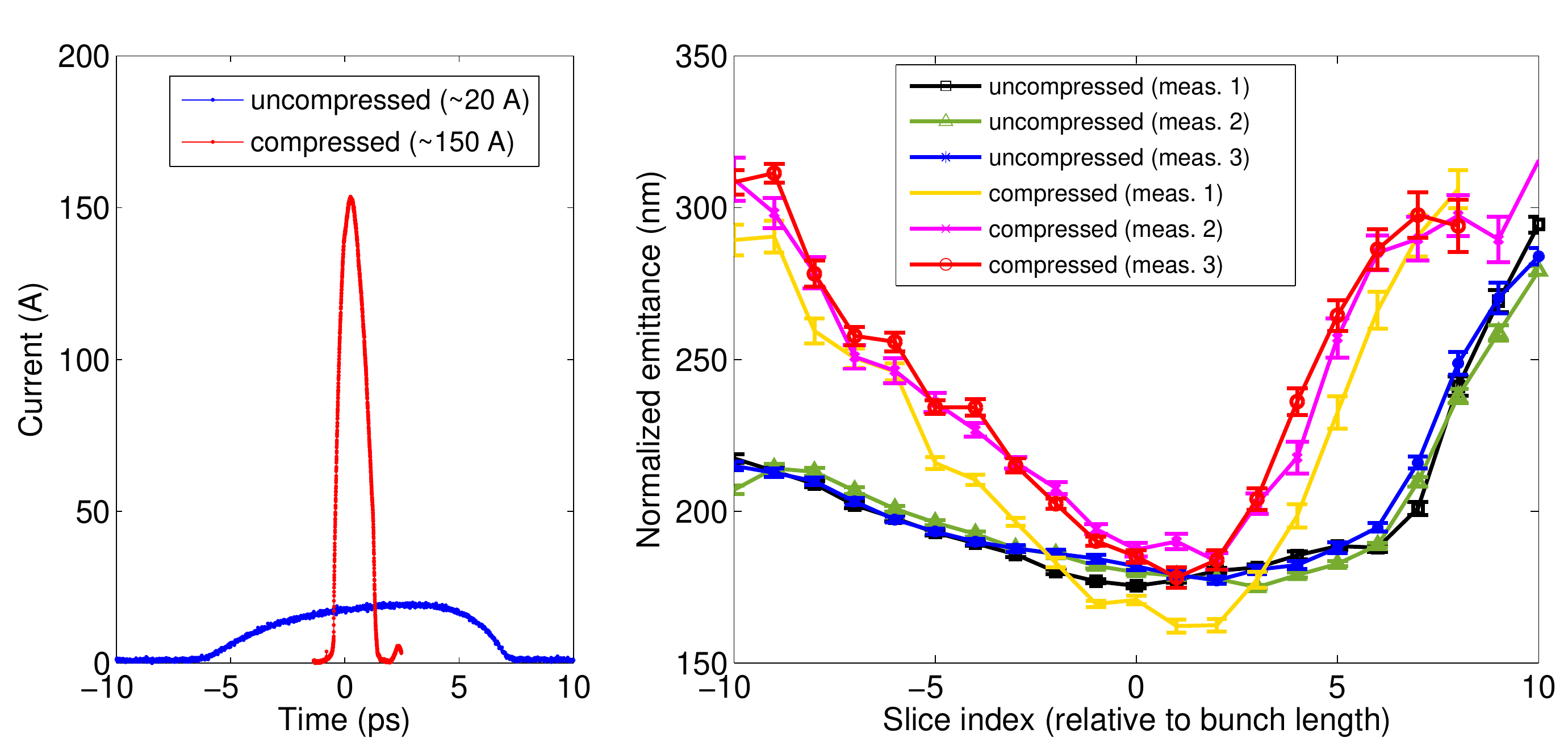}
    \caption{Current profiles (left plot) and three consecutive slice
      emittance measurements (right plot) for uncompressed and compressed
      beams (compression factor 7.5).
      The longitudinal extension of a slice is a fixed fraction of the 
      rms bunch length.}
    \label{fig:emit_comp}
\end{figure*}

Numerical simulations using the CSRTrack code~\cite{CSRTrack} for the tracking 
along the bunch compressor qualitatively reproduce the observed behavior, if a 
3D model based on the pseudo Green's function~\cite{Doh03} is used to describe 
the electron interactions, but not if the 1D model is selected.
From this we infer that the emittance growth is due to 3D effects originating
from coherent synchrotron radiation, but possibly also from other sources. 
A viable mitigation strategy therefore consists in minimizing the horizontal
$\beta$-function at the location of the last two dipoles of the chicane, where
the bunch length is reduced.
By careful tweaking of the beam optics along the bunch compressor, varying
the strengths of the last few quadrupoles in the matching section by a few
percent or less only, we managed to preserve the core slice emittance to within 
about 10\% for compression factors up to about 10.
Figure~\ref{fig:emit_comp} shows the result of three consecutive slice 
emittance measurements on a compressed beam, together with the current profile.
For comparison, the corresponding measurements before compression are also 
plotted.

Another conclusion from our studies, reported in detail in Ref.~\cite{Bet16a},
is that great care must be taken when matching the beam in transport sections 
where it carries a large energy chirp, as is the case in front of a bunch 
compressor:
chromatic effects acting on the energy-chirped bunch can give rise to a large
slice mismatch along the bunch, with the effect that only a short part of the
bunch fulfills the delicate beam optics conditions required to avoid emittance
degradation during compression.
A smooth optics in matching sections before bunch compressors, where focusing
of an energy-chirped beam is required, is therefore of vital importance.

The final results of our emittance minimization efforts for both compressed 
and uncompressed beams are summarized in Tables~\ref{tab:pars-high} and 
\ref{tab:pars-low}.

\subsection{\label{sec:bd-other}Further measurements}

Apart from the beam optimization efforts described above, the SITF was used
for a number of other beam dynamics studies as well as for beam-based tests of
beam diagnostics and manipulation techniques during its time of operation.
The detailed description of these additional developments is beyond the scope
of this report and the reader is referred to the corresponding publications for
more information.

Further studies performed at the SITF include:
\begin{itemize}
\item
  the evaluation and implementation of possible collimation strategies after the
  photocathode gun to limit the dark current transported further 
  downstream~\cite{Bet13};
\item
  a proof-of-principle experiment to test a new method for measuring 
  longitudinal bunch profiles by utilizing the beam's interaction with the 
  transverse wakefield self-induced by off-axis passage through a 
  dielectric-lined or corrugated waveguide~\cite{Bet16} (see also 
  Sec.~\ref{sec:long});
\item
  the test of a novel method to correct for beam tilts, which exploits 
  dispersive effects in the bunch compression chicane acting on an 
  energy-chirped beam~\cite{Gue15};
\item
  a systematic comparison between the transverse emittance measurement method 
  based on quadrupole scans and the method using multi-screen data  
  at fixed beam optics in the FODO section~\cite{Yan14}.
\end{itemize}

\section{\label{sec:undulator}Undulator experiment}

The availability of a well tuned, low-emittance electron beam at the SITF
afforded the opportunity to perform a test of the SwissFEL hard-X-ray undulator
design under realistic beam conditions.
(The design of the SwissFEL soft-X-ray undulators had not yet been finalized at
the time of SITF operation.)
To this end a prototype of the SwissFEL U15 undulator was installed in December
2013 and tested during the run period lasting from January to April 2014.

\subsection{\label{sec:und-motiv}Motivation and goals}

The undulator test at the SITF provided opportunities
\begin{itemize}

\item
to broadly verify---with respect to FEL performance---essential quality 
features of both the electron beam and the undulator design under real 
operating conditions (even if only at vacuum ultraviolet, VUV, and optical photon 
wavelengths),

\item
to validate the undulator magnetic model as well as the results of the magnetic
measurements directly with the help of the electron beam, and

\item
to test the undulator alignment concept based on permanent quadrupole magnets.

\end{itemize}

The primary goals of the experiment were therefore 
the observation and characterization of undulator radiation arising from the 
SASE process,
the systematic measurement of undulator induced orbit kicks as a function of 
the magnetic gap to uncover possible flaws in the undulator field, and 
the demonstration of the undulator alignment procedure.

\subsection{\label{sec:und-setup}Experiment setup}

\subsubsection{The U15 undulator}

The 4~m long hybrid in-vacuum U15 undulator features a period length of 15~mm 
and a tunable undulator parameter, $K$, between 0.1 and 1.8, achieved with a 
movable gap, variable between 3 and 20~mm by means of a wedge based gap drive
system. 
The U15 was designed for the SwissFEL hard-X-ray beam line based on 
the experience made at the SLS with hybrid in-vacuum 
undulators~\cite{Sch01,Sch15}, with appropriate modifications for the specific 
requirements of a linac driven FEL~\cite{Sch09}.
(The general concept of in-vacuum undulators was pioneered by 
SPring-8~\cite{Har98}.)
The relatively large number of identical units (13) planned for SwissFEL 
called for several design changes in order to benefit from an industrial
series production. 
To provide more rigidity to the magnetic structure, a closed mineral cast frame
design was chosen.
The mineral cast option was selected in place of the more common metallic 
frame solution to reduce costs, mechanical vibrations and disturbances from
stray fields.
The entire frame is equipped with a five-axis camshaft mover system for remote 
positioning with micrometer precision.

The new approach introduced additional constraints for the magnetic 
measurements and necessitated the development of a novel bench for the 
optimization and characterization of the undulator magnetic 
properties~\cite{Cal12}. 
The more modest requirements on the field quality for a single-pass machine
allows for a smaller good-field region (larger tolerance on multipole 
components), thereby reducing the magnetic forces and enabling higher $K$ 
values.
A more complete description of the U15 design can be found in 
Ref.~\cite{Sch12}. 

\subsubsection{Undulator preparation}

The U15 prototype undulator was subjected to the same preparation procedure as 
all U15 undulators built for SwissFEL.
The undulator elements are delivered to PSI in three distinct parts:
the frame, equipped with the gap drive system, the magnetic array and 
the vacuum components. 
For the first set of measurements the magnetic array is assembled inside the 
frame.
The trajectory and the phase errors are tuned adjusting the pole height for the
nominal $K$ value ($K = 1.2$).
After carefully securing the positions of the magnets within the magnetic 
array, the unit is removed and equipped with all vacuum components. 
Once the undulator is assembled in its final configuration, the phase error is 
remeasured and improved upon if necessary. 
We then characterize the magnetic array within the entire range of operation,
producing models to operate it in the accelerator ($K$ values and field errors 
as a function of gap height)~\cite{Cal14}.

Before removing the undulator from the measurement bench, we position two small
permanent quadrupole magnets along the measured undulator magnetic axis, one 
immediately before and one after the U15 module, to be later used as a 
reference for the alignment in the accelerator beam line.
These alignment quadrupoles are only used during the alignment procedure and
are retracted from the beam by a remotely controlled pneumatic system during
normal (FEL) operation.
Taking the magnetic axis as reference has the advantage of reducing the 
requirements on the mechanical alignment accuracy.

\subsubsection{Installation and instrumentation at the SITF}

To accommodate the U15 prototype in the SITF required substantial 
modification to the beam line in the diagnostics section.
Specifically, one of the 5~m girders in the FODO channel of the diagnostics 
section had to be removed and several accelerator components rearranged.

The principal instrumentation needed for the planned undulator measurements 
consisted of two appropriately positioned BPMs to monitor the effect of the 
undulator on the electron orbit and three YAG:Ce screens for the observation of 
the undulator radiation, located downstream of the undulator.
The first of these YAG:Ce screens included a void area in the lower half to 
allow the vertically deflected electron beam to pass through without disturbing
the photon signal registered in the upper screen area.
The scintillating screens did not allow for measurements of the absolute 
intensity of the photon beam.
Additional information was gained from an in-air spectrometer capable of 
measuring the photon energy in the optical spectrum range.

Throughout the undulator experiment, the radiation dose was constantly 
monitored by means of dose rate monitors described in Sec.~\ref{sec:drm}.
During the three month period of the experiment an integral dose of about
100~Gy was recorded.
The radiation had no observable effect either on the magnetic or on the 
electronic components of the undulator system, as was later confirmed with
lab measurements following the removal of the undulator from the beam line.

\subsection{\label{sec:und-results}Results}

\subsubsection{Undulator alignment}

With the one undulator installed, we define the reference orbit for its 
alignment simply by the centers of the nearest fixed quadrupoles upstream and 
downstream of the undulator and calibrate the adjacent BPMs to this orbit.
Once the reference orbit has been identified, an alignment procedure is applied
to ensure the coincidence of the undulator's magnetic axis with this 
orbit~\cite{Cal14a}.
For the measurements in the horizontal plane, the undulator gap is set to the 
nominal $K$ value, where field errors have been minimized by the magnetic
optimization.
For the measurements in the vertical plane, however, the undulator gap is kept
open to avoid the strong focusing effects occurring at the relatively low 
energies available at the SITF.

\begin{figure}[tb]  
  \includegraphics*[width=0.9\linewidth]{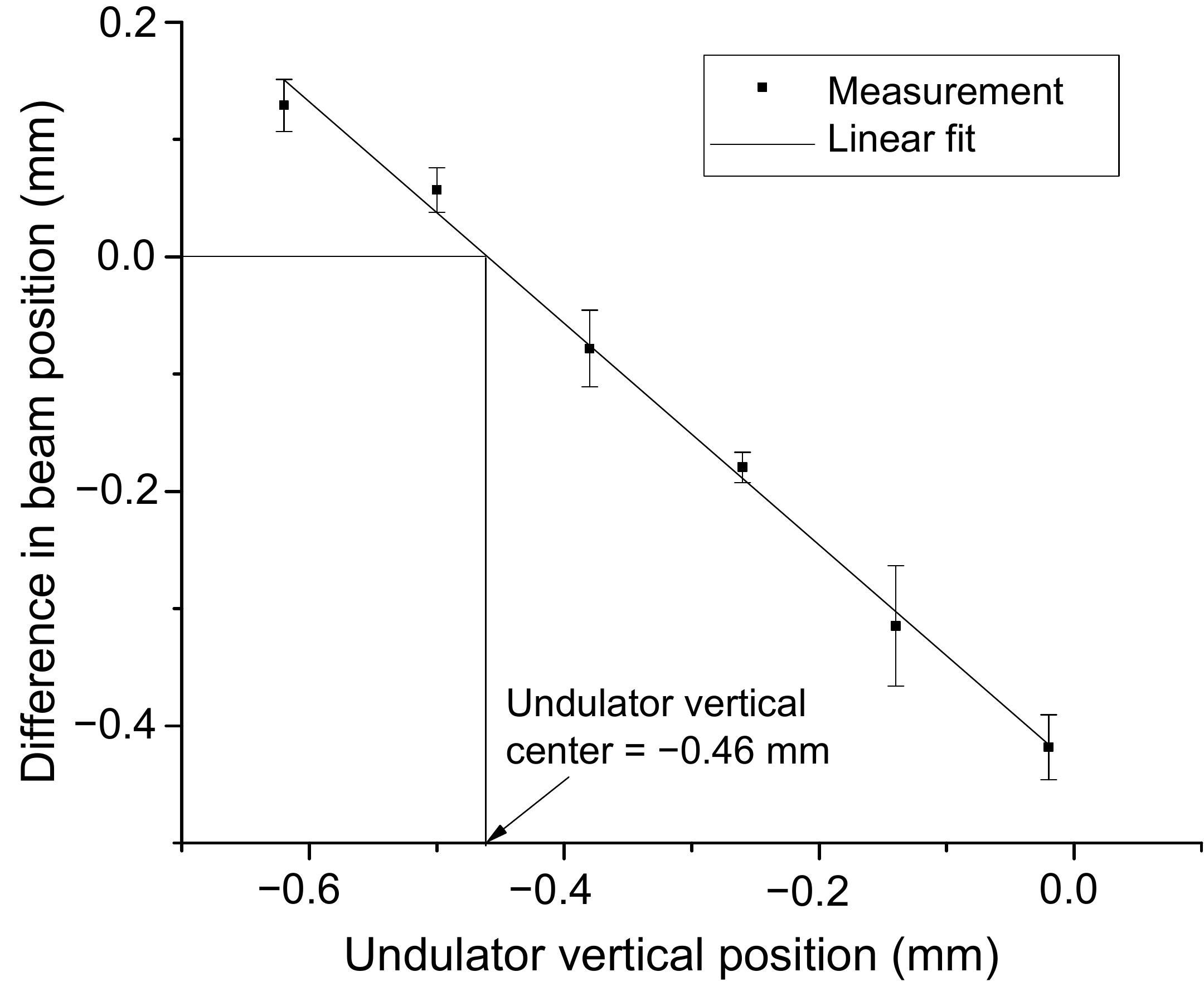}
  \caption{Example of an undulator alignment scan showing the measured beam
  positions relative to the reference orbit, $\Delta y$, versus vertical 
  undulator position.}
  \label{fig:und_align}
\end{figure}

Starting from the reference orbit with both alignment quadrupoles retracted,
the upstream alignment quadrupole is placed into the beam, and the undulator
is moved both horizontally and vertically.
Monitoring the downstream BPMs, we then find the position of the undulator for
which the reference orbit readings are recovered, i.e., where the upstream 
alignment quadrupole has no steering effect.
After removing the upstream alignment quadrupole, the downstream alignment
quadrupole is inserted into the beam and the undulator is adjusted again both
horizontally and vertically until the reference orbit is retrieved.
Once these two positionings have been determined, the final position and
orientation (pitch and yaw angle), for which the undulator is aligned with
the reference orbit, can be computed and applied to the mover system.
Figure~\ref{fig:und_align} shows an example of an undulator position scan 
performed as part of the alignment procedure.

\subsubsection{Residual field error measurements}

When working with a variable gap configuration the preservation of the
undulator's field integrals (governing the integral effect on the electron
orbit) for all gap settings is a demanding requirement.
It is therefore important to characterize the undulator's net effect on the
electron orbit as a function of the gap height to quantify possible residual
field errors. 

Such residual field errors mainly occur at the entrance and the exit of the 
undulator, giving rise to corresponding deflections (``kicks'') in the 
electron orbit.
If the electrons experience no focusing inside the undulator, these entrance 
and exit kicks can be inferred from the orbit modification at the undulator 
exit, both in position and angle ($\Delta x$ and $\Delta x'$), induced by the 
setting of the undulator gap to a certain $K$ value.
While this change in orbit cannot be measured directly, it can be derived from
the orbit response further downstream.
In our case we use two dedicated BPMs downstream of the undulator.
Moreover, all magnets between the undulator and the two BPMs are switched off.
Corrector magnets upstream of the undulator are used to ensure a constant 
injection orbit. 
The undulator gap is then reduced from the open position to the position 
corresponding to a certain $K$ value, and the associated change in the orbit 
downstream of the undulator is recorded by the BPMs.
The measurement is then repeated for different $K$ values.

\begin{figure}[b]  
  \includegraphics*[width=0.8\linewidth]{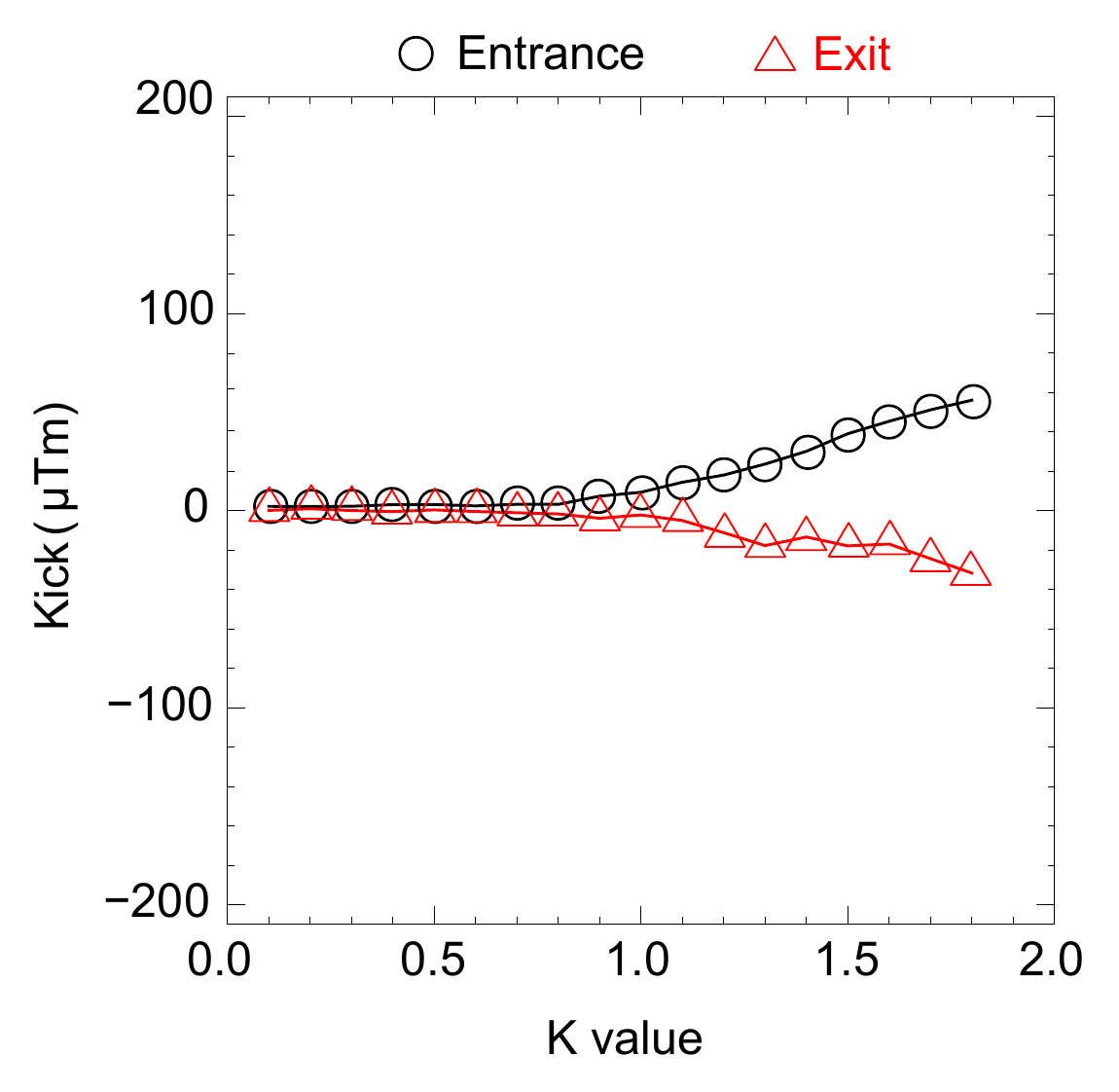}
  \caption{Undulator entrance (black circles) and exit (red triangles) kicks 
    measured as a function of the undulator $K$ value in the horizontal plane.}
  \label{fig:und_xkick}
\end{figure}

Figure~\ref{fig:und_xkick} shows the entrance and exit kicks in the horizontal 
plane as a function of the $K$ value, derived from downstream orbit 
measurements at a beam energy of 100~MeV.
The kicks, which are shown relative to the $K = 0$ reference case (not the 
nominal $K = 1.2$), exhibit the pattern expected from the central symmetry of 
the field profile, which is clearly not masked by field errors.
At the highest $K$ values (1.8) the kicks amount to about 60~\textmu Tm, 
close to predictions based on numerical calculations and magnetic measurements.

\begin{figure}[tb]  
  \includegraphics[width=1\linewidth]{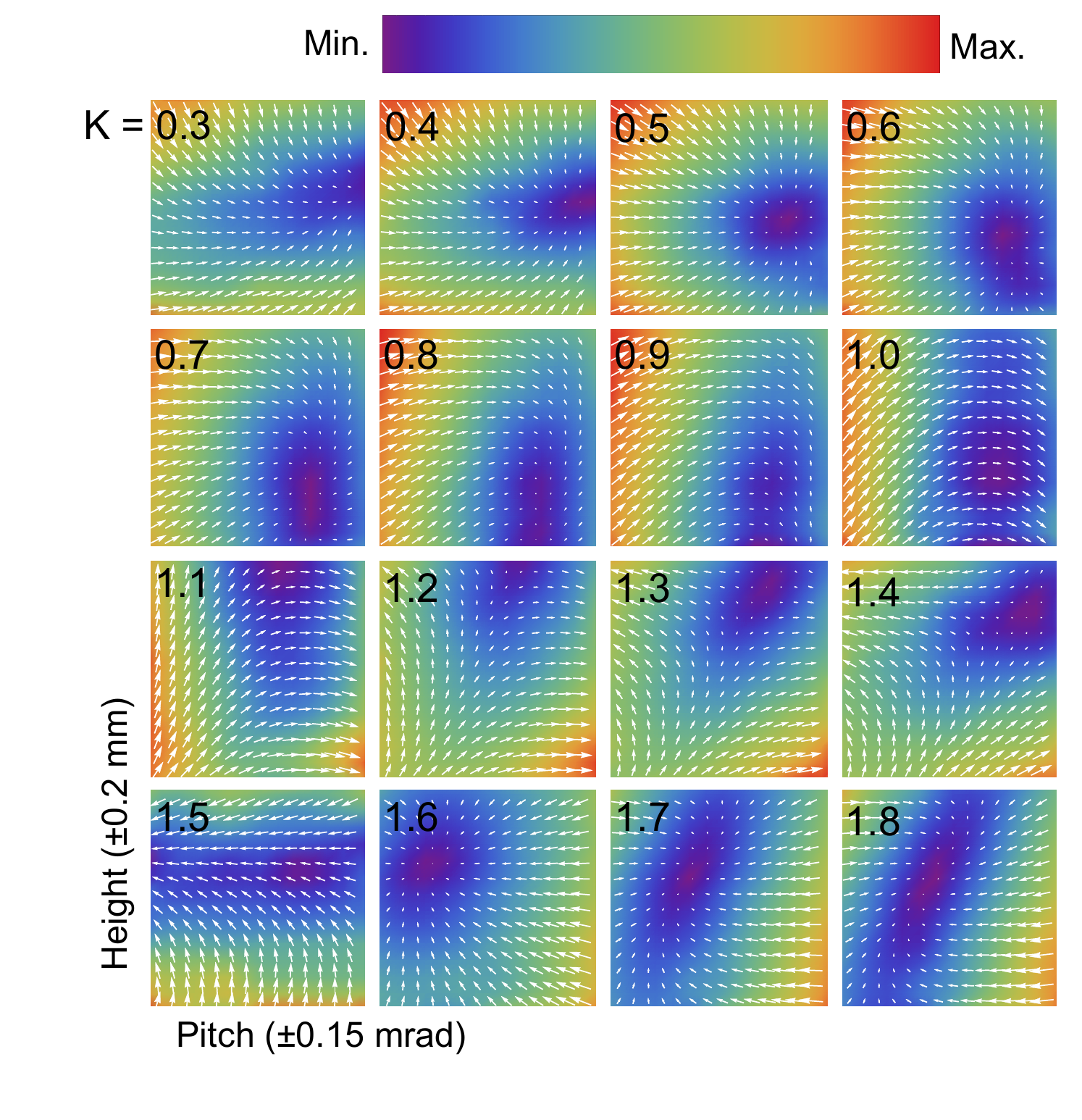}
  \caption{Position and angle changes of the electron orbit in the vertical
    plane at the undulator exit ($\Delta y$ and $\Delta y'$), represented as 
    vertical and horizontal components of small white arrows, as a function of 
    undulator pitch and height, individually plotted for undulator $K$ values 
    ranging from 0.3 to 1.8.
    The underlying colors correspond to the lengths of the arrows, i.e., the
    deviations from zero orbit change, and illustrate the dependence of the
    natural focusing strength and center on the $K$ value.
    The plots for the different $K$ values are individually normalized to 
    illustrate the focusing and bending effects.}
  \label{fig:und_yscan}
\end{figure}

The presence of natural focusing in the vertical plane 
prevents a straightforward interpretation
of the orbit response at the undulator exit (in this case $\Delta y$ and 
$\Delta y'$) in terms of entrance and exit kicks.
Nevertheless a comprehensive picture of the combined effects of focusing
and possible kicks from residual field errors may be obtained by determining the
orbit changes $\Delta y$ and $\Delta y'$ not only as a function of the $K$ 
value, but also, for each $K$ value, as a function of the undulator height and 
pitch.
The resulting set of two-dimensional plots, reproduced in 
Fig.~\ref{fig:und_yscan}, clearly reveals the quadrupole component, with axis
and phase advance both depending on the $K$ value, indicating that the field
errors change with the undulator gap.

The overall relatively large orbit deflections in the vertical plane 
highlighted the need for improvements in the assembly and measurement of the 
magnetic array.
Consequently a new sorting algorithm based on the measurement of individual
magnets has been planned and implemented for the next modules with the goal
of increasing the quality of the field in the horizontal plane (affecting
the vertical electron orbit), thereby reducing the imperfections below 
specifications.
Moreover, the SwissFEL undulators will be equipped with additional vertical
steering magnets to counteract any residual vertical orbit kicks.

\subsubsection{SASE performance}

Our first attempts to produce FEL radiation from the undulator were made at a 
beam energy of 220~MeV, corresponding to photon wavelengths between 61~nm (for 
$K$ = 1) and 106~nm (for $K$ = 1.8).
Soon after setting up the beam transport through the undulator, a bright signal
emerged on the first YAG:Ce screen downstream of the undulator, clearly 
related to FEL radiation.
But the asymmetric appearance of the beam spot in the form of a diffraction arc
and the fact that it could not be propagated beyond the first screen strongly 
suggested that only a reflection of the actual FEL signal was observed.
Later it became clear that the unexpectedly strong downward kick exerted by the
fringe field at the undulator entrance prohibited the proper propagation of 
the photon beam in our initial beam-line configuration.

After a modification of the beam line downstream of the undulator, increasing
the effective aperture, a round beam spot could be observed on three screens
located at distances between 1 and 4~m from the undulator exit~\cite{Rei14}.
Due to the temporary failure of the klystron driving the last two S-band 
structures, all subsequent measurements were performed at beam energies ranging
between 100 and 130~MeV, using, when needed, the X-band cavity to produce the 
energy chirp required for bunch compression.
At a beam energy of 100~MeV, a peak current of 20~A (obtained without 
compression) is sufficient to generate FEL radiation with wavelength around 
300~nm for $K$ = 1, i.e., in the optical (ultraviolet) range.
In addition to this ``long-bunch'' mode, we also compressed the beam to reach
a peak current of 200~A and characterized the FEL radiation resulting from this
``short-bunch'' mode.

We verified the SASE nature of the observed radiation by confirming two of its
salient characteristics, the beam divergence and the statistical distribution 
of the pulse intensities.
To this end we compare the measured properties to simulation results, 
obtained with the Genesis 1.3 code~\cite{Rei99}.

A measurement of the photon beam spot size on the first three screens 
downstream of the undulator yields a photon beam divergence of 0.30~mrad, about
an order of magnitude below the expected divergence for spontaneous radiation 
emanating from the undulator, but somewhat lower than the simulation value of 
0.38~mrad, computed for the same beam conditions (130~MeV beam energy, 
200~A peak current).
The difference between measurement and simulation is likely related to the
uncertainty in the undulator's gain length, which cannot be measured with 
only one module, and the associated variation of gain guiding effects.

\begin{figure}[b]  
  \includegraphics*[width=1\linewidth]{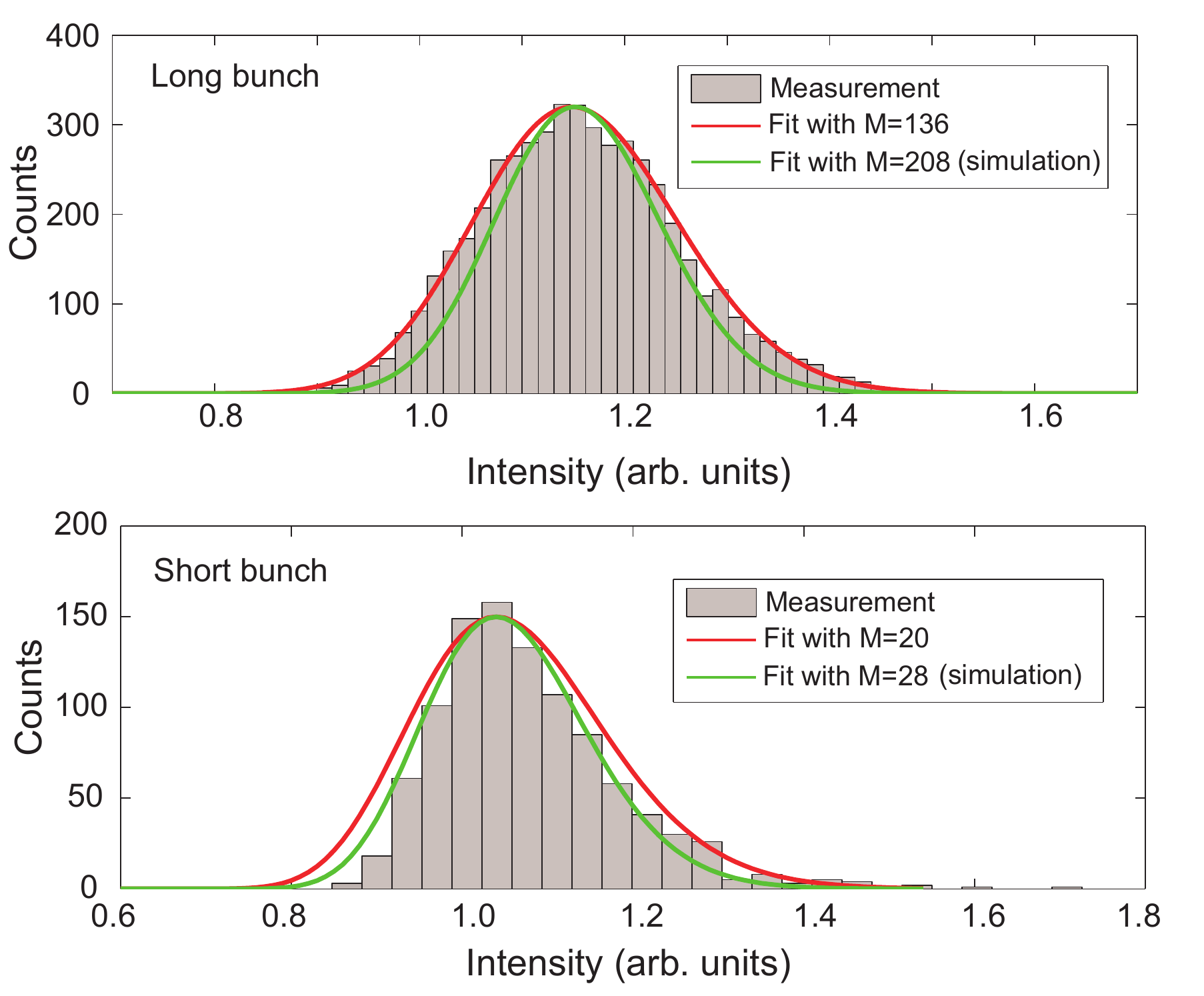}
  \caption{Intensity distributions of SASE FEL pulses measured in the
    long-bunch (top) and the short-bunch configuration (bottom), shown as
    histograms.
    The curves overlaid represent fits to gamma distributions with the $M$ 
    parameter left free (red) and fixed to simulation results obtained for
    equivalent beam conditions.}
  \label{fig:SASE}
\end{figure}

A further distinctive property of SASE radiation is the statistical distribution
of the intensities of individual pulses, which is known to follow a 
single-parameter gamma distribution with free parameter $M$ corresponding to 
the number of contributing modes~\cite{Sal98}.
A histogram based on 5000 pulses measured in the long-bunch configuration is 
shown in the upper plot of Fig.~\ref{fig:SASE}, along with a fit to the 
gamma distribution, which yields a mode parameter $M = 136$. 
Our simulation for this configuration results in $M = 208$, with the 
corresponding curve also indicated in the figure.
The measured pulse intensities are based on the signals recorded by the first
YAG:Ce screen after the undulator, corrected for background, and assuming
reasonable linearity of the YAG:Ce response in the relevant intensity range.
The lower plot of Fig.~\ref{fig:SASE} summarizes the analogous data for 1000
pulses obtained in the short-pulse configuration.
In this case, the $M$ parameters fitted to the measurement and simulation data
are 20 and 28, respectively.
The agreement between measured and simulated distributions, even if not 
perfect, further corroborates our interpretation of the observed radiation as
originating from the SASE FEL process.

\subsubsection{Measurement of $K$ values}

The limitation in beam energy due to the outage of the last S-band booster
rf station constrained the FEL spectral range to the optical regime, for which
an in-air spectrometer was already available at the SITF.
The spectrometer (Ocean Optics QE65000), which had previously been used for 
studies of optical transition radiation at the same facility~\cite{Smi15}, 
covers a spectral range between 220 and 950~nm wavelength.

\begin{figure}[t]  
  \includegraphics*[width=0.85\linewidth]{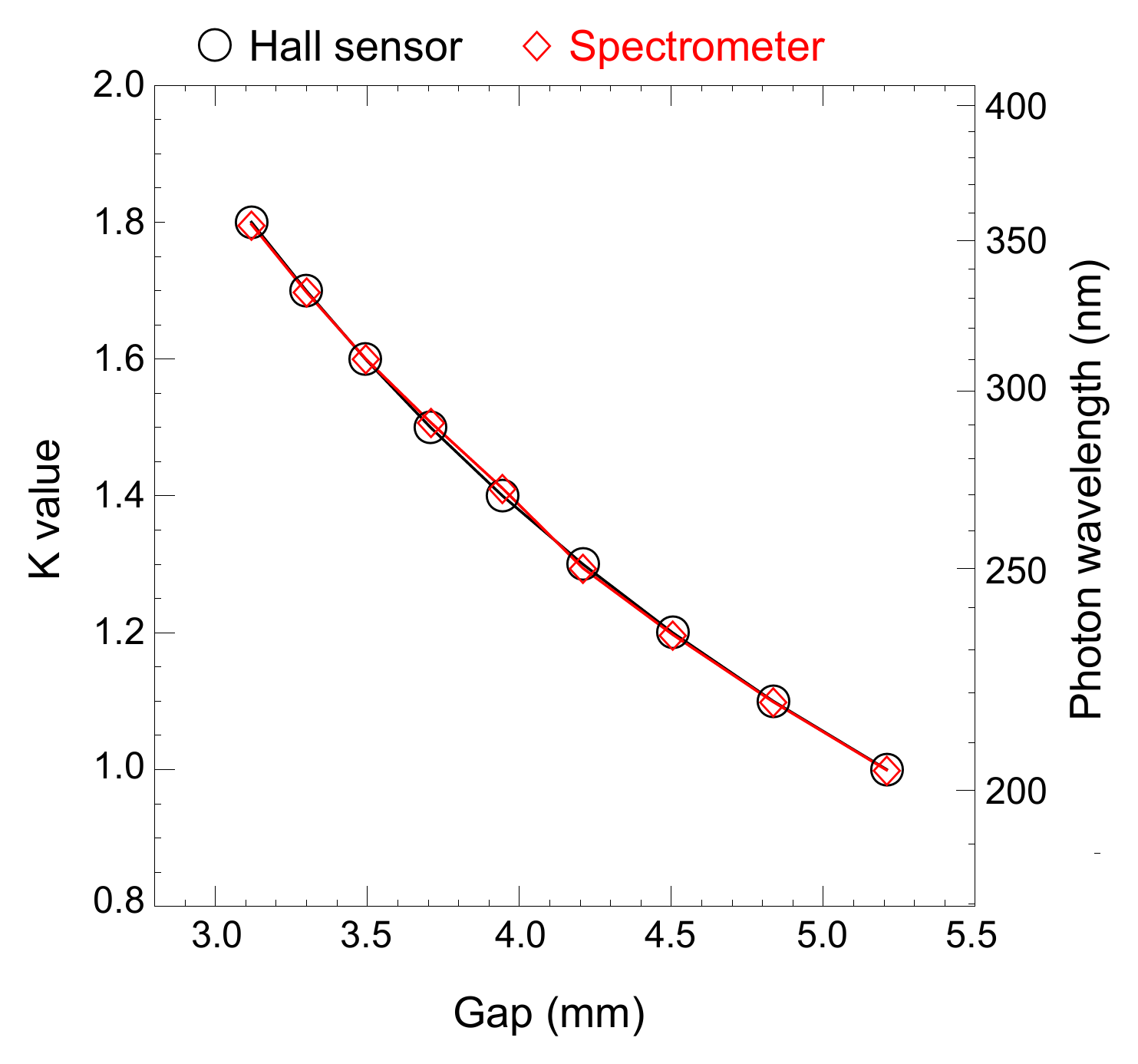}
  \caption{Undulator $K$ value as a function of undulator gap as derived from
    magnetic measurements using Hall sensors (black circles) and from 
    spectral measurements of the generated SASE FEL radiation (red diamonds).}
  \label{fig:Kval}
\end{figure}

At a beam energy of 120~MeV, the variation of the undulator gap between 3.0 and
5.5~mm gives access to a spectral tuning range of the SASE FEL wavelength 
extending from about 200 to 360~nm.
From the measured central wavelengths the corresponding $K$ values of the 
undulator can be retrieved and compared to the $K$ values derived from the
Hall probe measurements on the magnetic bench, see Fig.~\ref{fig:Kval}.
The observed concurrence bears out the magnetic design and validates the 
concept of the magnetic measurement bench.

\section{\label{sec:summary}Summary and outlook}

We have summarized our experience commissioning and operating the SwissFEL
Injector Test Facility during four years, with particular emphasis on system
developments and beam physics results relevant for the realization of the
SwissFEL facility.

In addition to demonstrating the beam parameters necessary to drive the first
stage of an FEL facility, the experience gathered at the SITF has proven 
invaluable for
  the evaluation and realization of a UV laser system to drive the 
    photoinjector rf gun,
  the implementation of a stable large-scale rf system, synchronized at the 
    tens of femtosecond level, and
  the development of numerous diagnostics devices essential for SwissFEL.
Furthermore it has enabled the demonstration of a complete undulator system 
qualified for FEL operation.
Beyond the establishment of essential linac and FEL capabilities at PSI
the most important advances in view of the larger accelerator community consist 
in 
  the development of a new profile monitor capable of measuring beams of very 
    low charge and emittance~\cite{Isc15},
  the systematic comparison of intrinsic emittances between cesium telluride and
    copper cathodes~\cite{Pra15}, 
  the demonstrated reduction of intrinsic emittance through modifications of 
    the wavelength of the laser driving the electron gun~\cite{Div15a} or the 
    gun rf gradient~\cite{Pra15a}, and 
  the development or refinement of various beam characterization 
    methods~\cite{Pra14,Pra14a,Pra14b,Aib14,Pra14c}.

In terms of beam emittance our efforts resulted in values as low as 100~nm
for bunch charges of 10--20~pC and 200~nm in the case of 200~pC~\cite{Pra14}.
These numbers refer to the normalized transverse emittance of individual slices,
as relevant for driving FEL facilities (see Sec.~\ref{sec:bd-results} for
details).
They have been shown to be reproducible within a few percent and stable
under longitudinal bunch compression~\cite{Bet16a}.
At the electron source, we have measured intrinsic emittance values close to 
the theoretical limit~\cite{Pra15,Pra15a,Div15a}.
Overall, the achieved beam quality represents a significant advance in 
photoinjector physics and technology.

The SITF has also fulfilled an important role as a training facility for 
machine operators as well as for students of accelerator or FEL-based sciences.
The commissioning of the SwissFEL main accelerator and the hard-X-ray undulator
and optical lines, scheduled for the years 2016 and 2017~\cite{Sch15a}, will
profit considerably from the experience gained at the SITF.

\begin{acknowledgments}

We acknowledge the valuable contributions of the entire SwissFEL project staff 
and all PSI support groups to the design, procurement, production, assembly, 
installation, validation and maintenance of the many components and systems 
that constituted the SwissFEL Injector Test Facility.
We also express our thanks to the PSI operations group for helpful assistance
with many commissioning tasks.
We are indebted to Albin Wrulich for initiating the injector project and for
supporting it throughout its earlier stages.
Both the SITF design and its operation have much profited from the valuable
advice and guidance generously offered by the SwissFEL machine advisory 
committee (FLAC), chaired by J\"org Ro\ss bach.
We further thank Jiaru Shi for help with the tuning of the X-band
structure and for providing the corresponding figure, and Nicole Hiller for
the careful proofreading of the manuscript.
Two detector developments have received financial support from the European 
Commission under its Seventh Framework Programme (FP7/2007--2013):
the X-band wake-field monitors through the FP7 Research Infrastructures project
EuCARD-2, Grant Agreement No.~312453, and the electro-optic bunch length
monitor under Grant Agreement No.~290605 (PSI-FELLOW/COFUND).
A part of the beam science program at the SITF was supported by the 
Swiss National Science Foundation through grant No.~200021\_140659.

\end{acknowledgments}

\end{document}